\documentclass[a4paper,fleqn,usenatbib]{mnras}

\usepackage{newtxtext,newtxmath}

\usepackage[T1]{fontenc}
\usepackage{ae,aecompl}

\usepackage{times}
\usepackage{amssymb}
\usepackage{amsmath}
\usepackage{color}
\usepackage{bm}
\usepackage{graphicx}
\usepackage{tabu}
\usepackage{epsfig}
\usepackage{longtable,tabu}
\usepackage{url}
\usepackage{xcolor}
\usepackage{ulem}
\usepackage{supertabular}
\usepackage{caption}
\usepackage[multidot]{grffile}
\usepackage{subcaption}
\DeclareSymbolFont{matha}{OML}{txmi}{m}{it}
\DeclareMathSymbol{\varv}{\mathord}{matha}{118}

\newcommand*{\cii}{\text{[C\,\textsc{ii}]}}
\newcommand*{\nii}{\text{[N\,\textsc{ii}]}}
\newcommand*{\oiii}{\text{[O\,\textsc{iii}]}}
\newcommand*{\oi}{\text{[O\,\textsc{i}]}}
\newcommand*{\hii}{\text{H\,\textsc{ii}}}

\newcommand{\HII}{H\,{\sc ii}}

\newcommand{\OIII}{[O\,{\sc iii}]}
\newcommand{\OV}{[O\,{\sc v}]}
\newcommand{\OII}{[O\,{\sc ii}]}
\newcommand{\OI}{[O\,{\sc i}]}

\newcommand{\NII}{[N\,{\sc ii}]}
\newcommand{\NIII}{[N\,{\sc iii}]}
\newcommand{\CII}{[C\,{\sc ii}]}

\newcommand{\LIR}{$L_{\rm IR}$}

\newcommand{\NIIab}{[N\,{\sc ii}]\,205-$\mu {\rm m}$}

\newcommand{\mum}{$\mu$m}
\newcommand{\herschel}{{\it Herschel}}

\newcommand*{\J}[2]{ \ensuremath{\text{\emph{J}}\! = \! #1\! \rightarrow\! #2}}
\newcommand*{\kms}{\text{km\,s\(^{-1}\)}}
\newcommand*{\lir}{\ensuremath{L_\text{IR}}}

\newcommand*{\lum}[1]{\ensuremath{L_\text{#1}}}
\newcommand*{\mass}[1]{\ensuremath{M_\text{#1}}}
\newcommand*{\lsun}{\ensuremath{\text{L}_{\odot}}}
\newcommand*{\msun}{\ensuremath{\text{M}_{\odot}}}

\newcommand*{\Td}{\ensuremath{T_\text{d}}}
\newcommand*{\um}{\ensuremath{\umu \hspace{0em} \text{m}}}

\newcommand*{\mJy}{\text{mJy}}

\renewcommand*{\fd}[1]{\ensuremath{S_\text{#1}}}
\newcommand*{\lineflux}{\ensuremath{10^{-18}\,\text{W}\,\text{m}^{-2}}}
\newcommand*{\xunits}{ \ensuremath{\msun \left(\text{K} \, \kms \text{pc}^2 \right)^{-1}} }
\newcommand*{\programme}[2]{\textsc{#1t}\scalebox{1}[0.8]{#2}}
\newcommand*{\programmefull}[4]{ \programme{#1}{#2}\textsc{\_#3\_}\scalebox{1}[0.8]{#4} }
\newcommand*{\otone}{\programme{o}{1}}
\newcommand*{\otonefull}{\programmefull{o}{1}{rivison}{1}}
\newcommand*{\ottwo}{\programme{o}{2}}
\newcommand*{\ottwofull}{\programmefull{o}{2}{rivison}{2}}

\newcommand*{\gtonefull}{\programmefull{g}{1}{ivaltcha}{1}}

\def\Htwo      {H$_2$}

\def\Jykms     {{ Jy}\,km\,s$^{-1}$}

\def\Msun      {$M_\odot$}
\def\Lsun      {$L_\odot$}
\def\Md        {$M_{\rm dust}$}
\def\cmt       {cm$^{-3}$}
\def\RNII      {$R_{122/205}$}
\def\ROIII     {$R_{52/88}$}
\def\mmin      {$M^{\rm min}_{\rm H^+}$}
\def\ne        {$n_{\rm e}$}
\def\te        {$T_{\rm e}$}

\def\purple#1  {{\textcolor{purple}{#1}}\ }
\def\red#1     {\textcolor{red}{#1}}
\def\new#1     {{\bf #1 }}

\title[Far-infrared spectroscopy of lensed starbursts]
{Far-infrared {\it Herschel} SPIRE spectroscopy of lensed starbursts reveals physical conditions of ionised gas}

\author[Zhang et al.]{%
\Large 
Zhi-Yu~Zhang,\(^{\! 1,2}\)
R.\,J.~Ivison,\(^{\! 2,1}\) 
R.\,D.~George,\(^{\! 1}\) 
Yinghe~Zhao,\(^{\!3}\) 
L.~Dunne,\(^{\! 4,1}\) 
R.~Herrera-Camus,\(^{\! 5}\) 
\newauthor \Large 
A.\,J.\,R.~Lewis,\(^{\! 1}\) 
Daizhong~Liu,\(^{\! 6,7}\) 
D.~Naylor,\(^{\! 8}\)
Ivan~Oteo,\(^{\! 1,2}\) 
D.~A.~Riechers,\(^{\!9}\)
Ian~Smail,\(^{\!10}\)  
Chentao~Yang,\(^{\! 11,12,13,6}\) 
\newauthor \Large 
Stephen~Eales,\(^{\! 4}\)
Ros~Hopwood,\(^{\! 14}\)
Steve~Maddox,\(^{\! 4,1}\) 
Alain~Omont\(^{12,13}\)   
and
Paul~van der Werf\(^{15}\)  
\vspace*{1mm}\\
\(^1\) Institute for Astronomy, University of Edinburgh, Blackford Hill, Edinburgh, EH9~3HJ, UK\\
\(^2\) European Southern Observatory, Karl-Schwarzschild-Strasse~2, D-85748 Garching, Germany\\
\(^3\) Yunnan Observatories, CAS, Kunming 650011, P.R. China\\
\(^4\) School of Physics and Astronomy, Cardiff University, The Parade, Cardiff, CF24~3AA, UK\\
\(^5\) Max-Planck-Institut f\"ur Extraterrestrische Physik (MPE), Giessenbachstr., D-85748 Garching, Germany\\
\(^6\) Purple Mountain Observatory, 2 West Beijing Road, Nanjing, 230000, P.R. China\\
\(^7\) MPIA, Koenigstuhl 17, 69117 Heidelberg, Germany\\
\(^8\) University of Lethbridge, Lethbridge, Alberta T1K~3M4, Canada\\
\(^9\) Cornell University, Space Sciences Building, Ithaca, NY 14853, USA\\
\(^{10}\) Department of Physics, Centre for Extragalactic Astronomy, Durham University, South Road, Durham DH1~3LE, UK\\
\(^{11}\) European Southern Observatory, Alonso de C{\'o}rdova 3107, Casilla 19001, Vitacura, Santiago, Chile \\
\(^{12}\) Institut d'Astrophysique Spatiale, CNRS UMR 8617, Universit\'{e} Paris-Sud, Universit\'{e} Paris-Saclay, 91405 Orsay, France \\
\(^{13}\) CNRS, UMR 7095, Institut d'Astrophysique de Paris, F-75014, Paris, France\\
\(^{14}\) Department of Physics, Imperial College London, Prince Consort Road, London SW7 2AZ, UK\\
\(^{15}\) Sterrewacht Leiden, Leiden University, PO Box 9513, 2300 RA, Leiden, The Netherlands\\
}

\date{ Submitted to MNRAS Main Journal, 2017 July ; Manuscript ID: MN-17-2572-MJ }

\pubyear{2018}

\begin{document}
\label{firstpage}
\pagerange{\pageref{firstpage}--\pageref{lastpage}}
\maketitle


\begin{abstract}
The most intensively star-forming galaxies are extremely luminous at
far-infrared (FIR) wavelengths, highly obscured at optical and ultraviolet
wavelengths, and lie at $z\ge 1$--3.  We present a programme of \herschel\ FIR
spectroscopic observations with the SPIRE FTS and photometric observations with
PACS, both on board \herschel, towards a sample of 45 gravitationally lensed,
dusty starbursts across $z\sim 1$--3.6. In total, we detected 27 individual
lines down to 3-$\sigma$, including nine \CII\ 158-\mum\ lines with confirmed
spectroscopic redshifts, five possible \CII\ lines consistent with their
far-infrared photometric redshifts, and in some individual sources a few \OIII\
88-\mum, \OIII\ 52-\mum, \OI\ 145-\mum, \OI\ 63-\mum, \NII\ 122-\mum, and
OH 119-\mum\ (in absorption) lines.  To derive the typical physical properties
of the gas in the sample, we stack all spectra weighted by their intrinsic
luminosity and by their 500-$\mu$m flux densities, with the spectra scaled to a
common redshift.  In the stacked spectra, we detect emission lines of \CII\
158-\mum, \NII\ 122-\mum, \OIII\ 88-\mum, \OIII\ 52-\mum, \OI\ 63-\mum, and
the absorption doublet of OH at 119-\mum, at high fidelity. We find that the
average electron densities traced by the \NII\ and \OIII\ lines are higher than
the average values in local star-forming galaxies and ULIRGs, using the same
tracers. From the \NII/\CII\ and \OI/\CII\ ratios, we find that the \CII\
emission is likely dominated by the photo-dominated regions (PDR), instead of
by ionised gas or large-scale shocks.
\end{abstract}
\begin{keywords}
galaxies: high-redshift --- 
galaxies: active ---
galaxies: starburst --- 
submillimetre: galaxies ---
infrared: galaxies ---
\end{keywords}



\section{Introduction}
\label{introduction}

The mean star-formation rate (SFR) density in the Universe was much higher in
the past, peaking around 10 billion years ago, at \(z \approx 2\)
\citep[e.g.][]{Hopkins2006, MD2014}, at which time the SFR per unit co-moving
volume peaked at levels 10--30$\times$ higher than the current rate.  Most of
the stars created at this time were located within low-mass (\(M_* <
10^{10.5}\,\msun\)) galaxies with moderate star-formation rates (SFR\,$\le 100
\, \msun\,\text{yr}^{-1}$) \citep[e.g.][]{Daddi2007, Hopkins2010, Sparre2015}.

Surveys at far-infrared (FIR) and sub-millimetre (submm) wavelengths revealed a
population of so-called submm galaxies or dusty star-forming galaxies
\citep[SMGs or DSFGs, e.g.][]{Smail1997,Eales2010}, mostly at $z=1$--3
\citep[e.g.][]{Chapman2005, Simpson2014,Danielson2017}, but with a smattering
at $z>4$ \citep[e.g.][]{Ivison2016,Asboth2016}, which can account for much of
the submm background.  These galaxies were forming stars at tremendous rates,
$\ge 300$\,M$_{\odot}$\,yr$^{-1}$ \citep[e.g.][]{Blain2002} -- intense
star-formation events that are thought to have been powered primarily by major
mergers \citep[e.g.][]{Ivison2007, Engel2010,Ivison2011, Oteo2016}, although
some fraction are likely isolated, fragmenting gas disks
\citep[e.g.][]{Hodge2012}.  The subsequent rapid period of physical evolution
likely passes through a quasar stage into a compact passive galaxy, which grows
via dry minor mergers into a massive elliptical galaxies at the present day
\citep[e.g.][]{Cote2007, Naab2009}.

Determining the physical conditions of the ionised gas, powered directly by
star formation, is one of the most important objectives that remain in our
study of DSFGs. The interstellar medium (ISM) is central to many galaxy-wide
physical processes, and is therefore critical to our understanding of the
gas-star-black hole interplay and evolution of galaxies
\citep[e.g.][]{Kormendy2013}. However, the large quantities of dust within
these starbursts obscure the most common ISM tracers: rest-frame optical
spectral lines from the recombination of hydrogen, and those from the most
abundant metal species, C, N, O.

Atomic fine-structure forbidden transitions in the FIR, such as [C\,{\sc ii}]
158-\mum, \NII\ 122-\mum, [O\,{\sc i}] 63-\mum\ and [O\,{\sc iii}] 88-\mum\ are
important coolants of the ISM, providing critical diagnostics of physical
conditions across all redshifts \citep[e.g.][]{Stacey1991, Lord1996, HC2016,
Zhao2016a, Wardlow2017,HC2018,HC2018b}.  Among these lines, [C\,{\sc ii}] 158-\mum\ 
is probably the most important, and the best studied, since it is the brightest
FIR line in most galaxies and often accounts for 0.1--1 per cent of the total
FIR luminosity \citep[e.g.][]{Stacey1991, DiazSantos2013}.  However, neutral
carbon (C) has an ionisation energy of 11.3\,eV, meaning that it co-exists in
both photo-dissociation regions (PDRs) and H\,{\sc ii} regions
\citep[e.g.][]{Stacey2010}. \CII\ can be excited by three independent
collisions excitation mechanisms, electrons, neutral hydrogen (H\,{\sc i}), and
molecular hydrogen (H$_2$). These mechanisms make the \cii\ emission arising
from nearly all ISM phases difficult to discriminate from each other.

The \NIIab\ transition provides complementary information on the origin of the
[C\,{\sc ii}] 158-\mum\ emission \citep[e.g.][]{Oberst2006,Walter2009,
Stacey2010, Decarli2014,Pavesi2016}.  The line ratio of \CII\ 158-\mum /\NIIab\
only depends on the abundances of N$^+$ and C$^+$ in the \HII\ region, and the
relative contributions of the neutral and ionised ISM phases, making the
observed \CII/\NII\ line ratio an excellent probe of the fraction of \CII\ from
the ionised gas phase \citep[e.g.][]{Oberst2011, Oberst2006}.

Unfortunately these lines are typically unobservable from the ground at low
redshifts due to poor atmospheric transmission. The Spectral and Photometric
Imaging REceiver (SPIRE) \citep{Griffin2010} instrument aboard the {\it
Herschel Space Observatory} \citep{Pilbratt2010} incorporated a Fourier
Transform Spectrometer (FTS), covering many of the brightest FIR lines.
However, with its 3.5-m aperture, the few $\times 10$-mJy flux densities
exhibited by \CII\ 158-\mum\ in typical high-redshift DSFGs were well below
the capabilities of {\it Herschel}'s SPIRE FTS, requiring prohibitively long
integration times.

An alternative solution, exploited since the earliest SCUBA observations
\citep{Smail1997}, is to use the flux boost provided by gravitational lensing
due to foreground galaxies, or clusters of galaxies.  Most occurrences grant a
factor of a few increase in brightness, but the most strongly lensed systems
enable very detailed study of the background object \citep[e.g.][]{Fu2012,
Bussmann2012, Bussmann2013, Messias2014, Dye2015,Spilker2016}, as epitomised by
SMM\,J2135$-$0102 -- the Cosmic Eyelash -- serendipitously discovered in the
neighbourhood of a massive cluster, and possessing a high average amplification
\citep[\(37.5 \pm 4.5\),][]{Swinbank2010, Swinbank2011}.  Such strongly lensed
DSFGs are rare \citep[\(\sim 0.26\,\text{deg}^{-2}\) --][]{Bussmann2013},
necessitating surveys covering large areas in order to assemble a statistically
significant sample.  This population can be selected efficiently at FIR/submm
wavelengths, where the number density of unlensed sources at high flux
densities drops quickly (after removal of local spiral galaxies and blazars),
with \(S_{500 \um} > 100\text{-mJy}\) sources being strongly lensed DSFGs
\citep{Negrello2010,Wardlow2013}, with just a smattering of hyperluminous IR
galaxies \citep{Ivison2013,Fu2013}.  Such FIR surveys have recently been
undertaken \citep[e.g.][]{Eales2010,Vieira2010, Oliver2012}, using the {\it
Herschel Space Observatory} and the South Pole Telescope \citep{Carlstrom2011},
which has resulted in hundreds to thousands of strongly lensed DSFGs candidates
\citep[e.g.][]{Negrello2017,Mocanu2013,GN2012}. Many of them have been
confirmed as such by follow-up observations
\citep[e.g.][]{Spilker2016,Negrello2014}.

In this paper we present the results of a \herschel\ Open Time programme
comprising FIR spectroscopic observations of 45 gravitationally-lensed DSFGs
using the SPIRE FTS aboard \herschel.  This paper is organised as follows.
Section~\ref{sec:sample} provides an overview of the sample selection and the
overall properties of the data. Section~\ref{sec:observations} details the
observations and data reduction for \herschel\ SPIRE FTS spectroscopy and PACS
photometry. Section~\ref{sec:results} presents the observed spectra and fitted
dust spectral energy distributions (SEDs). Section~\ref{sec:analysis_stacking}
presents the stacked spectra and associated analysis. We discuss caveats in
statistical biases, stacking methods, absorption contaminations and abundances
in Section~\ref{sec:caveats}. We summarise our results and draw conclusions in
Section~\ref{sec:summary}.  Throughout, we adopt a standard $\Lambda$-CDM
cosmology with $\Omega_{\rmn m} = 0.3$, $\Omega_\Lambda = 0.7$ and $H_0 = 70$
\kms\ Mpc$^{-1}$.

\section{Sample} \label{sec:sample}

\begin{table*}
  \centering
  \begin{tabular}{l l l l l l r c}
    \hline
    \hline
    IAU Name                          & Short Name       & R.A.         & Dec.             & Redshift                     & Notes\\
    \hline
    HATLAS\,J090740.0\(-\)004200      & SDP.9            &09h07m40.032s & $-$00d41m59.64s  & 1.577                        & \\                               
    HATLAS\,J091043.0\(-\)000322      & SDP.11           &09h10m43.056s & $-$00d03m22.68s  & 1.784                        & \\                               
    HATLAS\,J090302.9\(-\)014127      & SDP.17           &09h03m03.024s & $-$01d41m27.24s  & 2.3050                       & \\                               
    HATLAS\,J090311.6\(+\)003906      & SDP.81           &09h03m09.408s & $+$00d39m06.48s  & 3.0425                       & \\                               
    HATLAS\,J091305.0\(-\)005343      & SDP.130          &09h13m05.112s & $-$00d53m43.44s  & 2.6256                       & \\                               
    HATLAS\,J085358.9\(+\)015537      & G09-v1.40        &08h53m58.872s & $+$01d55m37.56s  & 2.0923                       & \\                               
    HATLAS\,J083051.0\(+\)013224      & G09-v1.97        &08h30m51.168s & $+$01d32m24.36s  & 3.634                        & \\                               
    HATLAS\,J084933.4\(+\)021443      & G09-v1.124       &08h49m33.336s & $+$02d14m44.52s  & 2.4101                       & \\                               
    HATLAS\,J091840.8\(+\)023047      & G09-v1.326       &09h18m40.92s  & $+$02d30m46.08s  & 2.5812                       & \\                               
    HATLAS\,J114638.0\(-\)001132      & G12-v2.30        &11h46m37.992s & $-$00d11m31.92s  & 3.2588                       & \\                               
    HATLAS\,J113526.3\(-\)014606      & G12-v2.43        &11h35m26.28s  & $-$01d46m06.6s   & 3.1275                       & \\                               
    HATLAS\,J115820.1\(-\)013753      & G12-v2.257       &11h58m20.04s  & $-$01d37m51.6s   & 2.1909                       & \\                               
    HATLAS\,J142935.3\(-\)002836      & G15-v2.19        &14h29m35.232s & $-$00d28m36.12s  & 1.026                        & \\                               
    HATLAS\,J141351.9\(-\)000026      & G15-v2.235       &14h13m52.08s  & $-$00d00m24.48s  & 2.4778                       & \\                               
    HATLAS\,J134429.4\(+\)303036      & NA.v1.56         &13h44m29.52s  & $+$30d30m34.2s   & 2.3010                       & \\                               
    HATLAS\,J133649.9\(+\)291801      & NA.v1.144        &13h36m49.992s & $+$29d17m59.64s  & 2.2024                       & \\                               
    HATLAS\,J132859.3\(+\)292317      & NA.v1.177        &13h28m59.256s & $+$29d23m26.16s  & 2.778                        & \\                               
    HATLAS\,J132504.4\(+\)311537      & NA.v1.186        &13h25m04.512s & $+$31d15m36s     & 1.8358                       & \\                               
    HATLAS\,J132427.0\(+\)284452      & NB.v1.43         &13h24m27.216s & $+$28d44m49.2s   & 1.676                        & \\                               
    HATLAS\,J133008.4\(+\)245900      & NB.v1.78         &13h30m08.52s  & $+$24d58m59.16s  & 3.1112                       & \\                               
    HATLAS\,J125632.7\(+\)233625      & NC.v1.143        &12h56m32.544s & $+$23d36m27.72s  & 3.565                        & \\                               
{\it HATLAS\,J223829.0\(-\)304148}    & {\it SA.v1.44 }  &22h38m29.472s & $-$30d41m49.2s   & 1.33$\pm$ 0.11$^\star$       & \\                               
{\it HATLAS\,J222536.3\(-\)295646}    & {\it SA.v1.53 }  &22h25m36.48s  & $-$29d56m49.56s  & 1.64$\pm$ 0.16$^\star$       & \\                               
{\it HATLAS\,J232531.4\(-\)302234}    & {\it SB.v1.143}  &23h25m31.608s & $-$30d22m35.76s  & 2.67$\pm$ 0.13$^\star$       & \\                               
{\it HATLAS\,J232623.0\(-\)342640}    & {\it SB.v1.202}  &23h26m23.064s & $-$34d26m43.8s   & 2.17$\pm$ 0.11$^\star$       & \\                               
{\it HATLAS\,J232419.8\(-\)323924}    & {\it SC.v1.128}  &23h24m19.944s & $-$32d39m28.08s  & 2.51$\pm$ 0.15$^\star$       & \\                               
{\it HATLAS\,J000912.6\(-\)300809}    & {\it SD.v1.70 }  &00h09m12.864s & $-$30d08m09.24s  & 1.19$\pm$ 0.10$^\star$       & \\                               
{\it HATLAS\,J000722.3\(-\)352014}    & {\it SD.v1.133}  &00h07m22.272s & $-$35d20m15s     & 1.38$\pm$ 0.11$^\star$       & \\                               
{\it HATLAS\,J002625.1\(-\)341737}    & {\it SD.v1.328}  &00h26m25.176s & $-$34d17m38.4s   & 2.70$\pm$ 0.16$^\star$       & \\                               
{\it HATLAS\,J004736.0\(-\)272953}    & {\it SE.v1.165}  &00h47m36.072s & $-$27d29m53.16s  & 2.03$\pm$ 0.15$^\star$       & \\                               
{\it HATLAS\,J010250.7\(-\)311721}    & {\it SF.v1.88 }  &01h02m50.88s  & $-$31d17m23.64s  & 1.57$\pm$ 0.13$^\star$       & \\                               
{\it HATLAS\,J012407.3\(-\)281435}    & {\it SF.v1.100}  &01h24m07.536s & $-$28d14m35.16s  & 2.00$\pm$ 0.13$^\star$       & \\                               
{\it HATLAS\,J014834.7\(-\)303532}    & {\it SG.v1.77 }  &01h48m34.704s & $-$30d35m32.64s  & 1.53$\pm$ 0.13$^\star$       & \\                               
    HERMES\, J004714.1+032453         &  HeLMS08         &00h47m14.136s & $+$03d24m55.44s  & 1.19$^\dagger$               & no FTS spectra  \\               
    HERMES\, J001626.0+042613         &  HeLMS22         &00h16m26.064s & $+$04d26m12.48s  & 2.5093$^\dagger$             & no FTS spectra  \\               
    HERMES\, J005159.4+062240         &  HeLMS18         &00h51m59.448s & $+$06d22m41.52s  & 2.392$^\dagger$              & no FTS spectra  \\               
    HERMES\, J233255.5-031134         &  HeLMS2          &23h32m55.584s & $-$03d11m36.24s  & 2.6899$^\dagger$             & \\                               
    HERMES\, J234051.3-041937         &  HeLMS7          &23h24m39.576s & $-$04d39m34.2s   & 2.473$^\dagger$              & \\                               
    HERMES\, J004723.3+015749         &  HeLMS9          &00h47m23.352s & $+$01d57m50.76s  & 1.441$^\dagger$              & no FTS spectra \\               
    HERMES\, J001615.8+032433         &  HeLMS13         &00h16m15.864s & $+$03d24m36.72s  & 2.765$^\dagger$              & no FTS spectra \\               
    HERMES\, J233255.7-053424         &  HeLMS15         &23h32m55.824s & $-$05d34m26.76s  & 2.4024$^\dagger$             & no FTS spectra \\               
    HERMES\, J234051.3-041937         &  HeLMS5          &23h40m51.528s & $-$04d19m40.8s   & 3.50$^\dagger$               & no FTS spectra \\               
    1HerMES S250 J142823.9\(+\)352619 & HBo\"otes03      &14h28m24.072s & $+$35d26m19.32s  & 1.325                        & \\                               
    1HerMES S250 J021830.5\(-\)053124 & HXMM02           &02h18m30.672s & $-$05d31m31.44s  & 3.39                         & \\                               
    SMM\,J213511.6\(-\)010252         & Eyelash          &21h35m11.64s  & $-$01d02m52.44s  & 2.32591                      & \\                               
    \hline                                                                                                                                         
  \end{tabular}
  \caption{ Summary of the Galaxy sample. \newline
    $^\dagger$: Spectroscopic redshifts adopted from \citet{Nayyeri2016}. 
    $^\star$: Redshift estimated from photometric data, using the SED of average ALESS galaxies as the template \citep{Ivison2016}.
    For galaxy without spec-$z$ information we label their names with italic fonts. 
    The order of the table is organised by different survey fields and then by the right ascensions of the galaxies. 
    }
  \label{tab:sources}
\end{table*}


\begin{table*}
  \centering
  \begin{tabular}{l l r@{\,}l r@{\,}l r@{\,}l r@{\,}l r@{\,}l r@{\,}l r@{\,}l r@{\,}l r@{\,}l}
    \hline
    \hline
    Source & $z$ & \multicolumn{2}{c}{\hspace{0.5em}Amplification}      & \multicolumn{2}{c}{\fd{70\,\um}}     & \multicolumn{2}{c}{\fd{100\,\um}} & \multicolumn{2}{c}{\fd{160\,\um}} & \multicolumn{2}{c}{\fd{250\,\um}} & \multicolumn{2}{c}{\fd{350\,\um}} & \multicolumn{2}{c}{\fd{500\,\um}} & \multicolumn{2}{c}{\fd{850\,\um}}& \multicolumn{2}{c}{\fd{880\,\um}}\\
           &   & &                                                    & \multicolumn{2}{c}{  mJy}            & \multicolumn{2}{c}{  mJy}         & \multicolumn{2}{c}{  mJy}         & \multicolumn{2}{c}{  mJy}         & \multicolumn{2}{c}{  mJy}         & \multicolumn{2}{c}{/ mJy}         & \multicolumn{2}{c}{  mJy}        & \multicolumn{2}{c}{  mJy}         \\
    \hline                                                                                                
    SDP.9       & 1.574                     &  8.8 & \(\pm\, 2.2\)           &    & \(-\)                                  & 307 & \(\pm\, 15\)                & 546 & \(\pm\, 20\)    & 478 & \(\pm\, 34\) & 328 & \(\pm\, 24\) & 171 & \(\pm\, 14\) &         &\(-\,       \)  &  24.8 & \(\pm\,  3.3\) \\
    SDP.11      & 1.786                     & 10.9 & \(\pm\, 1.3\)           &    & \(-\)                                  & 161 & \(\pm\, 10\)                & 363 & \(\pm\, 20\)    & 421 & \(\pm\, 30\) & 371 & \(\pm\, 26\) & 221 & \(\pm\, 17\) &  52     &\(\pm\,  1    \)  &  30.6 & \(\pm\,  2.4\) \\
    SDP.17      & 2.305                     &  4.9 & \(\pm\, 0.7\)           &    & \(-\)                                  &  66 & \(\pm\,  7\)                & 244 & \(\pm\, 19\)    & 354 & \(\pm\, 25\) & 339 & \(\pm\, 24\) & 220 & \(\pm\, 17\) &         &\(-\,       \)  &  54.7 & \(\pm\,  3.1\) \\
    SDP.81      & 3.040                     & 15.9 & \(\pm\, 0.7 ^\text{a}\) &    & \(<\,9\)                               &     & \(-\)                       &  58 & \(\pm\, 10\)    & 133 & \(\pm\, 11\) & 186 & \(\pm\, 14\) & 165 & \(\pm\, 14\) & 108     &\(\pm\, 10    \)  &  78.4 & \(\pm\,  8.2\) \\
    SDP.130     & 2.6256                    &  2.1 & \(\pm\, 0.3\)           &    & \(<\,9\)                               &     & \(-\)                       &  66 & \(\pm\, 10\)    & 118 & \(\pm\,  9\) & 137 & \(\pm\, 11\) & 104 & \(\pm\,  9\) & 67      &\(\pm\, 9     \)  &  36.7 & \(\pm\,  3.9\) \\
    G09-v1.40   & 2.093                     & 15.3 & \(\pm\, 3.5 \)          &    & \(-\)                                  &  70 & \(\pm\,  4\)                & 280 & \(\pm\, 13\)    & 396 & \(\pm\, 28\) & 368 & \(\pm\, 26\) & 228 & \(\pm\, 17\) &         &\(-\,       \)  &  61.4 & \(\pm\,  2.9\) \\
    G09-v1.97   & 3.634                     &  6.9 & \(\pm\, 0.6 \)          &    & \(-\)                                  &  53 & \(\pm\,  3\)                & 198 & \(\pm\, 10\)    & 249 & \(\pm\, 18\) & 305 & \(\pm\, 22\) & 269 & \(\pm\, 20\) & 121     &\(\pm\, 8     \)  &  85.5 & \(\pm\,  4.0\) \\
    G09-v1.124  & 2.410                     &  1.1 & \(\pm\, 0.1 \)          & 16 & \(\pm\, 4\)                            &  57 & \(\pm\,  4\)                & 169 & \(\pm\, 15\)    & 217 & \(\pm\, 16\) & 249 & \(\pm\, 18\) & 209 & \(\pm\, 16\) & 62      &\(\pm\, 10    \)  &  50.0 & \(\pm\,  3.5\) \\
    G09-v1.326  & 2.5812                    &  5.0 & \(\pm\, 1.0 ^\text{b}\) &    & \(-\)                                  &  41 & \(\pm\,  4\)                & 106 & \(\pm\, 10\)    & 126 & \(\pm\, 10\) & 151 & \(\pm\, 12\) & 128 & \(\pm\, 11\) & 61      &\(\pm\, 9     \)  &  18.8 & \(\pm\,  1.6\) \\
    G12-v2.30   & 3.259                     &  9.5 & \(\pm\, 0.6 \)          & 30 & \(\pm\, 4\)                            &  62 & \(\pm\,  4\)                & 235 & \(\pm\, 15\)    & 317 & \(\pm\, 23\) & 358 & \(\pm\, 25\) & 291 & \(\pm\, 21\) & 142     &\(\pm\, 8     \)  &  86.0 & \(\pm\,  4.9\) \\
    G12-v2.43   & 3.127                     & 17.0 & \(\pm\,11.0 ^\text{b}\) & 16 & \(\pm\, 3\)                            &  81 & \(\pm\,  5\)                & 196 & \(\pm\, 11\)    & 279 & \(\pm\, 20\) & 284 & \(\pm\, 21\) & 205 & \(\pm\, 16\) & 116     &\(\pm\, 9     \)  &  48.6 & \(\pm\,  2.3\) \\
    G12-v2.257  & 2.191                     & 13.0 & \(\pm\, 7.0 ^\text{b}\) & 15 & \(\pm\, 4\)                            &  43 & \(\pm\,  5\)                & 143 & \(\pm\, 11\)    & 119 & \(\pm\,  9\) & 124 & \(\pm\, 10\) & 101 & \(\pm\,  9\) & 40      &\(\pm\, 9     \)  &       & \(-\)          \\
    G15-v2.19   & 1.027                     &  9.7 & \(\pm\, 0.7 ^\text{c}\) & 316& \(\pm\, 16\)                           & 850 & \(\pm\, 10\)                &1190 & \(\pm\, 53\)    & 802 & \(\pm\, 56\) & 438 & \(\pm\, 31\) & 200 & \(\pm\, 15\) &         &\(-\,       \)  &       & \(-\)          \\
    G15-v2.235  & 2.479                     &  1.8 & \(\pm\, 0.3 \)          &    & \(-\)                                  &  48 & \(\pm\,  5\)                &  87 & \(\pm\,  6\)    & 189 & \(\pm\, 14\) & 217 & \(\pm\, 16\) & 176 & \(\pm\, 14\) & 104     &\(\pm\, 11    \)  &  33.3 & \(\pm\,  2.6\) \\
    NA.v1.56    & 2.301                     & 11.7 & \(\pm\, 0.9 \)          & 14 & \(\pm\, 3\)                            &  86 & \(\pm\,  4\)                & 308 & \(\pm\, 19\)    & 462 & \(\pm\, 33\) & 466 & \(\pm\, 33\) & 343 & \(\pm\, 25\) & 142     &\(\pm\, 8     \)  &  73.1 & \(\pm\,  2.4\) \\
    NA.v1.144   & 2.202                     &  4.4 & \(\pm\, 0.8 \)          & 11 & \(\pm\, 3\)                            &  47 & \(\pm\,  4\)                & 193 & \(\pm\, 10\)    & 294 & \(\pm\, 21\) & 286 & \(\pm\, 21\) & 194 & \(\pm\, 15\) &         &\(-\,       \)  &  36.8 & \(\pm\,  2.9\) \\
    NA.v1.177   & 2.778                     &      & \(-\)                   &    & \(-\)                                  &  40 & \(\pm\,  3\)                & 155 & \(\pm\, 14\)    & 268 & \(\pm\, 19\) & 296 & \(\pm\, 21\) & 249 & \(\pm\, 18\) & 149     &\(\pm\, 11    \)  &  50.1 & \(\pm\,  2.1\) \\
    NA.v1.186   & 1.839                     &      & \(-\)                   &    & \(-\)                                  &  60 & \(\pm\,  6\)                & 163 & \(\pm\,  9\)    & 241 & \(\pm\, 18\) & 227 & \(\pm\, 17\) & 165 & \(\pm\, 13\) & 39      &\(\pm\, 8     \)  &       & \(-\)          \\
    NB.v1.43    & 1.680                     &  2.8 & \(\pm\, 0.4 \)          &    & \(-\)                                  &  52 & \(\pm\,  4\)                & 170 & \(\pm\, 24\)    & 342 & \(\pm\, 25\) & 371 & \(\pm\, 27\) & 251 & \(\pm\, 19\) & 71      &\(\pm\, 10    \)  &  30.2 & \(\pm\,  2.2\) \\
    NB.v1.78    & 3.111                     & 13.0 & \(\pm\, 1.5 \)          & 40 & \(\pm\, 3\)                            &  87 & \(\pm\,  4\)                & 212 & \(\pm\, 16\)    & 271 & \(\pm\, 20\) & 278 & \(\pm\, 20\) & 203 & \(\pm\, 16\) & 108     &\(\pm\, 11    \)  &  59.2 & \(\pm\,  4.3\) \\
    NC.v1.143   & 3.565                     & 11.3 & \(\pm\, 1.7 \)          &    & \(-\)                                  &  25 & \(\pm\,  3\)                &  97 & \(\pm\,  8\)    & 209 & \(\pm\, 16\) & 289 & \(\pm\, 21\) & 264 & \(\pm\, 20\) & 160     &\(\pm\, 10    \)  &  97.2 & \(\pm\,  6.5\) \\
{\it SA.v1.44 } & 1.33$\pm0.11^\star$       &      & \(-\)                   &    & \(-\)                                  &  93 & \(\pm\,  5\)                & 225 & \(\pm\, 14\)    & 252 & \(\pm\, 18\) & 207 & \(\pm\, 15\) & 100 & \(\pm\,  9\) &         &\(-\,       \)  &       & \(-\)          \\
{\it SA.v1.53 } & 1.654$^\dagger$           &      & \(-\)                   &    & \(-\)                                  &  57 & \(\pm\,  3\)                & 163 & \(\pm\, 23\)    & 194 & \(\pm\, 16\) & 200 & \(\pm\, 17\) & 119 & \(\pm\, 14\) &         &\(-\,       \)  &       & \(-\)          \\
{\it SB.v1.143} & 2.42$^\dagger$            &      & \(-\)                   &    & \(-\)                                  &  35 & \(\pm\,  5\)                & 102 & \(\pm\, 15\)    & 176 & \(\pm\, 13\) & 227 & \(\pm\, 17\) & 176 & \(\pm\, 14\) & 100     &\(\pm\, 9     \)  &       & \(-\)          \\
{\it SB.v1.202} & 2.055$^\dagger$           &      & \(-\)                   &    & \(-\)                                  &  42 & \(\pm\,  4\)                & 130 & \(\pm\, 17\)    & 154 & \(\pm\, 12\) & 178 & \(\pm\, 13\) & 123 & \(\pm\, 11\) & 57      &\(\pm\, 11    \)  &       & \(-\)          \\
{\it SC.v1.128} & 2.574$^\dagger$           &      & \(-\)                   &    & \(-\)                                  &  35 & \(\pm\,  5\)                & 123 & \(\pm\, 13\)    & 213 & \(\pm\, 16\) & 244 & \(\pm\, 18\) & 169 & \(\pm\, 13\) & 73      &\(\pm\, 10    \)  &       & \(-\)          \\
{\it SD.v1.70 } & 1.19$\pm0.10^\star$       &      & \(-\)                   &    & \(-\)                                  & 156 & \(\pm\,  5\)                & 365 & \(\pm\, 23\)    & 353 & \(\pm\, 25\) & 273 & \(\pm\, 20\) & 156 & \(\pm\, 13\) &         &\(-\,       \)  &       & \(-\)          \\
{\it SD.v1.133} & 1.60$^\dagger$            &      & \(-\)                   &    & \(-\)                                  & 142 & \(\pm\,  5\)                & 267 & \(\pm\, 20\)    & 237 & \(\pm\, 17\) & 193 & \(\pm\, 15\) & 108 & \(\pm\, 10\) &         &\(-\,       \)  &       & \(-\)          \\
{\it SD.v1.328} & 2.70$\pm0.16^\star$       &      & \(-\)                   &    & \(-\)                                  &     & \(< 10\)                    &  70 & \(\pm\,  9\)    & 138 & \(\pm\, 11\) & 186 & \(\pm\, 14\) & 149 & \(\pm\, 12\) & 92      &\(\pm\, 13    \)  &       & \(-\)          \\
{\it SE.v1.165} & 2.03$\pm0.15^\star$       &      & \(-\)                   &    & \(-\)                                  &  27 & \(\pm\,  3\)                & 108 & \(\pm\,  9\)    & 171 & \(\pm\, 13\) & 197 & \(\pm\, 15\) & 146 & \(\pm\, 12\) &         &\(-\,       \)  &       & \(-\)          \\
{\it SF.v1.88 } & 1.57$\pm0.13^\star$       &      & \(-\)                   &    & \(-\)                                  &  73 & \(\pm\,  4\)                & 168 & \(\pm\,  9\)    & 268 & \(\pm\, 19\) & 253 & \(\pm\, 19\) & 168 & \(\pm\, 14\) &         &\(-\,       \)  &       & \(-\)          \\
{\it SF.v1.100} & 2.00$\pm0.13^\star$       &      & \(-\)                   &    & \(-\)                                  &  50 & \(\pm\,  4\)                & 135 & \(\pm\,  9\)    & 258 & \(\pm\, 19\) & 271 & \(\pm\, 20\) & 204 & \(\pm\, 16\) & 94      &\(\pm\, 10    \)  &       & \(-\)          \\
{\it SG.v1.77 } & 1.53$\pm0.13^\star$       &      & \(-\)                   &    & \(-\)                                  & 148 & \(\pm\,  9\)                & 344 & \(\pm\, 15\)    & 238 & \(\pm\, 18\) & 220 & \(\pm\, 17\) & 127 & \(\pm\, 13\) &         &\(-\,       \)  &       & \(-\)          \\
 HeLMS08        & 1.52$\pm0.11^\star$       &      & \(-\)                   &    & \(-\)                                  &  87 & \(\pm\, 10\)                & 227 & \(\pm\, 15\)    & 300 & \(\pm\, 22\) & 246 & \(\pm\, 18\) & 170 & \(\pm\, 15\) &         &\(-\,       \)  &       & \(-\)          \\
 HeLMS22        & 2.46$\pm0.18^\star$       &      & \(-\)                   &    & \(-\)                                  &  13 & \(\pm\,  3\)                &  65 & \(\pm\, 10\)    & 130 & \(\pm\, 15\) & 180 & \(\pm\, 18\) & 130 & \(\pm\, 15\) &         &\(-\,       \)  &       & \(-\)          \\
 HeLMS18        & 2.07$\pm0.13^\star$       &      & \(-\)                   &    & \(-\)                                  &  31 & \(\pm\,  3\)                &  91 & \(\pm\, 15\)    & 163 & \(\pm\, 13\) & 202 & \(\pm\, 15\) & 142 & \(\pm\, 12\) &         &\(-\,       \)  &       & \(-\)          \\
 HeLMS2         & 2.41$\pm0.19^\star$       &      & \(-\)                   &    & \(-\)                                  &  25 & \(\pm\,  4\)                & 146 & \(\pm\, 14\)    & 250 & \(\pm\, 18\) & 324 & \(\pm\, 23\) & 247 & \(\pm\, 19\) &         &\(-\,       \)  &       & \(-\)          \\
 HeLMS7         & 1.97$\pm0.14^\star$       &      & \(-\)                   &    & \(-\)                                  &  33 & \(\pm\,  4\)                & 129 & \(\pm\,  7\)    & 219 & \(\pm\, 16\) & 227 & \(\pm\, 17\) & 166 & \(\pm\, 13\) &         &\(-\,       \)  &       & \(-\)          \\
 HeLMS9         & 1.18$\pm0.11^\star$       &      & \(-\)                   &    & \(-\)                                  & 132 & \(\pm\,  4\)                & 340 & \(\pm\, 20\)    & 367 & \(\pm\, 25\) & 293 & \(\pm\, 21\) & 170 & \(\pm\, 14\) &         &\(-\,       \)  &       & \(-\)          \\
 HeLMS13        & 2.05$\pm0.16^\star$       &      & \(-\)                   &    & \(-\)                                  &  39 & \(\pm\,  3\)                & 168 & \(\pm\, 15\)    & 176 & \(\pm\, 13\) & 210 & \(\pm\, 15\) & 134 & \(\pm\, 11\) &         &\(-\,       \)  &       & \(-\)          \\
 HeLMS15        & 2.66$\pm0.17^\star$       &      & \(-\)                   &    & \(-\)                                  &  14 & \(\pm\,  3\)                &  44 & \(\pm\,  8\)    & 153 & \(\pm\, 12\) & 186 & \(\pm\, 14\) & 152 & \(\pm\, 13\) &         &\(-\,       \)  &       & \(-\)          \\
 HeLMS5         & 2.73$\pm0.21^\star$       &      & \(-\)                   &    & \(-\)                                  &   7 & \(\pm\,  3\)                &  68 & \(\pm\,  7\)    & 149 & \(\pm\, 12\) & 197 & \(\pm\, 15\) & 188 & \(\pm\, 15\) &         &\(-\,       \)  &       & \(-\)          \\
    HBo\"otes03 & 1.325                     & 3.0  & \(\pm\, 1.5 \)          &    & \(-\)                                  & 104 & \(\pm\,  5\)                & 279 & \(\pm\, 16\)    & 323 & \(\pm\, 23\) & 244 & \(\pm\, 18\) & 140 & \(\pm\, 34\) &         &\(-\,       \)  &  18.4 & \(\pm\,  2.5\) \\
    HXMM02      & 3.395                     & 4.4  & \(\pm\, 1.0 \)          &    & \(-\)                                  &  29 & \(\pm\,  3\)                &  93 & \(\pm\, 15\)    &  92 & \(\pm\, 10\) & 122 & \(\pm\, 12\) & 113 & \(\pm\, 11\) &         &\(-\,       \)  &  66.0 & \(\pm\,  5.4\) \\
    Eyelash     & 2.32591                   & 37.5 & \(\pm\, 4.5 ^\text{d}\) &    & \(-\)                                  &  25 & \(\pm\,  3\)                & 147 & \(\pm\,  7\)    & 366 & \(\pm\, 55\) & 429 & \(\pm\, 64\) & 325 & \(\pm\, 49\) &     115 &\(\pm\,     13\)  & 106.0 & \(\pm\, 12.0\) \\
    \hline
  \end{tabular}
  \caption{
FIR continuum flux densities of the sample.  250, 350 and 500\,\um{} data from
\textit{H}-ATLAS and \textit{Her}MES, uncertainties include a 7\,\% calibration
uncertainty \citep{Swinyard2010, Bendo2013}. 
Most of the 70 \mum\ fluxes are from \citet{Wardlow2017}, except for SDP.81 and
SDP.130, which are measured with our PACS observations.  100 and 160\,\um{}
uncertainties include 2.75\,\% and 4.15\,\% calibration uncertainties
respectively, following the PACS Photometer - Point-Source Flux Calibration
document. The 850 \mum\ data are from the SCUBA2 observations \citep{Bakx2018}.
 The 880\,\um{} data is from \citet{Bussmann2013} and
\citet{Swinyard2010}. We notice that the SCUBA2 850 \mum\ fluxes are higher
than the SMA 880 \mum\ fluxes, likely due to the interferometric filtering
issue.  Where not otherwise noted, amplification values are taken from
\citep{Bussmann2013}. We also notice that for HXMM02, the new ALMA flux density
is $63.33\pm 0.58$ \citep[i.e.,][]{Bussmann2015}, consistent with the SMA
value.
\(^\star\): Redshift estimate from photometric data, using the SED of Eyelash as the template.
\(^\dagger\): Redshift is estimated from both photometric data and a possible (low significance) spectral feature.
\(^\text{a}\)~\citet{Dye2015} 
\(^\text{b}\)~Estimate from CO line luminosity and FWHM from \citet{Harris2012} 
\(^\text{c}\)~\citet{Messias2014}
\(^\text{d}\)~\citet{Swinbank2011}
  }
  \label{tab:flux_densities}
\end{table*}


In Table~\ref{tab:sources} we present basic information for our sample of 45
DSFGs, observed in \otonefull\ and \ottwofull. The majority of the targets were
selected from the \herschel\ Astrophysical Terahertz Large Area Survey
(\textit{H}-ATLAS; \citealp{Eales2010}) and \herschel\ Multi-Tiered
Extragalactic Survey (\textit{Her}MES; \citealp{Oliver2012}) Large Mode Survey
(HeLMS). \herschel\ SPIRE 250-, 350- and 500-\um\ images were used to identify
strongly lensed DSFG candidates, from which we selected those satisfying
\(\fd{350} \gtrsim 200 \, \mJy\), with no indication that they could be a
blazar or a \(z \lesssim 0.1\) spiral, and with a color cut attempting to
remove the highest redshift objects such that the \cii\,158-\um\ line would
remain within the FTS spectral range. The SPIRE images show point sources, or
only show marginally resolved features. The sample was supplemented with
several objects for which substantial ancillary data existed, including
SMM\,J2135$-$0102. About half of the sources have been confirmed to be lensed targets
from submm continuum observations \citep[e.g.][]{Bussmann2013}. 

\section{Observations and data reduction}\label{sec:observations}

\subsection{PACS observations and flux density measurements}
\label{sec:pacs_measurements}

The original \textit{H}-ATLAS parallel PACS imaging data at 100  and 160\,\um\
have noise levels of $\sim 25$ -- $50 $ mJy \citep[almost ten times higher than
our new targted observations; see][]{Ibar2010,Eales2010}. They were
insufficiently deep to detect many of our sample, and the targets outside
\textit{H}-ATLAS had no coverage at these wavelengths. To complement the SPIRE
photometric measurements across the flux-density peak of their SEDs and to
provide stronger constraints on their rest-frame mid-infrared (MIR) emission,
we obtained deep imaging observations at 100 and 160\,\um\ with \herschel\
PACS.

For each galaxy we obtained two cross-linked mini-scans with PACS, recording
data at 100 and 160\,\um\ simultaneously. Each mini-scan covers an area of
$\sim 10' \times 3'$, with two orientations of 70 and 110\degr, resulted to
three minutes on-source and a total of 791 seconds observing time including
overhead.  On average, we reach  1-\(\sigma\) depths of \(\sim 3\) and 7\,mJy at
100 and 160\,\um, respectively.  Archival mini-scan imaging data covering
SMM\,J2135$-$0102 was combined with our two scans to produce a deeper image,
which is particularly useful as this field has a large number of FIR sources
visible in the region around the target lensed galaxy.  Information about the
\herschel\ PACS observations is listed in Appendix~\ref{observingdate}.

We adopted the {\it Herschel} Interactive Processing Environment \citep[{\sc
hipe} v12;][]{Ott2010} to process and combine the mini-scan data using the
standard pipeline scripts.  To remove the 1/\(f\) noise, a high-pass filter was
applied, after masking visible sources \citep[e.g.][]{Popesso2012}.
Corrections were adopted to the measured flux densities to compensate for the
losses due to this filter. We also applied a color correction to take into
account the spectral index within the bandpass of the PACS
spectrometer\footnote{\url{http://herschel.esac.esa.int/twiki/bin/view/Public/PacsCalibrationWeb}}.

To measure the flux density, we first removed all visible background sources,
masking with 8-10$''$ diameter circles, then fitted the global background and
subtracted it from the masked image. The global background level was on the
order of $10^{-6}$\,Jy\,pixel$^{-1}$, making a negligible contribution to the
final flux density measurements. We performed aperture photometry to remove the
local background.  Most targets displayed compact 100-\um\ emission within an
aperture 5-7$''$ in radius. Therefore, we adopted aperture corrections
according to the encircled energy fraction (EEF) curves in the \herschel\ PACS
manual\footnote{\url{http://herschel.esac.esa.int/twiki/bin/view/Public/PacsCalibrationWeb}},
assuming point-like sources. A few targets displayed extended 100-\um\ emission
that was clearly resolved by the PACS beam. For these sources, we used an
aperture of $\sim 25$--$30''$ (FWHM) to ensure all the emission was included.
Around a third of the targets were resolved by PACS into two components in our
100- and 160-\mum\ maps, with separations ranging from $5''$ to $15''$, too
close to be resolved by SPIRE. For these, we first identify the peak of
individual emitting structure by fitting 2-D Gaussian distribution and
performed aperture photometry on the two components separately and summed their
flux densities within the SPIRE beam size. Among the sample, around half have
been confirmed as lensed galaxies in previous studies
\citep[e.g.][]{Bussmann2013,Bussmann2015}.  A few targets had components which
displayed different $S_{100}/S_{160}$ or $S_{70}/S_{100}$ colors, perhaps
indicative that they are not two segments from the same background lensed
galaxy, but rather different galaxies along the line of sight.  In
Appendix~\ref{stamps}, we show postage-stamp images of the PACS observations of
all of our targets.

The calibration uncertainties for the 100- and 160-\um\ images were $\sim 3\%$ and
$\sim~4\%$ \footnote{According to the PACS Photometer - Point-Source Flux
Calibration document 2011.
\url{http://herschel.esac.esa.int/twiki/pub/Public/PacsCalibrationWeb/pacs\_bolo\_fluxcal\_report\_v1.pdf}},
respectively. We tested different high-pass filters, photometric aperture radii
and mask radii, finding that these choices in total contribute uncertainties of
$\lesssim$10\%. In the end, we combine all these into our final estimate of the
flux uncertainty. Typical noise levels are $\sim$0.1 mJy\,pix$^{-1}$ and
$\sim$0.15 mJy\,pix$^{-1}$ for the  100- and 160-\um\ images, which are  around
20 -- 30 times deeper than the PACS maps of the \textit{H}-ATLAS survey
\citep{Ibar2010b,Smith2017}.  For the common sources in \cite{Wardlow2017}, we
have re-measured fluxes at 100- and 160-\um, and found consistent results
differing by $<$5\%.  Measured flux densities are shown in
Table~\ref{tab:flux_densities}, along with their SPIRE 250-, 350- and 500-\um\
and Submillimeter Array (SMA) 880-\um\ flux density measurements, taken from
\citet{Bussmann2013}.

\subsection{SPIRE FTS observations and data reduction} \label{sec:spire}

The SPIRE FTS instrument \citep[i.e.,][]{Griffin2010} is comprised of two
bolometer arrays (SSW and SLW), covering the wavelength ranges
\(\lambda_\text{obs} = \text{191--318} \) and 295--671\,\um, which corresponds
to \cii\ redshift ranges of 0.2 to 1.0 and 0.85 to 3.2, respectively.  The
observations were performed in the high-resolution mode, in which each mirror
scan takes 66.6\,s, producing a maximum optical path difference of 12.56\,cm,
resulting a uniform spectral resolution of $\sim1.2$\,GHz.

The profile of the SPIRE FTS beam varies with observing frequency in a complex
manner \citep[][]{Makiwa2013}.  The effective angular resolution varies from
$\sim 17''$ at the highest SSW frequency to a maximum of $\sim 42''$ at the
lowest SLW frequency \citep[][]{Swinyard2010,Makiwa2013}. The pointing accuracy
was within $2''$ \citep[][]{Pilbratt2010}. Our target sizes are mostly less
than 2--$3''$, as revealed by high-resolution submm and radio imaging
\citep[e.g.][]{Bussmann2013}, except for G09.124 which consists of multiple
galaxies at $z=2.4$ with separations of up to
10.5$''$ \citep[e.g.][]{Ivison2013}.

We obtained spectra of 38 sources, including five repeat observations of
SMM\,J2135$-$0102 \citep[see][]{George2014}. Each observation consisted of 100
repetitions (100 forward and 100 reverse scans of the SMEC mirror),
corresponding to 13320\,s of on-source integration time. The SLW detectors are
separated by $51''$, and the SSW detectors by $33''$, sufficiently far to avoid
sidelobe contamination by emission from the targets.  The central detectors of
the arrays were centred on the target in each case.

We reprocessed the data from each observing epoch using the \herschel\ data
processing pipeline \citep{Fulton2010} within HIPE \citep{Ott2010} v14.2. The
version of the calibration data is SPIRE\_CAL\_14\_3.  Due to the weakness of
the emission from the sample, in most galaxies we could not obtain a good
detection of the continuum level using the pipeline
\citep[e.g.][]{Hopwood2014,Hopwood2015,Fulton2016}, so their SLW and SSW
spectra could not be matched to each other. This likely generates systematic
continuum offsets as a function of frequency \citep[e.g.][]{Hopwood2015}, which
adds uncertainties to the spectra.  The continuum levels detected with
SPIRE/FTS have a good agreement with the SPIRE photometry, in general
\citep{Makiwa2016,Valtchanov2018}.  For rare cases ($\la$ 30\%) where the
baselines between SLW and SSW match each other, the continuum levels estimated
from the FTS spectra at $\sim 250$ \mum\ are consistent with those measured
from SPIRE photometric fluxes within $\sim 30$\%.

For SLW and SSW, especially at the high-frequency end of the SLW
spectra, about half of the spectra show a high level of fringing at the band
edges (at the levels of $\ge 0.5$\,Jy), in particular at the high frequency end
of SLW spectra. Lower amplitude fringing is observed in all spectra.
The fringing can be traced to resonant cavities that exist within the
spectrometer (for example the air-gaps between the different band defining filters
and/or field lenses) and their effect on continuum calibration. The imperfect
subtraction of the warm background generated by the Herschel telescope has its
root in the derived relative  spectral response function \citep[RSRF, e.g.\,
{\it Herschel}
workshop\footnote{https://nhscsci.ipac.caltech.edu/sc/index.php/Workshops/Fall2014Talks}
2014,][]{Swinyard2014, Fulton2014}.  These fringes represent correlated pink
(1/\(f\)) noise, which is the dominant noise contribution to our spectra and
to the overall shape of the continuum.  Moreover, the ripples not only vary
with time, which can be seen in the six separate scans of SMM\,J2135$-$0102,
but they also vary between adjacent scans of different targets. It is not
possible to fit polynomial baselines to subtract these ripples.

We tried baseline subtraction to remove the fringes, by subtracting spectra
obtained on dark sky on the same observational days (OD) or subtracting an
average spectra from off-centre pixels. However, due to small differences
in the thermal environment of the telescope, spectrometer fringes remain the
dominant source of spectral noise.  This is not only because the weak level of
the continuum response is not well calibrated, for the science target, for the
dark sky, and for the spectra observed with the off-centre pixels, but also
because subtracting another spectrum -- with a different continuum and
telescope model calibration error -- adds noise.

Fortunately, these fringes appear as relatively low spatial frequencies which
do not seriously impede extraction of parameters from narrow spectral lines.
Real line emission displays a relatively narrow width (FWHM) of $\sim
2$--3\,GHz, which is the convolution of a typical linewidth of $\sim
500$\,\kms\ and the 1.2-GHz Sinc width at the lowest observing frequency, $\sim
500$\,GHz).

We therefore performed baseline fitting designed to subtract the low-frequency
features of the spectra. To this end, we masked 2-GHz frequency ranges
(corresponding to $\sim 500$ \kms\ at $z = 2$ for \CII\ 158-\mum)  around the
few strong lines (\CII\ 158-\mum, \OIII\ 88-\mum, and OH 119-\mum), then we
convolved the masked spectra with a Gaussian profile with 15-GHz FWHM in order
to derive an overall local baseline profile which fits the fringes very well
and does not contaminate the narrow line features.  We then subtracted the
baseline profiles from the original spectra to obtain the final spectra for
each target. We tested this method by inserting artificial spectral signals
into the raw data, subtracting the baseline, then fitting the signal, in order
to determine whether we can recover the line flux.  The tests demonstrated that
a robust flux measurement can be recovered after the baseline subtraction.
We present the detailed tested results in
Appendix~\ref{app:baseline_subtraction} and show the final spectra of our
targets in Appendix~\ref{app:All_spectra}.

The default spectral response of \herschel\ FTS is a Sinc function with a
uniform width of $\sim 1.2$\,GHz. We present the raw spectra at the original
spectral resolution.  Measurements of the \cii\ spectral line flux densities
were made by fitting a Sinc+Gaussian function -- considering the large line
width of a typical high-redshift DSFG --
\citep[e.g.][]{1998ApJ...506L...7F,Bothwell2013, Yang2017} -- to each spectra,
and a Gaussian function to the stacked spectra.  Several FTS spectra from this
dataset have been published previously
\citep[e.g.][]{Valtchanov2011,George2013, George2014}. Here, we include
re-reductions of those data with the latest pipeline (HIPE v14.2). The line
fluxes and uncertainties are shown in Table~\ref{tab:line_fluxes} and cut-out
spectral regions around the \cii\,158-\um\ line are shown in
Fig.~\ref{fig:sed_fits}.

\subsection{APEX observations and data reduction}
\label{sec:apex}

To check the \CII\ flux and to resolve the line profiles of galaxies detected
in our \herschel\ observations, we obtained ground-based observations of SDP.11
and NA.v1.186 with the 12-m Atacama Pathfinder EXperiment (APEX) telescope on
the Chajnantor Plateau in Chile, in good (pwv $<$ 0.6\,mm) weather conditions.
We conducted the observations during the science verification phase of the
band-9 Swedish ESO PI receiver for APEX \citep[i.e., SEPIA B9,][]{Belitsky2018}
during 2016 August and September. The project number is E-097.F-9808A-2016.

All observations were performed in a wobbler-switching mode. Beam throws were
$2'$, offset in azimuth. The focus was determined using Mars. Pointing was
corrected every 30\,min using nearby carbon stars, resulting in a typical
uncertainty of $2''$ (r.m.s.).  Typical system temperatures were 600--1000\,K.
The Fast Fourier Transform Spectrometer backend provided a bandwidth of 4\,GHz.
The beamsize was $\sim 8''$ at 670\, GHz.  All data were reduced with the {\sc
class} package in {GILDAS}\footnote{\url{http://www.iram.fr/IRAMFR/GILDAS}}.

We first combine the spectral data of each sub-scan obtained in two different
spectrometers, which have 2.5 GHz bandwidth each, and an overlap of 1 GHz. The
data from two spectrometers can not be simply added since sometimes they have
different continuum levels (even after the wobbler-switch). The source spectra
are centralised in the overlapped region, so the spectrum obtained in each
spectrometer only covers one side of the line-free channels (baseline). If the
baseline is fitted (and subtracted) to the spectrum from each spectrometer
individually, the slope of the order-1 polynomial baseline is little
constrained. 

To combine the data from the two spectrometer for each individual sub-scan, we
fit the spectra in the overlapped frequency range and combine the two spectra
with a uniform weighting. We also removed $\sim 50$ MHz at both spectral edges,
to avoid poor responses there.  Then, we checked the line profiles of the CO
transitions in the literature \citep[see][for SDP.11]{Oteo2017} and establish
the velocity range over which a linear baseline fit will be applied. For
NA.v1.186, we set the velocity range to $\pm$ 400 \kms\,  for its emission.
Linear baselines were subtracted after inspecting each individual combined
spectrum. An automatic quality control is made to get rid of the spectra whose
noise is much higher than the theoretical one \citep[][Details are described in
the James Clerk Maxwell Telescope MALATANG survey 
\footnote{\url{https://github.com/malatang-jcmt-survey/auto_qualification}}]{Zhang2018b} .
Only $<$ 5\% of the data are thrown away for poor baseline qualities.  We
converted the antenna temperatures ($T_{\rm A}^\star$) to flux density using
the telescope efficiency of $150\pm 30$ Jy\,K$^{-1}$, which was determined
using Callisto and Mars during the science verification tests. The flux
uncertainty is estimated to be $\sim 25$\%.

\section{Results and analysis}\label{sec:results}

\subsection{ \herschel\ photometry}

\begin{figure*}
   \centering
    \begin{minipage}[b]{0.48\textwidth}
               \centering
         \includegraphics[width=1.0\textwidth]{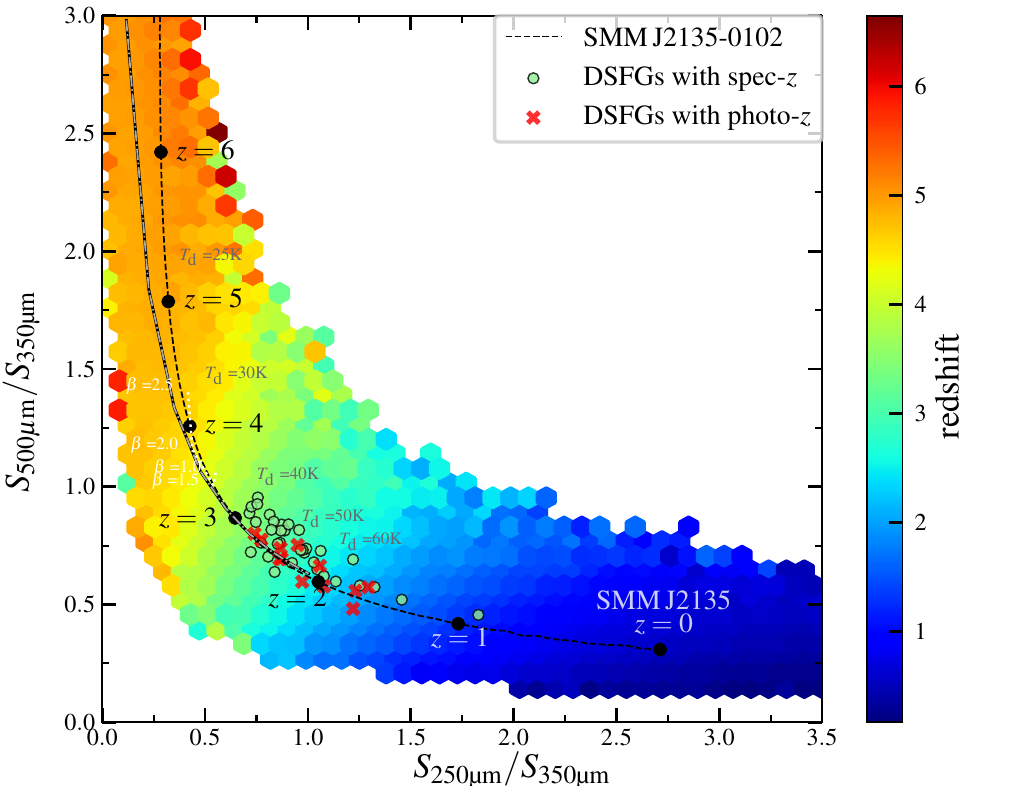} 
         \caption{\herschel\ SPIRE flux density ratios for the sample.  Sources
                 with known redshifts are displayed as green circles, those
                 without as red crosses.  Underlaid are a dashed line
                 displaying the expected flux density ratios of
                 SMM\,J2135$-$0102 with changing redshift, determined from the
                 SED presented in \citet{Ivison2010}, and shading displaying
                 the predicted colors of \(10^6\) modified blackbody sources
                 with a range of redshifts, dust temperatures, dust spectral
                 emissivity indices and a flux density measurement uncertainty
                 of 10\%, based on Figure~1 of \citet{Amblard2010AA}. White
                 dotted line indicates the color track of various $\beta$ at
                 \Td= 30 K. Gray dash dotted line indicates the color track of
                 various \Td, with $\beta$= 1.8. }
           \label{fig:colorcolor}
    \end{minipage}%
       \hspace{0.02 \textwidth}
    \begin{minipage}[b]{0.48\textwidth}
        \includegraphics[width=1.0\textwidth]{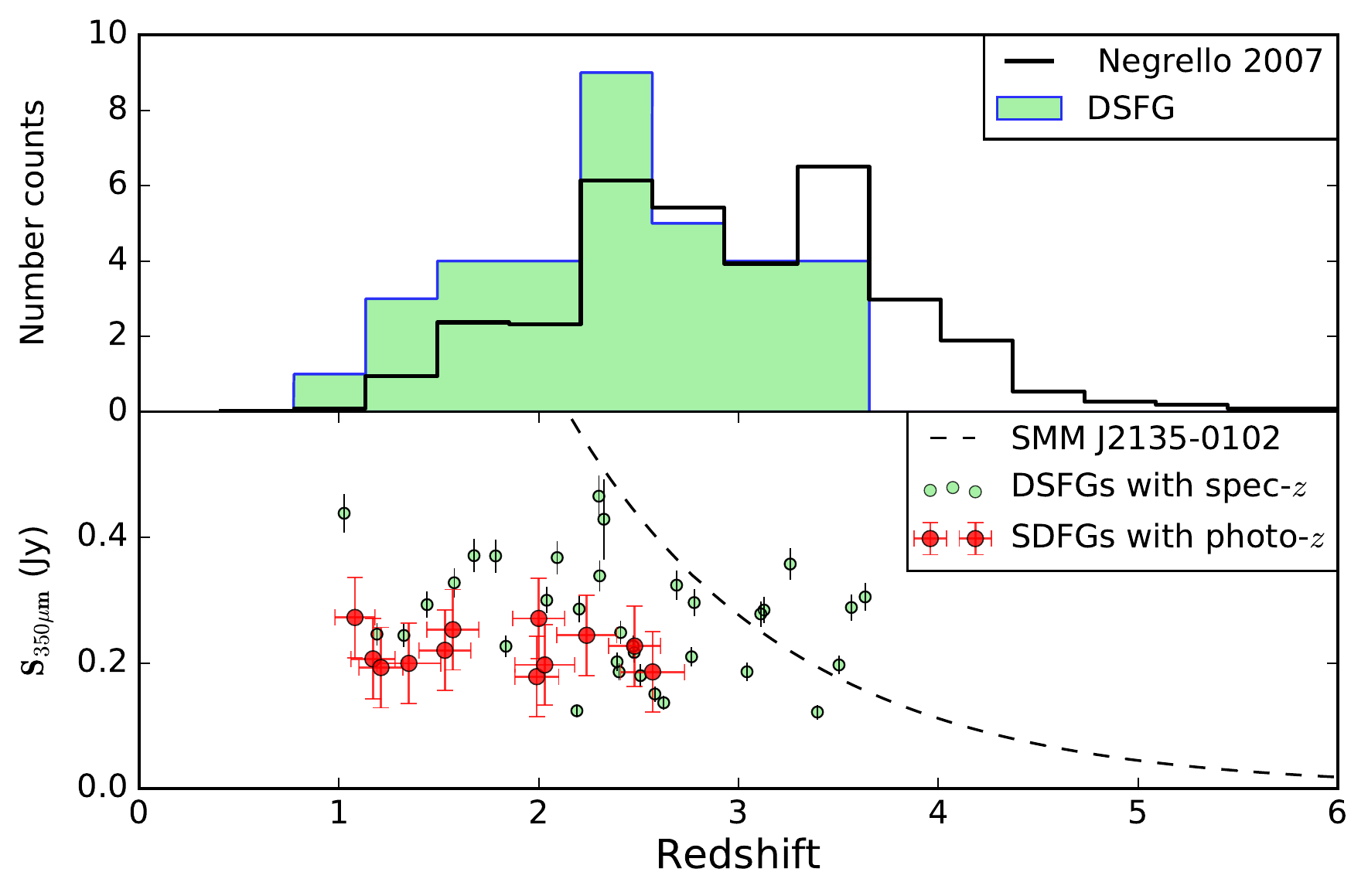} 
        \caption{ {\bf \it Upper:} Redshift distribution of sources in the
        sample. The shaded histogram displays the number of sources within our
        sample in redshift bins of \(\Delta z = 0.36\), and the area under the
        solid line displays the predicted redshift distribution of strongly
        lensed DSFGs with \(\{ S_{250} \text{, } S_{350} \text{, } S_{500} \} >
        100 \, \mJy \) from \citet{Negrello2007}, scaled to the same number of
        sources.\\
          {\bf \it Lower:} 
          \herschel\ SPIRE 350-\um\ flux densities as a function of the
          redshift distribution. Vertical lines show the uncertainties in flux
          density. Horizontal lines show uncertainties in photometric
          redshifts. The dashed line represents the 350-\um\ flux density
          expected to be observed from a SMM\,J2135$-$0102-like galaxy at
          different redshifts. }
        \label{fig:hisogram}
       \end{minipage}
\end{figure*}

Fig.~\ref{fig:colorcolor} presents a color-color plot of the \herschel{} SPIRE
flux densities of the sample. Following \cite{Amblard2010AA}, we generate $>2
\times 10^6$ SEDs of single-temperature modified blackbodies (MBB) and fill
their colors as the background. The MBBs are generated with a flux density
$F_\nu$:

\begin{equation}
        F_\nu = \varepsilon_\nu B_\nu \propto \frac{\nu^{3+\beta}}{ \exp(\frac{ h \nu }{k T_{\rm d}}) -1}\, 
\end{equation}  

\noindent
where $\varepsilon_\nu$ is the frequency-dependent emissivity, $\varepsilon_\nu
\propto \nu^{\beta}$, $T_{\rm d}$ is the dust temperature and $\beta$ is the
dust emissivity index.

To generate these models, we randomly sample a uniform range of \Td\ from 15 to
60\,K, of redshift from 0.1 to 7.0, of $\beta$ from 1.0 to 2.5. We limit the
plots with extreme color limits of $S_{500}/S_{350} > 3$ and $S_{250}/S_{350} >
3.5$.  For computing the colors, we also randomly add an extra noise to each
modeled data point with a Gaussian standard deviation of 10\% of its flux.
Then we bin the modeled data points with a hexagonal binning method
and plot the average colors in a hexagonal box. The SPIRE colors of the sample
are well within the limits defined by the models we have considered. We adopt
the SED template of SMM J2135$-$0102, shift it to different redshifts and
measure the `observed' $S_{500}/S_{350}$ and $S_{250}/S_{350}$ colors. Then we
plot the simulated colors as a dashed line, with their redshifts marked, to
compare with the observed values.  As displayed, the sample appears relatively
similar to the track of SMM\,J2135$-$0102 in the observed frame, although the
average colors of \(\left.\fd{250}\middle/\fd{350}\right.\) and
\(\left.\fd{350}\middle/\fd{500}\right.\) are a little higher than those of
SMM\,J2135$-$0102.  This is expected due to the very high magnification factor
experienced by SMM\,J2135$-$0102, which allows a less-extreme galaxy with a
relatively lower SFR surface density, and therefore a likely lower dust
temperature, to reach flux densities comparable to the others in our sample.
The distribution and the trend also seem to be consistent with those shown by
\citet{Yuan2015AA}, who use different templates to model the SPIRE color-color
plots in high-redshift DSFGs and show that higher \Td\ and/or smaller $\beta$
could produce the observed colors. We also present two tracks to demonstrate
the physical parameters that drive the spread of values in
Fig.\, \ref{fig:colorcolor}.  The white dotted line presents the color track for
\Td= 30 K, with $\beta$ varying from 1.0 to 2.5. The gray dash dotted line
presents the track for $\beta$=1.8, with \Td\ ranges from 20 K to 60 K. The
degeneracy between \Td\ and $\beta$ is clearly seen from the similarity of the
two tracks, whilst the temperature seems to be more sensitive in shifts along
the redshift axis.

Fig.~\ref{fig:hisogram} shows the spectroscopic redshift distribution of the
sample. The redshift distribution of strongly-lensed DSFGs predicted by
\citet{Negrello2007} has a higher mean value than observed in our sample.  This
is due partly to the different flux-density cuts employed, with our primarily
\(\fd{350} > 200\)-mJy subsample of the lens candidates preferentially
selecting lower redshift objects than the \(\fd{500} > 100\,\mJy\) used by
\citet{Negrello2007}. This is compounded by the methods and instruments used
to determine spectroscopic redshifts for our galaxy sample. Further details of
specific galaxies are given in Appendix~\ref{sec:comments}.

Galaxies at $z>$ 3 show higher flux densities than that scaled from Eyelash,
indicating that the \(z > 3\) galaxies in our sample are on average not only
brighter after lensing but much more intrinsically luminous than
SMM\,J2135$-$0102, which has the highest lensing amplification \(37.5 \pm 4.5\)
in the known sample.

\subsection{Dust SED modeling}\label{sec:seds}


\begin{figure*}
\includegraphics[]{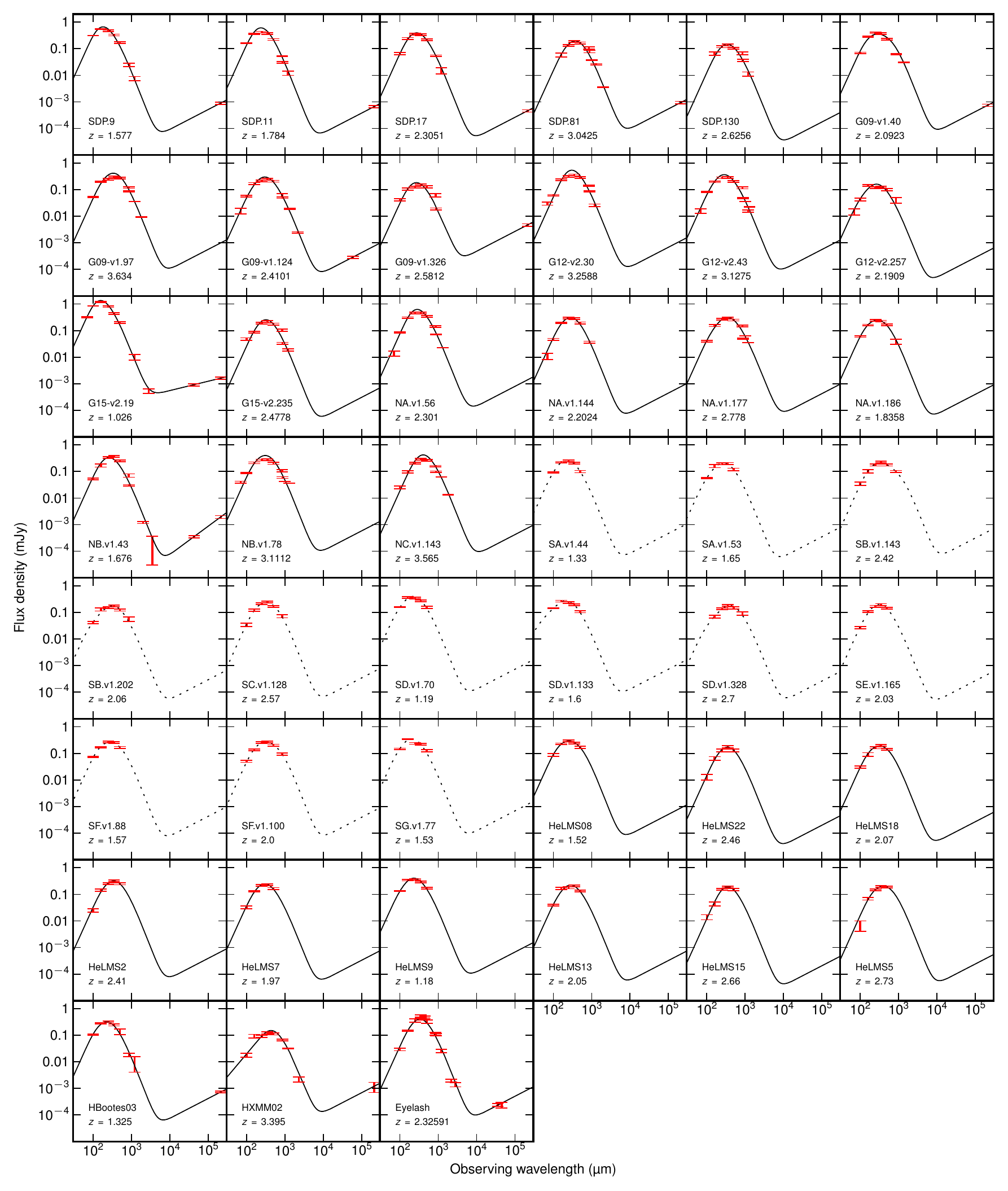}
\caption[Source SED fits]{ SED fits to available FIR--radio photometry for
        objects in the sample, using the power-law dust temperature
        distribution model of \citet{Kovacs2010}.  Sources selected from the
        \textit{H}-ATLAS SGP field do not have secure redshifts, and fits for
these sources are performed assuming a best-fit photometric redshift and are
shown as a dotted line.  }
\label{fig:sed_fits}
\end{figure*}

\begin{table*}
\centering
\begin{tabular}{
l          l                            l                            l                          l                           l                           l                           l                           l     }
      \hline
      \hline
Source      & \CII\ 158 \mum          & \NII\ 122 \mum             & \NII\ 205 \mum\            & \OIII\ 88 \mum\            & \OIII\ 52 \mum\           & \OI\ 63\mum\              & \OI\ 145 \mum\            & OH\ 119 \mum                     \\
            &\scriptsize $\rm 10^{-18}\,W\,m^{-2}$ &\scriptsize $\rm 10^{-18}\,W\,m^{-2}$ &\scriptsize $\rm 10^{-18}\,W\,m^{-2}$ &\scriptsize $\rm 10^{-18}\,W\,m^{-2}$ &\scriptsize $\rm 10^{-18}\,W\,m^{-2}$ &\scriptsize $\rm 10^{-18}\,W\,m^{-2}$ &\scriptsize $\rm 10^{-18}\,W\,m^{-2}$ &\scriptsize $\rm 10^{-18}\,W\,m^{-2}$    \\
\hline
SDP.9            & 3.5$\pm$0.6             & $<$2.8                   & $<$2.5                   & 4.8$\pm$1.5             & --                       & --                      & $<$1.5                   & $<$4.6         \\
SDP.11           & 6.1$\pm$0.4             & $<$1.5                   & $<$2.1                   & 4.9$\pm$1.0             & --                       & --                      & $<$1.4                   & $-$2.0$\pm$0.7   \\
SDP.17           & $<$2.1                  & $<$1.0                   & --                       & $<$2.88                 & --                       & $<$3.6                  & 2.1$\pm$0.5              & $<$1.4    \\
SDP.81           & 2.9$\pm$0.6             & $<$2.6                   & --                       & 2.9$\pm$0.5             & $<$3.7                   & 2.1$\pm$0.7             & $<$2.7                   & $<$2.3         \\
SDP.130          & $<$1.3                  & $<$2.6                   & --                       & $<$1.6                  & --                       & $<$3.9                  & $<$1.7                   & $<$1.4         \\
G09-v1.40        & 2.5$\pm$0.9             & $<$1.4                   & $<$2.1                   & $<$2.0                  & --                       & $<$6.7                  & $<$1.0                   & $<$1.3         \\
G09-v1.97        & --                      & $<$1.7                   & --                       & $<$1.1                  & 3.5$\pm$0.6              & $<$2.7                  & --                       & $<$2.2         \\
G09-v1.124       & 3.0$\pm$1.0             & $<$1.3                   & --                       & $<$3.0                  & --                       & 2.1$\pm$0.7             & $<$1.5                   & $<$1.1         \\
G09-v1.326       & $<$1.7                  & $<$1.2                   & --                       & $<$2.0                  & --                       & $<$4.3                  & $<$1.9                   & $<$0.8         \\
G12-v2.30        & --                      & $<$1.5                   & --                       & $<$1.1                  & $<$2.6                   & $<$2.3                  & $<$1.4                   & $<$1.7         \\
G12-v2.43        & $<$2.9                  & $<$2.1                   & --                       & $<$1.1                  & $<$4.3                   & $<$4.3                  & $<$2.8                   & $<$2.3         \\
G12-v2.257       & 3.4$\pm$0.6             & $<$1.1                   & --                       & $<$2.6                  & --                       & $<$4.0                  & $<$1.5                   & $<$1.1         \\
G15-v2.19        & 7.9$\pm$1.0             & $<$5.2                   & $<$1.8                   & --                      & --                       & --                      & $<$6.6                   & $<$8.0         \\
G15-v2.235       & $<$2.2                  & $<$1.5                   & --                       & $<$4.1                  & --                       & --                      & 3.5$\pm$1.2              & $<$1.4         \\
NA.v1.56         & $<$1.5                  & $<$0.7                   & --                       & 2.3$\pm$0.6             & --                       & $<$3.9                  & 3.0$\pm$0.6              & $<$0.9         \\
NA.v1.144        & $<$1.8                  & $<$1.2                   & --                       & $<$2.6                  & --                       & $<$3.1                  & $<$1.4                   & $<$1.4         \\
NA.v1.177        & $<$1.2                  & $<$0.8                   & --                       & $<$1.5                  & $<$5.1                   & $<$3.3                  & $<$1.5                   & $<$0.8         \\
NA.v1.186        & 4.2$\pm$0.4             & $<$1.7                   & $<$1.5                   & $<$2.0                  & --                       & --                      & $<$1.0                   & $<$1.7         \\
NB.v1.43         & 8.8$\pm$0.5             & $<$2.6                   & $<$2.6                   & $<$3.4                  & --                       & --                      & $<$1.2                   & $<$3.3         \\
NB.v1.78         & $<$2.5                  & $<$2.6                   & --                       & $<$1.6                  & $<$3.9                   & $<$3.1                  & $<$2.5                   & $<$2.2         \\
NC.v1.143        & --                      & $<$1.7                   & --                       & $<$1.3                  & $<$2.2                   & $<$1.7                  & --                       & $<$5.5          \\
{\it SA.v1.44 }  & $<$2.0                  & $<$3.0                   & $<$1.8                   & $<$5.3                  & --                       & --                      & $<$2.0                   & $<$4.0         \\
{\it SA.v1.53 }  & 3.7$\pm$0.4             & $<$2.8                   & $<$2.7                   & $<$5.1                  & --                       & --                      & $<$1.6                   & $<$2.8         \\
{\it SB.v1.143}  & 3.9$\pm$1.0             & $<$0.8                   & --                       & $<$4.4                  & --                       & $<$2.8                  & $<$1.6                   & $<$0.9         \\
{\it SB.v1.202}  & 3.1$\pm$0.6             & $<$1.0                   & $<$2.1                   & --                      & --                       & $<$9.2                  & $<$0.9                   & $<$0.8         \\
{\it SC.v1.128}  & 1.5$\pm$0.5             & $<$1.2                   & --                       & $<$3.2                  & --                       & $<$3.3                  & $<$2.5                   & $<$1.2         \\
{\it SD.v1.70 }  & $<$2.5                  & $<$2.4                   & $<$1.8                   & $<$2.2                  & --                       & --                      & $<$0.9                   & $<$1.2         \\
{\it SD.v1.133}  & 4.1$\pm$0.6             & $<$2.7                   & $<$2.1                   & $<$4.6                  & --                       & --                      & $<$1.6                   & $-$3.4$\pm$1.1   \\
{\it SD.v1.328}  & $<$1.2                  & $<$1.6                   & --                       & $<$1.6                  & --                       & $<$2.7                  & $<$1.4                   & $<$1.3         \\
{\it SE.v1.165}  & --                      & $<$1.7                   & --                       & $<$1.1                  & $<$3.6                   & $<$2.0                  & $<$1.3                   & $<$1.3         \\
{\it SF.v1.88 }  & $<$1.4                  & $<$1.4                   & $<$2.4                   & $<$3.1                  & --                       & --                      & $<$1.0                   & $<$1.3   \\
{\it SF.v1.100}  & $<$2.1                  & $<$0.9                   & --                       & $<$2.0                  & $<$5.1                   & $<$2.9                  & $<$2.0                   & $<$0.9         \\
{\it SG.v1.77 }  & $<$1.3                  & $<$1.5                   & --                       & $<$2.1                  & --                       & $<$3.8                  & $<$1.5                   & $<$1.3         \\
     HeLMS2      & $<$2.2                  & $<$0.6                   & --                       & $<$2.2                  & --                       & $<$2.3                  & $<$1.5                   & $<$0.9         \\
     HeLMS7      & $<$1.3                  & $<$1.0                   & --                       & $<$2.3                  & --                       & $<$4.2                  & $<$1.8                   & $<$0.9         \\
HBootes03        & 3.1$\pm$1.4             & $<$4.0                   & $<$3.3                   & 3.1$\pm$1.6             & --                       & --                      & --                       & $<$3.8         \\
HXMM02           & --                      & $<$1.6                   & --                       & $<$0.7                  & --                       & $<$2.0                  & $<$2.6                   & $<$1.8         \\
Eyelash          & 5.0$\pm$0.7             & 1.8$\pm$0.2              & --                       & --                      & --                       & --                      & $<$1.5                   & $-$2.8$\pm$0.4   \\
\hline
\end{tabular}
\caption{
Far-IR line fluxes of individual galaxies, measured by fitting Sinc-Gaussian
profiles to each line. The uncertainties are from the propagation of the
fitting error and from bootstrapping in the neighbouring $\pm$5000 \kms\
velocity range around the targeted lines, to get the `local' noise level.}
\label{tab:line_fluxes}
\end{table*}



As a first step towards understanding the physical properties of these
galaxies, we start by fitting their SEDs using broad-band continuum flux
densities from our multi-wavelength imaging observations. The SEDs are
constructed by combining our recently obtained PACS photometric data with the
250-, 350- and 500-\um\ photometric measurements obtained with SPIRE, the
850-\um\ flux densities measured with the Submillimetre Common-User Bolometer
Array 2 onboard JCMT \citep{Bakx2018,Holland2013}, the 880-\um\ flux densities
measured with SMA \citep{Bussmann2013}, the 1.2- and 2-mm data measured with
the Institut de Radioastronomie Millim\'{e}trique's NOrthern Extended
Millimeter Array \citep[NOEMA][]{Yang2016} and the 1.4-GHz radio continuum data
from the literature \citep{Becker1995}.  We list the measured FIR flux
densities at \herschel\ wavelengths in Table~\ref{tab:flux_densities}.

We first used a single-temperature MBB model to fit the
observed dust SEDs.  However, more than half of the sources could not be fitted
adequately with a single MBB, indicating that either multiple excitation
components are needed, or the assumption of a single-MBB dust emission does not
hold. Also, the dust emission is assumed to be optically thin, which may not be
valid for our extreme targets. The MBB fitting is also biased by the data in
the Wien regime, meaning that the luminosities are systematically
under-estimated.  On the other hand, there are not enough data points at the
longer wavelengths to fit two independent MBB models for most of the sources.

Instead, we estimate the dust properties with a power-law temperature
distribution method introduced by \citet{Kovacs2010}.  We model the FIR SEDs
with dust emission following a thermally-motivated power-law distribution of
dust masses, \(M_\text{d}\), with temperature components \(T\):
\(\frac{\text{d}\,M_\text{d}}{\text{d} T} \propto T^{-\gamma}\), and a
low-temperature cutoff. This model does not assume optically thin dust emission
everywhere and this simple prescription can reproduce both the Wien and
Rayleigh-Jeans sides of the FIR peak.  For consistency, all sources are
modelled using this method, regardless of previous independent
determinations of their SEDs.
              
\begin{figure*}
\includegraphics[width=0.49\textwidth,trim=0bp 0bp 0bp 0bp ,clip]{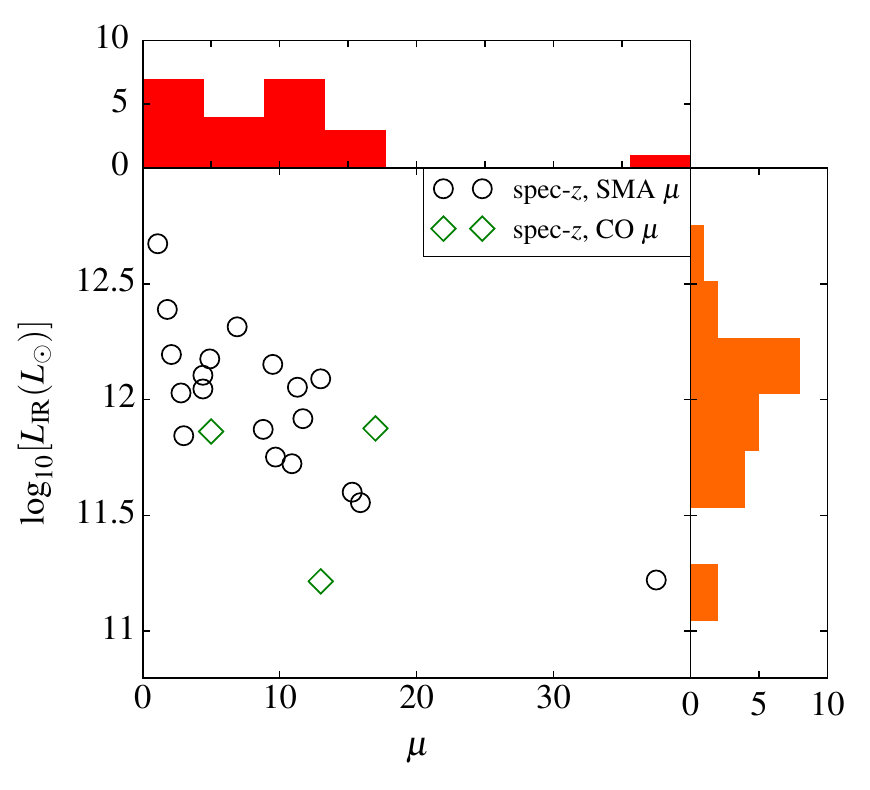}
\includegraphics[width=0.49\textwidth,trim=0bp 0bp 0bp 0bp ,clip]{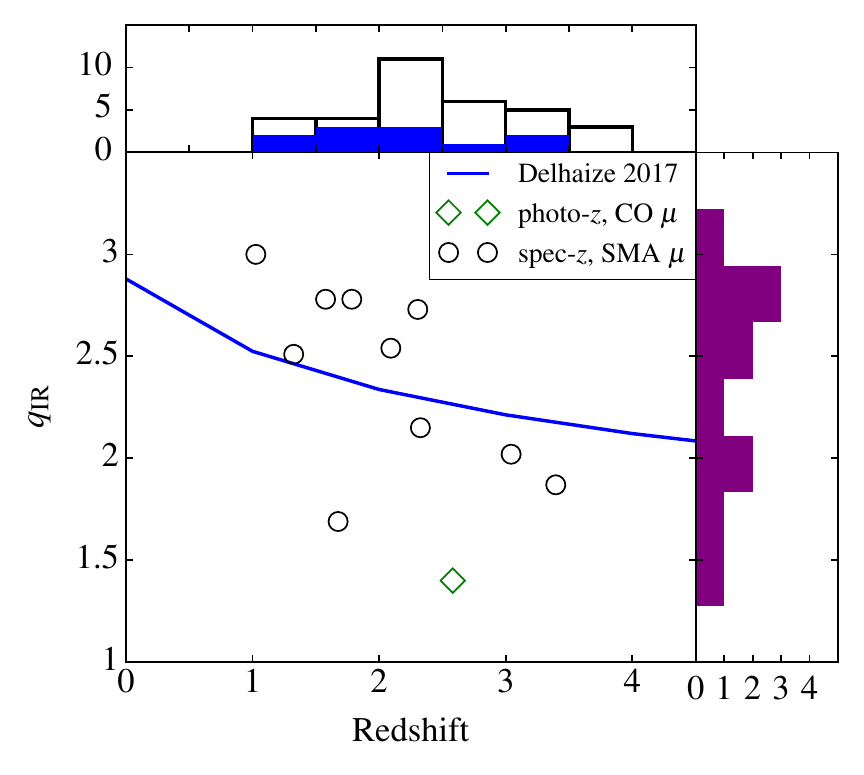}
\includegraphics[width=0.49\textwidth,trim=0bp 0bp 0bp 0bp ,clip]{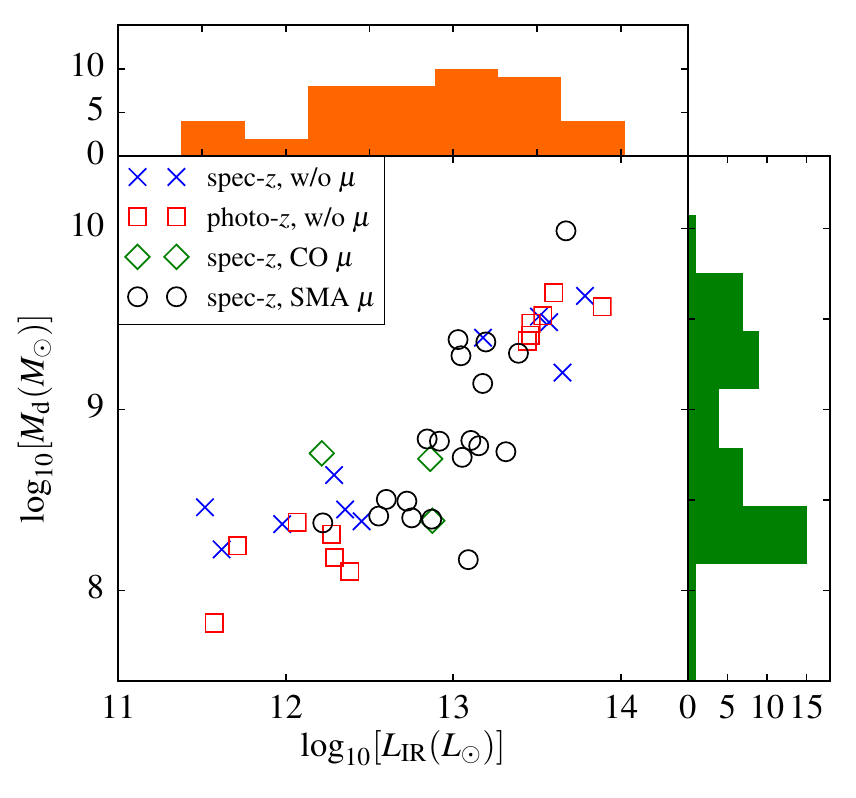}
\includegraphics[width=0.49\textwidth,trim=0bp 0bp 0bp 0bp ,clip]{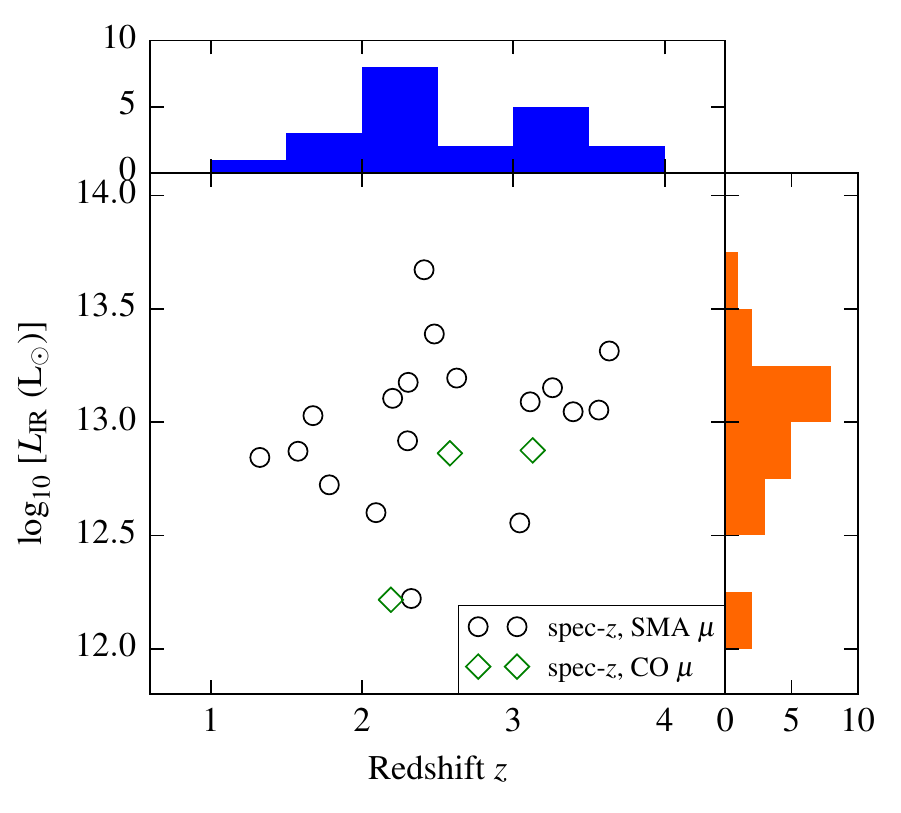}
\caption{Distributions and correlations of SED fit parameters. 
{\it Upper left:} 
The distributions of gravitational amplification factor $\mu$ and the total
infrared luminosity, \LIR\ (8-1000 $\mu$m). 
{\it Upper right:} 
The distribution of redshift and the total infrared to radio correlation factor
$q_{\rm IR}$. 
The blue line shows the infrared-radio correlation fitted
in star-forming galaxies in the COSMOS field \citep[i.e.][]{Delhaize2017}:
$q_{\rm TIR}(z) = (2.88 \pm 0.03)(1 + z)^{- 0.19 \pm 0.01}$.  
The open histogram (black) shows the redshift  distribution of the whole sample. 
{\it Lower left:} 
The distributions of total infrared luminosity and the fitted dust mass $M_{\rm
d}$, after lensing amplification corrections.  For targets without
amplification factors (blue cross and red squares), we have divided by a factor
of five for both axes, in order to compare with the de-lensed results. 
{\it Lower right:} 
The distributions of redshift and the total infrared luminosity after lensing
correction. Filled histograms (in color) show the distribution of the sample.
Open
circles represent systems with known redshifts and lens models, primarily from
\citet{Bussmann2013}.  Green diamonds represent targets with known redshifts,
but with estimates of gravitational amplification factors only from their CO
line characteristics \citep{Harris2012}. Blue crosses represent sources with
known redshifts, with no existing lens model. Red squares represent sources
with photometric redshifts, as shown in Table~\ref{tab:flux_densities}.
} 
\label{fig:sed_fit_properties}
\end{figure*}

For the SED modelling for the high-redshift DSFGs, we fit the black-body model
and a power-law synchrotron emission component simultaneously. Thermal
free-free emission is unlikely to contaminate strongly in our fitting
wavelengths \citep[e.g.,][]{Aravena2013}, so we excluded it from the fitting.
The power-law index, \(\gamma\), of the dust is fixed to be 7.2, the
best-fitting value found in local starbursts \citep[see][]{Kovacs2010}; the
dust spectral emissivity index, \(\beta\), is fixed at 1.8, the true value of
which likely varies inversely with temperature \citep{Knapp1993, Mennella1998,
Dunne2000}; the characteristic photon cross-section to mass ratio of particles
$\kappa_0$ is assumed to be $\kappa_{850 \mu m} = 0.15\, \rm
m^2\,kg^{-1}$ \citep[see][]{Kovacs2010}; the conversion from \LIR\ to star
formation rate follows \citet{Kennicutt2012} with a Kroupa initial mass
function (IMF) \citep{Kropua2003}; the synchrotron spectral index, $\alpha$, is
fixed as $-0.75$ for objects with less than two photometric measurements at
\(\lambda_\text{rest} > 4\,\text{mm}\) \citep[e.g.][]{Ibar2010b}.

Total infrared luminosities and dust masses are comparable to those estimated
for these galaxies via other methods, such as \textsc{magphys}
\citep{Negrello2014}, or a single grey-body component \citep{Bussmann2013}.
However, the fitted SFRs are higher than those suggested by \textsc{magphys},
which is likely due to the different adopted calibrations. To compare the total IR
luminosity ($S_{\rm IR}^{8-1000\mu \rm m}$) with the radio continuum
observations, we also calculate $q_{\rm IR}$, following
\citet{Ivison2010MNRAS,Ivison2010AA}, as $q_{\rm IR} = {\rm log}_{10} [(S_{\rm
IR} / 3.75 \times 10^{12} {\rm W\, m^{-2}}) / (S_{\rm 1.4 GHz}/ {\rm W\, m^{
-2} Hz^{-1}})]$, 
where
$S_{\rm 1.4 GHz} $ is $K$-corrected assuming
$S_{\nu} \propto \nu^{\alpha}$, again with $\alpha = -0.75$.
Finally, we list the derived properties, such as \LIR, SFR, \Md\ and
\Td$^{\rm cutoff}$ in Table~\ref{tab:sed_fit_properties}.

In Fig.~\ref{fig:sed_fit_properties} we plot correlations and distributions for
a few of the parameters we have fitted. It is not surprising that the
gravitational amplification factor, $\mu$, is weakly anti-correlated with the
intrinsic infrared luminosity, \(\lum{IR}\),  after correcting the lensing
magnification.  This is because correspondingly higher \(\mu \lum{IR}\) is
required for higher-redshift sources to reach our \(\fd{350}{\um} >
200\,\text{mJy}\) flux-density threshold.

We find a tentative trend of decreasing $q_{\rm IR}$ with increasing redshift,
indicating a potential variation of the IR-radio correlation with redshift.
This is consistent with the decreasing trend found recently in the 3-GHz survey
in the COSMOS field \citep{Delhaize2017}. However our results indicate much
shallower slope index, whose significance is very limited by the current sample
size. 

The total infrared luminosity \lir\ shows an good increasing correlation with
fitted dust mass, $M_{\rm d}$, over two orders of magnitude, which is also
expected since all targets in the sample show similar dust temperatures and
far-IR colors. The correlation shown in the targets with lensing correction
seems consistent with that shown in targets without lensing corrections. 

The median redshift (including both photo$-z$ and spec$-z$) for the whole
sample is 2.5$\pm 0.6$, with a redshift range spanning $z=1.0$--3.6. The
intrinsic \lir\ spans from $10^{12}$ to $10^{13.5}$ \Lsun, with a median of
$\sim 10^{13}$ \Lsun, indicating their starburst natures. 

In the end, we find that the redshift does not show significant correlation
with the intrinsic \lir, indicating that our sample is not potentially
biased to more luminous targets at higher redshifts, although they are selected
with flux cutoffs. 

\subsection{Magnification factors}

Gravitational amplification values are primarily taken from
\citet{Bussmann2013} that are derived from SMA data in preference to values
derived from Near-IR imaging \citep[e.g][]{Calanog2014,Dye2014}, due to the
likelihood of a physical separation (and hence amplification difference) of the
starburst and existing stellar population \citep{Fu2012,
Dye2015,Hodge2016,Oteo2017}.  Most galaxies in the sample are strongly lensed,
with a median amplification of around $10\times$, distributed between
1--$15\times$. SMM\,J2135$-$0102 is an outlier with its very high
amplification factor.

When fitting dust SED and measuring lines ratios, we neglect the possibility of
differential lensing at different wavelengths, which may bring in extra
uncertainties \citep[e.g.][]{Hezaveh2012,Serjeant2012,Yang2017}. The lensed
status of this sample could complicate the modelling of their SEDs,
particularly the low-temperature dust components which may not be co-spatial
with hotter dust near the starburst.  This similar effect could also be
important for optical and near-IR flux density measurements and modeling
\citep[e.g.][]{Negrello2014, Ma2015}. 

The different origins of the fine-structure ionised lines make differential
lensing more severe. Resolved observations have shown that the \nii\ emission
is more extended than the \oiii\ emission in local galaxies
\citep[e.g.,][]{Hughes2015}. Besides, \cii\ originates from multiple gas
phases, making it more extended than the \nii\ emission, which has been seen in
the Milky Way galaxy, local galaxies and galaxies at high redshift
\citep[e.g.,][]{Goldsmith2015,Lapham2017,Pavesi2016}.  However, the line ratios
of the same species at the same energy level, e.g., the two \nii\ lines and the
two \oiii\ lines, are expected to be not as severe as those of species from
different origins.  Detailed studies on the differential lensing effect need
very high-quality high-resolution images at all studies wavelengths, and are
highly dependent on the model interpretation, which is beyond the scope of our
study.

\begin{table*}
\centering
\begin{tabular}{l r@{\,}l c r@{\,}l r@{\,}l r@{\,}l r@{\,}l r@{\,}l r@{\,}l}
\hline
\hline
Source &\multicolumn{2}{c}{Amplification}& \(\mu \mass{H\(_2\)}\) & \multicolumn{2}{c}{\(\mu\)\lum{IR}}   & \multicolumn{2}{c}{\lum{IR}}             & \multicolumn{2}{c}{SFR}                  & \multicolumn{2}{c}{\(\mass{d}^\text{tot}\)}         & \multicolumn{2}{c}{\(T_\text{d}^\text{cutoff}\)} & \multicolumn{2}{c}{\(q_\text{\,IR}\)}  \\
       &                                 &                        &   \(10 ^ {11}\,\msun\)                & \multicolumn{2}{c}{  \(10^{12}\,\lsun\)} & \multicolumn{2}{c}{  \(10^{12}\,\lsun\)} & \multicolumn{2}{c}{  \(\msun\,\text{yr}^{-1}\)} & \multicolumn{2}{c}{  \(10^{8}\,\msun\)}   & \multicolumn{2}{c}{  K}                   &                               \\
    \hline
SDP.9            &  8.8 & \(\pm\, 2.2\)          & 1.9 \(\pm\, 0.5\)   &  71.5 & \(\pm\, 2.4\) &      8.1 & \(\pm\, 1.9\) &   1200 & \(\pm\, 300\) &      2.2 & \(\pm\,  1.0\) &     36.1 & \(\pm\, 3.4\) &     2.78 & \(\pm\, 0.10\) \\
SDP.11           & 10.9 & \(\pm\, 1.3\)          & 2.4 \(\pm\, 0.3\)   &  68.9 & \(\pm\, 2.3\) &      6.3 & \(\pm\, 0.7\) &   900  & \(\pm\, 100\) &      3.8 & \(\pm\,  0.9\) &     34.9 & \(\pm\, 1.6\) &     2.78 & \(\pm\, 0.10\) \\
SDP.17           &  4.9 & \(\pm\, 0.7\)          & 4.3 \(\pm\, 0.6\)   &  76.7 & \(\pm\, 3.1\) &     15.7 & \(\pm\, 2.2\) &   2300 & \(\pm\, 300\) &     13.5 & \(\pm\,  4.2\) &     27.8 & \(\pm\, 1.7\) &     2.73 & \(\pm\, 0.12\) \\
SDP.81           & 15.9 & \(\pm\, 0.7\)          & 3.8 \(\pm\, 0.2\)   &  58.0 & \(\pm\, 4.0\) &      3.7 & \(\pm\, 0.3\) &   550  & \(\pm\,  50\) &      2.6 & \(\pm\,  0.5\) &     38.9 & \(\pm\, 3.8\) &     2.02 & \(\pm\, 0.13\) \\
SDP.130          &  2.1 & \(\pm\, 0.3\)          & 1.7 \(\pm\, 0.2\)   &  31.8 & \(\pm\, 2.8\) &     15.1 & \(\pm\, 2.5\) &   2300 & \(\pm\, 400\) &     29.6 & \(\pm\,  6.6\) &     24.1 & \(\pm\, 1.4\) &          & \(-\)          \\
G09-v1.40        & 15.3 & \(\pm\, 3.5\)          & 3.3 \(\pm\, 0.8\)   &  65.1 & \(\pm\, 2.1\) &      4.3 & \(\pm\, 0.9\) &   600  & \(\pm\, 130\) &      7.4 & \(\pm\,  1.0\) &     43.5 & \(\pm\, 2.9\) &     2.54 & \(\pm\, 0.11\) \\
G09-v1.97        &  6.9 & \(\pm\, 0.6\)          & 5.0 \(\pm\, 0.4\)   & 212.0 & \(\pm\, 6.0\) &     30.8 & \(\pm\, 2.0\) &   4600 & \(\pm\, 300\) &      5.3 & \(\pm\,  1.9\) &     50.7 & \(\pm\, 1.6\) &          & \(-\)          \\
G09-v1.124       &  1.1 & \(\pm\, 0.1\)          & 2.2 \(\pm\, 0.2\)   &  64.5 & \(\pm\, 4.9\) &     58.6 & \(\pm\, 5.9\) &   8800 & \(\pm\, 900\) &     82.0 & \(\pm\, 12.6\) &     24.4 & \(\pm\, 1.6\) &     2.40 & \(\pm\, 0.22\) \\
G09-v1.326       &  5.0 & \(\pm\, 1.0^\text{H}\) & 2.7 \(\pm\, 0.5\)   &  47.2 & \(\pm\, 1.9\) &      9.4 & \(\pm\, 1.5\) &   1400 & \(\pm\, 200\) &      4.5 & \(\pm\,  1.7\) &     30.8 & \(\pm\, 4.4\) &     1.40 & \(\pm\, 0.12\) \\
G12-v2.30        &  9.5 & \(\pm\, 0.6\)          & 7.1 \(\pm\, 0.5\)   & 202.0 & \(\pm\, 7.0\) &     21.3 & \(\pm\, 1.2\) &   3200 & \(\pm\, 170\) &      5.6 & \(\pm\,  1.8\) &     37.3 & \(\pm\, 1.2\) &          & \(-\)          \\
G12-v2.43        & 17.0 & \(\pm\,11.0^\text{H}\) & 1.3 \(\pm\, 0.8\)   & 150.0 & \(\pm\, 3.8\) &      8.8 & \(\pm\, 4.8\) &   1300 & \(\pm\, 700\) &      2.8 & \(\pm\,  1.6\) &     31.7 & \(\pm\, 3.0\) &          & \(-\)          \\
G12-v2.257       & 13.0 & \(\pm\, 7.0^\text{H}\) & 1.4 \(\pm\, 0.8\)   &  34.0 & \(\pm\, 1.2\) &      2.6 & \(\pm\, 0.9\) &   400  & \(\pm\, 100\) &      2.7 & \(\pm\,  3.7\) &     24.8 & \(\pm\, 2.3\) &          & \(-\)          \\
G15-v2.19        &  9.0 & \(\pm\, 1.0\)          & 2.9 \(\pm\, 0.3\)   &  58.1 & \(\pm\, 2.0\) &      6.0 & \(\pm\, 0.5\) &   900  & \(\pm\, 100\) &      2.4 & \(\pm\,  0.8\) &     35.2 & \(\pm\, 1.9\) &     3.0. & \(\pm\, 0.07\) \\
G15-v2.235       &  1.8 & \(\pm\, 0.3\)          & 3.5 \(\pm\, 0.6\)   &  49.8 & \(\pm\, 1.8\) &     27.7 & \(\pm\, 4.2\) &   4100 & \(\pm\, 600\) &     23.3 & \(\pm\,  6.0\) &     35.3 & \(\pm\, 1.8\) &          & \(-\)          \\
NA.v1.56         & 11.7 & \(\pm\, 0.9\)          & 5.8 \(\pm\, 0.4\)   & 109.0 & \(\pm\, 4.4\) &      9.3 & \(\pm\, 0.8\) &   1400 & \(\pm\, 110\) &      7.3 & \(\pm\,  1.8\) &     31.1 & \(\pm\, 1.7\) &          & \(-\)          \\
NA.v1.144        &  4.4 & \(\pm\, 0.8\)          & 1.8 \(\pm\, 0.3\)   &  54.8 & \(\pm\, 2.1\) &     12.5 & \(\pm\, 2.4\) &   1900 & \(\pm\, 300\) &      6.9 & \(\pm\,  2.2\) &     40.1 & \(\pm\, 3.7\) &          & \(-\)          \\
NA.v1.177        &      & \(-\)                  & 4.2                 &  94.1 & \(\pm\, 3.0\) &     18.8 & \(\pm\, 2.8\) &   2800 & \(\pm\, 600\) &      1.6 & \(\pm\,  0.4\) &     34.7 & \(\pm\, 3.2\) &          & \(-\)          \\
NA.v1.186        &      & \(-\)                  & 3.6                 &  32.7 & \(\pm\, 1.3\) &      6.5 & \(\pm\, 1.4\) &   1000 & \(\pm\, 200\) &      7.4 & \(\pm\,  2.1\) &     28.8 & \(\pm\, 2.2\) &          & \(-\)          \\
NB.v1.43         &  2.8 & \(\pm\, 0.4\)          & 3.6 \(\pm\, 0.5\)   &  31.2 & \(\pm\, 1.3\) &     11.1 & \(\pm\, 1.6\) &   1700 & \(\pm\, 400\) &     24.7 & \(\pm\,  4.9\) &     23.8 & \(\pm\, 1.2\) &     1.67 & \(\pm\, 0.09\) \\
NB.v1.78         & 13.0 & \(\pm\, 1.5\)          & 4.6 \(\pm\, 0.5\)   & 156.0 & \(\pm\, 7.2\) &     12.0 & \(\pm\, 1.4\) &   1800 & \(\pm\, 400\) &      2.5 & \(\pm\,  0.5\) &     49.7 & \(\pm\, 4.4\) &          & \(-\)          \\
NC.v1.143        & 11.3 & \(\pm\, 1.7\)          & 6.2 \(\pm\, 0.9\)   & 153.0 & \(\pm\, 6.1\) &     13.5 & \(\pm\, 1.9\) &   2000 & \(\pm\, 300\) &      9.6 & \(\pm\,  2.3\) &     35.2 & \(\pm\, 3.7\) &          & \(-\)          \\
{\it SA.v1.44 }  &      & \(-\)                  & \(-\)               &  18.2 & \(\pm\, 0.9\) &      3.6 & \(\pm\, 0.8\) &   540  & \(\pm\, 120\) &      5.5 & \(\pm\,  1.7\) &     26.1 & \(\pm\, 2.1\) &          & \(-\)          \\
{\it SA.v1.53 }  &      & \(-\)                  & \(-\)               &  22.9 & \(\pm\, 0.8\) &      4.6 & \(\pm\, 0.8\) &   680  & \(\pm\,  90\) &      5.7 & \(\pm\,  1.2\) &     27.3 & \(\pm\, 1.9\) &          & \(-\)          \\
{\it SB.v1.143}  &      & \(-\)                  & \(-\)               &  55.5 & \(\pm\, 3.2\) &     11.1 & \(\pm\, 1.5\) &   1700 & \(\pm\, 250\) &     14.6 & \(\pm\,  4.3\) &     30.9 & \(\pm\, 3.7\) &          & \(-\)          \\
{\it SB.v1.202}  &      & \(-\)                  & \(-\)               &  35.1 & \(\pm\, 3.1\) &      7.0 & \(\pm\, 2.5\) &   1000 & \(\pm\, 200\) &      5.5 & \(\pm\,  2.1\) &     29.9 & \(\pm\, 4.5\) &          & \(-\)          \\
{\it SC.v1.128}  &      & \(-\)                  & \(-\)               &  54.4 & \(\pm\, 2.4\) &     10.9 & \(\pm\, 1.3\) &   1600 & \(\pm\, 230\) &      9.1 & \(\pm\,  2.3\) &     31.7 & \(\pm\, 3.2\) &          & \(-\)          \\
{\it SD.v1.70 }  &      & \(-\)                  & \(-\)               &  23.5 & \(\pm\, 2.8\) &      4.7 & \(\pm\, 1.1\) &   700  & \(\pm\, 200\) &      6.7 & \(\pm\,  1.1\) &     27.0 & \(\pm\, 2.6\) &          & \(-\)          \\
{\it SD.v1.133}  &      & \(-\)                  & \(-\)               &  25.3 & \(\pm\, 3.9\) &      5.1 & \(\pm\, 0.4\) &   760  & \(\pm\, 200\) &      3.8 & \(\pm\,  0.7\) &     28.9 & \(\pm\, 2.0\) &          & \(-\)          \\
{\it SD.v1.328}  &      & \(-\)                  & \(-\)               &  42.4 & \(\pm\, 2.7\) &      8.5 & \(\pm\, 1.3\) &   1300 & \(\pm\, 240\) &     13.1 & \(\pm\,  1.5\) &     29.3 & \(\pm\, 3.4\) &          & \(-\)          \\
{\it SE.v1.165}  &      & \(-\)                  & \(-\)               &  37.2 & \(\pm\, 3.3\) &      7.4 & \(\pm\, 1.7\) &   1100 & \(\pm\, 300\) &      7.3 & \(\pm\,  0.9\) &     29.6 & \(\pm\, 2.9\) &          & \(-\)          \\
{\it SF.v1.88 }  &      & \(-\)                  & \(-\)               &  34.8 & \(\pm\, 1.4\) &      7.0 & \(\pm\, 1.3\) &   1000 & \(\pm\, 300\) &      7.0 & \(\pm\,  1.3\) &     29.3 & \(\pm\, 2.3\) &          & \(-\)          \\
{\it SF.v1.100}  &      & \(-\)                  & \(-\)               &  51.2 & \(\pm\, 2.8\) &     10.2 & \(\pm\, 2.4\) &   1500 & \(\pm\, 400\) &     12.4 & \(\pm\,  2.6\) &     30.6 & \(\pm\, 2.8\) &          & \(-\)          \\
{\it SG.v1.77 }  &      & \(-\)                  & \(-\)               &  49.4 & \(\pm\, 1.7\) &      9.9 & \(\pm\, 1.3\) &   1500 & \(\pm\, 400\) &      3.3 & \(\pm\,  0.8\) &     33.6 & \(\pm\, 2.4\) &          & \(-\)          \\
 HeLMS08         &      & \(-\)                  & \(-\)               &  14.1 & \(\pm\, 0.6\) &      2.8 & \(\pm\, 0.3\) &   420  & \(\pm\, 100\) &     10.7 & \(\pm\,  2.4\) &     23.3 & \(\pm\, 1.9\) &          & \(-\)          \\
 HeLMS22         &      & \(-\)                  & \(-\)               &  32.6 & \(\pm\, 2.2\) &      6.5 & \(\pm\, 1.1\) &   970  & \(\pm\, 180\) &      9.1 & \(\pm\,  1.8\) &     28.3 & \(\pm\, 3.8\) &          & \(-\)          \\
 HeLMS18         &      & \(-\)                  & \(-\)               &  40.9 & \(\pm\, 1.8\) &      8.2 & \(\pm\, 1.2\) &   1200 & \(\pm\, 200\) &      6.5 & \(\pm\,  1.4\) &     30.5 & \(\pm\, 3.0\) &          & \(-\)          \\
 HeLMS2          &      & \(-\)                  & \(-\)               &  75.3 & \(\pm\, 4.4\) &     15.1 & \(\pm\, 2.0\) &   2300 & \(\pm\, 300\) &     11.5 & \(\pm\,  1.9\) &     33.5 & \(\pm\, 3.0\) &          & \(-\)          \\
 HeLMS7          &      & \(-\)                  & \(-\)               &  53.9 & \(\pm\, 1.9\) &     10.8 & \(\pm\, 2.5\) &   1600 & \(\pm\, 340\) &      6.8 & \(\pm\,  1.3\) &     32.3 & \(\pm\, 2.8\) &          & \(-\)          \\
 HeLMS9          &      & \(-\)                  & \(-\)               &  33.1 & \(\pm\, 1.0\) &      6.6 & \(\pm\, 1.1\) &   1000 & \(\pm\, 290\) &      6.8 & \(\pm\,  1.3\) &     29.1 & \(\pm\, 2.0\) &          & \(-\)          \\
 HeLMS13         &      & \(-\)                  & \(-\)               &  70.7 & \(\pm\, 2.6\) &     14.1 & \(\pm\, 2.1\) &   2100 & \(\pm\, 300\) &      3.8 & \(\pm\,  1.0\) &     27.2 & \(\pm\, 2.9\) &          & \(-\)          \\
 HeLMS15         &      & \(-\)                  & \(-\)               &  30.1 & \(\pm\, 1.6\) &      6.0 & \(\pm\, 9.6\) &   900  & \(\pm\, 140\) &     13.0 & \(\pm\,  3.4\) &     35.1 & \(\pm\, 4.1\) &          & \(-\)          \\
 HeLMS5          &      & \(-\)                  & \(-\)               &  82.3 & \(\pm\, 7.5\) &     16.5 & \(\pm\, 3.3\) &   2500 & \(\pm\, 500\) &      7.5 & \(\pm\,  2.0\) &     37.9 & \(\pm\, 3.3\) &          & \(-\)          \\
HBo\"otes03      & 3.0  & \(\pm\, 1.5\)          & 1.0 \(\pm\, 0.5\)   &  20.8 & \(\pm\, 0.9\) &      6.9 & \(\pm\, 1.4\) &   1000 & \(\pm\, 200\) &      6.9 & \(\pm\,  1.9\) &     23.8 & \(\pm\, 6.2\) &     2.51 & \(\pm\, 0.09\) \\
HXMM02           & 5.33 & \(\pm\, 0.19^\text{B}\)& 3.4 \(\pm\, 0.8\)   &  66.2 & \(\pm\, 6.5\) &     15.1 & \(\pm\, 3.3\) &   2300 & \(\pm\, 500\) &     24.1 & \(\pm\,  7.0\) &     23.0 & \(\pm\, 2.3\) &     1.87 & \(\pm\, 0.22\) \\
Eyelash          & 37.5 & \(\pm\, 4.5\)          & 5.6 \(\pm\, 0.7\)   &  62.3 & \(\pm\, 2.9\) &      1.7 & \(\pm\, 0.2\) &    250 & \(\pm\,  30\) &      2.4 & \(\pm\,  0.6\) &     33.3 & \(\pm\, 3.4\) &     2.15 & \(\pm\, 0.26\) \\
\hline
\end{tabular}
\caption{
    Properties from the SED fits described in Section \ref{sec:seds}.
    In addition, \(\text{H}_2\) masses from CO measurements are shown,
    taken from
    \citet{2006PASJ...58..957I, 2012PASJ...64L...2I,
      2011ApJ...726L..22F, Fu2012, Harris2012,
      2012ApJ...757..135L, Danielson2013, Ivison2013,
      Messias2014}
      These were converted to $L^\prime_{\rm CO 1-0}$ where necessary
    by the brightness temperature ratios given in
    \citet{Bothwell2013}, and then to a molecular gas mass via
    an $\alpha_{\rm CO}$ conversion factor of \(0.8\,\xunits\).
    \(^\text{H}\)~Amplification estimate from CO line luminosity
      and FWHM from \citet{Harris2012}.\newline
    \(^\text{B}\)~Amplification estimate from ALMA 870-\mum\ data \citep{Bussmann2015}.
} \label{tab:sed_fit_properties}
\end{table*}


\section{Spectroscopy results} \label{sec:spectroscopy}

\herschel\ spectra of all our targets are presented in
Appendix~\ref{app:All_spectra}. In galaxies with known spectroscopic redshifts,
we report individual detections of nine \CII\ 158-\mum\ emission lines, four
\OIII\ 88-\mum\ lines, three \OI\ 145-\mum\ lines, two \OI-63\mum\ lines,
one \OIII-52-\mum\ line, one \NII-122-\mum\ line, and one OH 119-\mum\
line in absorption. For galaxies without  spectroscopic redshifts, we 
estimate their photometric redshift based on the FIR-based photometric
redshifts fitting method \citep[for details see][]{Ivison2016}, which fits
FIR templates of different high-$z$ galaxies. We adopt the best fits using
the ALESS template and add the difference between different templates to the
final error. We then search for lines within a range of $z_{\rm phot}\pm0.5$
($\sim 3 \sigma$), and find five possible \CII\ emission lines at $z_{\rm
phot}\pm0.5$ (one of these, SD.v1.133, with a possible OH 119-\mum\ absorption
feature), thus yielding five plausible new spectroscopic redshifts.  These
redshifts need to be confirmed with follow-up observations before they are
considered robust.

\begin{figure}
\centering
\includegraphics[width=0.5\textwidth]{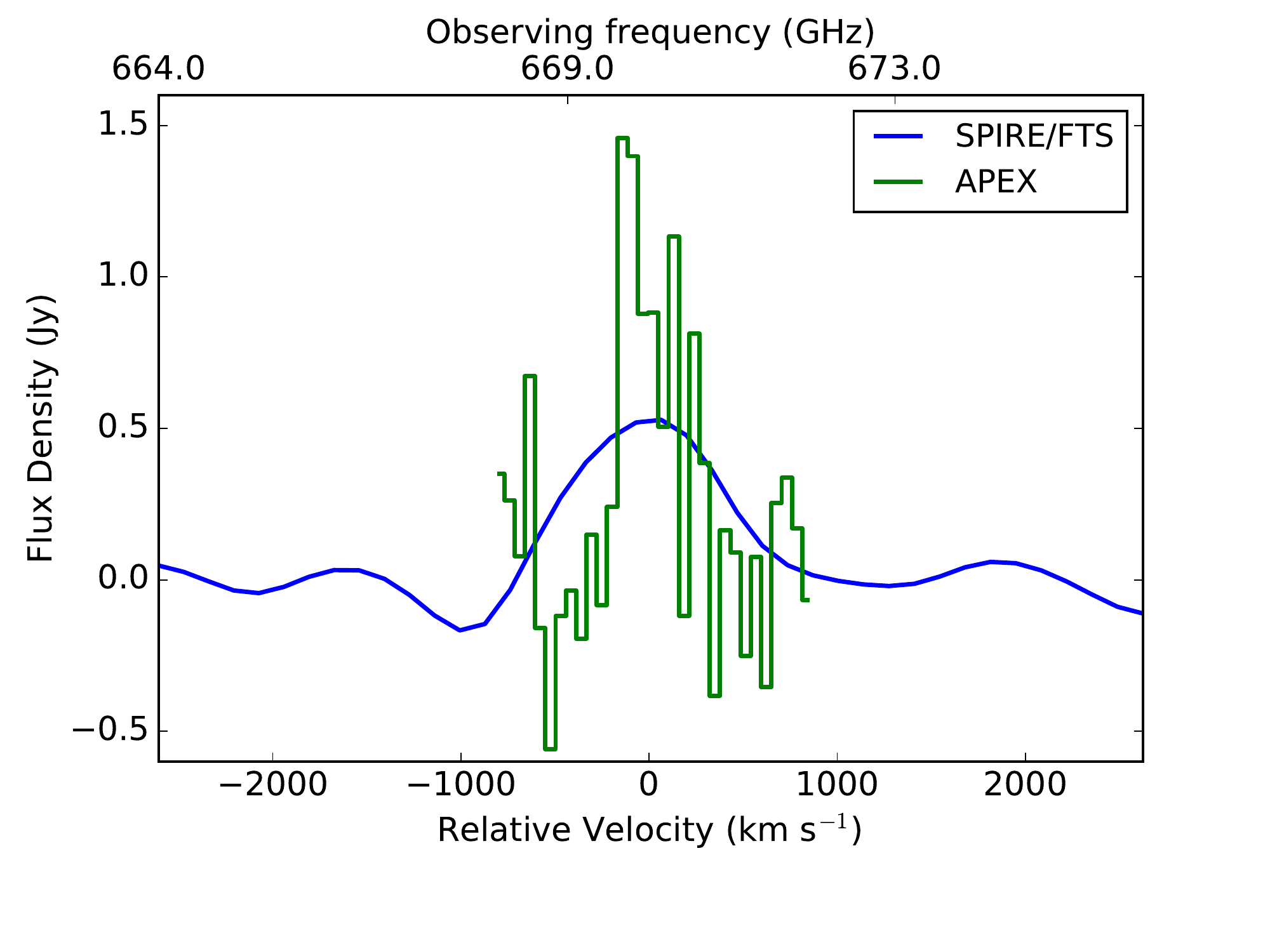}
\includegraphics[width=0.5\textwidth]{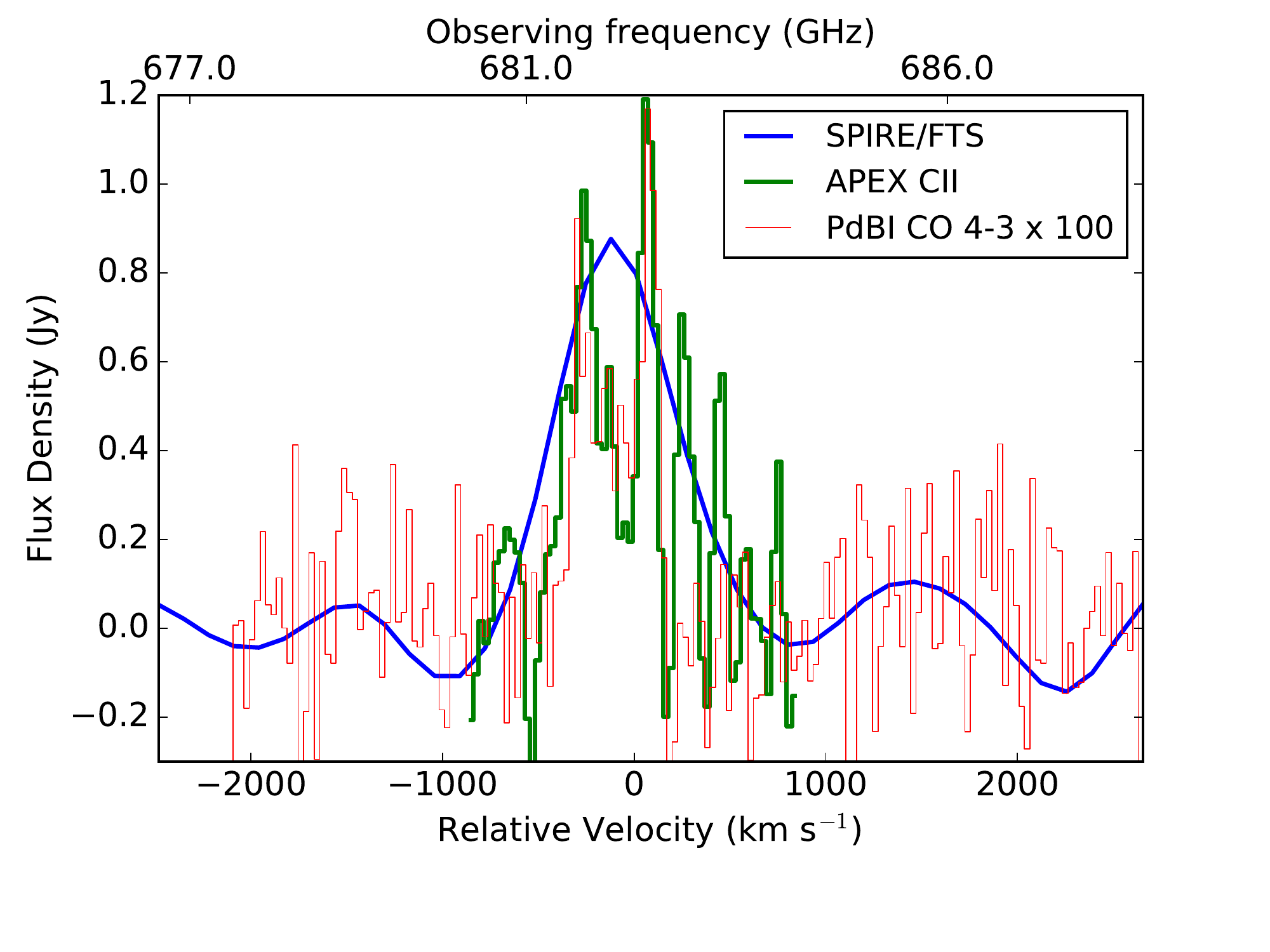}
\caption{{\it Upper:} \CII\ spectra of NA.v1.186. The blue line shows the un-apodised 
\herschel\  FTS spectrum, and the green line shows the APEX
spectrum. {\it Lower:} \CII\ and CO $J$=4-3 spectra of SDP.11.  The blue line
shows the un-apodised \herschel\ FTS spectrum, the green line shows the APEX
spectrum, and the thin red line shows the CO $J$=4-3 spectrum (multiplied by
100$\times$) obtained with
the PdBI \citep{Oteo2016}.} 
\label{fig:apex}
\end{figure}

We also compare the \CII\ spectra of SDP.11 and NA.v1.186 obtained with
\herschel\ and those observed with APEX, which are shown in
Fig.~\ref{fig:apex}.  For SDP.11, the velocity-integrated \CII\ flux obtained
with APEX is $265\pm 65$\,Jy\,\kms, fully consistent with our \herschel\
detection, $269\pm 30$\,Jy\,\kms. This is also close to the \CII\ flux measured
with the second-generation $z$(Redshift) and Early Universe Spectrometer
(ZEUS-2) on APEX \citep[][]{Ferkinhoff2014}. We note that the \CII\ line
profile of SDP.11 is resolved into two velocity components, separated by $\sim
300$\,\kms, which were not resolved by either the \herschel\ or ZEUS-2
observations. We also overlay the CO $J=4\rightarrow3$ spectrum with the \CII\
lines, and find that the twin-peaked profile of the APEX \CII\ line is
consistent with CO (also for HCN) detections in \citet{Oteo2017}, likely
indicative of a merger.  For NA.v1.186, our APEX observations give a \CII\ flux
of $310\pm 90$\,Jy\,\kms, around 60\% higher than that obtained from \herschel\
($190\pm 40$\,Jy\,\kms), but in agreement within the uncertainties.  

\subsection{\cii\,158-\um\ as a diagnostic of ISM properties} \label{sec:cii_diagnostic}

\begin{figure*}
\centering
\includegraphics[width=1.0\textwidth,trim=30bp 100bp 50bp 0 ]{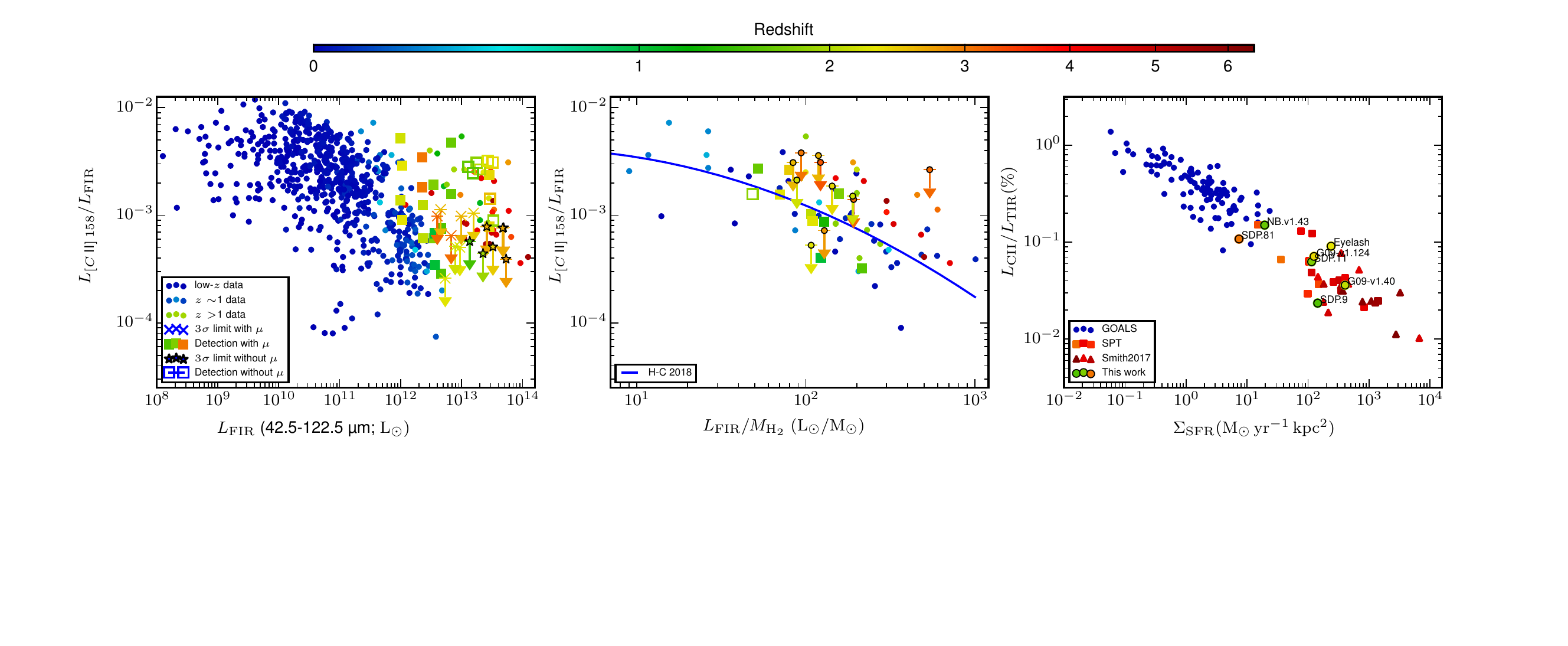}
\caption{ {\it Left}: 
The well-known \(\lum{\cii\,158\,\um} / \lum{FIR}\) deficit, locally and at
high redshift in starburst-dominated galaxies.  Data from this work are shown
as squares, or downward arrows indicating \(3\sigma\) upper limits.  Filled
points have FIR luminosities corrected for magnification, non-filled points use
magnification estimates from SED fitting.  Low-redshift data are taken from
\citet{Brauher2008, DiazSantos2013, Sargsyan2014, GA2015, Ibar2015,
Rosenberg2015}.  Data at \(z \sim 1\) are from \citet{Stacey2010,
Farrah2013,Magdis2014}.  Higher redshift data are taken from
\citet{Cox2011,Swinbank2012, Wagg2012, Riechers2013, DB2014A, Riechers2014,
Rawle2014, Gullberg2015}. Where necessary, luminosities have been scaled from
literature measurements by the mean of the values from our SED fits
\lum{FIR}(42.5--122.5\,\um) = 0.60 \lum{IR}(8--1000\,\um). 
{\it Middle}: The \cii/$L_{\rm FIR}$ ratio as a function of
\lum{FIR}/\mass{\(\text{H}_2\)}, which should correlate more strongly than the
continuum luminosity alone \citep[e.g.][]{GC2011}.  Gas masses are determined
from integrated-galaxy CO observations, converted to CO $J=1-0$ luminosities
where necessary. The blue line shows the scaling relation for \cii/FIR -
FIR/$M_{\rm H_2}$ found in local star-forming galaxies \citep{HC2018}. 
{\it Right}:  The \cii/$L_{\rm TIR}$ ratio as a function of surface density of
SFR, $\Sigma_{\rm SFR}$. This expands the KINGFISH-GOALS-High-$z$ compilation
of \citet{Smith2017a}, with additional data from \citet{Spilker2016}. There are
six new measurements from this work, which have both \cii\ detection and size
estimate, labeled with inner-colourised black circles. This plot includes data
from the Great Observatories All-sky LIRG Survey (GOALS) sample \citep{DS2013},
South Pole Telescope (SPT) sample \citep{Spilker2016,Gullberg2015}, and high
redshift galaxies collected from the literature
\citep{Walter2009b,Carniani2013,Riechers2013,Wang2013b,DB2014,Neri2014,Riechers2014,Yun2015,DS2016,Oteo2016,Smith2017a}.}
\label{fig:cii_lfir}
\end{figure*}

The 158-\um\ \cii\ fine-structure transition is often the brightest observed
FIR emission line.  With the Atacama Large Millimeter/submillimeter Array
(ALMA) now operational, interest in the interpretation of \cii\ 158-\um\
emission from high-redshift galaxies has increased in recent years.

\cii\ has been reported to have an apparent relative decrease in the gas
cooling efficiency with increasing radiation field intensity, dust temperature,
and/or SFR
\citep[e.g.][]{Abel2009,Stacey2010,DiazSantos2013,DiazSantos2014,HC2018,HC2018b}.
The fraction of energy transmitted via this line is not constant, however,
exhibiting a well-known decrease in the \(L_\text{line} / \lum{FIR}\) ratio
with increasing \lum{FIR} in the local Universe \citep{Luhman2003}. This
results in the brightest (ultra-)luminous infrared galaxies (ULIRGs) exhibit
the highest deficits.  Compared with nearby normal quiescent star-forming
spirals, the luminosity ratio between \cii\ and the FIR continuum emission
decreases by more than an order of magnitude in extreme star-forming systems
\citep[e.g.\ LIRGs and ULIRGs][]{DiazSantos2013,DiazSantos2014}.  The
relationship also hold in spatially-resolved measurements on scales of a few
hundreds pc, e.g.  surveys of KINGFISH \citep{Smith2017a},
GOALS\citep{DiazSantos2014}, SHINING \citep{HC2018},  with lower values found
in the nuclei of galaxies than in the extended disks, and the trend may be even
more pronounced with \(\lum{FIR} / M_{\text{H}_2}\) \citep[e.g.][]{GC2011}.

Several theories have been proposed to explain this deficit. The ionisation
parameter, \(U\), and hence dust temperature and SFR surface density, appear to
be strongly correlated with the line-to-continuum ratio \citep{DiazSantos2013},
with the most extreme values of all three parameters found in the dense merging
nuclei of low-redshift ULIRGs. Resolved observations and detailed modelling
\citep{HC2018,HC2018b} show that not only the high ionisation parameter, but also  
the reduction in the photoelectric heating efficiency make the \cii\ line not
able to trace the FUV radiation field. Dust grain charging in higher radiation
fields, leading to a lower photoelectric gas-heating efficiency, may
additionally play a role \citep{Malhotra1997}.  Column density will also have
an effect as \cii\,158-\um\ is primarily produced in PDRs, with a small
contribution from \hii\ regions for local galaxies \citep[e.g.][]{Abel2006};
the column density at which the \cii\ becomes luminous is then governed by the
ionisation parameter and dust extinction.  FIR continuum radiation, however,
can also be produced by non-PDR sources. Lines of sight through high optical
depth molecular material to a PDR and starburst will contain a substantial
continuum contribution from the molecular region, as well as from the \hii\
region close to the ionising source.  This increases the FIR continuum emission
and, at very high column densities, the optical depth to the lines may become
large enough to reduce their observed flux. \citet{Croxall2012} found that the
\cii/\lir\ ratio decreases with increasing dust temperature traced by the
$F_{\rm 70 \mu m}/F_{\rm 100 \mu m}$ color, while the \nii/\lir\, ratio keeps
unchanged. The \cii\ deficit was  also found to have a relationship with the
ionization state of small grains, revealed from polycyclic aromatic
hydrocarbons (PAH) features \citep[i.e.][]{Croxall2012}.  

\begin{table*}
  \centering
  \begin{tabular}{l l l l l l l l}
    \hline
    \hline
    Transition                       & rest wavelength & rest frequency  & ionisation energy              & critical density             & critical density & $E_{\rm up}$ & ref \\
                                     & $\mu$m          & GHz             &   eV\,(low-up)                 &  cm$^{-3}$ (H$_2$)           & cm$^{-3}$ ($e$)  &        K       &   \\
    \hline                                                              
    $\rm[OIII]^3 P_2-^3 P_1$         & 51.81           & 5786            &    35.12 -- 54.94              &                              &  3.6$\times10^3$ &                & \citet{Draine2011book}\\      
    $\rm[NIII]^2 P_{3/2}- ^2P_{1/2}$ & 57.32           & 5229            &    29.60 -- 47.45              &                              &  3.0$\times10^3$ &                & \citet{Malhotra2001}  \\          
    $\rm[OI]  ^3 P_1-^3 P_2$         & 63.18           & 4744.8          &    0     -- 13.6               & 4$\times10^5(T/100)^{-0.34}$ &                  & 227.7          & \citet{Draine2011book}\\ 
    $\rm[OIII]^3 P_1-^3 P_0$         & 88.36           & 3393            &    35.12 -- 54.94              &                              &  5.1$\times10^2$ &                & \citet{Draine2011book}\\      
    OH\,$\Pi_{\rm 3/2}- \Pi_{\rm 3/2} \frac{5}{2}^- - \frac{3}{2}^+$ & 119.23          & 2514.32         &    0     -- 4.4                &$\sim1\times10^8$             &                  & 120.7          & LAMBDA \\
    OH\,$\Pi_{\rm 3/2}- \Pi_{\rm 3/2} \frac{5}{2}^+ - \frac{3}{2}^-$ & 119.44          & 2509.95         &    0     -- 4.4                &$\sim1\times10^8$             &                  & 120.5          & LAMBDA \\
    $\rm[NII] ^3 P_2- ^3 P_1$        & 121.90          & 2459.4          &    14.53 -- 29.6               &                              &  3.1$\times10^2$ & 188.1          & \citet{Draine2011book}\\            
    $\rm[OI]  ^3 P_0-^3 P_1$         & 145.53          & 2060.1          &    0     -- 13.6               & 8$\times10^4(T/100)^{-0.34}$ &                  & 326.6          & \citet{Draine2011book}\\    
    $\rm[CII] ^2 P_{3/2}-^2 P_{1/2}$ & 157.74          & 1900.4          &    11.26 -- 24.4               &   3$\times10^3$              &     50           & 91.21          & \citet{Malhotra2001} \\       
    $\rm[NII] ^3 P_1- ^3 P_0$        & 205.18          & 1461.1          &    14.53 -- 29.6               &                              &  44              & 70.10          & \citet{Draine2011book}\\            
    \hline                                                                                                                                         
  \end{tabular}
  \caption{ Physical properties of the observed lines. 
LAMBDA database: \url{http://home.strw.leidenuniv.nl/~moldata/}
The critical densities are calculated with a kinetic temperature of 10,000\, K for electrons, and a kinetic temperature of 100\, K for collisions with the \Htwo\ gas.
    }
  \label{tab:ncrit}
\end{table*}



This effect appears to continue at high redshift, albeit potentially to a lower
extent \citep[e.g.][]{Stacey2010}. While the global SFRs in high-redshift DSFGs
often equal or exceed those found in local ULIRGs, significant differences may
exist in the distribution of gas and star formation within these two
populations. In some systems, a larger volume of gas may be illuminated by a
lower flux of ionising photons, producing both a smaller ionisation parameter,
\(U\), and a lower optical depth.  \citet{DiazSantos2013} and
\citet{Gullberg2015} discuss the deficit and correlations with other parameters
in further detail, in particular with emission area and molecular gas mass, for
low- and high-redshift galaxies, respectively.

Our \(\lum{\cii{}\,158\, \um{}} / \lum{FIR}\) measurements are plotted in
Fig.~\ref{fig:cii_lfir} along with values from local and high-redshift
starburst galaxies. Gravitational lensing has allowed us to push
intrinsic luminosities towards or below those of local ULIRGs, and our sources
occupy much of the space between the highest redshift ALMA detections
(primarily with \(\lum{FIR} > 10^{13}\,\lsun\)) and low-redshift systems
(typically with \(\lum{FIR} < 3 \times 10^{12}\,\lsun\)).  We see a spread in
\(\lum{\cii{}\,158\, \um{}} / \lum{FIR}\), from \(\sim 3 \times 10^{-3}\) to \(3
\times 10 ^{-4}\), suggesting that the local relation does not hold at higher
redshifts; however, the non-detections may have lower ratios.  These values
appear to confirm other observations indicating that -- for starburst-dominated
systems at least -- the deficit is lower for high-redshift ULIRGs than for
their low-redshift cousins, likely resulting from a lower intensity of ionising
radiation due to their similar total star formation occurring over a larger
volume.

The line-to-continuum ratio is expected to correlate more strongly with
\lum{FIR}/\mass{\(\text{H}_2\)} \citep{GC2011} and our measurements -- plotted
in Fig.~\ref{fig:cii_lfir} using gas masses derived from CO luminosities --
agree with this theory. Fig.~\ref{fig:cii_lfir} also shows that the overall
scaling relation derived from local star-forming galaxies \citep{HC2018} and
our data are also fully consistent with the same trend.

In addition to the total IR luminosity, the FIR lines -- as coolants of
star-formation-heated gas -- may be used to trace the instantaneous SFR. Since
it is often the brightest line, \cii{}\,158-\um\ is of primary interest.  The
conversion is non-linear, however, due to the changing efficiency of these
lines with increasing SFR.  \citet{Sargsyan2014} and \citet{HC2015} discuss the
effectiveness of \cii{}\,158-\um\ as a tracer of SFR in the local Universe,
noting that \cii{}\,158-\um\ can provide measurements consistent with other
star-formation tracers, such as PAH emission and MIR emission lines, but that
any calibration may require additional corrections, applicable to more strongly
star-forming sources, such that SFRs derived from continuum measurements may be
superior.

Literature calibrations from these local samples are presented in the form
\(\frac{\text{SFR}}{M_\odot \text{yr}^{-1}} = A (
\frac{\lum{\cii{}\,158-\um{}}}{L_\odot} )^B \), where \(A\) =
\(1.22\times 10^{-8}\), \(1\pm0.2\times 10^{-7}\) and
\(1.00\times 10^{-7}\), and \(B\) = 1.034, 1 and 0.983 from
\citet{DL2011}, \citet{Sargsyan2014} and \citet{HC2015}, respectively.
For our highest signal-to-noise \cii{}\,158-\um\ detection, that of
SMM\,J2135$-$0102, these suggest SFRs of 40, 165 and 120 \(M_\odot
\,\text{yr}^{-1}\), the range reflecting the inherent dispersion in the
observed \lum{\cii{}\,158\,\um{}}/SFR ratios.  An equivalent calibration
derived from high-redshift data has \(A = 3.02 \times 10^{-9}\) and \(B =
1.18\) \citep{DL2014}, with the higher exponent value accounting for the
decreasing line-to-continuum ratio in galaxies with the high infrared
luminosities discovered at such redshifts.  For SMM\,J2135$-$0102, this
calibration suggests a SFR of 230 \(\msun \text{yr}^{-1}\), fully consistent
with the 250 \(\msun \text{yr}^{-1}\) that we estimate from our photometric
\lum{IR} fit.

Recently, \citet{HC2018, HC2018b} present new scaling relations based on
several physical properties of galaxies and classifications, including
separation from the main-sequence of star-forming galaxies, star-formation
efficiency, and AGN/LINER/pure starburst categories. Using the calibration for
galaxies above the main-sequence \citep{HC2018}, the SFR of Eyelash is $\sim$
160 \(\msun \text{yr}^{-1}\), slightly below our fitted SFR, but consistent
with the values obtained in \citet{Sargsyan2014} and \citet{HC2015}.

An extension to these \lum{\cii\,158\,\um{}}--SFR relationships exploits local
resolved galaxies to consider surface densities.  \citet{DiazSantos2013} derive
a relationship for the nuclei of a large sample of LIRGs, which appears to hold
for high-redshift objects \citep{DiazSantos2014}, and which can be re-arranged
as:

\begin{equation}
      \text{log}_{10} \! \left( \frac{\text{area}}{\text{kpc}^2} \right)
      =
      \text{log}_{10} \! \left( \frac{\lum{IR}}{L_\odot} \right)
      +
      \frac{
        \text{log}_{10} \! \left( \frac{\lum{\cii{}\,158\,\um{}}}{\lum{FIR}} \right)
          - 1.21 \pm 0.24}
        {0.35 \pm 0.03} 
 \end{equation}

The resulting surface areas suggested for our sample are typically within a
factor of a few of the demagnified areas we have adopted (these can be fairly
uncertain), primarily from the lens modelling of \citet{Bussmann2013}, and are
compatible with the low-redshift scatter. As shown in Fig.~\ref{fig:cii_lfir}
(right panel), the \lum{\cii\,158\,\um{}}/\LIR\ ratio shows a decreasing trend
with surface density of SFR, $\Sigma_{\rm SFR}$, following exactly the same
trend found in local starbursts and high redshift galaxies in the literature
\citep{Smith2017a, Spilker2016}. The compactness of the star-formation
activities seems to be a very tight correlation with the
\lum{\cii\,158\,\um{}}/\LIR\ ratio. There seems to have a slight trend with
redshift as well, possibly due to the selection biases -- only the most
intensive, high surface-density starbursts can be selected in the more distant
Universe. Highest lensing magnification occurs within a small angular area, so
compact starbursts allow a higher fraction of the total luminosity to be
located within the lensed area.

\section{Rest-frame stacking} \label{sec:analysis_stacking}

\begin{figure*}
  \centering
  \includegraphics[width=\textwidth]{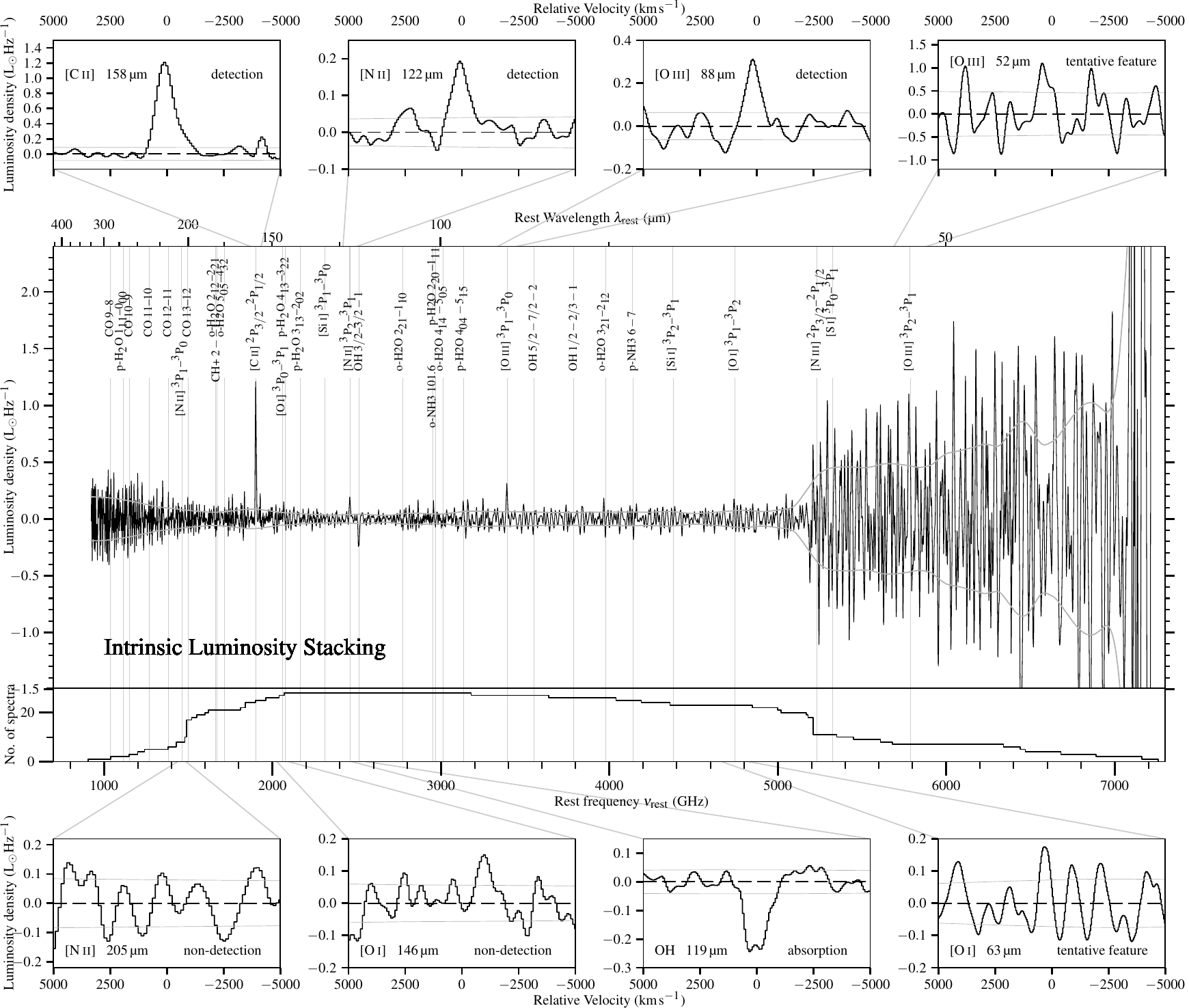}
  \caption{
    {\it Upper centre}: Rest-frame intrinsic (de-lensed) luminosity stack of
    all spectra where secure redshifts are in hand. Markers denote the
    positions of the primary spectral features expected.
    {\it Lower centre}: The number of spectra contributing to the
      stacked spectrum.
    {\it Upper and lower}: \(\pm 5000 \, \kms\) cuts around the positions
      of the primary atomic and ionic spectral lines analysed elsewhere in
      this work, plus the OH \(119 \, \um\) doublet.}
  \label{fig:intrinsicstack}
\end{figure*}

\begin{figure*}
  \centering
  \includegraphics[width=\textwidth]{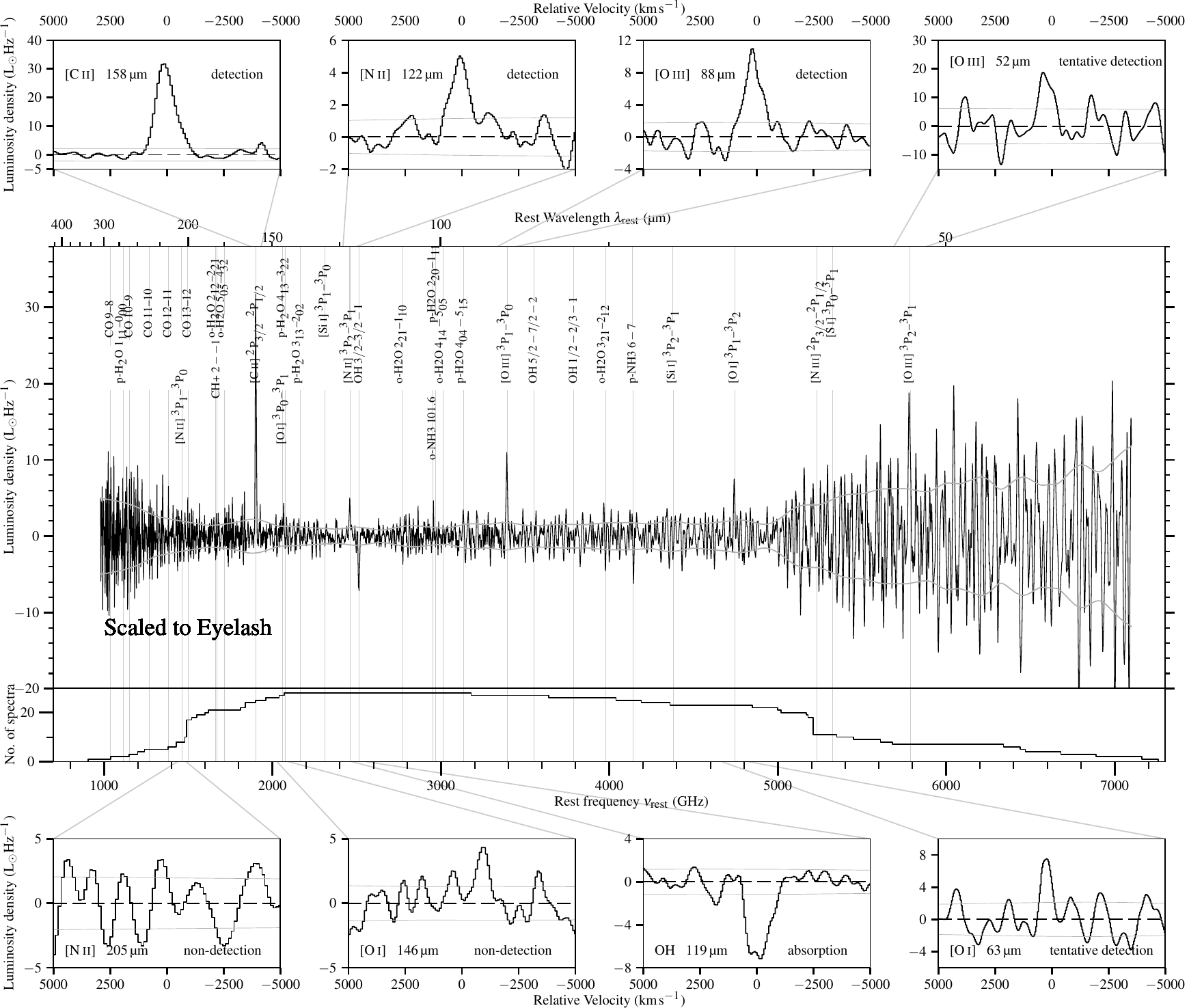}
  \caption{
    {\it Upper centre}: Scaling stacked spectrum, with all spectra scaled to
    the 500 \mum\ flux of SMM\,J2135$-$0102. Markers denote the positions of
    the primary expected spectral features.
    {\it Lower centre}: The number of spectra contributing to this
      stacked spectrum.
    {\it Upper and lower}: \(\pm 5000 \, \kms\) cuts around the positions
      of the primary atomic and ionic spectral lines analysed elsewhere in
      this work, plus the OH \(119 \, \um\) doublet.
  }
  \label{fig:scalingstack}
\end{figure*}

\begin{figure*}
  \centering
  \includegraphics[width=\textwidth]{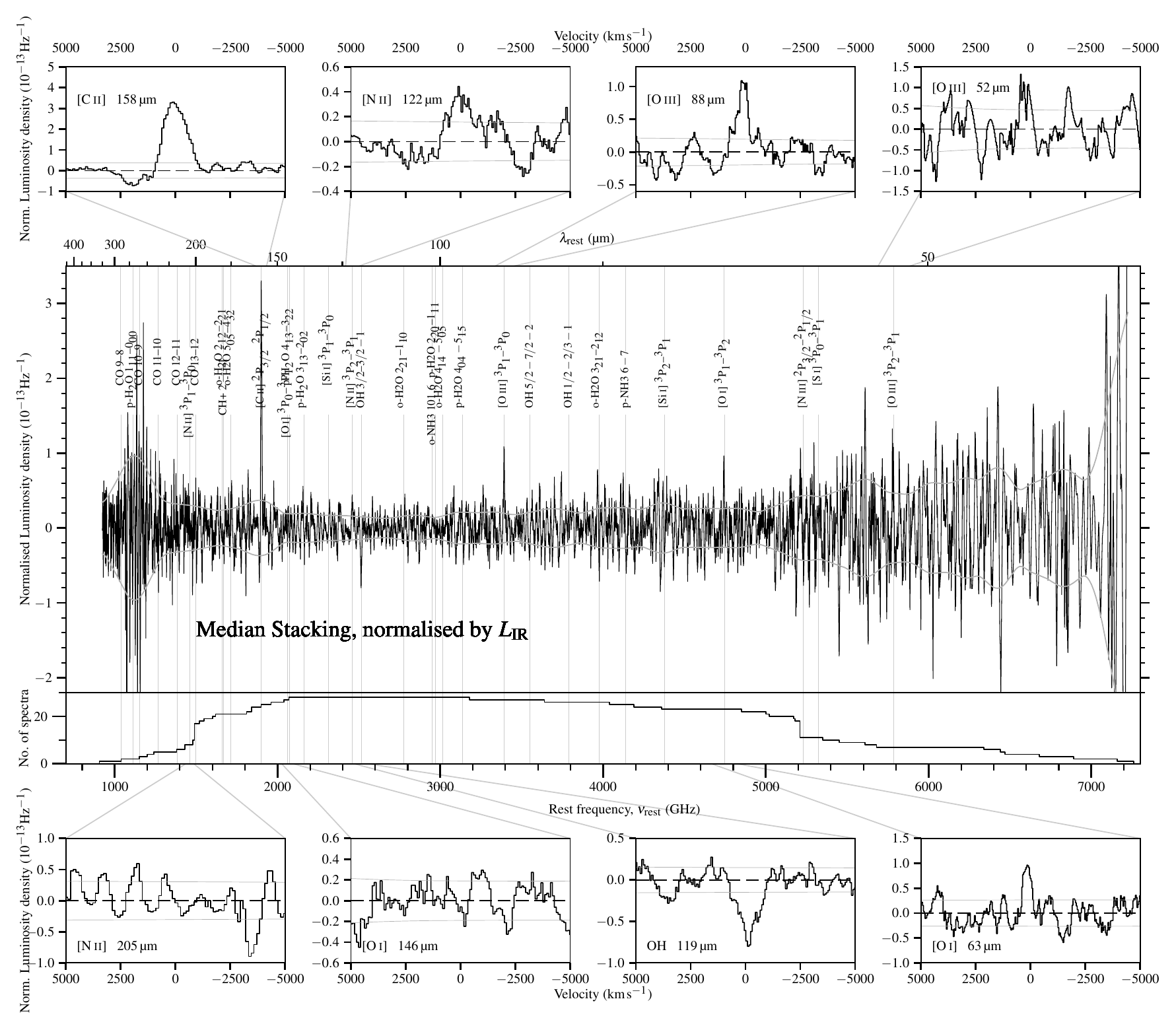}
  \caption{
    {\it Upper centre}: Median stacked spectrum, with all spectra normalised to
    their own \LIR.  Markers denote the positions of the primary expected
    spectral features.
    {\it Lower centre}: The number of spectra contributing to this stacked
    spectrum.
    {\it Upper and lower}: \(\pm 5000 \, \kms\) cuts around the positions of
    the primary atomic and ionic spectral lines analysed elsewhere in this
    work, plus the OH \(119 \, \um\) doublet.  }
  \label{fig:median_stacking}
\end{figure*}


\begin{figure}
  \centering
  \includegraphics[]{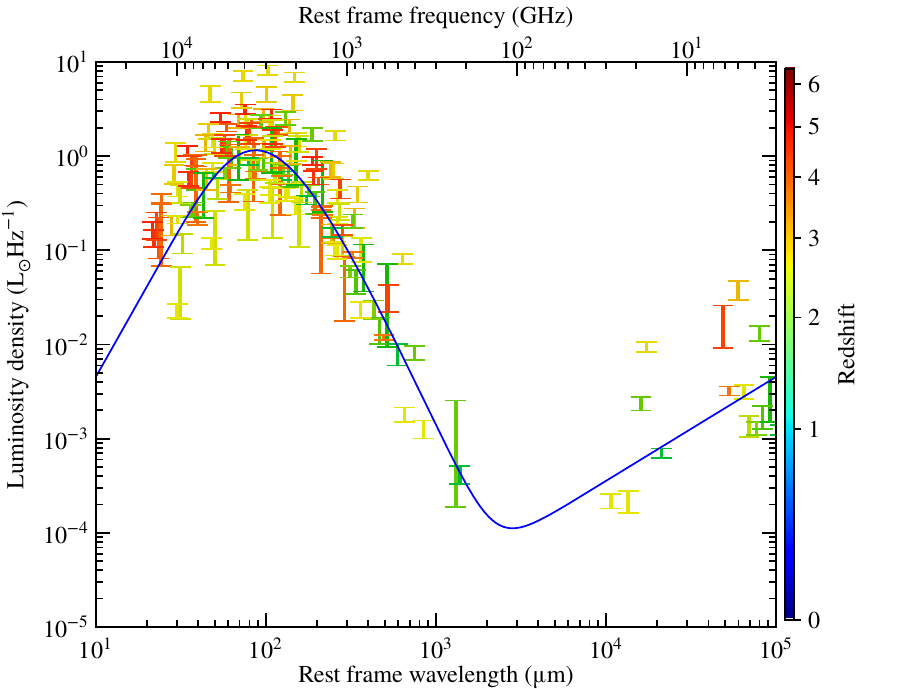}
  \caption[Mean demagnified SED]{ The rest-frame stacked SED (blue line)
          derived from fitting the power-law dust temperature distribution
          model \citep[i.e.,][]{Kovacs2010} to all demagnified photometric
          points (colourful dots) of the members of the sample for which
          lensing model and spectroscopic redshift are available. Colours show
          the different redshifts of the sources.}
  \label{fig:stacked_sed}
\end{figure}

In individual spectra of the sample, only the brightest FIR cooling lines may
be detected. Recently \citet{Wardlow2017,Wilson2017} stacked \herschel-PACS and
\herschel-SPIRE spectra of DSFGs at high redshift and detected ionised
fine-structure lines, such as [O {\sc iv}] 26-\mum, [S {\sc iii}] 33-\mum,
[O{\sc iii}] 52-\mum, and neutral lines, such as C{\sc i} 158-\mum, [O{\sc i}]
63-\mum.  Here, we stack our \herschel-SPIRE spectra to probe the faint lines. 

\subsection{Spectral stacking}\label{sec:spectral_stacking}

To search for fainter lines, and to determine the evolution of the average
properties of the population as a whole, we stacked the observed spectra in the
rest frame. In principle the noise should reduce as \(\sqrt{n}\), where $n$ is
the number of stacked spectra, when weightings are not considered. Different
stacking techniques have been adopted in the literature
\citep[e.g.][]{Spilker2014,Wardlow2017,Wilson2017} for various scientific
purposes, which give different weightings to the data. In the following
sections, we present three different stacking methods to explore the average
physical properties in the sample and the potential biases of the stacking
techniques.

Before stacking, we applied apodisation to convolve the spectral response from
a Sinc function to reduce sidelobes and generate a more ``Gaussian-like''
profile \citep[i.e.,][ see also HIPE manual]{Naylor2007}. To stack spectra
with different redshifts (see Section \ref{sec:analysis_stacking}), the spectra need
to be blueshifted to the rest frame, where the widths of the Sinc function for
each galaxy increases with redshift by $1+z$, i.e., from 2.6\,GHz at $z
= 1.2$ to 5.5\,GHz at $z = 3.6$, compared to the 1.2-GHz width of the system
response of the \herschel-SPIRE. If we simply add/average together the same
line from different redshifted galaxies without apodisation, the sidelobes of
the Sinc functions (of different widths) tend to cancel each other, which
biases the stacked signal.

To match the velocity resolution across the sample at the rest frequency, we
vary the Gaussian width of the apodisation with redshift, where width $=
((z-1.5)/2 + 1.1) \times 1.20671$\,GHz, which ensures a roughly uniform
velocity resolution at any given rest frequency. For the one galaxy at $z<1.5$,
G15-v2.19, we use a Gaussian width of 1.1\,GHz.  Converting from the Sinc
function to the Gaussian function may raise the flux by 5\%
\citep[e.g.][]{Hopwood2015}, which we consider as an extra source of error in
the final flux estimate.  To avoid the noisy edges of the spectra in both SLW
and SSW, which are heavily affected by ripples, we do not include the very edge
of the spectral ends in the stacking. For SLW, we exclude 30 channels at the
low-frequency end and 120 channels at the high-frequency end; for SSW, we
exclude 30 channels at both ends of the spectra.

The first method stacks spectra using their intrinsic de-lensed line
luminosities. This can only be applied to the sources with both known
spectroscopic redshifts and amplification factors. This method focuses on the
intrinsic properties of the galaxies and mitigates the influence of the highly
non-uniform lensing factors for different targets (and thus frequency coverage,
since the targets all lie at different redshifts). However, the sample size is
limited by the numbers of galaxies with lensing models (23 out of 38 galaxies
with \herschel-SPIRE spectra). 

We blue shift all apodised spectra to the rest frame, then convert flux
densities to luminosity densities (spectral luminosity) with their luminosity
distances. To arrive at de-lensed intrinsic properties, we divide the
luminosity by their amplification factors \citep[from e.g.][]{Bussmann2013}.
Then we create an empty output spectrum covering the total frequency range in
the rest frame, with a frequency sampling step equivalent to the minimum of all
individual spectra (to avoid under-sampling the high-redshift end). After
re-sampling the rest-frame de-lensed spectra (in luminosity density) into the
template output spectrum, we average all of them for each channel using
\(1/\sigma^2\) weighting, which is controlled by the noise level appropriate
for this specific channel. The final output is a spectrum in de-lensed mean
line luminosity density. The final noise level is a channel-based function.
The spectra resulting from this intrinsic stacking method is displayed in
Fig.~\ref{fig:intrinsicstack}.

The second method uses a similar stacking technique as \citet{Spilker2014},
taking advantage of the negative $K$ correction in the submm wavelength regime.
We use SMM\,J2135$-$0102 as a `master' template, and blue-shift (or red-shift
for galaxies with $z< 2.326$) all measured spectra to its redshift
($z=2.32591$). We then scale all spectra by their 500-$\mu$m continuum flux
relative to the value measured for SMM\,J2135$-$0102.  In this way, the lensing
factors for all of our spectra are scaled to a value similar to that of
SMM\,J2135$-$0102 at 500\,\mum, so the stacked spectrum should have a similar
amplification factor as SMM\,J2135$-$0102.

The blue-shifted and scaled spectra are then stacked together using
\(1/\sigma^2\) weighting. This stacking method does not require amplification
factors, and thus avoids the associated uncertainties. All spectra with known
redshifts can be used, regardless of lensing models, so this method can be
applied for a relatively large sample of galaxies.  Comparing to the results
from the first stacking technique, this approach can reveal any potential bias
in the different stacking methods. We present the spectrum obtained using this
scaling method in Fig.~\ref{fig:scalingstack}.

The third method is based on median, instead of the mean, to avoid weightings
that may bias to the more luminous or higher signal-to-noise spectra.  We
normalise all spectra with their infrared luminosities, \LIR, blueshift spectra
to their rest-frames, and then calculate the median value across all spectra
for each channel bin. When calculating the median values, we do not adopt
weightings, i.e., equal weighting. Such a stacking method can avoid systematic
biases by a few strong targets and is robust to test if the weakly detected
lines are common (more than 50\%) in the sample.  We present the spectrum
obtained using the median stacking method in Fig.~\ref{fig:median_stacking}. In
principle, an straight (non-weighted) stacking would also be little biased by
noise, but such method could only results in a detection of the \cii\ line and
a marginal detection of the \oiii\ 88\mum\ line.

From the line ratios of multiple far-IR lines, calculated between the intrinsic
stacked spectrum and the scaling stacked spectrum, the amplification factor of
the scaling stacking method is $\sim$ 15--30 (depending on the specific line),
higher than the mean value of the amplification factors.

This is because the scaling stacking method has a systematic bias.  We have
scaled all spectra to the 500-\um\ flux of Eyelash,  which actually has the
highest flux in the sample. This would actually artificially bias all spectra
to a higher amplification factor (compared to the mean value) for the final
stacked spectrum. Furthermore, the higher S/N of Eyelash and its very high
amplification factor \citep[$\sim$37.5][]{Swinbank2011} may further biases
the final amplification to high values.

\subsection{SED stacking}\label{sec:sed_stacking}

To probe the average dust properties and the mean IR luminosity, we also stack
the IR photometry data and generate a mean intrinsic SED. We combine all the
IR photometry data for each target, blue shift the PACS, SPIRE and 1.4-GHz
radio fluxes to their rest frequencies, derive the luminosity for each band,
and correct for the lensing magnifications. Then we fit a single MBB with a
power-law synchrotron emission. The stacked mean SED and the rest-frequency
luminosities of the continuum data are shown in Fig.~\ref{fig:stacked_sed}.  We
fit the SED with a single MBB with a dust emissivity slope index
of $\beta=1.8$, resulting a dust temperature of $\sim 45 \pm 5$\,K, close to
the SED modeling and dust temperatures measured in DSFGs at similar redshifts
\citep[e.g.][]{Swinbank2014}. The average dust mass is $3.7\pm0.5 \times
10^{8}$\,\msun, which corresponds to an H$_2$ gas mass of $\sim 3.7 \pm0.5
\times 10^{10}$\,\msun, when a typical dust-to-ISM mass ratio of $\sim 100$ is
adopted for metal-rich galaxies \citep[e.g.][]{Swinbank2014,Scoville2017}.

\begin{table}
  \centering
  \begin{tabular}{l l l l l l r c}
    \hline
    \hline
    Transitions                      & $L^{\rm intrin}$&  $M^{\rm min}_{\rm H^+}$       &$L^{\rm scale}$   &  $\mu M^{\rm min}_{\rm H^+}$\\  
    Unit                             & $10^9$\Lsun     &  $10^8$\Msun                   &$10^9$\Lsun       &    $10^8$\Msun    \\                                   
    \hline
    \OIII\ 52-\mum                   &14$\pm$2         &  0.36$\pm$0.1                  & 200$\pm$25       &    5   $\pm$1.4         \\         
     \OI\  63-\mum                   &1.7$\pm$0.3      &                                & 79 $\pm$8        &                         \\        
    \OIII\ 88-\mum                   &3.3$\pm$0.3      &  0.9$\pm$0.1                   & 87$\pm$5         &    24  $\pm$1.4         \\       
     OH 119-\mum$^\dagger$           &$-$2.9$\pm$0.6   &                                & $-$66 $\pm$8     &                         \\          
    \NII\ 122-\mum                   &1.5$\pm$0.3      &  7.4$\pm$1.5                   &36$\pm$5          &    180 $\pm$25          \\         
    \CII\ 158-\mum                   &7.6$\pm$0.2      &                                &216$\pm$4         &                         \\                               
    \NII\ 205-\mum                   &$<$0.9           &                                & $<$18            &                         \\        
    \hline                                                                                                                                         
  \end{tabular}
  \caption{
Measured line properties in the stacked spectra.  $M^{\rm min}_{\rm H^+}$  is
the estimated minimum ionised gas mass calculated from \NII\ 122-\mum\ and
\OIII\ 88-\mum\ lines, following the method used in \citet{Ferkinhoff2010}.
$^{\dagger}$: The OH flux is combined from two velocity components. $\mu$ is
the stacked amplification factor for each line, which ranges about 15--30,
depending on the specific line and weighting adopted (see Section
\ref{sec:spectral_stacking}).  
}
  \label{tab:stack}
\end{table}


%


\subsection{Molecular absorption features}

In the two weighted stacked spectra, we find clear absorption corresponding to
the OH 119-\um\ feature, with robust individual detections seen in the spectra
of G09-v2.19 and SMM\,J2135$-$0102, the latter presented first by
\citet{George2014}. If we remove SMM\,J2135$-$0102 and G09-v2.19 from the
sample then the OH absorption feature remains after re-stacking. In the median
stacked spectrum, the OH 119-\um\, absorption feature is still clearly
detected. This indicates that this OH 119-$\mu$m absorption feature is likely
common for the DSFGs in our sample, and is not dominated by a few strong
targets.

Blueshifted OH 163-\um\ line has also been observed in emission towards
high-redshift galaxies \citep{Riechers2014}. This feature provides further
evidence that out-flowing molecular gas may be common within the high-redshift
DSFG population, despite being difficult to observe.  However, further analysis
of this detection is beyond the scope of this work.

\begin{figure*}
  \centering
  \includegraphics[width=0.33\textwidth]{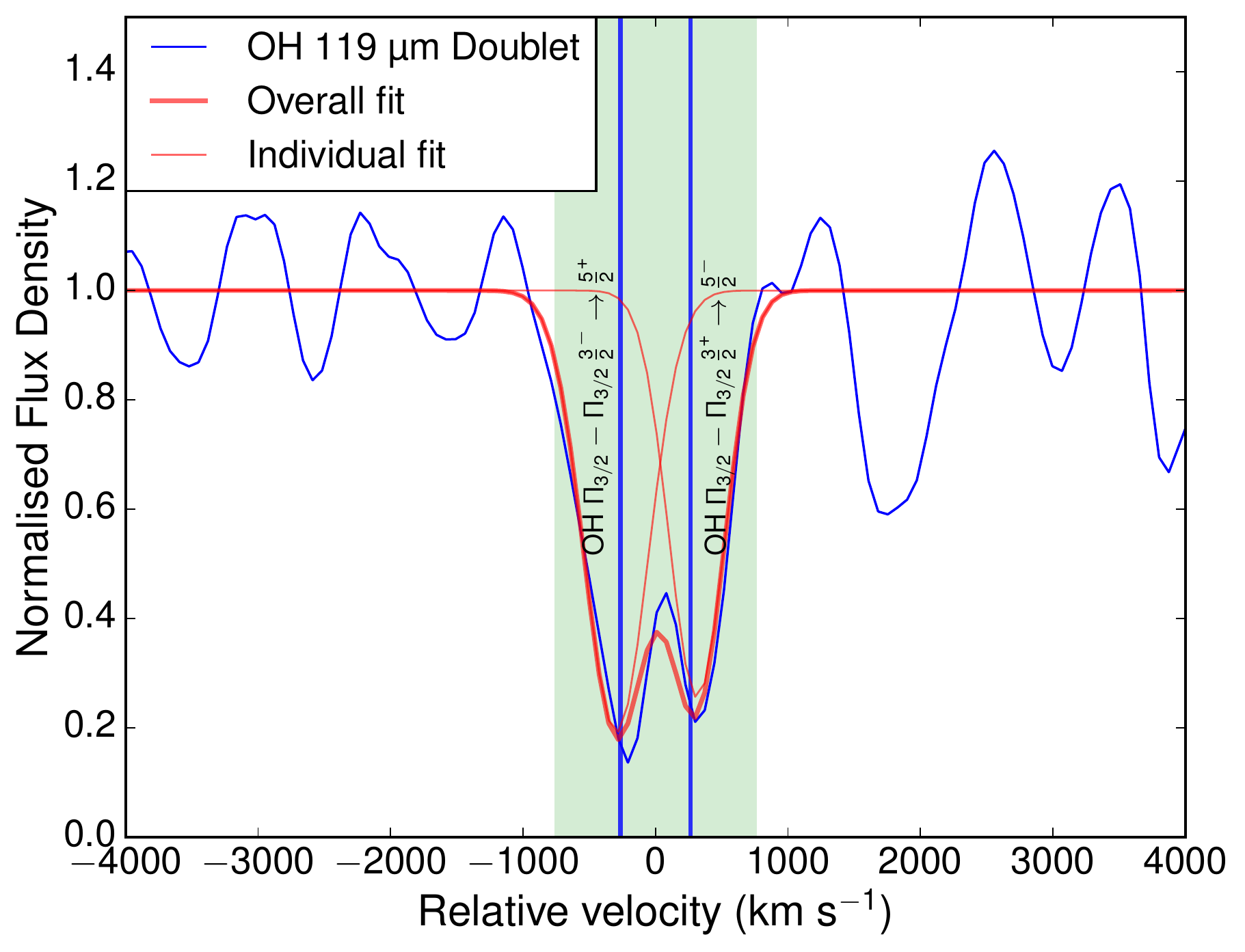}
  \includegraphics[width=0.33\textwidth]{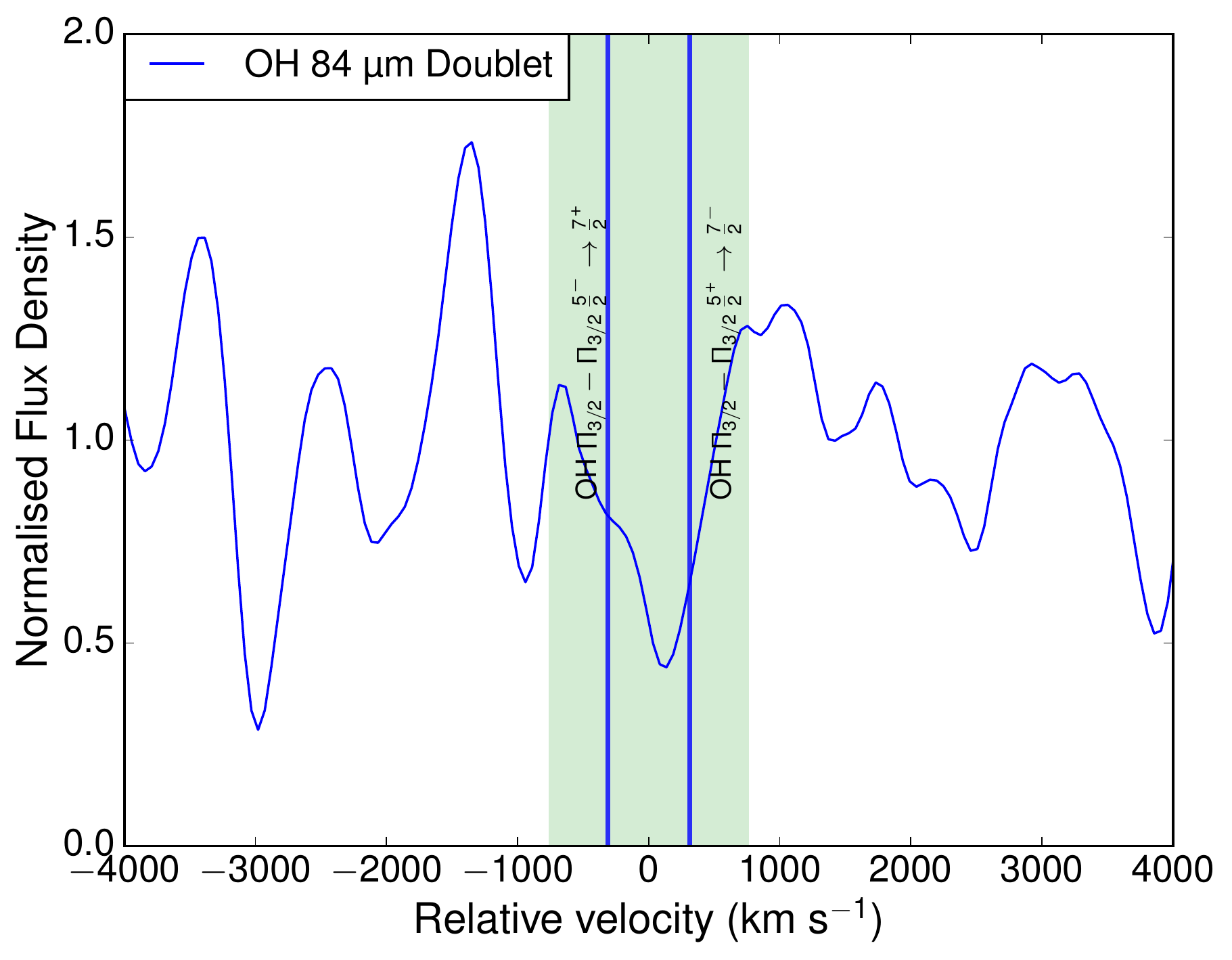}
  \includegraphics[width=0.33\textwidth]{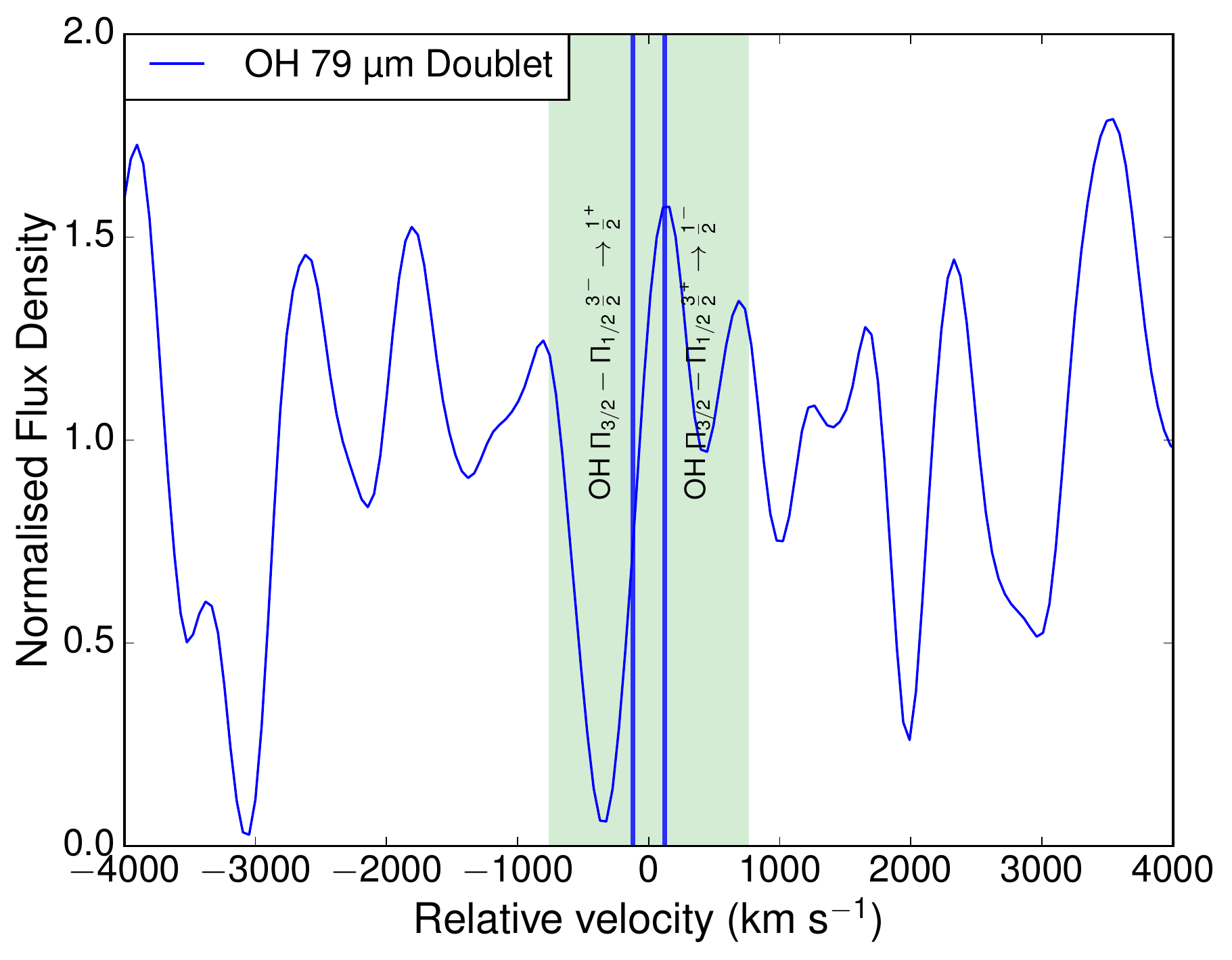}
  \caption{Normalised OH line profile of the stacked spectrum, using their
  intrinsic luminosity and continuum contribution interpolated from SED
  fitting. \textit{Left:} stacked OH 119 $\mu$m spectrum. The absorption
  feature is fitted with two Gaussian components, centered at the rest
  frequencies of the OH doublet transitions, $\nu_{\rm rest}=$2514.31 GHz for
  OH $\Pi_{\rm 3/2}- \Pi_{\rm 3/2} \frac{5}{2}^- - \frac{3}{2}^+$  and
  $\nu_{\rm rest}=$ 2509.95  GHz for  OH $\Pi_{\rm 3/2}- \Pi_{\rm 3/2}
  \frac{5}{2}^+ - \frac{3}{2}^-$. \textit{Middle:} stacked OH 84-\um\ 
  spectrum. \textit{Right:} stacked OH 79-\um\ spectrum.   }
\label{fig:OH_absorption} 
\end{figure*}

To derive the equivalent width of the OH absorption line, we create another
stack designed to include both the line emission and the continuum.  We take
the SED models derived in Section \ref{sec:pacs_measurements}, interpolate the SED to
the observed frequency of the absorption line, then add the continuum
intensity. We then stack the spectra with the added continuum contribution,
adopting the same weighting scheme as in Section \ref{sec:analysis_stacking} for
consistency. Finally, we fit a second-order polynomial baseline to the
continuum and normalise the spectra to unity to derive the equivalent width.
These continuum-adjusted, stacked spectra are shown in
Fig.~\ref{fig:OH_absorption}. The uncertainties on the integrated opacities
were estimated as $\Sigma_\tau \sqrt{\delta V \Delta V_{1/2} }$ where
$\Sigma_\tau$ is the r.m.s.\ uncertainty on the opacity for the velocity
resolution $\delta V$ and $\Delta V_{1/2} $ is the half-maximum velocity width.

The critical densities of the two OH $^2\Pi_{3/2}$ $J =5/2$--$3/2$ transitions
are very high (Table~\ref{tab:ncrit}) and their upper level energies are $\sim
120$\,K. This makes it difficult to excite these transitions to high-$J$ levels
-- we can anticipate that most of the OH molecules are in the ground state and 
well mixed in the low-density gas along the line of sight. This assumption is
supported by the fact that we do not detect any obvious signals of another
ground transition of OH $J=3/2$--$2/1$ at 79-$\mu$m, nor the other high lying 
doublet of OH $^2\Pi_{3/2}$--$^2\Pi_{1/2}$ $J=3/2$--$2/1$ at 84.6-$\mu$m (see
~Fig. \ref{fig:OH_absorption}).  These transitions were detected in absorption
or emission in nearby local compact (U)LIRGs, e.g., Mrk\,231,
NGC\,4418 and Arp\,220, where multiple transitions of OH have been detected
with absorption depths comparable to their 119-$\mu$m OH features
\citep[e.g.][]{Sturm2011,GA2012,Spoon2013}.

We follow the method in \citet[][]{GA2014} to calculate the theoretical ratios
of the equivalent width between OH 119-$\mu$m, OH 84-$\mu$m and OH 79-$\mu$m.
For optically thin absorption and ignoring any possible emission components and
assuming a uniform filling factor of unity (i.e. recovering all of the
continuum emission in the line of sight), the equivalent width of the
absorption lines is given by:

\begin{equation}
W_{\rm eq} = \lambda^3 g_{\rm u} A_{\rm ul} 
\frac{N_{\rm OH}^{\rm l}}
     {8 \pi g_{\rm l}},
\end{equation} \label{eq:Weq}

\noindent
where $\lambda$ is the wavelength, $A_{\rm ul}$ is the Einstein coefficient for
spontaneous emission, $g_{\rm u}$ and $g_{\rm l}$ are the statistical weighting
factors of the upper and lower levels and $N_{\rm OH}^{l}$ is the OH
column density in the lower transition, which is $^2\Pi_{3/2}\, J= 3/2$ for OH
119-$\mu$m.

In the optically thin limit, the OH119-to-OH84 equivalent width ratio is $\sim$
0.84. Although the OH 84-$\mu$m absorption line is often seen in local (U)LIRGs
at similar equivalent width as the 119-$\mu$m line \citep{GA2015,GA2017}, we
have only a non-detection for the OH 84-$\mu$m line in our stacked spectra.
This indicates that the OH population at the $^2\Pi_{3/2}$ $J =5/2$ energy
level is much less than that on the ground level, $^2\Pi_{3/2}$ $J =3/2$. This
means that the average excitation of OH in our lensed DSFGs is low, with most
OH molecules in the ground energy level, and there are much fewer OH molecules
in the $J=5/2$ level than the $J=3/2$ level, compared to what has been found in
local ULIRGs.

On the other hand, the OH119-to-OH79 equivalent width ratio is $\sim$ 39.3,
which is consistent with the non-detection of the OH 79\,$\mu$m absorption line
in our stacked spectrum, supporting the relatively optically thin assumptions.
However, in local (U)LIRGs, both lines are often detected at similar equivalent
widths, which show optically thick conditions \citep[e.g.][]{GA2017}. 

Unfortunately, our spectral resolution (around $\sim 400$\,\kms\ at
119-$\mu$m) hinders attempts to identify any P-Cygni profiles, or to set
constraints on the maximum velocity of the OH lines, given the signal-to-noise
ratio of our stacked spectrum, even without apodisation applied.  For nearby
local ULIRGs and QSOs, \citet[][]{2013ApJ...776...27V} found a median outflow
velocity of around $-200$\,\kms.  Such a velocity for the DSFGs would not be
resolved by our data.  We fit identical Gaussian profiles to both transitions
of the OH absorption doublet and find their line centres to be consistent with
their rest frequencies.

Following \citet{George2014}, we can estimate an upper limit for the OH column
density assuming optically thin OH absorption.  To derive the optical depth of
the absorption lines, we assume that the filling factor of the continuum source
and the foreground molecular gas is unity. The line optical depth,
$\tau(\varv)$, is given by:

\begin{equation}
        \tau(\varv) = - {\rm ln} ( \frac { I(\varv) - I_{\rm cont}}{ I_{\rm cont}}),
\end{equation}

\noindent
where $I(\varv)$ is the depth of the absorption relative to the continuum level
as a function of velocity $\varv$ and $I_{\rm sat}$ is the continuum intensity.
The equivalent width of the OH 119-$\mu$m absorption features is estimated to
be $\sim$ 1050 \kms.  We calculate the total OH column density following
\citet{Mangum2015} and \citet{George2014}: 

\begin{equation}
        N_{\rm total}^{\rm OH} = \frac{4 \pi } {h \nu} 
        Q(T_{\rm ex}) 
        \frac{1}{B_{\rm 21} g_{\rm u}}
        \frac{\exp(\frac{E_{\rm u}}{k_{\rm B} T_{\rm ex}}) } 
        { \exp(- h\nu/k_{\rm B} T_{\rm ex}) - 1} 
        \int \tau d V
\end{equation}

\noindent
where we assume optically thin lines and a filling factor of unity. $E_{\rm u}$
is the upper level energy of this transition, $Q(T_{\rm ex})$ is the partition
function (where we adopt the $Q$ values given by the JPL
database\footnote{\url{http://spec.jpl.nasa.gov/ftp/pub/catalog/doc/d017001.cat}}),
$g_{\rm I}=3$ for an ortho level and $g_{\rm I}=1$ for all any other cases.
$B_{21} = A_{21}/( 2 h\nu^3 / c^2 ) $, where $A_{\rm ul} = \frac{ 64 \pi ^4
\nu^3 }{ 3 h c^3} |\mu|^2 $ is the Einstein coefficient for spontaneous
emission, $\mu$ is the electric dipole moment and $\nu$ is the rest
frequency of the line.

As shown in aforementioned discussion, the non-detection of 84-$\um$ and
79-$\um$ lines indicates that the OH excitation is low. We therefore adopt a
low excitation temperature of 9\,K (comparable to the cosmic microwave
background temperature, $T_{\rm cmb}$, at $z=2.5$) and derive an estimate of
the OH column density: $\sim2\times 10^{15}$\,cm$^{-2}$. Assuming the OH
abundance is the same as the Galactic value, $X_{\rm OH} = 5 \times 10^{-6}$
\citep[e.g.][]{2002ApJ...576L..77G}, the average H$_2$ column density is $\sim
4 \times 10^{20}$\,cm$^{-2}$, which is considerably lower than the typical
H$_2$ column densities found in SMGs \citep[e.g.][]{Simpson2017}, which
indicates that either the OH abundance in DSFGs is much lower than that found
in local galaxies or, more likely, that the observed OH is associated with
out-flowing gas and does not trace the total column of H$_2$.

\section{Atomic and ionised lines}

The \cii\,158-\um\ and \oi\,63-\um\ lines often dominate the cooling of
neutral gas in starburst galaxies: and ionic fine-structure lines like \NII\
and \OIII\ normally dominate the FIR cooling of the ionised gas, which is
mostly in \HII\ regions \citep[e.g.][]{DL2014,Hughes2015}. The \OIII, \NII,
and \OII\ lines at optical wavelengths have the strongest cooling in \HII\
regions \citep[e.g.][]{Osterbrock2006}, but most of the optical cooling lines
are absorbed by dust gains in the dusty starburst galaxies. Dust grains re-emit
most of the incident FUV and optical radiation in the FIR and as a result the
luminosity ratio, \OI/\(\text{FIR}\), \OIII/\(\text{FIR}\), and \CII
/\(\text{FIR}\), typically \(\sim 0.01-0.1\,\%\) in high-redshift DSFGs (see
Fig.~\ref{fig:cii_lfir}) provides an estimate of the efficiency of gas
heating.  This value is, however, affected by the non-negligible fraction of
\cii\,158-\um\ produced in \hii{} regions, the strong dependence of the carbon
ionisation front upon metallicity \citep[e.g.][]{Croxall2017} and the optical
depths of the lines, in particular \(\tau_{\oi\,63\,\um} \sim 1\text{--}3\)
\citep[e.g.][]{Liseau2006, Hughes2015}. The \oiii\ and \nii\ lines come only
from the ionised gas phase -- the \NII\ lines originate from both diffuse and
dense \HII\ gas phases -- offering an excellent estimate of the total mass of
ionised gas \citep[e.g.][]{Liseau2006, Zhao2016a, Zhao2016b}. With similar
critical densities, but different ionisation potentials, the ratio
\oiii\,88-\um/\nii\,122-\um\ provides a good tracer of the stellar effective
temperature of the ionising source.

\subsection{Ionised gas mass}

The extreme starburst conditions in the SMG sample supplies enormous amount of
heating power to ionise neutral gas, mostly due to the feedback of
star-formation through UV radiation, X-rays, outflows, and shocks, etc. It has
been claimed that some high redshift starburst galaxies have very high mass
ratios between ionised gas and molecular gas, $M(\rm H^+)$/$M(\rm H_2)$
$\gtrsim 10 - 20\%$, in SMMJ02399 and Cloverleaf \citep{Ferkinhoff2011}, by
calculating the minimum ionized gas mass, \mmin, traced by the \NII\ and \OIII\
lines.  Following the method used in \citet{Ferkinhoff2010, Ferkinhoff2011}, we
assume that nitrogen in \HII\ regions is singly ionised and the \NII\ emission
could represent all the diffuse low-density ionised gas, neglecting other
energy levels (e.g.\ \NIII\ or  \OII,  \OV\ and higher). The minimum ionised
gas mass can be estimated from:

\begin{equation}
M^{\rm min}_{\rm H^+} = 
F_{\rm line} \cdot 
\frac{4 \pi \cdot D_{\rm L}^2 \cdot m_{\rm H}} 
     {g_{\rm upper} / g_{\rm tot} A_{\rm 10} h \nu_{\rm line} X_{\rm line} } ,
\end{equation}

\noindent
where $F_{\rm line}$ is the line flux in \Jykms, $D_{\rm L}$ is the luminosity
distance in Mpc, $m_{\rm H}$ is the mass of a Hydrogen atom, $A_{\rm 10}$ is
the Einstein spontaneous A coefficient, $g_{\rm upper}$ is the statistic
weighting of the upper energy level ($=3$ for \NII\ 122-\mum\ and \OIII\
88-\mum\ lines), $g_{\rm tot} (= \Sigma g_{\rm i} \exp(-\Delta E_{\rm i}/k T)$)
is the partition function, $h$ is the Planck constant, $\nu_{\rm line}$ is the
rest frequency of the emission line and $X_{\rm line}$ is the relative
abundance of the studied ionised atoms (N or O) relative to ionised hydrogen
(H$^+$). For the minimum mass estimate, we assume that $X_{\rm N_\textsc{II}} =
\rm N/H$, $X_{\rm O_\textsc{III}} = \rm O/H$, and we adopt the relative
abundances (comparing to H) of N and O to be $9.3 \times 10^{-5}$ and
$5.9\times 10^{-4}$ \citep[i.e.\ the \HII\ region abundances in][]{Savage1996}.

We calculate \mmin\, using the \NII\ 122-\mum\, \OIII\ 88-\mum\ and \OIII\
52-\mum\ lines for both kinds of stacked spectrum and list them in
Table~\ref{tab:stack}. We find that \mmin\ derived using only the \OIII\
88-\mum\ lines are about 12--13\% of those derived using the \NII\ lines,
likely because the \OIII\ 88-\mum\ lines trace highly ionised gas and the \NII\
lines probe almost all phases of H$^+$, including the diffuse phase as well as
gas ionised by less energetic UV photons. Similarly the \mmin\ derived using
\OIII\ 52-\mum\ lines are 20\%--30\% of that derived using \OIII\ 88 \mum,
likely due to the much higher critical density of \OIII\ 52-\mum, which is
almost solely contributed by the dense gas phase (\ne $> 10^3$\,\cmt).

We derive a \mmin\ of $\sim 7.4 \pm 1.5\times 10^8$\,\msun\ for the intrinsic
stacked spectrum and $180\pm 25/\mu \times 10^8$\,\msun\ for the scaled
spectral stack. The relative ratio between the \OIII{}-derived \mmin\ and the
\NII{}-derived \mmin\ are similar for the two different stacked spectra. We
list their \mmin\ values in Table~\ref{tab:stack}.

\subsection{ \NII\ lines}\label{nii} 

The two lines of \NII\ are particular interesting since they are excited solely
in ionised gas (\HII) given their ionisation potential of 14.53\,eV, i.e. a
value just above the ionisation energy of hydrogen, 13.6\,eV, which indicates that 
they are potentially related to star-forming activity. Both lines have
relatively low critical densities (44\,\cmt\ for \NII\ 205-\mum\ and
300\,\cmt\ for \NII\ 122-\mum, see Table~\ref{tab:ncrit}), meaning they are
easily excited in the diffuse ionised ISM.  Both lines are normally optically
thin at FIR wavelengths, even for high column densities and extreme conditions
\citep[e.g.][]{Goldsmith2015}.  Given their long wavelengths, they are much
less affected by dust extinction compared to their optical/NIR transitions. As
such, these lines are independent extinction-free indicators of the current SFR
\citep[e.g.][]{Zhao2013}.

Both \NII\ lines lie at high frequencies that are inaccessible from
ground-based telescopes. Only FIR space telescope missions (e.g.\ {\it ISO} and
\herschel) could detect the line in the Milky Way and some local galaxies
\citep[e.g.][]{wright1991, Goldsmith2015, Zhao2016a, HC2016,Croxall2017}. The
\NII\ lines, especially the \NII\ 205-\mum\ line, are found to be usually
10--$50\times$ weaker than the \CII\ 158-\mum\ line, making it difficult to
detect from redshifted galaxies. After some early searches in the distant
Universe \citep[e.g.][]{Ivison1996}, ALMA has begun to observe these lines in
high-redshift galaxies \citep[e.g.][]{Ferkinhoff2015,Pavesi2016}.
Self-absorption may also play a role in the \NII\ line ratio, which will be
discussed in Section~\ref{sec:absorptiong}. 

\begin{figure}
  \centering
  \includegraphics[width=0.5\textwidth]{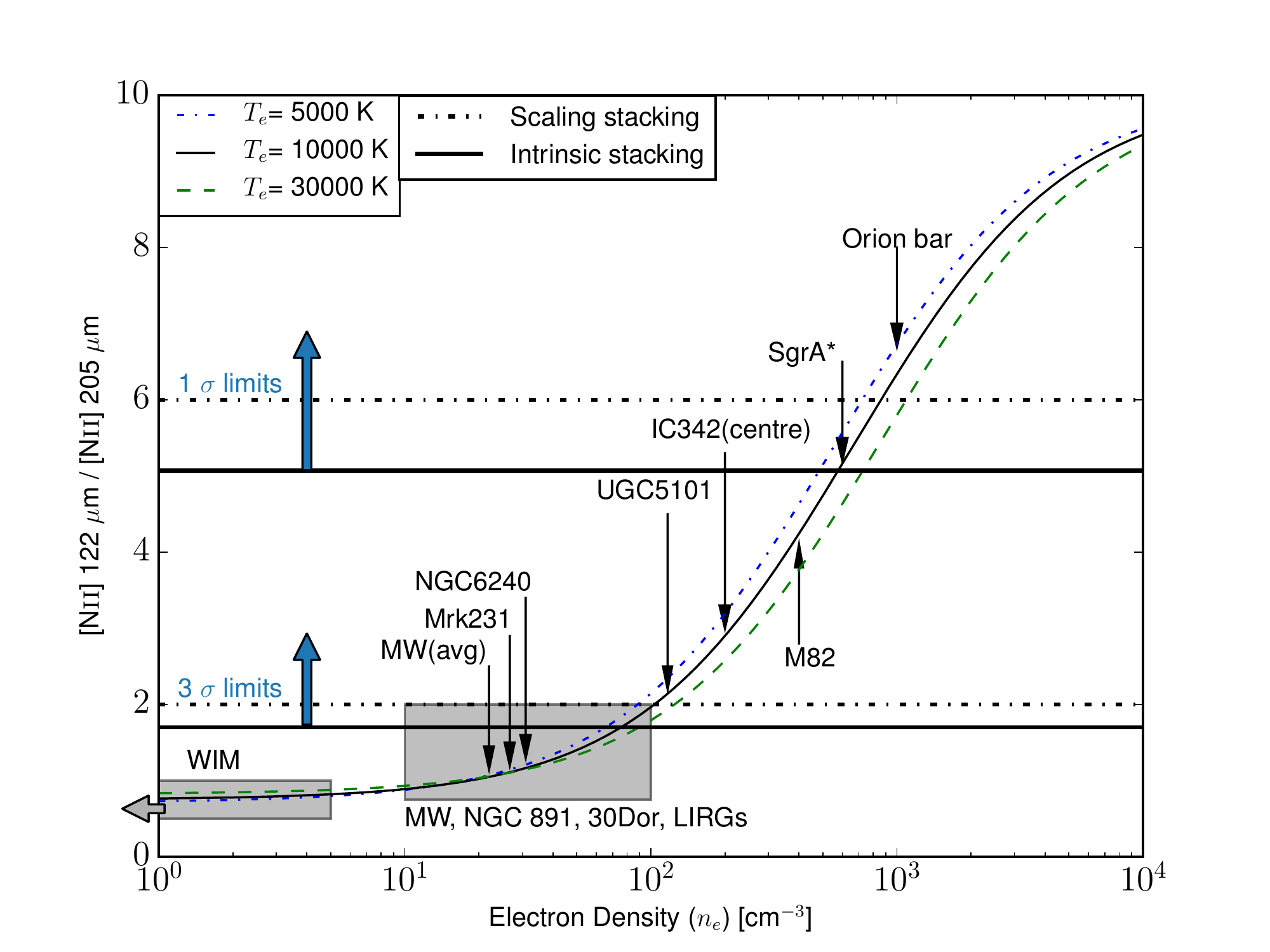}
  \caption{ Theoretical line luminosity (in \Lsun) ratio between the \NII\
          122-\mum\ and \NII\ 205-\mum\ lines (\RNII) as a function of
          electron density (\ne), following \citet{Rubin1985} and
          \citep{Draine2011book}.  We present \RNII\ models for electron
          temperatures \te\ $= 5,000$\,K (blue dotted), 10,000\,K (black solid)
          and 30,000\,K (green dashed).  The thick solid and dash dot lines
          show the 1-$\sigma$ and 3-$\sigma$ lower limits of the observed line
          ratios of \RNII, derived from intrinsic stacking and scaling
          stacking, respectively. We also overplot typical ranges of \ne\ measured with 
          the same \RNII, in the disks of nearby galaxies, in the central regions
          of galaxies, in the Galactic Centre, and for a few Galactic \HII\ regions.  }
\label{fig:NII_ne}
\end{figure}

In Fig.~\ref{fig:NII_ne}, we plot the theoretical line ratio between the \NII\
122-\mum\ and \NII\ 205-\mum\ lines (\RNII) following \citet{Rubin1985} and
\citep{Draine2011book}. We find that the minimum value of \RNII\ is around
unity at relatively low electron densities ($\sim 1$--10\,\cmt); at high
electron densities ($\ge 10^{4-5}$\,\cmt), \RNII\ saturates at around 10.  We
present \RNII\ models for three electron temperatures (\te\ = 5,000, 10,000 and
30,000\,K) in Fig.~\ref{fig:NII_ne}, showing \RNII\ as a monotonically
increasing function of electron density for each temperature. \RNII\ is not
sensitive to \te, but is very sensitive to \ne.  Clearly, this makes the \RNII\
ratio an excellent tracer of \ne, with little degeneracy in \te.  Furthermore,
the ratio of \RNII\ seems sensitive across a wide range of \ne\ between $10$
and $10^4$\,\cmt.

We over-plot a few representative data points measured from the same far-IR
\NII\ pair from the literature, trying to compare different \ne\ results
obtained using the same tracer. In our Milky Way, an average \ne\ of 22\,\cmt\
was found using observations made by the {\it Cosmic Background Explorer}
\citep[COBE,][]{wright1991,Bennett1994}.  Using \herschel, \citet{Goldsmith2015}
observed \NII\ lines along roughly 100 sight lines across the Galactic disk,
mostly close to the inner parts of our Galaxy ($\pm50^\circ$ of Galactic
longitude), and they found most to have \ne\ between 10 and 100\,\cmt (see
their Figure~19). In Fig.~\ref{fig:NII_ne}, the average conditions of the warm
ionised medium (WIM) are plotted as a gray box in the bottom-left corner. The
electron density of the WIM is quite uncertain, and is often taken to be $\sim
0.01$--0.1\,\cmt\ \citep[e.g.][]{Taylor1993, Goldsmith2015,Persson2014}.
\citet{Heiles1996} found a low volume filling factor and a relatively high \ne\
of 5\,\cmt, using 1.4-GHz radio recombination lines in the Galactic plane. In
our plot we set this value as the highest plausible \ne\ value for the WIM, and
use an arrow showing that it might also be much lower than the grey box.

We also plot the \ne\ range measured in NGC\,891 \citep[][]{Hughes2015}, as the
representative value found for nearby quiescent spirals and in the Milky Way
\citep[][]{Goldsmith2015}. A similar \ne\ range has been found in the starburst
region, 30 Doradus, in the Large Magellanic Cloud \citep[LMC,][]{Chevance2016}. For
local normal galaxies \citep[e.g.][]{HC2016} and LIRGs
\citep[e.g.][]{Zhao2016a,DS2017} most \ne\ values range between between 10 and
100\,\cmt, too. For most local ULIRGs, although higher SFR densities are
expected, they lie in the same region as spiral galaxies with the maximum, \ne\
$\sim 120$\,\cmt\ found in UGC\,05101 \citep[][]{Spinoglio2015} via \NII\
lines. Higher \ne\ values have also been found in a few spatially resolved
regions in our Galaxy and in some nearby starburst regions. In the central
regions of IC\,342 \citep[][]{Rigopoulou2013} and M\,82
\citep[][]{Petuchowski1994}, \ne\ values are much higher than the average for
the disks. Interestingly, the \ne\ measured in Sgr\,A$^\star$
\citep[][]{Goicoechea2013} and the ionising peak of the Orion bar
\citep[][]{BS2012} show the highest values, indicating high-density \HII\
regions compared to the lower densities found to represent the average
conditions in galaxies.

Unfortunately, our stacked spectra only give an upper limit for the \NII\
205-\mum\ line flux, so we can provide only a lower limit for \RNII ($>2$, for a
3-$\sigma$ limit), which corresponds to an \ne\ of $>100$\,\cmt, i.e.,
considerably higher than the global average conditions of most local
star-forming galaxies.  In Section~\ref{sec:sed_stacking}, we find an average
intrinsic IR colour of $S_{70}/S_{100}\sim 0.8$, which corresponds to $\sim
1.15$ in $\nu\,  S_{70}/\nu\, S_{100}$. Our \ne\ limit is roughly consistent
with the correlation found between $\nu\, S_{70}/\nu\, S_{100}$ and \ne\ by
\citet{HC2016}. Derived from Equation~6 in \citet{HC2016}, \ne\ is $\sim
90$\,\cmt, which lies at the high end of their galaxy sample, and has a wide
dynamical range of \ne\ and \RNII\ for comparison. Most local star-forming
galaxies, active star-forming regions in the LMC (e.g., 30~Dor), normal spirals
(e.g., NGC~891), LIRGs and ULIRGs (e.g., Mrk~231 and NGC~6240), have a global
average \ne\ between 10 \cmt\ and 100 \cmt, and show no obvious relations with
the global SFR, or \lir.  The high \ne\ found for our stacked DSFG spectrum
indicates higher pressure conditions for the ionised gas in their
star-formation regions, more like M~82, SgrA$^\star$, IC 342 centre and that
found in the Orion bar.  On the other hand, the electron temperature is not
expected to change dramatically (see also Sect.  \ref{radiationfield}), so
higher \ne\ values indicate higher pressures of the ionised gas phase.

In our Milky Way, younger \HII\ regions (hyper-compact \HII\ and ultra-compact
\HII, UC\HII) are found to have higher \ne\ \citep[e.g.][]{Churchwell2002},
however the high \ne\ likely cannot be attributed to the age of the \HII\
regions because the typical lifetime of an UC\HII\ region is only $\sim 10^5$
years, too brief to play an important role here. Moreover, the extremely high
SFRs found in DSFGs produces a prodigious flux of cosmic rays, $\sim
100$--$1000 \times$ the Galactic value, which efficiently heats up the dense
cores that are protected by the dust from UV photons, thereby increasing the
Jeans mass of the dense cores \citep[e.g.][]{PPP2010}. The most massive dense
cores can maintain their pressure via self-gravity, helping to confine \HII\ to
relatively small regions, such that the average \ne\ is systematically higher.

\subsection{ [O\,\textbf{\sc iii}] lines}\label{oiii} 
 
The \OIII\ 88-\mum\ and 52-\mum\ lines are important coolants of the ionised
phase of ISM in DSFGs because of the high abundance of oxygen. The ionisation
potential of \OIII\ is 35.12\,eV, which corresponds to very energetic physical
conditions, e.g.\ the UV radiation from very hot massive stars (O and B stars),
X-rays or shocks \citep[e.g.][]{stasiska2015}. As with the \NII\ lines, \OIII\
line ratios are not sensitive to the electron temperature. The critical
densities of the \OIII\ 88-\mum\ and 52-\mum\ transitions are $\sim 500$ and
3600\,\cmt, respectively, making them a sensitive probe of \ne\ in relatively
dense \HII\ regions.

\begin{figure}
  \centering
  \includegraphics[width=0.5\textwidth]{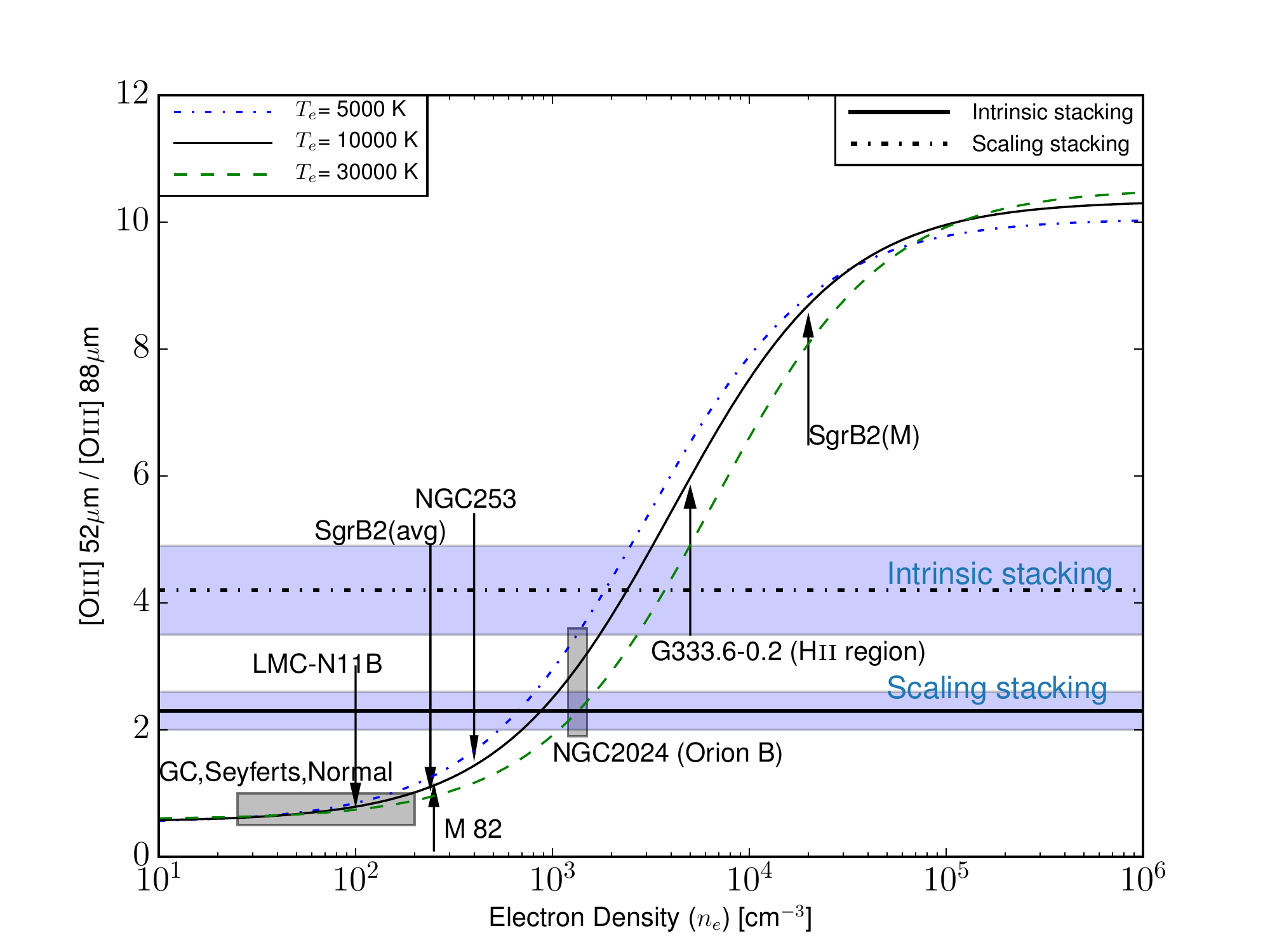}
  \caption{Theoretical line luminosity (in \Lsun) ratio between the \OIII\
          52-\mum\ and \OIII\ 88-\mum\ lines  (\ROIII), as a function of
          $n_{\rm e}$, following \citet{Rubin1985}.  We present \ROIII\ models
          for electron temperatures, \te\ = 5,000\,K (blue dotted line),
          10,000\,K (black solid line) and 30,000\,K (green dashed line).  The
          thick solid line and dash dot line show the observed \ROIII\ ratios
          derived from intrinsic stacking and scaling stacking, respectively.
          The blue shadow regions show the 1$\sigma$ error bar of the observed
          ratios.  We also overplot typical ranges of \ne\ found in the disks
          of nearby normal galaxies, the LMC, the central regions of our
  Galaxy, and a few Galactic \HII\ regions.}
\label{fig:OIII_ne}
\end{figure}

In Fig.~\ref{fig:OIII_ne}, we plot the theoretical line ratio of the \OIII\
52-\mum\ and 88-\mum\ (\ROIII) lines, following \citet{Rubin1985}.  The range
where these lines are relatively sensitive to electron density is between
$10^2$ and $10^5$\,\cmt, corresponding to line ratios, \ROIII, of $\sim 0.7$
and $\sim 10$, respectively. Similar to Fig.~\ref{fig:NII_ne}, we plot \ROIII\
with \ne\ of 5,000, 10,000 and 30,000\,K, which shows that \ROIII\ is not
sensitive to $T_{\rm e}$.

The \OIII\ 52-\mum\ and 88-\mum\ transitions have been observed from both
Galactic targets and external galaxies. The grey box in the bottom left corner
of Fig.~\ref{fig:OIII_ne} shows the \ne\ range measured with the same \OIII\
pair in the central region of the Milky Way \citep[i.e.][]{RF2005}, nearby
Seyfert galaxies \citep[e.g.][]{Spinoglio2015} and normal galaxies
\citep[e.g.][]{Negishi2001}.  These values are consistent with the \ne\ range
measured from the \NII\ lines, though \OIII\ lines are not very sensitive to
\ne\ below a few hundred \cmt. We also label the measured \ROIII\ in the
centres of a few typical nearby star-forming galaxies \citep[e.g.\ M\,82,
NGC\,253; see][]{Duffy1987,Carral1994} and the average \ROIII\ measured in
Sgr\,B2 \citep[][]{Goicoechea2004}, which are somewhat higher than the average
value measured in galaxies as a whole. This indicates a higher average \ne\ in
the centre of galaxies, consistent with the trend found in M\,51 and NGC\,4449
using optical observations \citep[e.g.][]{Gutierrez2010}.

We also overplot the \ROIII\ values found in Galactic sources, including the
range found in the \HII\ regions of NGC\,2024 \citep[Orion~B][]{Giannini2000},
G333.6$-$0.2 \citep[][]{Colgan1993}, and Sgr\,B2(M) \citep[][]{Goicoechea2004}.
These targets show the highest \ne, especially Sgr\,B2(M) where several
ultra-compact \HII\ regions reside in a very small volume.

In Section~\ref{nii} we estimated \ne\ using \NII\ 122-\mum\ and the upper
limit from \NII\ 205-\mum, finding that the \ne\ found in our DSFGs
is higher than 100\,\cmt. From the \mmin\ and \ne\ derived from \OIII\
lines, at least 10\% of the ionised gas has a very high \ne,
$\sim 10^3$--$10^4$\,\cmt. The \ne\ derived from \NII\ and \OIII\ are
both higher than the densities found in local star-forming galaxies
\citep[e.g.\ 30\,\cmt][see also Figs~\ref{fig:NII_ne} and
\ref{fig:OIII_ne}]{HC2016}, where the derived \ne\ values are
consistent with measurements using optical lines
\citep[e.g.][]{Zaritsky1994}.

More recent evidence from optical and near-IR studies suggest that \ne\ in
star-forming galaxies at $z\sim 2$ seem to be systematically higher than in
local star-forming galaxies, by up to two orders of magnitude
\citep[e.g.][]{Bian2016, Sanders2016}. These studies find a median \ne\ of
$\sim 250$\,\cmt, consistent with our lower limit for \ne\ from the \NII\
lines, indicating an increase of \ne\ with redshift for both normal
star-forming galaxies and DSFGs.

As revealed in resolved studies of nearby galaxies, \ne\ tends to be higher in
the centres of galaxies, likely due to the strength of the radiation field and
the density of the local ISM \citep[e.g.][]{HC2016}. High-redshift DSFGs may
have both a higher radiation field, revealed by the \OIII\ to \NII\ ratio, and
a denser ISM, which confines the expansion of \HII\ regions and keeps \ne\
relatively high.  However, optical measurements in both galaxy populations from
the Sloan Digital Sky Survey (SDSS) and local analogs of high-redshift
star-forming galaxies find that those with higher SFR surface densities
($\Sigma_{\rm SFR}$) tend to have lower \ne\ \citep[][]{Bian2016}; this is not
consistent with far-IR \NII\ results, which show a tight correlation between
$\Sigma_{\rm SFR}$ and \ne\ \citep[][]{HC2016}.
 
\subsection{ \CII\ emission from ionised gas}\label{cii_ionised} 

To trace the origins of the \CII\ 158-\mum\ emission, it is important to
separate between the \CII\ luminosity fraction contributed by the neutral gas
and that contributed by the ionised gas. The fraction from ionised gas is
normally small $\sim 20-40\%$ \citep{Croxall2017}, but is not negligible,
especially given the uncertainty in the C/N abundance ratios.  The \CII/\NII\
line ratio is sensitive to the ionised gas fraction that contributes the \CII\
emission, because \NII\ lines come only from the ionised gas, and should pick
up the majority of it, given their low critical densities.

In the ionised gas phase, both the \NII\ and \CII\ lines are excited by
collisions with electrons, and neither line is sensitive to electron
temperature. Using the line ratio between \NII\ lines and \CII\ 158-\mum, one
can probe the contribution to the observed \CII\ intensity from the ionised gas
phase with little dependence on \ne\ and \te\
\citep[e.g.][]{Oberst2006,Oberst2011,Hughes2015,Pavesi2016}.

\begin{figure}
  \centering
  \includegraphics[width=0.5\textwidth]{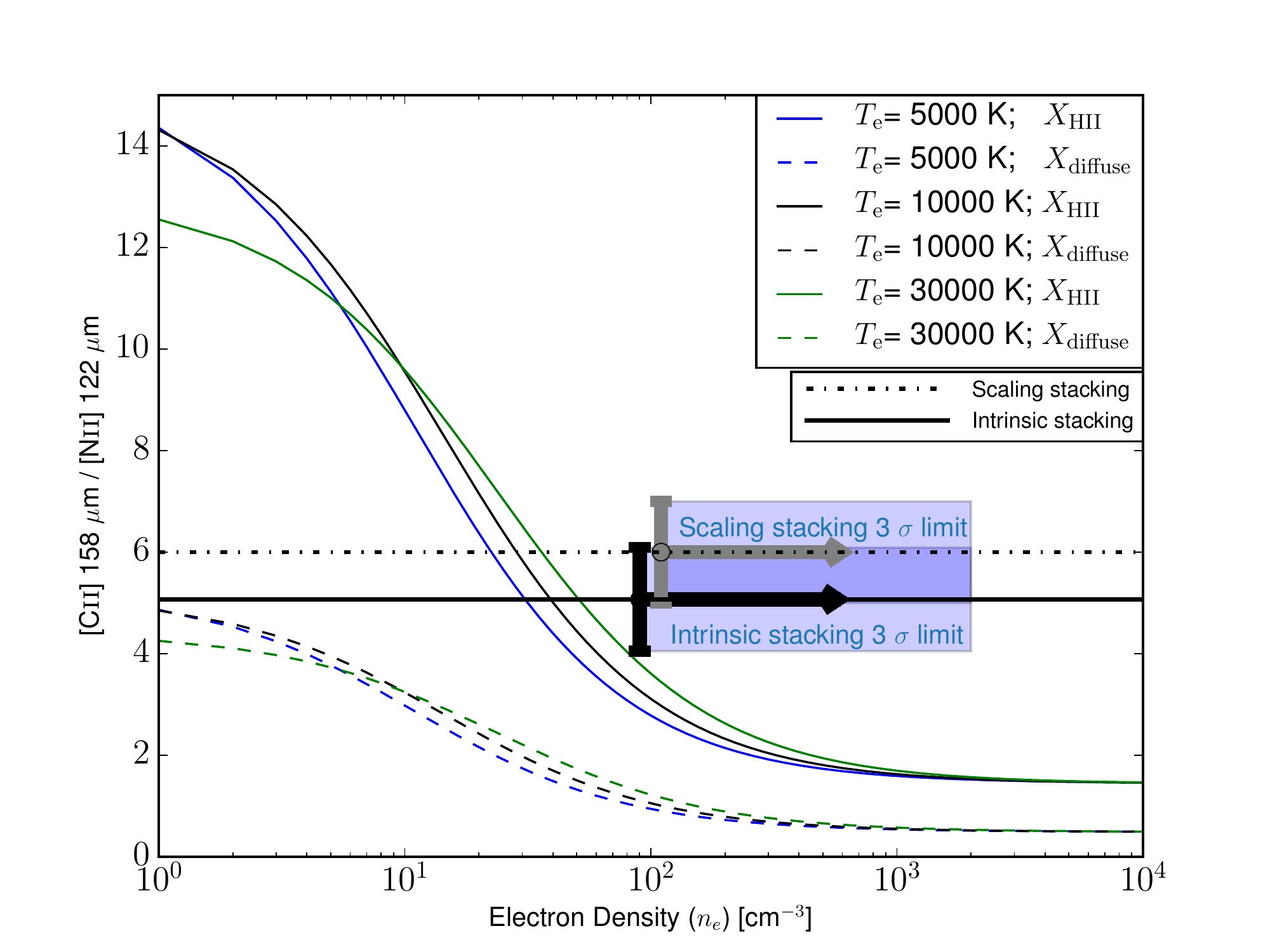}
  \caption{Theoretical line luminosity (in \Lsun) ratio of \CII\ 158-\mum\ and
          \NII\ 122-\mum\ as a function of electron density. This calculation
          considers the \CII\ 158- \mum\ emission that is excited by
          collisions with electrons only. We plot the \CII\ 158-\mum/\NII\
          122-\mum\  line ratios for 5000\,K (blue), 10,000\,K (black) and
          30,000\,K (green). Solid curves show the ratio using the Carbon and
          Nitrogen abundances measured in Galactic \HII\ regions; dashed curves
          show the ratio with abundances in diffuse gas
          \citep[e.g.][]{Savage1996}.  The observed ratios between \CII\
          158-\mum\ and \NII\ 122-\mum\ in our stacked DSFG spectrum are
          plotted as thick black line and dash dotted line, for intrinsic
          stacking and scaling stacking, respectively. The blue shadow regions
          show the 1$\sigma$ error bar of the observed \CII\ 158/\NII\ 122-
          ratios.  The two arrows in the x-axis show the lower 3-$\sigma$
          limits of electron density derived from \RNII\ ratios derived from
          the two stacking methods.  The maximum allowed electron density is
  2000\,\cmt, which is the \ne\ derived from \OIII\ lines. }
\label{fig:cii_nii}
\end{figure}

Although the \NII\ 122-\mum\ line has a higher critical density ($\sim
300$\,\cmt) than that of the \CII\ line, the ratio dependency on the electron
density  can be well modeled.  We can use the upper limit of \ne\ derived from
the two \NII\ lines to set a further constraint. In Fig.~\ref{fig:cii_nii}, we
plot the theoretical line ratio between \CII\ 158-\mum\ and \NII\ 122-\mum\ in
the ionised gas phase. The observed \CII/\NII\ ratios include \CII\ emission
with contribution from neutral gas phase, so the ratio is always above the
theoretical curves (for the ionised gas phase only).  The line ratio of \CII\
158-\mum\ to \NII\ 122-\mum\ does depend on \ne.  However, if the electron
density is higher than a few hundred \cmt (well above the critical densities of
both lines), both the \NII\ 122-\mum\ and the \CII\ 158-\mum\ lines will be
excited efficiently.  For the abundances of N/H and C/H, we adopt $7.76 \times
10^{-5}$ and $3.98 \times 10^{-4}$, respectively -- typical \HII\ region
values\citep{Savage1996}. Because the \CII/\NII\ ratio is sensitive to
abundances, we overplot the ratio using typical Galactic abundances found in
diffuse clouds \citep[$7.94 \times 10^{-5}$ for N; $1.38 \times 10^{-4}$ for C
--][]{Savage1996} for comparison, and to present the uncertainty. More detailed
discussion on the abundances will be presented in Section~\ref{sec:abundances}.

We plot the \CII\ 158-\mum/\NII\ 122-\mum\ line ratio curves at \te =
5,000\,K, 10,000\,K and 30,000\,K, which are consistent with each other,
showing that these ratios are not sensitive to the electron temperature.  The
\CII\ 158-\mum/\NII\ 122-\mum\ ratio has a stronger dependency on the
electron density, which can be constrained from the line ratio of the two \NII\
lines.  All line ratios decrease monotonically with electron density, and
saturate at the high density end ($\ge 1000$\,\cmt), where \ne\ is much higher
than the critical densities of both lines. 

The observed ratio of \CII\ 158-\mum\ and \NII\ 122-\mum\ in our stacked DSFG
spectrum is $5.1\pm 1.0$, which is plotted as the shadowed area.  We find a
lower limit for the electron density of 100\,\cmt, derived from the $3\sigma$
limit for the \NII\ line ratios; the upper limit for \ne\ is 2000\,\cmt,
derived using the \OIII\ lines, which are from the more energetic and denser
\HII\ regions. If we use the Galactic diffuse gas abundances \citep{Savage1996},
the contribution from ionised gas to the \CII\ line is $\sim 10$--15\%.
However, if \HII\ region abundances are adopted \citep{Savage1996}, the ionised
gas could contribute up to 60\% of our \CII\ emission.  The ratios derived from
the diffuse gas abundances is about $3\times$ less than those derived from
abundances in \HII\ regions. If an electron density of 100\,\cmt\ and an
electron temperature of 10,000\,K are assumed; this contribution drops to
30--40\% for an \ne\ of 1000\,\cmt. 

\subsection{Radiation field}\label{radiationfield}

The \NII\ 122-\mum\ and \OIII\ 88-\mum\ lines have similar critical densities
$\sim 300$--500\,\cmt, but their ionisation potentials differ by $\times 2$
from 14.5 to 35\,eV. Both lines are normally optically thin, and are
insensitive to the electron temperature. Their ratio is sensitive only to the
abundance and radiation field, which is mainly controlled by the relative
fraction of UV photons at different energies \citep[e.g.][]{Ferkinhoff2011}.

Following \citet{Ferkinhoff2011}, we can probe the stellar effective
temperature using the ratio between the \NII\, 122-\mum\  and \OIII\ 88-\mum\
lines in our stacked DSFG spectrum.  We adopt the \HII\ region models
calculated by \citet{Rubin1985}, who derived the theoretical intensities of the
ionised lines with different metal abundances, electron densities and stellar
effective temperatures.  Model `K' in \citet{Rubin1985} has elemental
abundances close to the values in Galactic \HII\ regions
\citep[e.g.][]{Savage1996}. The line intensity is proportional to abundance for
optically thin lines, so we correct the line ratio map with the Galactic \HII\
region abundances \citep[][see also Section~\ref{cii_ionised}]{Savage1996}, to
provide a more consistent and realistic comparison.  With the \HII\ region
abundances, the \OIII/\NII\ ratio is about $1.15\times$ higher than those in
\citet{Rubin1985} -- this does not influence our conclusions. The ratios
between \OIII\ 88-\mum\ and \NII\ 122-\mum\ from both our stacking methods are
$\sim 2.2\pm0.5$ (intrinsic stacking) and 2.4$\pm0.3$ (scaling stacking), fully
consistent with each other.  These values correspond to a stellar effective
temperature of $\sim 35,000$\,K, which corresponds to O8--O9 stars, according
to \citet{Vacca1996}.  These values are less than that obtained from the
resolved compact starburst center in a local ULIRG, Arp299, but are more
consistent with the resolved results in most local LIRGs and AGN hosts on
(sub-)kpc scales\citep{HC2018,HC2018b}.

However, the global ratio of \NII/\OIII\ is not just affected by abundances and
the hardness of the radiation fields, but also sensitive to variation in the
filling factors of the observing beam. The lower ionisation potential of \NII\
makes it expected to be more extended than the \OIII\ 88 \mum\ line. Only O
stars can emit powerful UV photons to excite this line. This means that our
\NII\ emission is contributed from both O stars and B stars.  On the other
hand, \nii\ lines have a very low critical densities, so its collisional
excitation saturates in \hii\ regions well before \OIII\ lines. If we zoom into
smaller \OIII\ emitting regions, the \OIII/\NII\ ratio would be higher than
that obtained from global average. The combination of the two leads to \nii\
dominating the extended, diffuse, ionized gas. So, our derived stellar
effective temperature is actually a lower limit for that in the dense \HII\
regions with \OIII\ emission.

To generate the observed \NII\ 122-\mum\ emission, $\sim 3 \times 10^7$ or
$\sim 5 \times 10^7$ O8 or O9 stars are needed, respectively, assuming an \ne\
of 1000\,\cmt\ in the \citet{Rubin1985} model. From the same models we can also
derive the ionised gas mass from the \NII 122-\mum\ emission.  We find that
different values of \ne\ give similar ionised gas masses (intrinsic), ranging
between 1--$2 \times 10^9$\,\msun, consistent with the \mmin\ derived in the
previous section. This is about $10\times$ less than that estimated for
SMM\,J2135$-$0102 by \citet{Ferkinhoff2011}. However, for the galaxies with CO
detections, the H$_2$ gas mass ranges from $1\times 10^{11}$ to $5\times
10^{11}$\,\msun. The ionised gas mass fraction is estimated to be less than 2\%
of the molecular gas mass, assuming an $\alpha _{\rm CO}$ conversion factor of
0.8 \citep[typical value for ULIRGs][]{2013ARA&A..51..207B}. This is consistent
with the ionised gas fraction found in local galaxies
\citep[e.g.][]{Brauher2008,Ferkinhoff2011}, but much less than the fractions
found in two high-redshift DSFGs \citep[35\% for Cloverleaf quasar  and 16\%
for SMM\,J2135$-$0102;][]{Ferkinhoff2011}. On the other hand, active galactic
nuclei (AGNs) can increase the UV radiation and bias the hardness of the
radiation field \citep[e.g.][]{Groves2004}, which shows that a high \OIII/\NII\
ratio can be easily achieved in the high-pressure narrow-line region of AGNs.

Using the [Si\,{\sc ii}] 34-\mum\ and [S\,{\sc iii}] 33-\mum\ lines in the
stacked PACS spectra for a similar sample of high redshift lensed starbursts,
\citet{Wardlow2017} found that SMGs on average contain AGNs, followed the
method proposed by \citet{Dale2006,Dale2009}. However, statistical studies
based on X-rays and mid-infrared spectroscopy studies find that the fraction of
DSFGs harbouring powerful AGN is small, thus they are unlikely to dominate the
gas heating \citep[e.g.][]{Alexander2005,Coppin2010,Wang2013}.  On the other
hand, in the stacked PACS spectrum \citet{Wardlow2017} found non-detection for
[O {\sc iv}], which is expected to arise from AGNs due to the high ionisation
potential. Thus we conclude that the AGN contamination to the far-IR lines
should be negligible for our stacked results.

\begin{figure}
  \centering
  \includegraphics[width=0.5\textwidth]{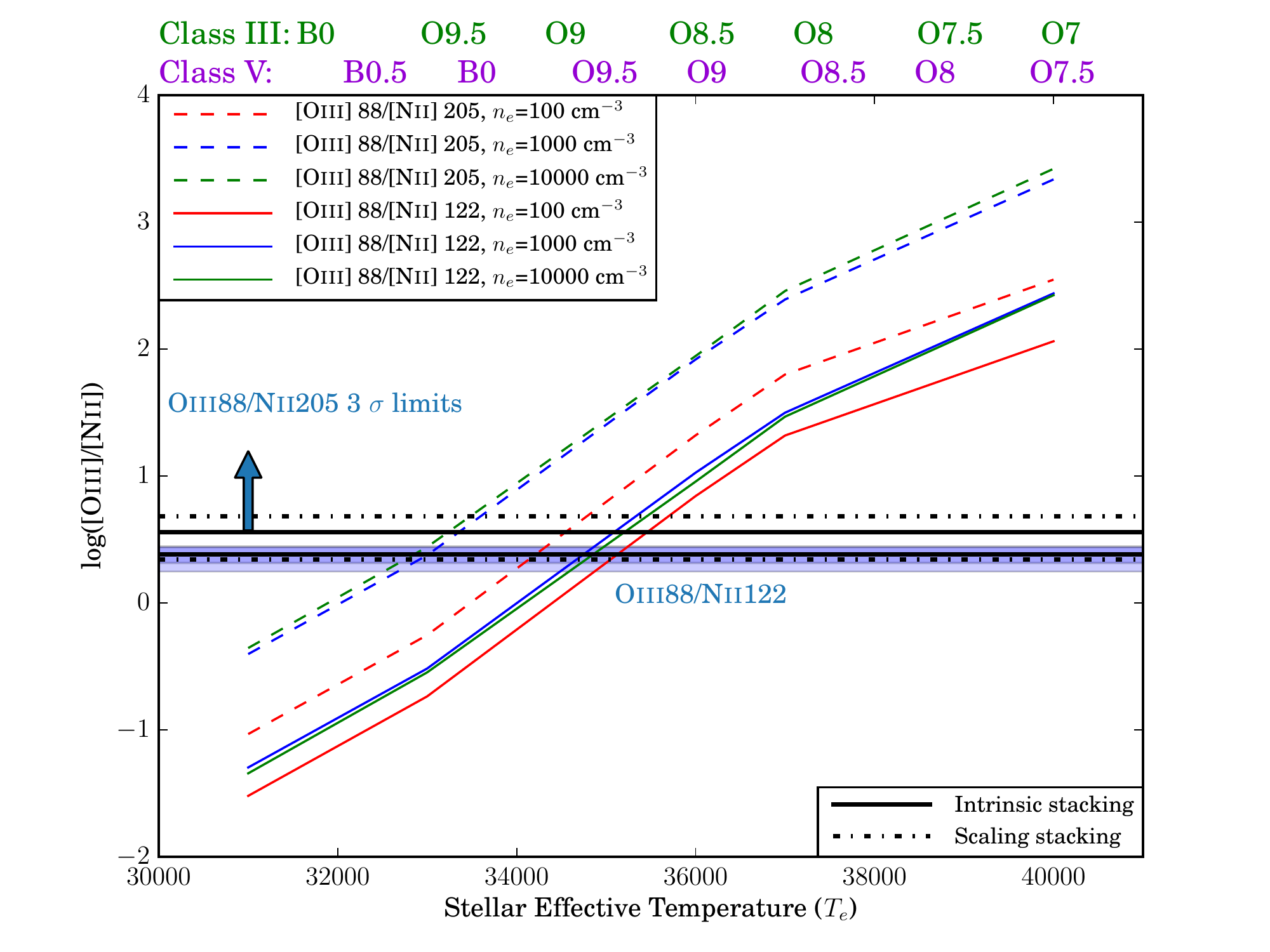}
  \caption{Theoretical line luminosity (in \Lsun) ratio of \OIII\ 88-\mum\ and
          the \NII\ lines as a function of the effective stellar temperature,
          derived from the \citet{Rubin1985} \HII\ models.  We plot the
          observed \OIII\ 88-\mum\ to \NII\ 122-\mum\ ratios and the upper
          limits of the observed \OIII\ 88-\mum\ to \NII\ 205-\mum\ ratios,
          derived from for intrinsic stacking (black thick line) and scaling
          stacking (dash dotted line), respectively. The upper axis labels show
          stellar types of OB stars of corresponding electron temperature for
          luminosity class {\sc iii} (Green) and luminosity class {\sc v}
          (Purple) stars, according to the classification of \citet{Vacca1996}.}
\label{fig:oiii_nii_ratio}
\end{figure}

\subsection{[O\,\textbf{\sc i}] lines} 

We detect \OI\ 63-\mum\ emission in our stacked spectrum, with an upper limit
for the \OI\ 145-\mum\ transition. These two lines have identical energy
potentials for further ionisation, but they have different upper level
energies, $\Delta\,E/k$, of 228 and 327\,K above the ground state. The critical
densities of the 145- and 63-\mum\ transitions are $\sim 10^4$ and
$\sim10^5$\,cm$^{-3}$, respectively, so both lines are sensitive to dense and
warm neutral gas. Their line ratio is a powerful diagnostic of the density and
temperature of the emitting regions, under the optically thin assumption
\citep[e.g.][]{Liseau2006,Tielens1985ApJ}. The heating of \OI\ lines could be
powered by radiative shocks \citep[e.g.][]{Hollenbach1989, Draine1993} or FUV
photons from massive stars, given the high SFRs of these galaxies.

Given the non-detection of the \OI\ 145-\mum\ transition, we can get an upper
limit for the ratio of \OI\ 63/145 $> 6$ ($1\sigma$) or $> 1.3$ ($3\sigma$).
These exclude a cold ($< 50$\,K) and optically thick scenario (see, e.g.,
Figure~4 of \citealt{Liseau2006}). If \OI\ 63-\mum\ is indeed optically thin,
then the \OI\ 145-\mum\ emission is expected to be even further below our
detection limit.

If the \OI\ 63-\mum\ emission is dominated by dissociative shocks, its
intensity is proportional to the mass-loss rate, which is caused by the
dominance of the cooling via \OI\ 63-\mum\, \citep{Draine1993,Hollenbach1989}.
However, the excitation from PDRs is another strong power source for \OI\
lines.  The \CII${158-\um}$/\OI${63-\um}$\, luminosity ratio provides a way to
discriminate between PDRs and shocks: this ratio is generally $\le 10$ in PDRs
and $\ge 10$ in shock-dominated regions \citep{Hollenbach1989}. Our stacked
DSFG spectrum shows \OI${63-\um}$/\CII${158-\um}$ $\sim 4$, on the high end of
the resolved local starburst regions or AGN centres, and much higher than the
values measured in normal galactic disks\citep{HC2018}.  Our global average
ratios indicate that the \OI\ emission must have some contribution from shocks.
but shock is also not dominant.  Moreover, the \OI\ 63-\mum\ emission often
suffers from self-absorption effects (see Section~\ref{sec:absorptiong}), which
has been seen both in the Milky Way molecular clouds
\citep[e.g.][]{Ossenkopf2015}, and in local starbursts
\citep[][]{Luhman2003,Rosenberg2015}.  If the \OI\ self-absorption also plays a
role, the intrinsic ratio of \OI${63-\um}$/\CII${158-\um}$ should be even
higher.

\section{Caveats}\label{sec:caveats}

\subsection{Statistics in the line ratios}

In this paper, we have stacked \herschel\ SPIRE FTS spectra and derived the
average physical conditions of the fine-structure lines using their relative
ratios. We have to note that the average of ratios (e.g., $<$ \NII\ 205-\um/
\NII\ 122-\um\, $>$ )  are not necessarily the same as the ratio of the
averages $<$ \NII\ 205-\um\,$>$/$<$ \NII\ 122-\um\,$>$, which introduces an
extra uncertainty in the statistics \citep[e.g.,][]{Brown2011}. 

For most lines, we do not have individual detections of the lines in the SPIRE
spectra, so it is not possible to directly measure individual line ratio and
get the average.  Although a ratio of line flux averages (observational
constraint) might be similar to an average of ratios (physical condition), the
expectation -- the average of the ratio --  would always be larger than the
ratio of the averages. Thus, our line ratios from stacked spectra are
underestimated \citep{Brown2011}. The derived \ne\ and radiation hardness are
likely more extreme than those we derived. Furthermore, this speculation is
especially true when the undetected lines are upweighted when adopting average
weighted by noise levels. 

\subsection{Biases in stacking } 

We have stacked a sample of $\sim$ 40 galaxies at a redshift range between 1.5
to 3.6. The \herschel\ SPIRE FTS has a fixed observing frequency range so the
final line coverage is not uniform. Thus, the stacked averaged lines are
actually from different galaxies, which makes another potential bias to the
final line ratios.  This effect is obvious at both the low- and high-frequency
ends, especially for the \oiii\ 52-\mum\ and \oi\ 63-\mum\ lines, which are
more contributed by galaxies at higher redshifts.  Moreover, the spectra are
noisier in the band edges, which makes the S/N even lower for these lines. 

Our first two stacking methods (intrinsic luminosity and scaling to Eyelash)
both adopt weights of $1/\sigma^2$, which preferentially weight galaxies with
better S/N levels (Eyelash) and smaller redshifts. The intrinsic luminosity
weighting is relatively more biased to higher S/N targets, whilst the scaling
method is more biased to higher luminosity targets. Although a straight
(equally weighted) stacking could avoid weighting biases, such stacking will be
highly dominated by non-Gaussian noises, especially the edges of both bands
that cover across the whole frequency range. A median stacking of the \LIR\
normalised spectra is less biased by galaxy properties, and it can test if some
weak features are common for the sample, e.g., the OH 119-\um\ absorption line.
However, the median stacked spectrum is noisier and seems still contaminated by
the noisy edges of individual band. This method also does not allow to
calculate median of the dust continuum or \LIR\ using the same weighting from
spectral line stacking.

\subsection{Absorptions to the fine-structure lines }\label{sec:absorptiong}

During our analysis, we accounted for the integrated line luminosities and
neglected self-absorptions. This might be severe for the \oi\ lines, which have
been found in various Galactic and extra-galactic
conditions \citep{Fischer1999,Liseau2006,GA2012,Fischer2014}. The \CII\
158-\mum\ line also often shows self-absorption features in Galactic studies
\citep[e.g.][]{Gerin2015,Graf2012}, and it might also affect extra-galactic
studies as well \citep{Malhotra1997}, which is very difficult to identify with
limited angular resolution \citep{Ibar2015}.

We also examine the possibility of the differential self-absorption to the
\NII\ line pair, due to the relatively lower  excitation energy of the
205-\mum\ line compared to the 122-\mum\ line. Following Equation \ref{eq:Weq},
the equivalent width ratio between the two lines are $W_{\rm eq}(205) / W_{\rm
eq}(122) \sim  1.026$, indicating that their theoretical absorption depths are
similar. Only a high optical depth can produce biased self-absorption to
eliminate the observed \NII\ 205-\mum\ line flux. However, due to the small
abundance of nitrogen, it is unlikely that the \NII\ 205-\mum\ line can be
optically thick \citep{Langer2016}.  Furthermore, self-absorption has not been
found in local galaxies in our Milky Way galaxy, where \CII\ and \OI\ have been
found to have self-absorption in some cases \citep{Gerin2015}. We conclude that
our high \RNII\  is unlikely to be caused by self-absorption, but more detailed
observations are needed to further support this conclusion. 

Another, possible contamination that may bias the line measurement is an
absorption against strong background continuum emission. This has been seen in
\CII\ absorption in high-redshift galaxies \citep[e.g.,][]{Nesvadba2016}
against strong continuum sources powered by intense star-formation or AGNs.
Although this may not be a dominant bias for the fine-structure lines, such
an effect could contribute very high level of `line-deficit' \citep{Nesvadba2016}.

On the other hand, the upper energy levels of the \NII\ lines are 70-K and
188-K for the $\rm ^3 P_1-^3 P_0$  (205-\mum) and $\rm ^3 P_2-^3 P_1$
(122-\mum) lines, respectively. Therefore, the WIM component of them can be
easily absorbed against background sources with high brightness temperatures.
Using high-spectral-resolution observations from the HIFI instrument onboard
\herschel, \citet{Persson2014} detected absorption features of the \NII\
205-\mum\ lines toward a few massive Galactic star-forming regions. They found
that \NII\ emission from ionised gas in the dense and hot \HII\ regions is
likely absorbed by widespread low-density and relatively low-temperature WIM
gas in the foreground. 

Because the upper level temperature of \NII\ 205-\mum\ is relatively lower than
that of \NII\ 122-\mum, the former is easily to get absorbed. The \RNII\ ratio
obtained from global volume average observations can be biased by the
differential absorptions, and may further bias the final derived \ne\ to lower
values (suggestive of more diffuse gas) and this has possibly already
influenced observational results in local galaxies \citep[e.g.,][]{Zhao2016a,
HC2016}.  Observations at high-angular and high-spatial resolution of the \NII\
lines are needed to avoid the confusion from the absorption of the \NII\ line
in low-density gas and any WIM contribution.

We have neglected the optical depth of dust throughout this study. Dust can
have non-negligible optical depths at mid- to far-IR wavelengths, especially
for the compact starbursts, e.g. Arp220 has an optical depth of $\sim$ 1 even
at the 3-mm band \citep{Scoville2017}, i.e. it is optically thick at far-IR.
\citet{Lutz2016} found that local ULIRGs generally have compact sizes and their
optical depths in the FIR can be close to optically thick on average.  It is
difficult to estimate the accurate dust attenuation of the lines, but this
effect may contribute to the line deficit over the sample \citep{Fischer2014}.

\subsection{Elementary abundances}\label{sec:abundances}

Our study adopted line ratios between different elements to derive physical
conditions, e.g., \CII/\NII\ for the contribution of ionised gas, and
\OIII/\NII\ for the hardness of the radiation field. These are based on the
assumption of the elementary abundances, which are actually fairly uncertain,
due to unclear nucleosynthesis and galactic chemical evolution processes
\citep[e.g.,][]{Matteucci2001}.  

The C/N abundance ratio is still largely under debate. Based on Galactic
stellar determinations for the variations with metallicity \citep{Nieva2012},
\citet[][]{Croxall2017} find a decreasing C/N abundance ratio with increasing
metallicity traced by [O/H]. This is partially consistent with the results
found in dwarf galaxies, which show a C/N--[O/H] curve similar to a negative
parabola shape \citep{Garnett1995}. Furthermore, recent work by \citet{PG2017}
shows that the C/N abundance ratio linearly increases with [O/H] in starburst
galaxies with top-heavy IMFs, contradictory to the findings of
\citet{Croxall2017}. This indicates that the primary element channel for the N
production is not negligible in metal poor conditions, and this effect is more
sever for the top-heavy IMF conditions, which have more massive stars to supply
the primary N yields \citep{Coziol1999,PG2017}. It seems that it is likely not
appropriate to adopt the Galactic chemical evolution and Galactic chemical
abundances to starburst galaxies, which have systematically different evolution
tracks and IMFs \citep{Zhang2018}.

On the other hand, the O/N abundance ratio that adopted in the \OIII/\NII\ line
ratios also have large uncertainties. Using Galactic \HII\ regions,
\citet{Carigi2005} find a solo increasing trend of N/O ratio with [O/H], which
is consistent with the results found using optical spectra measured from
Galactic B-stars \citet{Nieva2012}. However, local starburst galaxies have
systematically biased N/O ratios from the trend found in the Milky Way
\citep{Coziol1999}.  More confusingly, \citet{Contini2017} shows that the N/O
abundance ratios in gamma-ray and supernova host galaxies at $z<4$ can not be
explained by stellar chemical evolution models calculated for starburst
galaxies, nor for the Milky Way, which suggests that different evolutionary
tracks need to be applied for various galaxy types and redshifts.

\section{Summary and concluding remarks} \label{sec:summary}

We present \herschel\ SPIRE FTS spectroscopy and PACS photometry of a sample of
45 gravitationally lensed DSFGs at $z = 1.0$--3.6, targeting the \CII\
158-$\mu$m, \NII\ 205- and 122-\mum, \OIII\ 88 and 52-\mum, \OI\ 63- and
145-\mum, and OH 119-\mum\ lines.  We obtained 17 individual detections of
\CII\ 158-\mum, five detections of \OIII\ 88-\mum, three detections of \OI\
63-\mum, and three detections of OH 119-\mum\ in absorption.

We find that the \CII/\LIR-\LIR\ ratio shows a deficit at high FIR
luminosities, high star-formation efficiency (\LIR/$M_{\rm H_2 }$) and higher
surface densities of star formation rates, consistent with the trends found in
local star-forming galaxies, ULIRGs, SPT sources and high-redshift starburst
galaxies found in the literature. 

To determine the average conditions of the ionised gas in our high-redshift
DSFG sample, we stack the SPIRE spectra using three different methods, each
with a different bias.  We derive physical properties of the ionised gas from
the stacked spectra using un-lensed intrinsic luminosity, and from the spectra
scaled to a common redshift.  In the stacked spectrum we detected emission
lines of \CII\ 158-\mum, \NII\ 122-\mum, \OIII\ 88-\mum, \OIII\ 52-\mum, \OI\
63-\mum, and OH in absorption at 119-\mum. Median stacking has lower S/N but
provides further evidences for the weak line detections.

Using the \NII\ 122-\mum\ detection and the upper limit for \NII\ 205-\mum,
we derive a lower limit for the electron density of $>100$\,\cmt, which is
higher than the average conditions found in local star-forming galaxies and
ULIRGs using the same \NII\ line pair. From the \OIII\ 63- and 145-\mum\
detections, the electron density is found to be $10^3$--$10^4$\,\cmt, which is
one-two orders of magnitude higher than that found in local star-forming
galaxies using the same \OIII\ lines. We also use \NII\ 122-\mum\ to derive
the ionised gas contribution to \CII\ 158-\mum\ and find the fraction of
ionised gas to be 10--15\%.  If we adopt the N and C abundances found in
Galactic \HII\ regions, this fraction can be as high as 60\%.  The \OI/\CII\
ratio indicates that the \CII\ emission is likely dominated by PDRs rather than
by large-scale shocks. Finally, we derived the OH column density from the OH
119-\mum\ absorption feature, which we find likely traces outflows driven by
the star formation.  

The physical conditions derived for the sample seem to be systematically more
extreme than for local star-forming galaxies and ULIRGs, i.e., in
star-formation rates, in electron densities derived from both total ionised gas
and dense ionised gas. We use the\NII/\OIII\ ratio to derive an average
radiation hardness of the sample, which indicates that the star-formation is
dominated by massive stars with masses higher than O8.5, modulo the potential
biases from diffuse ionised gas and uncertainties in the N/O abundance ratio.

\section*{Acknowledgements}

The authors are grateful to the referee for the constructive suggestions and
comments. 
ZYZ thanks J.D. Smith, Justin Spilker for sharing data. 
ZYZ thanks Donatella Romano for very helpful discussion. 
ZYZ, RJI, LD, SM, IO, RDG and AJRL acknowledge support from the European
Research Council (ERC) in the form of Advanced Grant,321302, \textsc{cosmicism}.  
RDG also acknowledges an STFC studentship. 
YZ thanks the support from the NSFC grant numbers 11673057 and 11420101002. 
CY was supported by an ESO Fellowship. 
US participants in {\it H}-ATLAS acknowledge support from NASA through a
contract from JPL.  
\herschel\ was an ESA space observatory with science instruments provided by
European-led Principal Investigator consortia and with important participation
from NASA.
PACS has been developed by a consortium of institutes led by MPE (Germany) and
including UVIE (Austria); KU Leuven, CSL, IMEC (Belgium); CEA, LAM (France);
MPIA (Germany); INAF-IFSI/OAA/OAP/OAT, LENS, SISSA (Italy); IAC (Spain). This
development has been supported by the funding agencies BMVIT (Austria),
ESA-PRODEX (Belgium), CEA/CNES (France), DLR (Germany), ASI/INAF (Italy), and
CICYT/MCYT (Spain).
SPIRE has been developed by a consortium of institutes led by Cardiff
University (UK) and including Univ. Lethbridge (Canada); NAOC (China); CEA, LAM
(France); IFSI, Univ. Padua (Italy); IAC (Spain); Stockholm Observatory
(Sweden); Imperial College London, RAL, UCL-MSSL, UKATC, Univ. Sussex (UK); and
Caltech, JPL, NHSC, Univ. Colorado (USA). This development has been supported
by national funding agencies: CSA (Canada); NAOC (China); CEA, CNES, CNRS
(France); ASI (Italy); MCINN (Spain); SNSB (Sweden); STFC, UKSA (UK); and NASA
(USA).
We thank Drew Brisbin for providing machine readable forms of the \HII\ models
presented in \citet{Rubin1985}.




\bibliographystyle{mnras}
\bibliography{stack} 




\begin{appendix}

\section{Observing dates}\label{observingdate} 

{Basic information of the \herschel{} observations}
{\scriptsize 
\tablehead{
\hline
Source & Instrument & Time  & Date & Proj. & Notes\\[-1ex]
                  &            & s     &      &           & \\ 
\hline }
\begin{supertabular}{llllll}
  SDP.9               & SPIRE & 13320 & 2011-11-07 & {\sc ot1} & \\
                      & PACS  &    90 & 2011-12-02 & {\sc ot1} & \\
                      & PACS  &    90 & 2011-12-02 & {\sc ot1} & \\
   SDP.11             & SPIRE & 13320 & 2012-11-18 & {\sc ot1} & \\
                      & PACS  &    90 & 2011-10-18 & {\sc ot1} & \\
                      & PACS  &    90 & 2011-10-18 & {\sc ot1} & \\
  SDP.17              & SPIRE & 13320 & 2012-11-18 & {\sc ot2} & \\
                      & PACS  &    90 & 2011-12-02 & {\sc ot1} & \\
                      & PACS  &    90 & 2011-12-02 & {\sc ot1} & \\
  SDP.81              & SPIRE & 13320 & 2010-06-01 & {\sc gt1} & \cite{Valtchanov2011} \\
                      & PACS  &    90 & 2010-06-03 & {\sc gt1} & 70- and 160-{\mum} maps\\
                      & PACS  &    90 & 2010-06-03 & {\sc gt1} & 70- and 160-{\mum} maps\\
  SDP.130             & SPIRE & 13320 & 2010-06-01 & {\sc gt1} & \cite{Valtchanov2011}\\
                      & PACS  &    90 & 2010-06-03 & {\sc gt1} & 70- and 160-{\mum} maps \\
                      & PACS  &    90 & 2010-06-03 & {\sc gt1} & 70- and 160-{\mum} maps \\
  G09-v1.40           & SPIRE & 13320 & 2011-11-06 & {\sc ot1} & \\
                      & PACS  &    90 & 2012-04-24 & {\sc ot1} & \\
                      & PACS  &    90 & 2012-04-24 & {\sc ot1} & \\
  G09-v1.97           & SPIRE & 13320 & 2011-11-08 & {\sc ot1} & \\
                      & PACS  &    90 & 2011-11-10 & {\sc ot1} & \\
                      & PACS  &    90 & 2011-11-10 & {\sc ot1} & \\
  G09-v1.124          & SPIRE & 13320 & 2012-11-18 & {\sc ot2} & \\
                      & PACS  &    90 & 2012-04-13 & {\sc ot1} & \\
                      & PACS  &    90 & 2012-04-13 & {\sc ot1} & \\
  G09-v1.326          & SPIRE & 13320 & 2011-11-07 & {\sc ot1} & \\
                      & PACS  &    90 & 2011-11-28 & {\sc ot1} & \\
                      & PACS  &    90 & 2011-11-28 & {\sc ot1} & \\
  G12-v2.30           & SPIRE & 13320 & 2012-07-07 & {\sc ot1} & \\
                      & PACS  &    90 & 2011-11-27 & {\sc ot1} & \\
                      & PACS  &    90 & 2011-11-27 & {\sc ot1} & \\
  G12-v2.43           & SPIRE & 13320 & 2012-07-07 & {\sc ot1} & \\
                      & PACS  &    90 & 2011-07-14 & {\sc ot1} & \\
                      & PACS  &    90 & 2011-07-14 & {\sc ot1} & \\
  G12-v2.257          & SPIRE & 13320 & 2012-07-07 & {\sc ot1} & \\
                      & PACS  &    90 & 2011-07-14 & {\sc ot1} & \\
                      & PACS  &    90 & 2011-07-14 & {\sc ot1} & \\
  G15-v2.19           & SPIRE & 13320 & 2012-02-04 & {\sc ot1} & \\
                      & PACS  &    90 & 2011-07-09 & {\sc ot1} & \\
                      & PACS  &    90 & 2011-07-09 & {\sc ot1} & \\
  G15-v2.235          & SPIRE & 13320 & 2012-02-04 & {\sc ot1} & \\
                      & PACS  &    90 & 2012-01-12 & {\sc ot1} & \\
                      & PACS  &    90 & 2012-01-12 & {\sc ot1} & \\
  NA.v1.56            & SPIRE & 13320 & 2012-07-22 & {\sc ot1} & \\
                      & PACS  &    90 & 2011-11-27 & {\sc ot1} & \\
                      & PACS  &    90 & 2011-11-27 & {\sc ot1} & \\
  NA.v1.144           & SPIRE & 13320 & 2012-08-03 & {\sc ot1} & \\
                      & PACS  &    90 & 2011-12-03 & {\sc ot1} & \\
                      & PACS  &    90 & 2011-12-03 & {\sc ot1} & \\
  NA.v1.177           & SPIRE & 13320 & 2012-08-03 & {\sc ot1} & \\
                      & PACS  &    90 & 2011-11-27 & {\sc ot1} & \\
                      & PACS  &    90 & 2011-11-27 & {\sc ot1} & \\
  NA.v1.186           & SPIRE & 13320 & 2012-06-18 & {\sc ot1} & \\
                      & PACS  &    90 & 2011-07-11 & {\sc ot1} & \\
                      & PACS  &    90 & 2011-07-11 & {\sc ot1} & \\
       NB43           & SPIRE & 13320 & 2012-08-02 & {\sc ot1} & \cite{George2013}\\
                      & PACS  &    90 & 2011-11-27 & {\sc ot1} & \\
                      & PACS  &    90 & 2011-11-27 & {\sc ot1} & \\
  NB.v1.78            & SPIRE & 13320 & 2012-08-02 & {\sc ot1} & \\
                      & PACS  &    90 & 2011-12-03 & {\sc ot1} & \\
                      & PACS  &    90 & 2011-12-03 & {\sc ot1} & \\
  NC.v1.143           & SPIRE & 13320 & 2012-07-21 & {\sc ot1} & \\
                      & PACS  &    90 & 2011-07-11 & {\sc ot1} & \\
                      & PACS  &    90 & 2011-07-11 & {\sc ot1} & \\
  SA.v1.44            & SPIRE & 13320 & 2012-12-03 & {\sc ot2} & \\
                      & PACS  &    90 & 2012-11-27 & {\sc ot2} & \\
                      & PACS  &    90 & 2012-11-27 & {\sc ot2} & \\
  SA.v1.53            & SPIRE & 13320 & 2012-12-02 & {\sc ot2} & \\
                      & PACS  &    90 & 2012-11-27 & {\sc ot2} & \\
                      & PACS  &    90 & 2011-11-27 & {\sc ot2} & \\
  SB.v1.143           & SPIRE & 13320 & 2012-12-29 & {\sc ot2} & \\
                      & PACS  &    90 & 2012-12-12 & {\sc ot2} & \\
                      & PACS  &    90 & 2012-12-12 & {\sc ot2} & \\
  SB.v1.202           & SPIRE & 13320 & 2012-12-03 & {\sc ot2} & \\
                      & PACS  &    90 & 2012-12-12 & {\sc ot2} & \\
                      & PACS  &    90 & 2012-12-12 & {\sc ot2} & \\
  SC.v1.128           & SPIRE & 13320 & 2012-12-03 & {\sc ot2} & \\
                      & PACS  &    90 & 2012-12-12 & {\sc ot2} & \\
                      & PACS  &    90 & 2012-12-12 & {\sc ot2} & \\
  SD.v1.70            & SPIRE & 13320 & 2012-12-28 & {\sc ot2} & \\
                      & PACS  &    90 & 2012-12-25 & {\sc ot2} & \\
                      & PACS  &    90 & 2012-12-25 & {\sc ot2} & \\
  SD.v1.133           & SPIRE & 13320 & 2012-12-28 & {\sc ot2} & \\
                      & PACS  &    90 & 2012-12-11 & {\sc ot2} & \\
                      & PACS  &    90 & 2012-12-11 & {\sc ot2} & \\
  SD.v1.328           & SPIRE & 13320 & 2012-12-29 & {\sc ot2} & \\
                      & PACS  &    90 & 2012-12-11 & {\sc ot2} & \\
                      & PACS  &    90 & 2012-12-11 & {\sc ot2} & \\
  SE.v1.165           & SPIRE & 13320 & 2012-12-29 & {\sc ot2} & \\
                      & PACS  &    90 & 2012-12-25 & {\sc ot2} & \\
                      & PACS  &    90 & 2012-12-25 & {\sc ot2} & \\
  SF.v1.88            & SPIRE & 13320 & 2012-07-08 & {\sc ot2} & \\
                      & PACS  &    90 & 2012-07-15 & {\sc ot2} & \\
                      & PACS  &    90 & 2012-07-15 & {\sc ot2} & \\
  SF.v1.100           & SPIRE & 13320 & 2012-07-08 & {\sc ot2} & \\
                      & PACS  &    90 & 2012-07-15 & {\sc ot2} & \\
                      & PACS  &    90 & 2012-07-15 & {\sc ot2} & \\
  SG.v1.77            & SPIRE & 13320 & 2012-07-17 & {\sc ot2} & \\
                      & PACS  &    90 & 2012-06-25 & {\sc ot2} & \\
                      & PACS  &    90 & 2012-06-25 & {\sc ot2} & \\
  HeLMS05             & PACS  &    90 & 2013-01-06 & {\sc ot2} & \\
                      & PACS  &    90 & 2013-01-06 & {\sc ot2} & \\
  HeLMS06             & PACS  &    90 & 2013-01-06 & {\sc ot2} & \\
                      & PACS  &    90 & 2013-01-06 & {\sc ot2} & \\
  HeLMS09             & PACS  &    90 & 2013-01-11 & {\sc ot2} & \\
                      & PACS  &    90 & 2013-01-11 & {\sc ot2} & \\
  HeLMS44             & SPIRE & 13320 & 2013-01-08 & {\sc ot2} & \\
                      & PACS  &    90 & 2013-01-01 & {\sc ot2} & \\
                      & PACS  &    90 & 2013-01-01 & {\sc ot2} & \\
  HeLMS45             & SPIRE & 13320 & 2012-12-29 & {\sc ot2} & \\
                      & PACS  &    90 & 2013-01-01 & {\sc ot2} & \\
                      & PACS  &    90 & 2013-01-01 & {\sc ot2} & \\
  HeLMS49             & PACS  &    90 & 2013-01-06 & {\sc ot2} & \\
                      & PACS  &    90 & 2013-01-06 & {\sc ot2} & \\
  HeLMS51             & PACS  &    90 & 2013-01-06 & {\sc ot2} & \\
                      & PACS  &    90 & 2013-01-06 & {\sc ot2} & \\
  HeLMS61             & PACS  &    90 & 2013-01-01 & {\sc ot2} & \\
                      & PACS  &    90 & 2013-01-01 & {\sc ot2} & \\
  HeLMS62             & PACS  &    90 & 2012-12-26 & {\sc ot2} & \\
                      & PACS  &    90 & 2012-12-26 & {\sc ot2} & \\
  HBo\"otes03         & SPIRE & 13320 & 2012-02-05 & {\sc ot1} & \\
                      & PACS  &    90 & 2011-11-27 & {\sc ot1} & \\
                      & PACS  &    90 & 2011-11-27 & {\sc ot1} & \\
  HXMM02              & SPIRE & 13320 & 2012-02-05 & {\sc ot1} & \\
                      & PACS  &    90 & 2012-02-14 & {\sc ot1} & \\
                      & PACS  &    90 & 2012-02-14 & {\sc ot1} & \\
            Eyelash   & SPIRE & 13320 & 2011-04-24 & {\sc ot1} & \cite{George2014}\\
                      & SPIRE & 13320 & 2012-11-17 & {\sc ot2} & as above\\
                      & SPIRE & 13320 & 2012-11-24 & {\sc ot2} & as above\\
                      & SPIRE & 13320 & 2012-11-24 & {\sc ot2} & as above\\
                      & SPIRE & 13320 & 2012-12-01 & {\sc ot2} & as above\\
                      & SPIRE & 13320 & 2012-12-02 & {\sc ot2} & as above\\
                      &  PACS &    90 & 2011-11-17 & {\sc ot1} & \\
                      &  PACS &    90 & 2011-11-17 & {\sc ot1} & \\
\hline
\end{supertabular} 
\bigskip\\
Herschel  observations of the sample of strongly lensed DSFGs.  SPIRE
spectrometer observations consist of 100 forward and 100 reverse scans of the
SMEC mirror.  The two PACS photometer observations of each source consist of
\(10 \times 3'\) scan legs oriented at 70\degr\ and 110\degr respectively, and
combined during processing into a single mini scan map.   
}


\section{PACS imaging} \label{stamps}

    \begin{figure*}
    \centering
    \includegraphics[width=0.33\textwidth,trim=60bp 0bp 140bp 40bp ,clip]{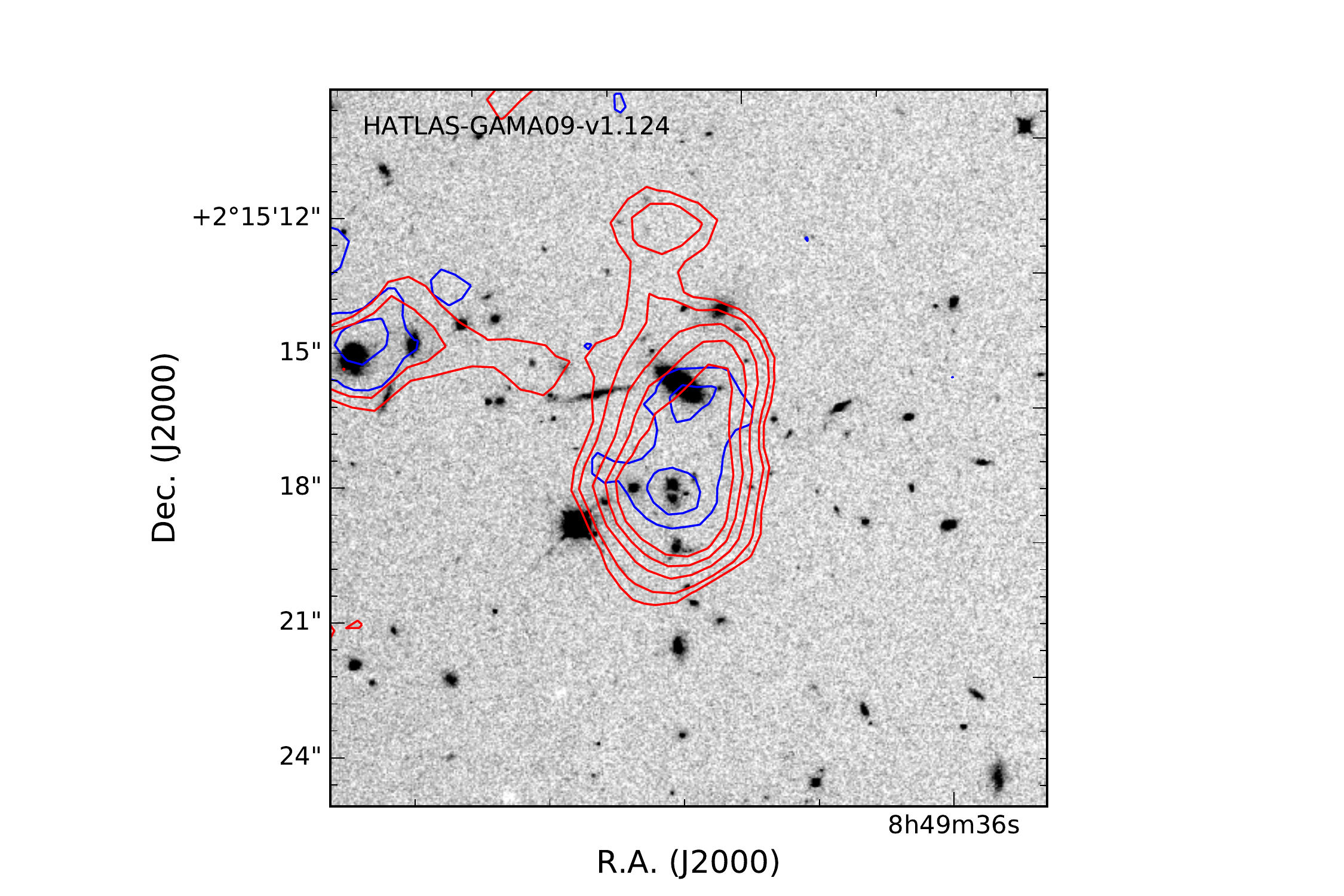}
    \includegraphics[width=0.33\textwidth,trim=60bp 0bp 140bp 40bp ,clip]{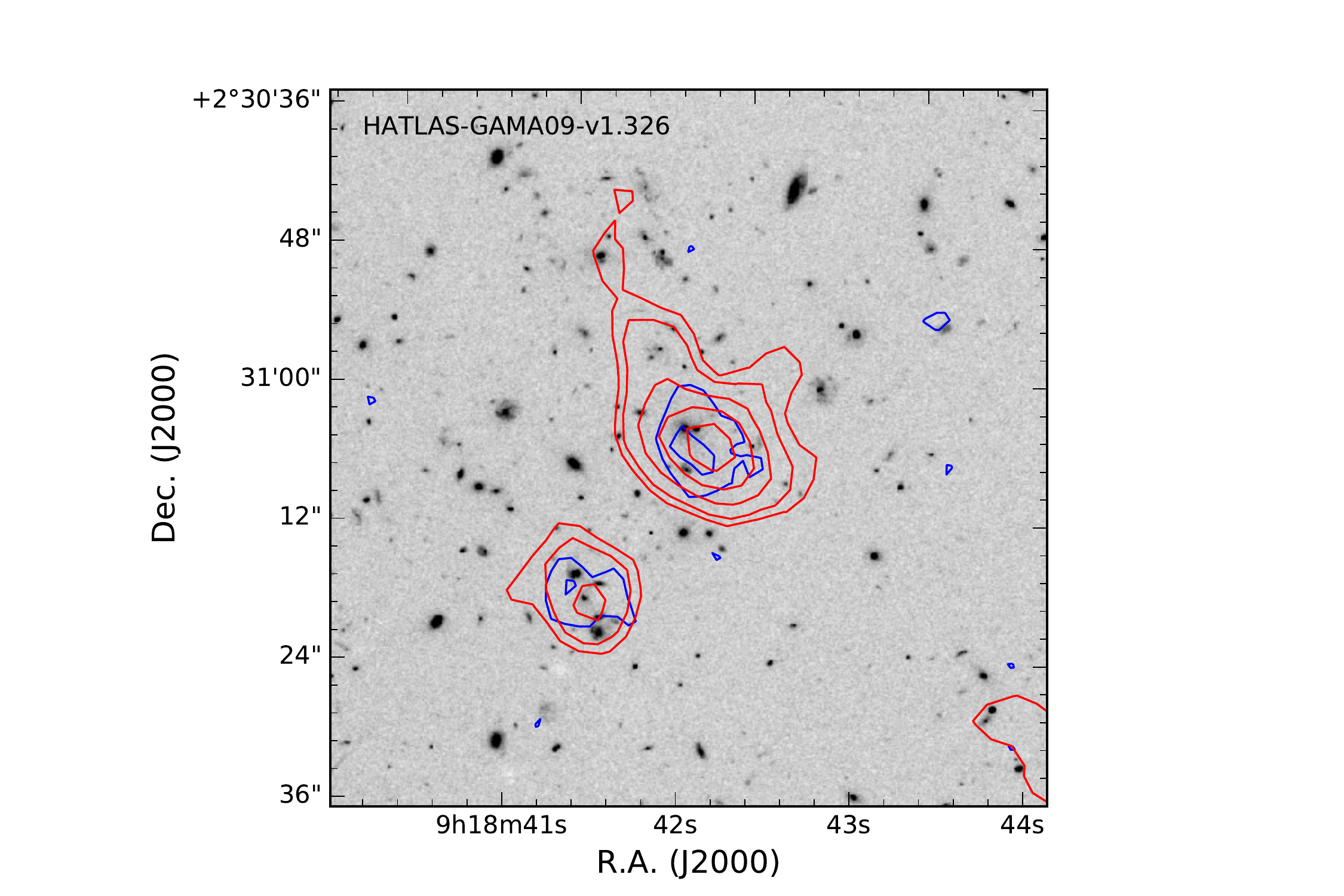}
    \includegraphics[width=0.33\textwidth,trim=60bp 0bp 140bp 40bp ,clip]{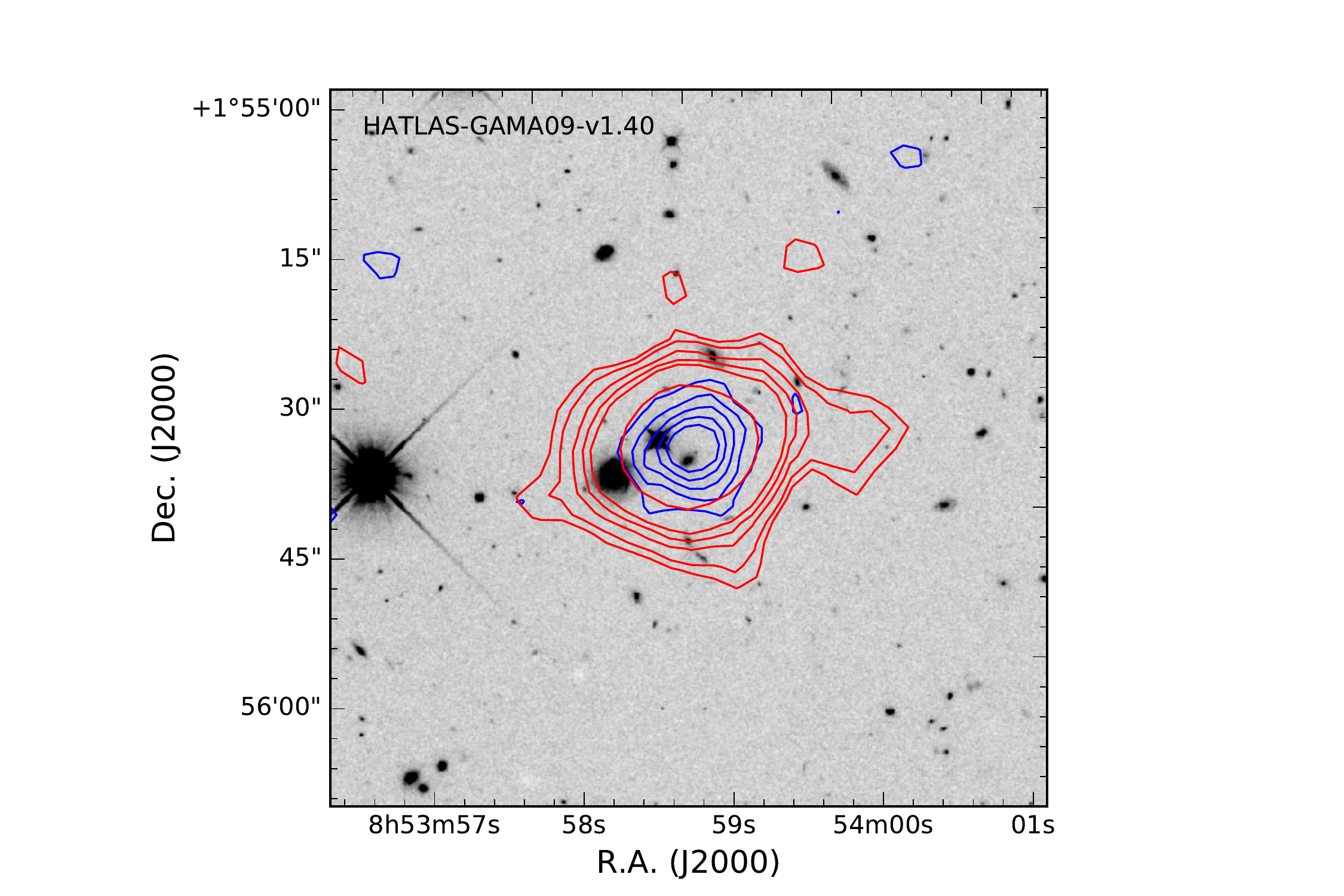}
    \includegraphics[width=0.33\textwidth,trim=60bp 0bp 140bp 40bp ,clip]{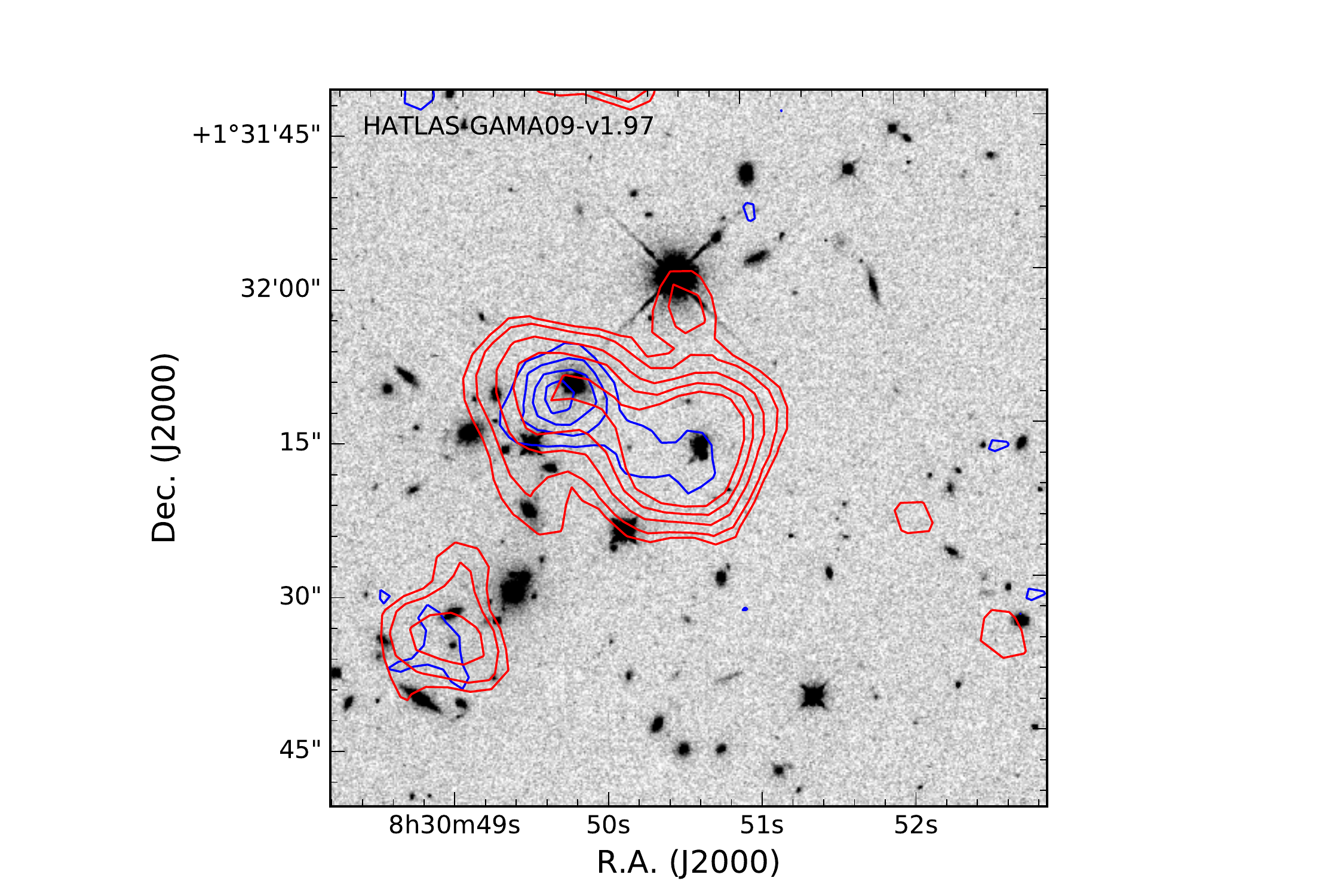}
    \includegraphics[width=0.33\textwidth,trim=60bp 0bp 140bp 40bp ,clip]{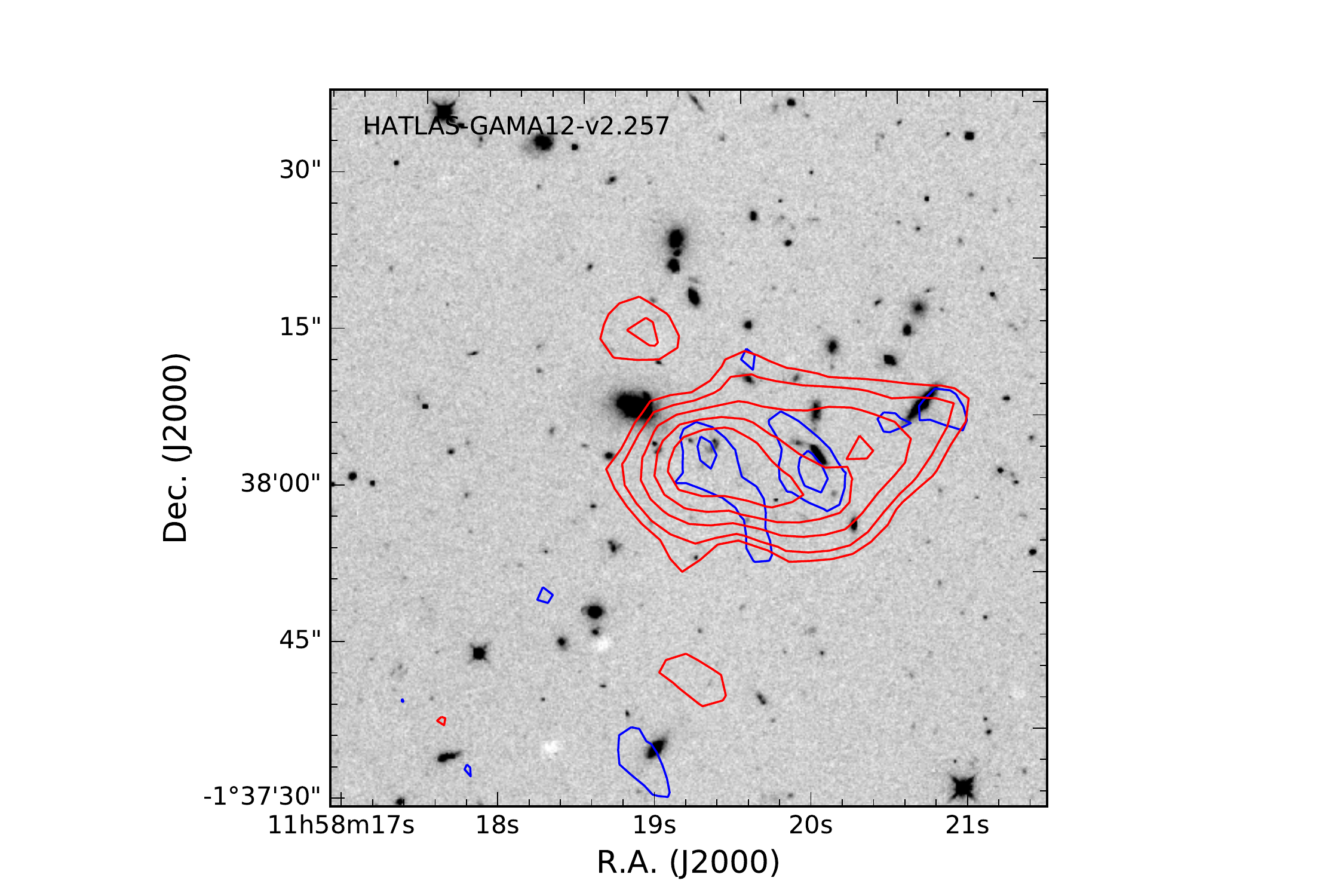}
    \includegraphics[width=0.33\textwidth,trim=60bp 0bp 140bp 40bp ,clip]{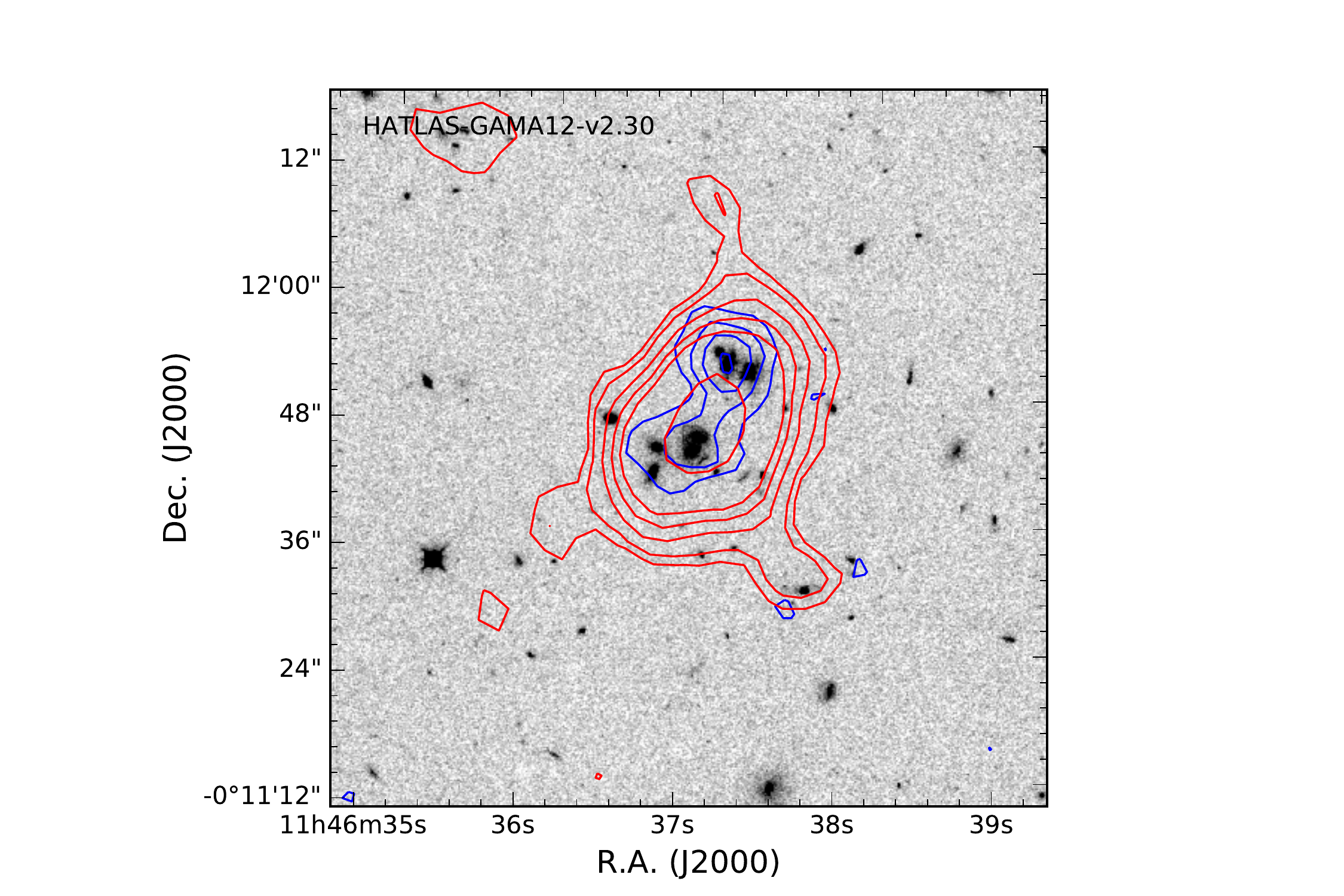}
    \includegraphics[width=0.33\textwidth,trim=60bp 0bp 140bp 40bp ,clip]{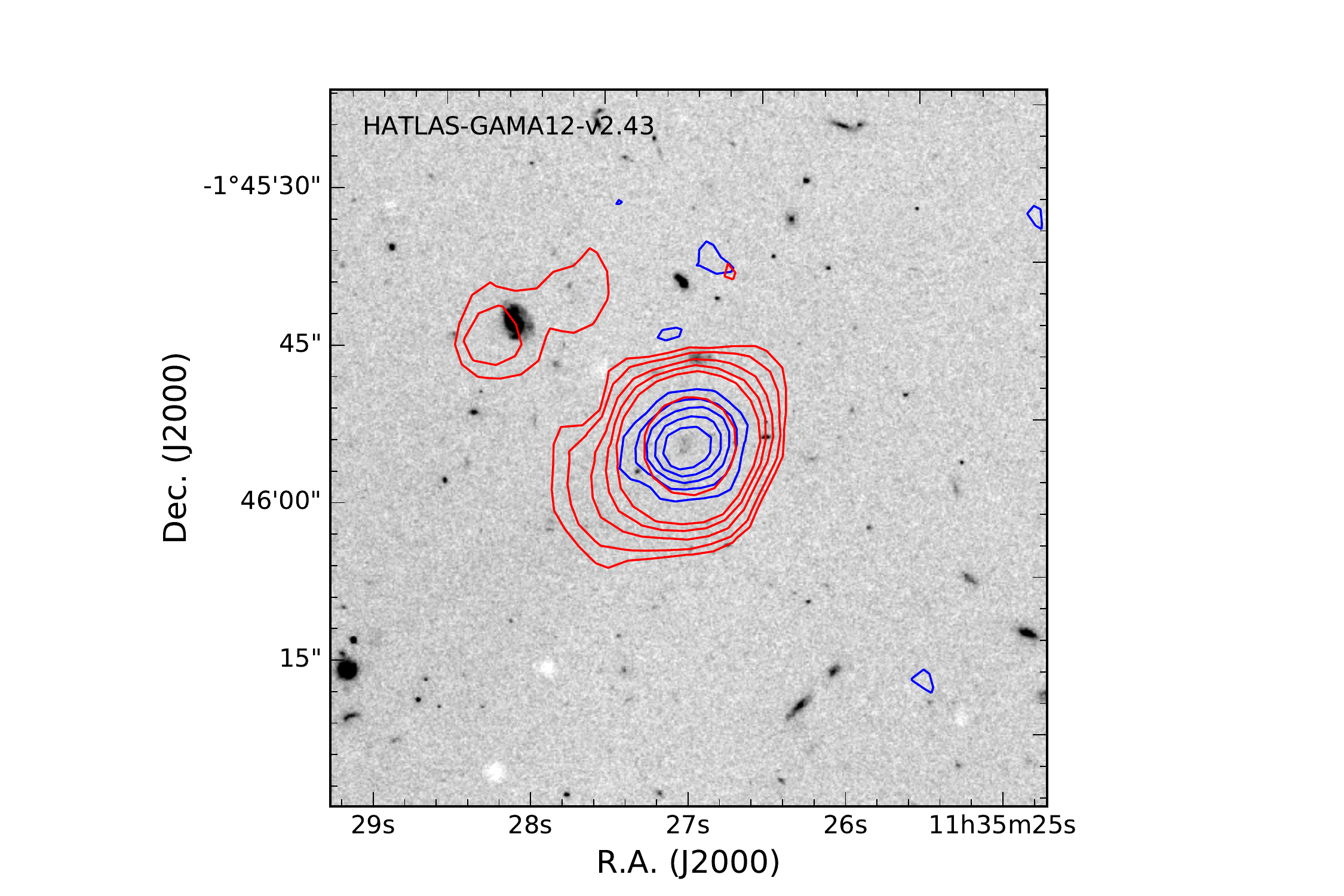}
    \includegraphics[width=0.33\textwidth,trim=60bp 0bp 140bp 40bp ,clip]{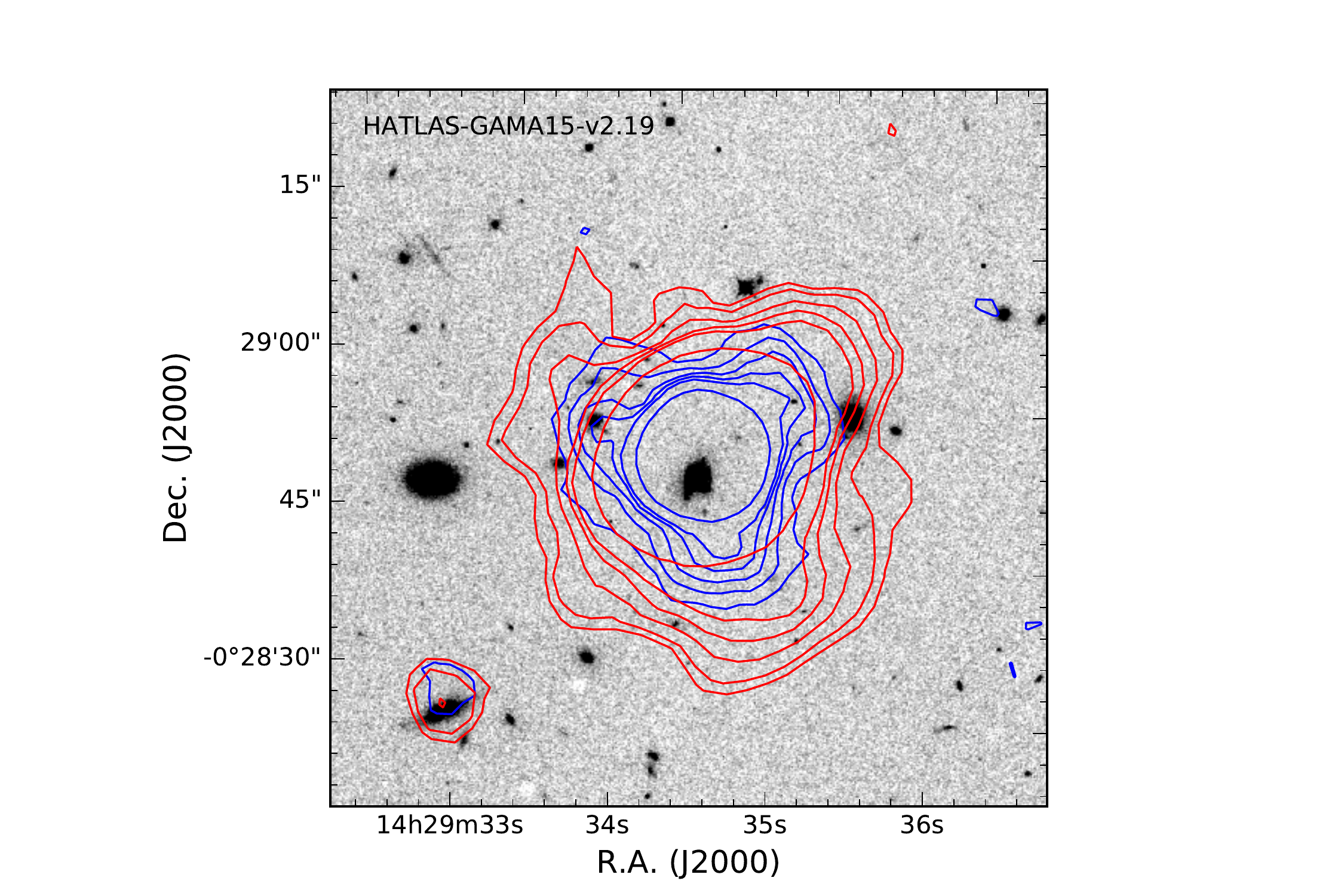}
    \includegraphics[width=0.33\textwidth,trim=60bp 0bp 140bp 40bp ,clip]{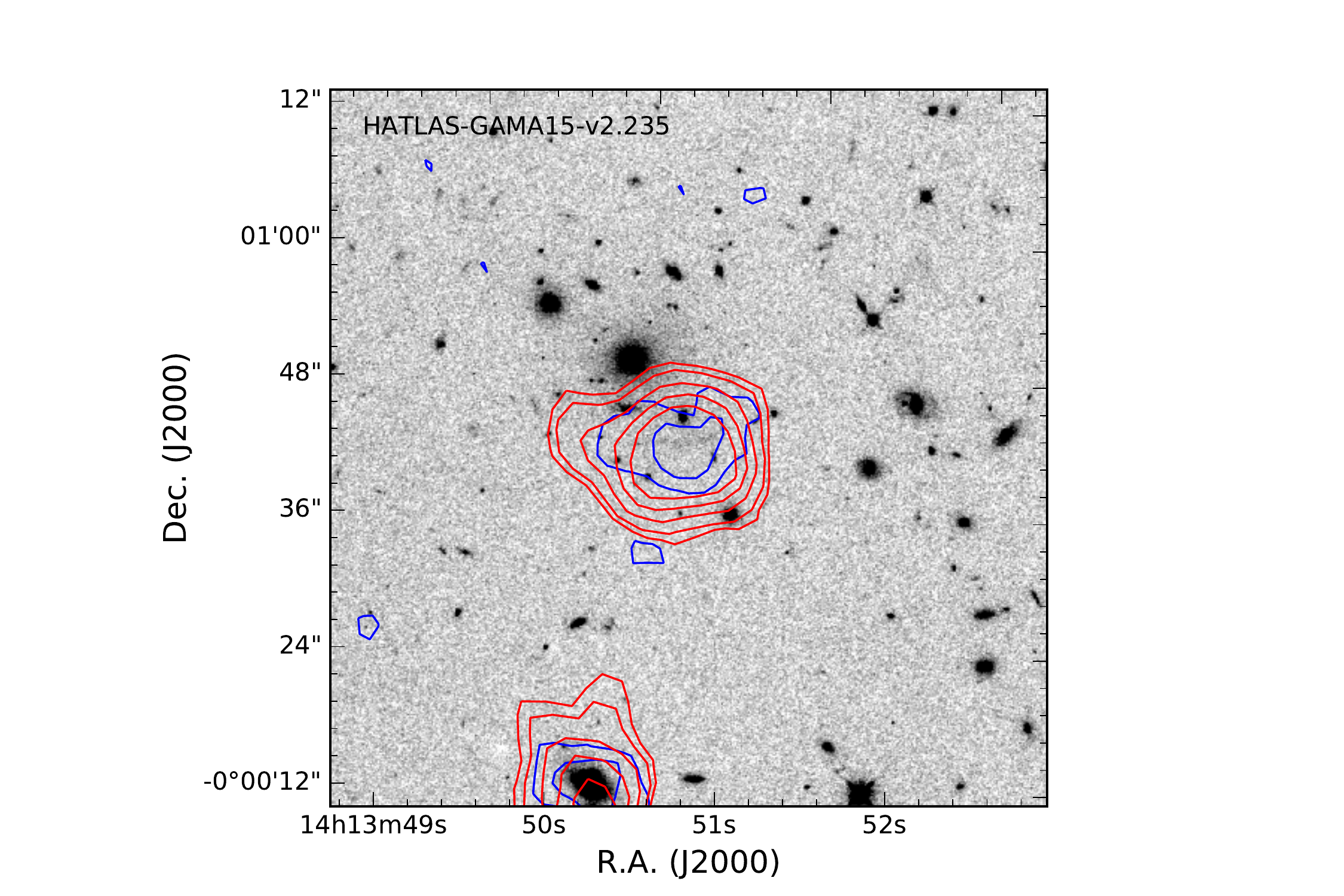}
    \includegraphics[width=0.33\textwidth,trim=60bp 0bp 140bp 40bp ,clip]{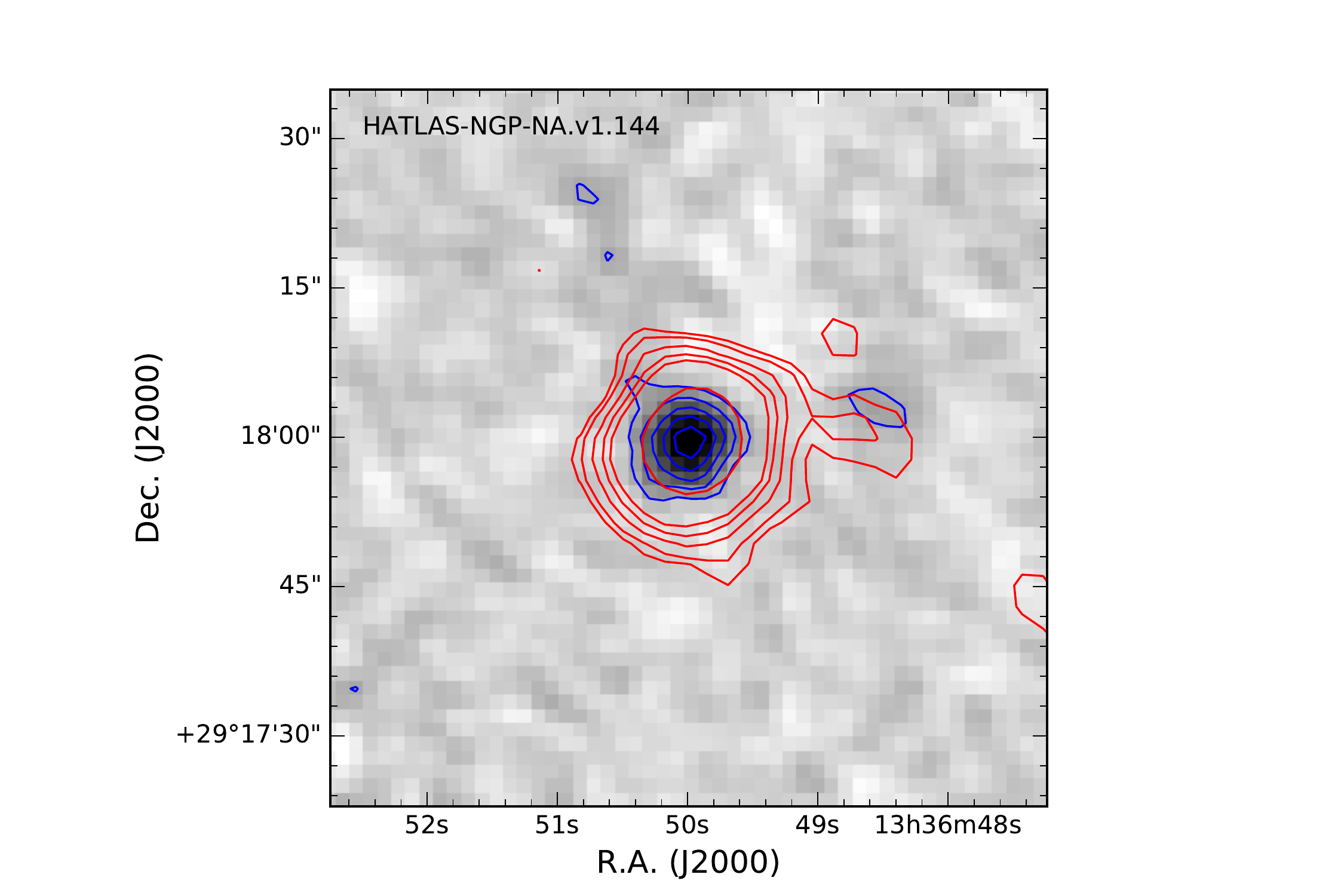}
    \includegraphics[width=0.33\textwidth,trim=60bp 0bp 140bp 40bp ,clip]{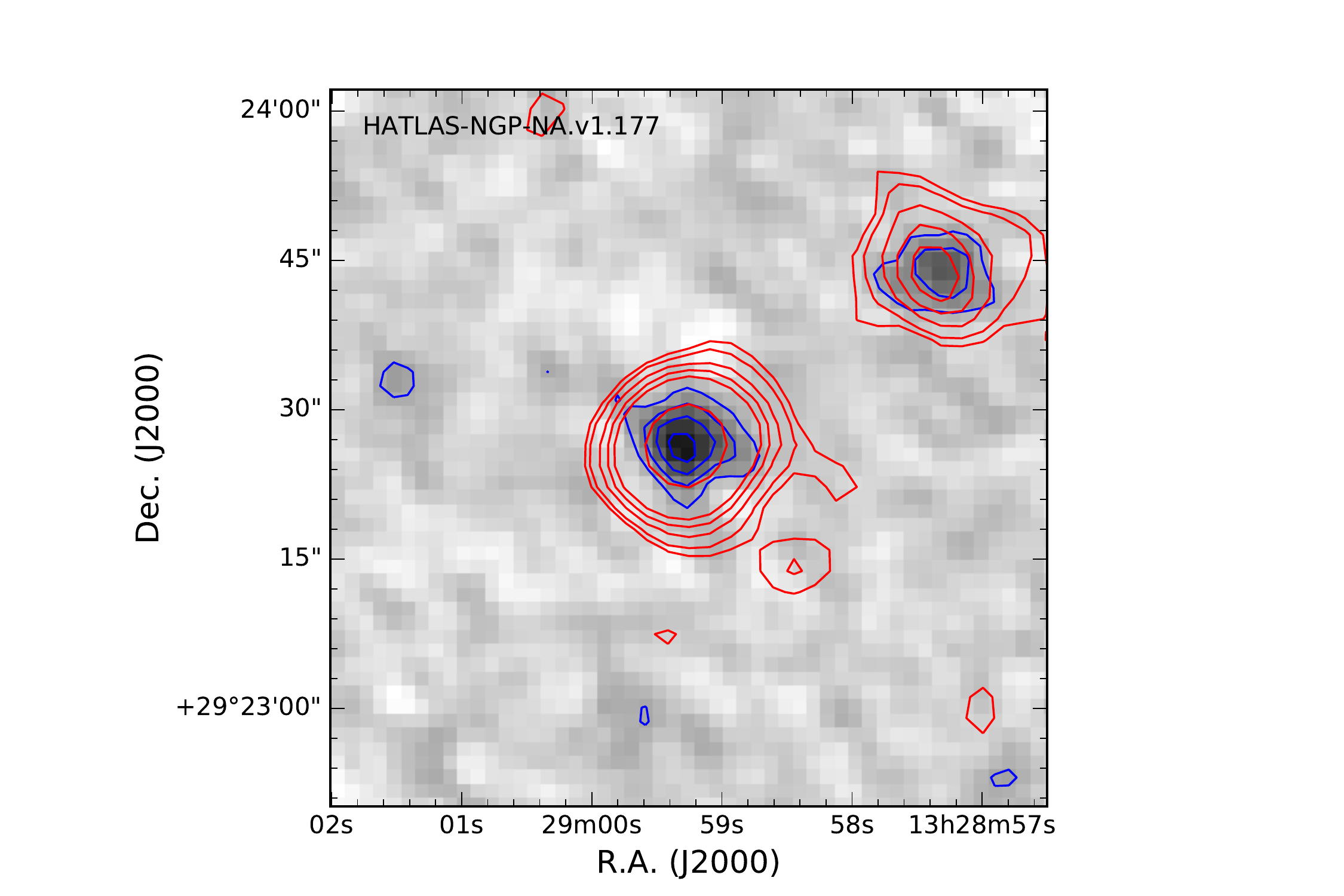}
    \includegraphics[width=0.33\textwidth,trim=60bp 0bp 140bp 40bp ,clip]{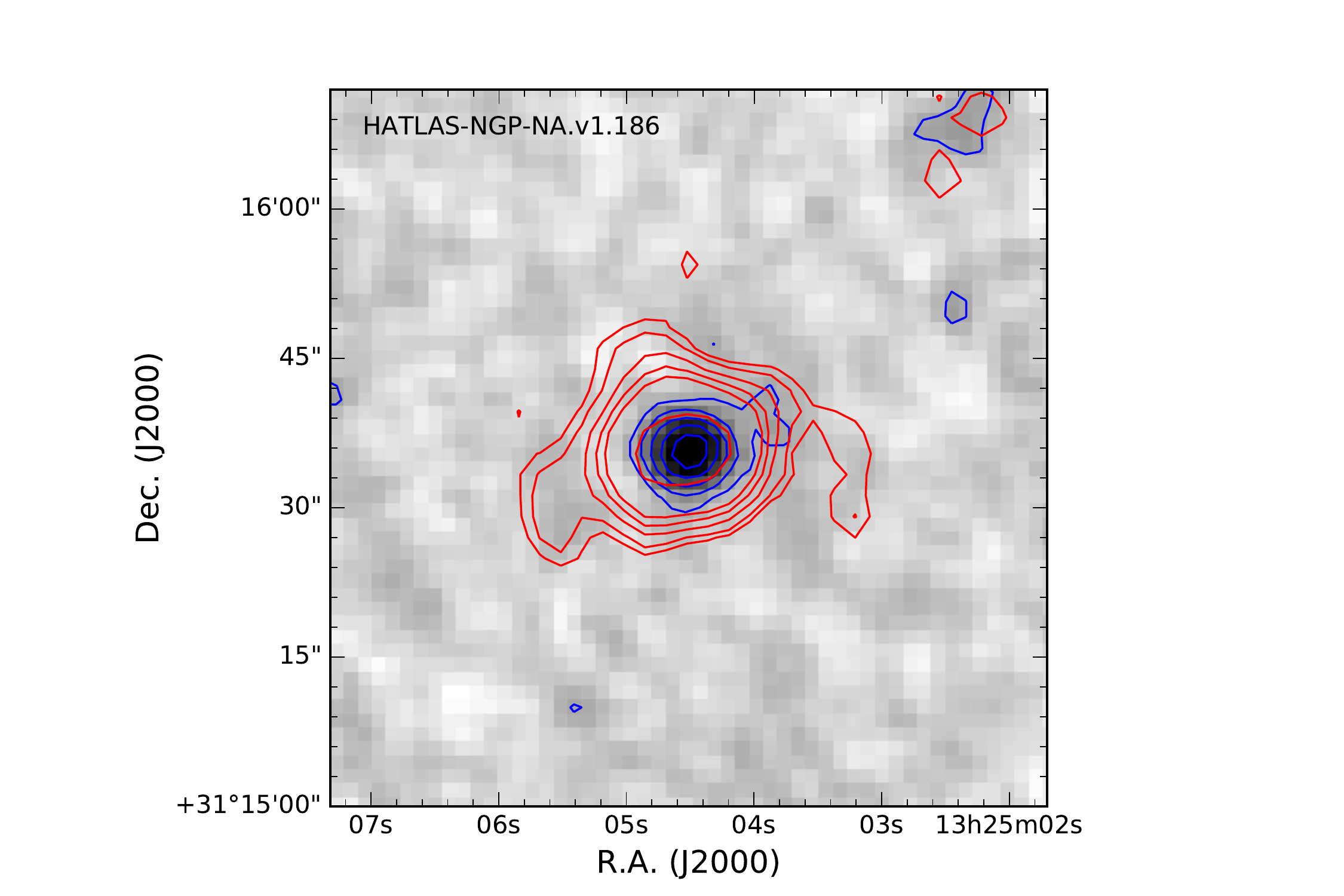}
  \caption{
\herschel\ PACS images of the sample. The high resolution background images are
HST F110W archival data. The low resolution background images are PACS 100
\mum\ data. Red contours are PACS 160 \mum\ data, with levels of
0.3,0.4,0.6,0.8,1,2 mJy/pixel. Blue contours are PACS 100 \mum\ data, with
levels of 0.2, 0.4, 0.6, 0.8, 1, and 2 mJy/pixel. }
    \label{fig:all_stamp1}
    \end{figure*}

    \begin{figure*}
    \centering
    \includegraphics[width=0.33\textwidth,trim=60bp 0bp 140bp 40bp ,clip]{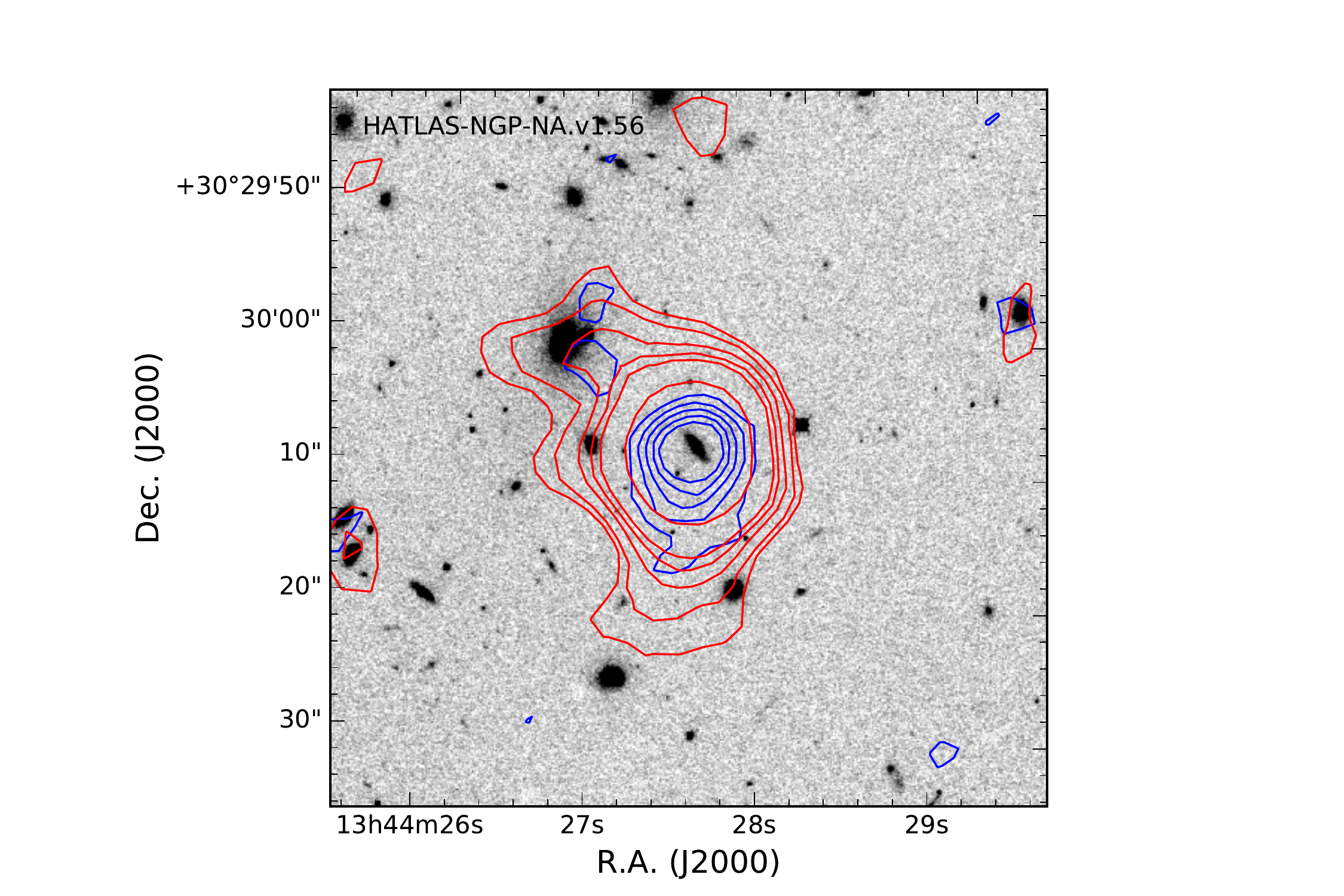}
    \includegraphics[width=0.33\textwidth,trim=60bp 0bp 140bp 40bp ,clip]{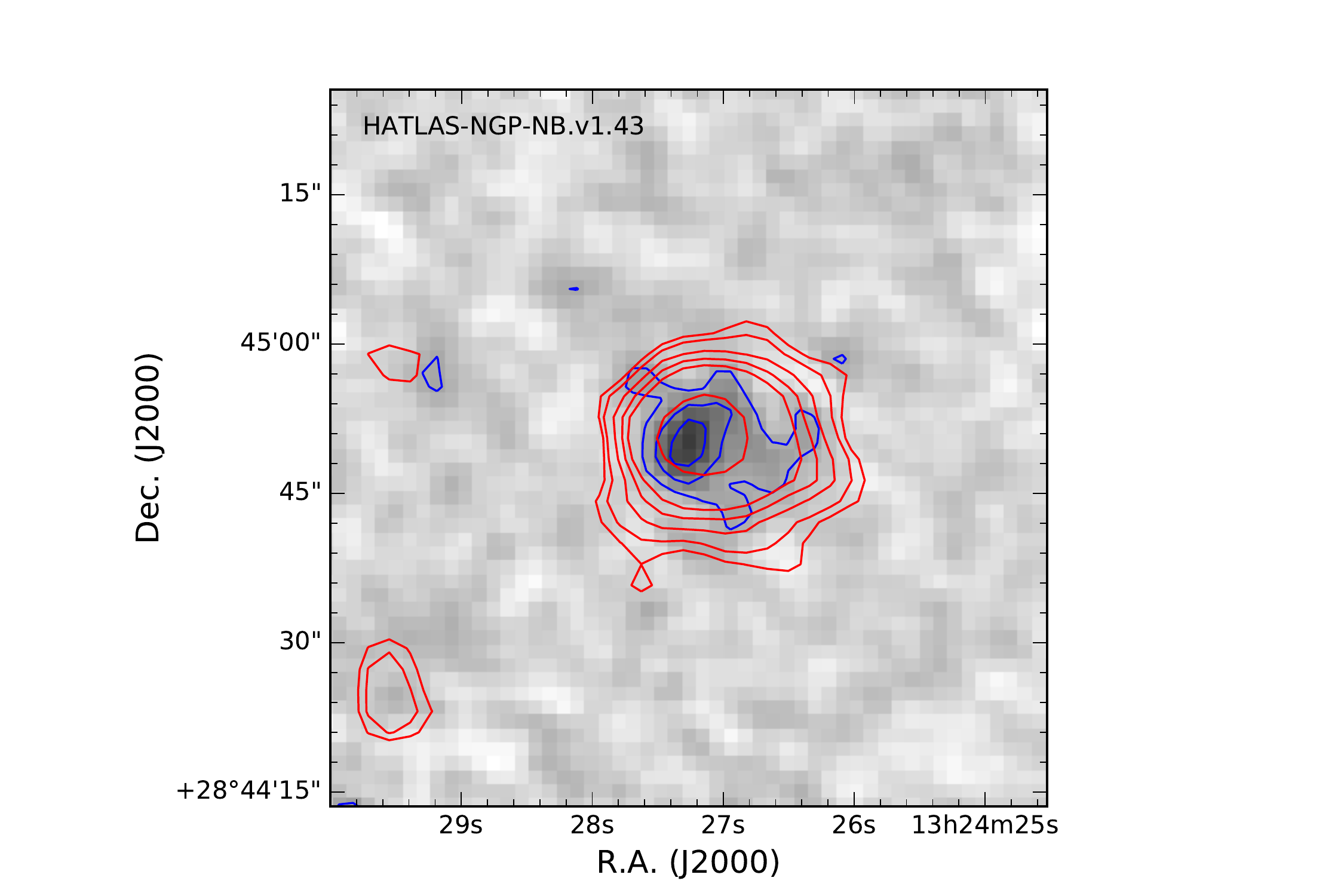}
    \includegraphics[width=0.33\textwidth,trim=60bp 0bp 140bp 40bp ,clip]{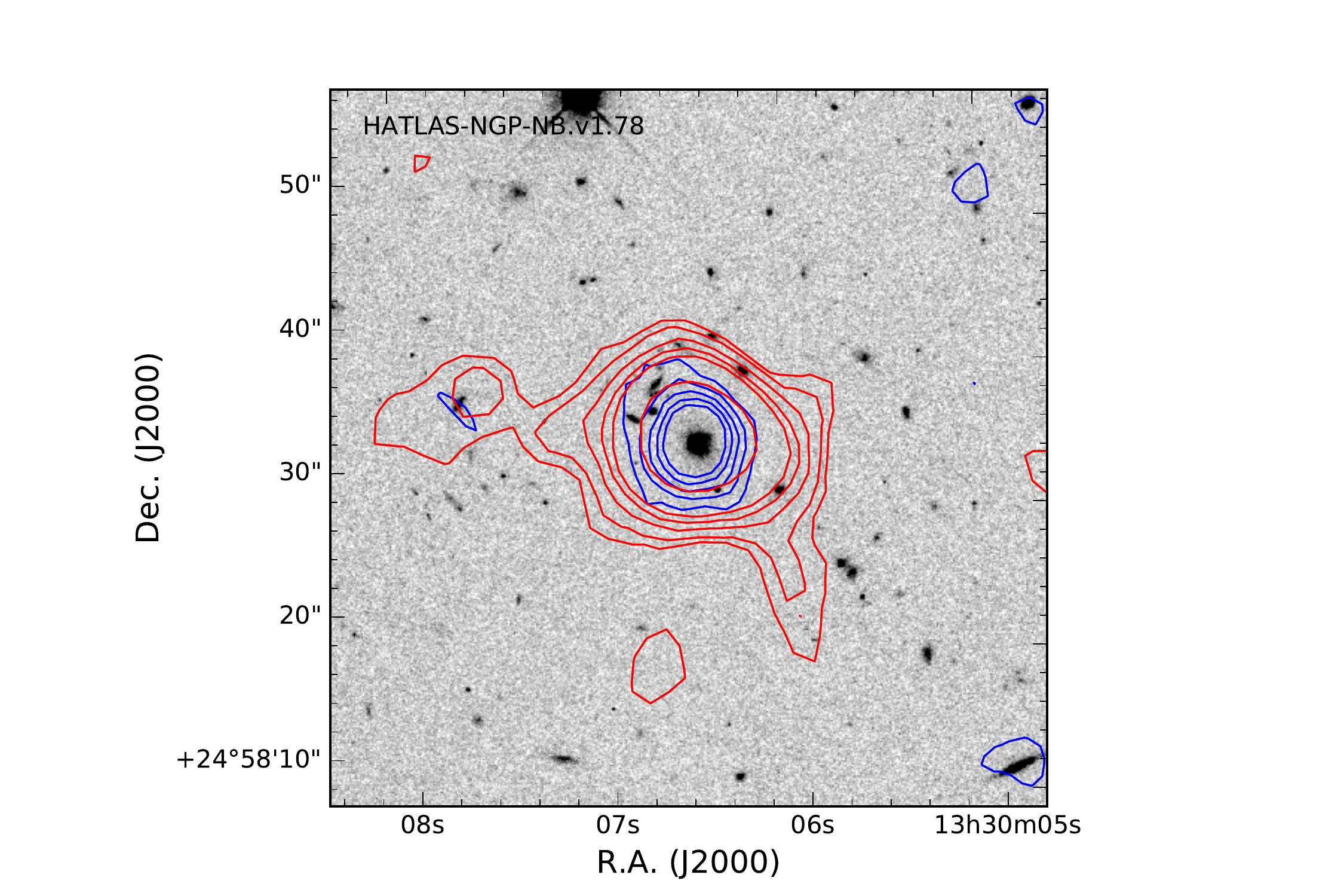}
    \includegraphics[width=0.33\textwidth,trim=60bp 0bp 140bp 40bp ,clip]{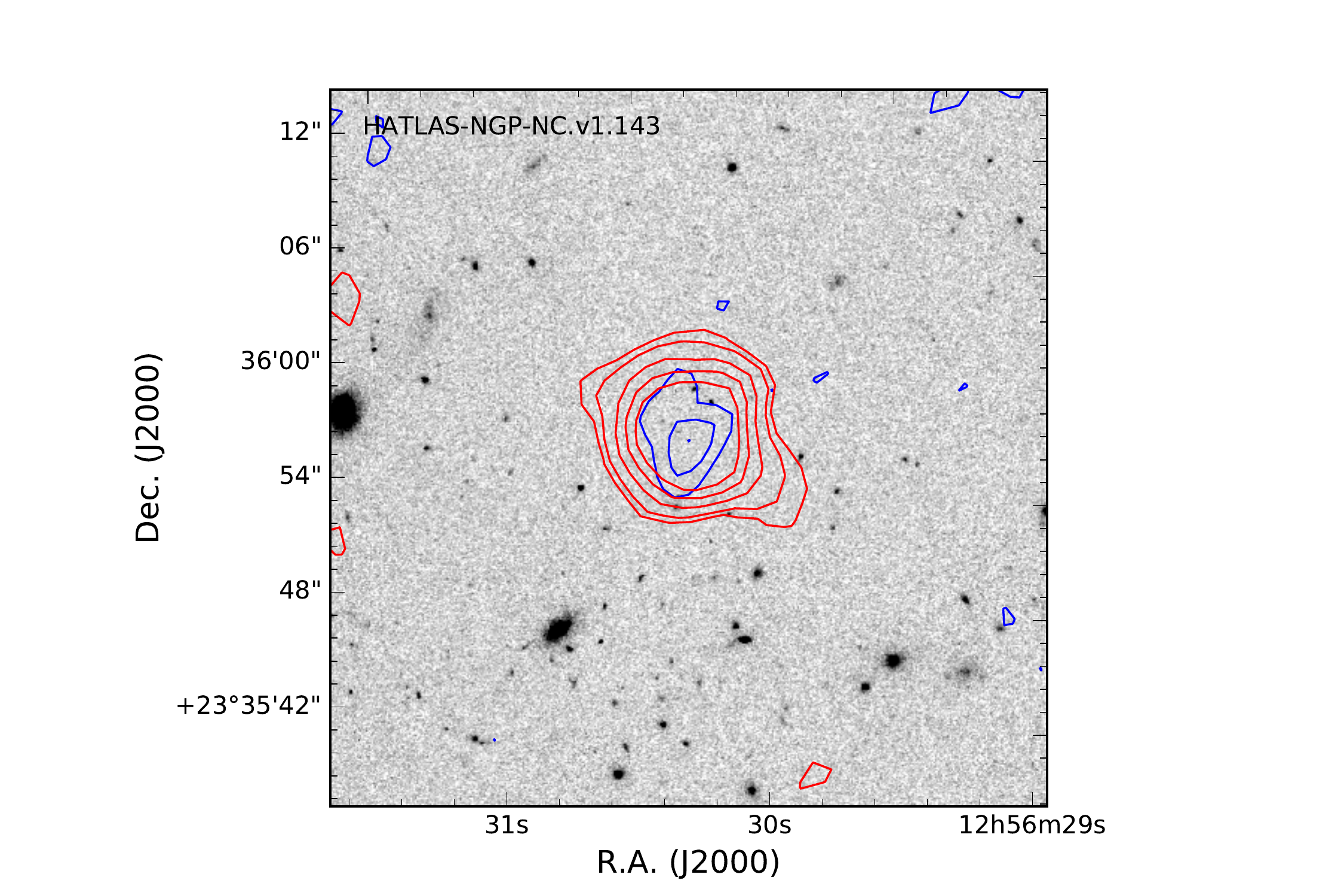}
    \includegraphics[width=0.33\textwidth,trim=60bp 0bp 140bp 40bp ,clip]{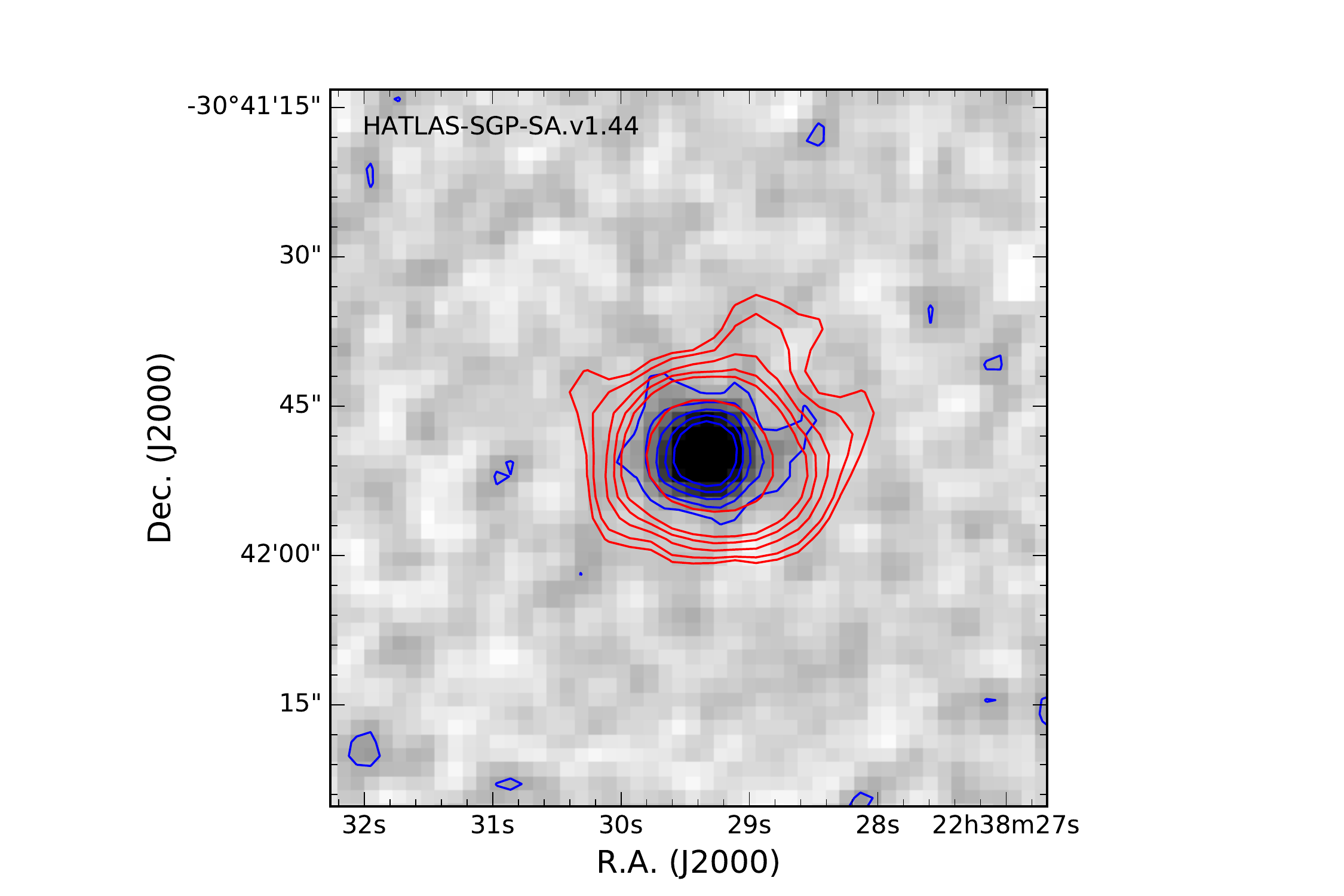}
    \includegraphics[width=0.33\textwidth,trim=60bp 0bp 140bp 40bp ,clip]{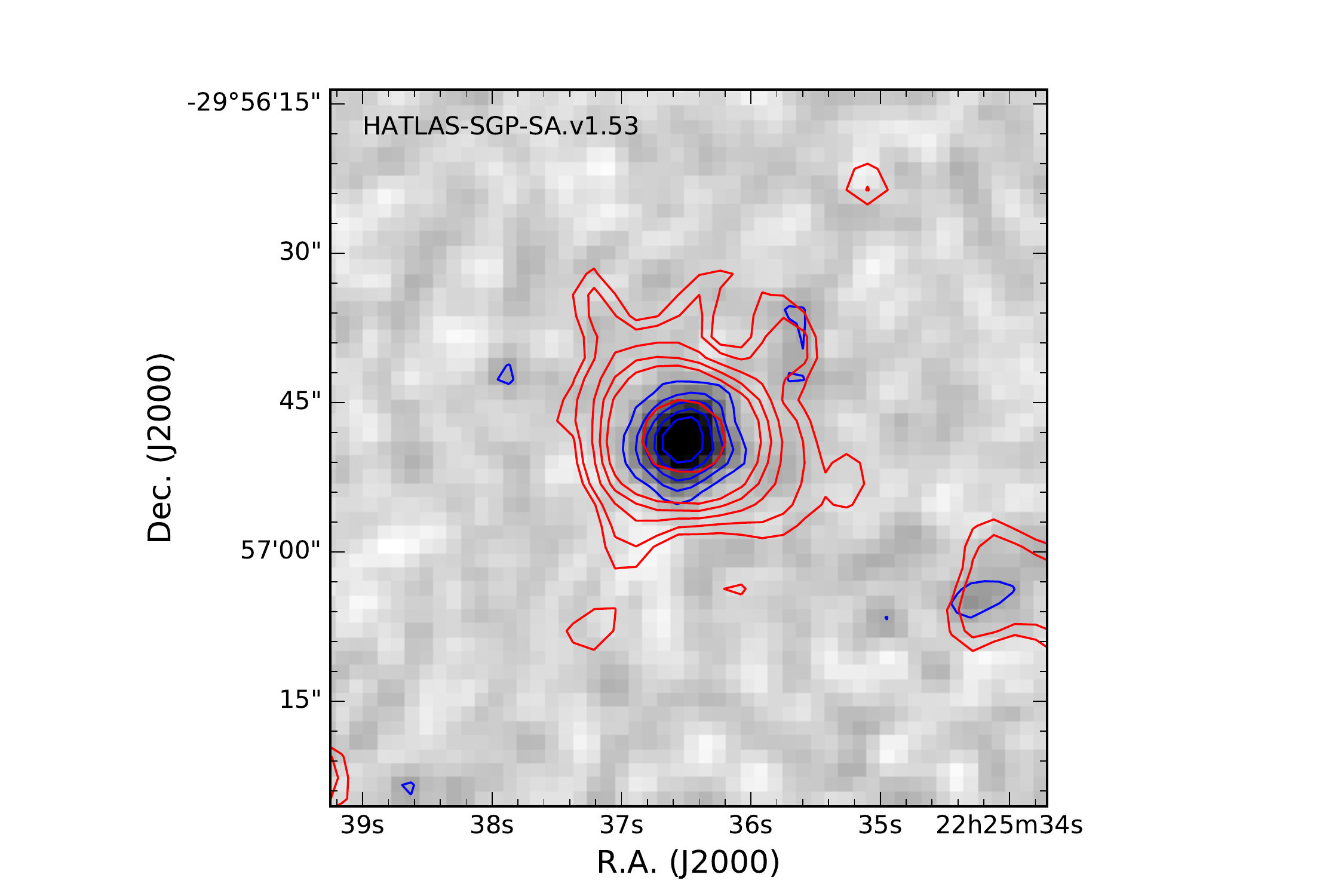}
    \includegraphics[width=0.33\textwidth,trim=60bp 0bp 140bp 40bp ,clip]{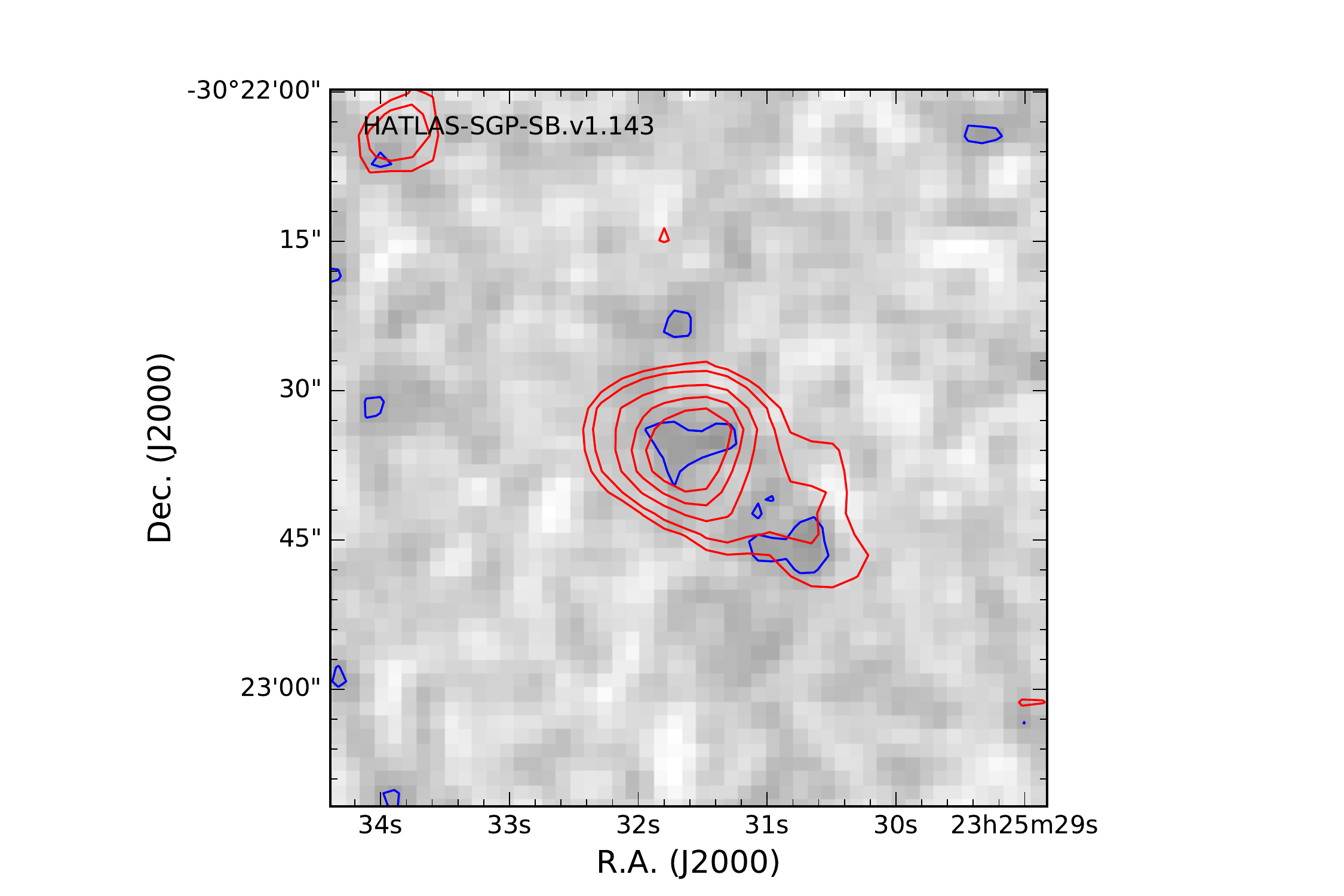}
    \includegraphics[width=0.33\textwidth,trim=60bp 0bp 140bp 40bp ,clip]{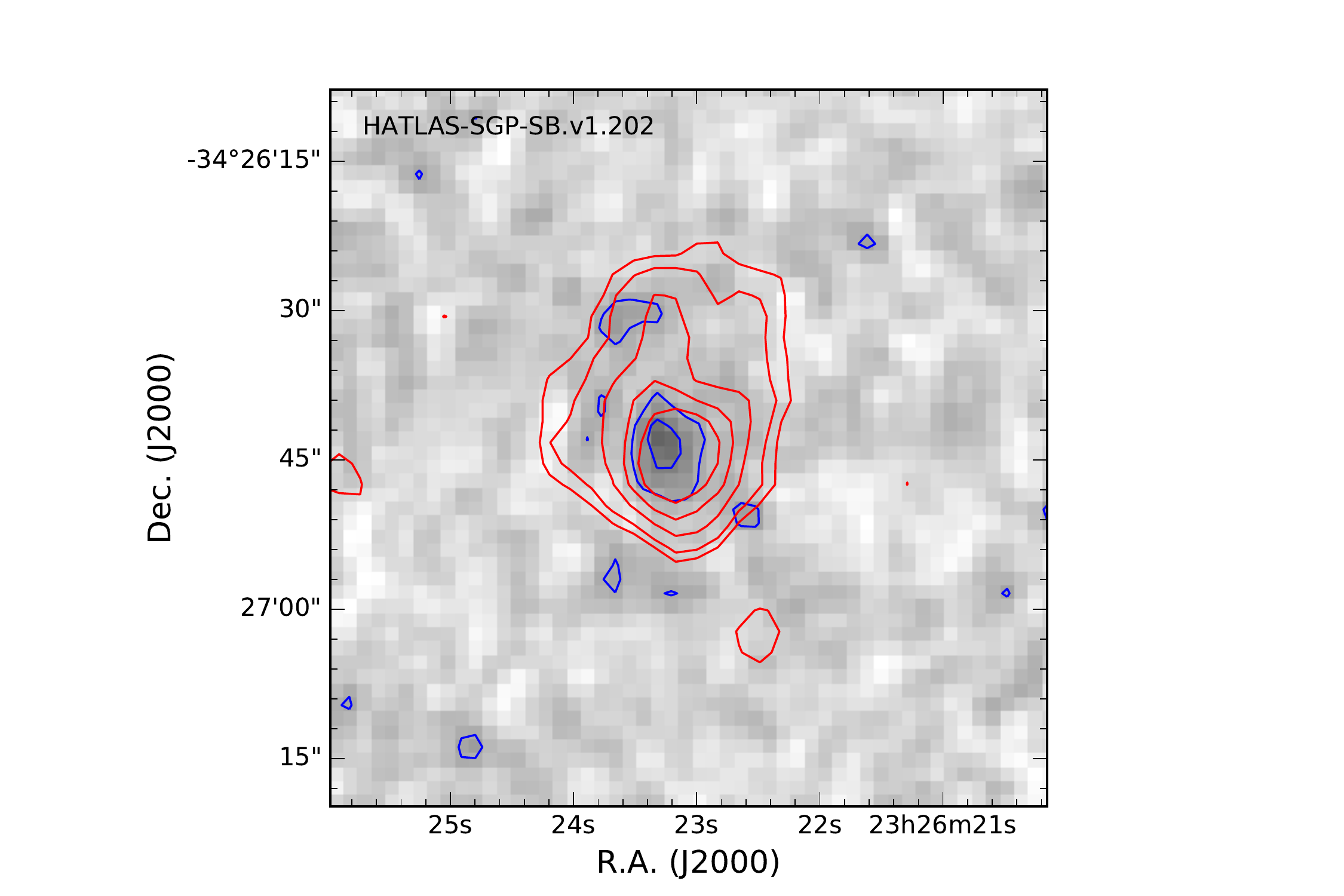}
    \includegraphics[width=0.33\textwidth,trim=60bp 0bp 140bp 40bp ,clip]{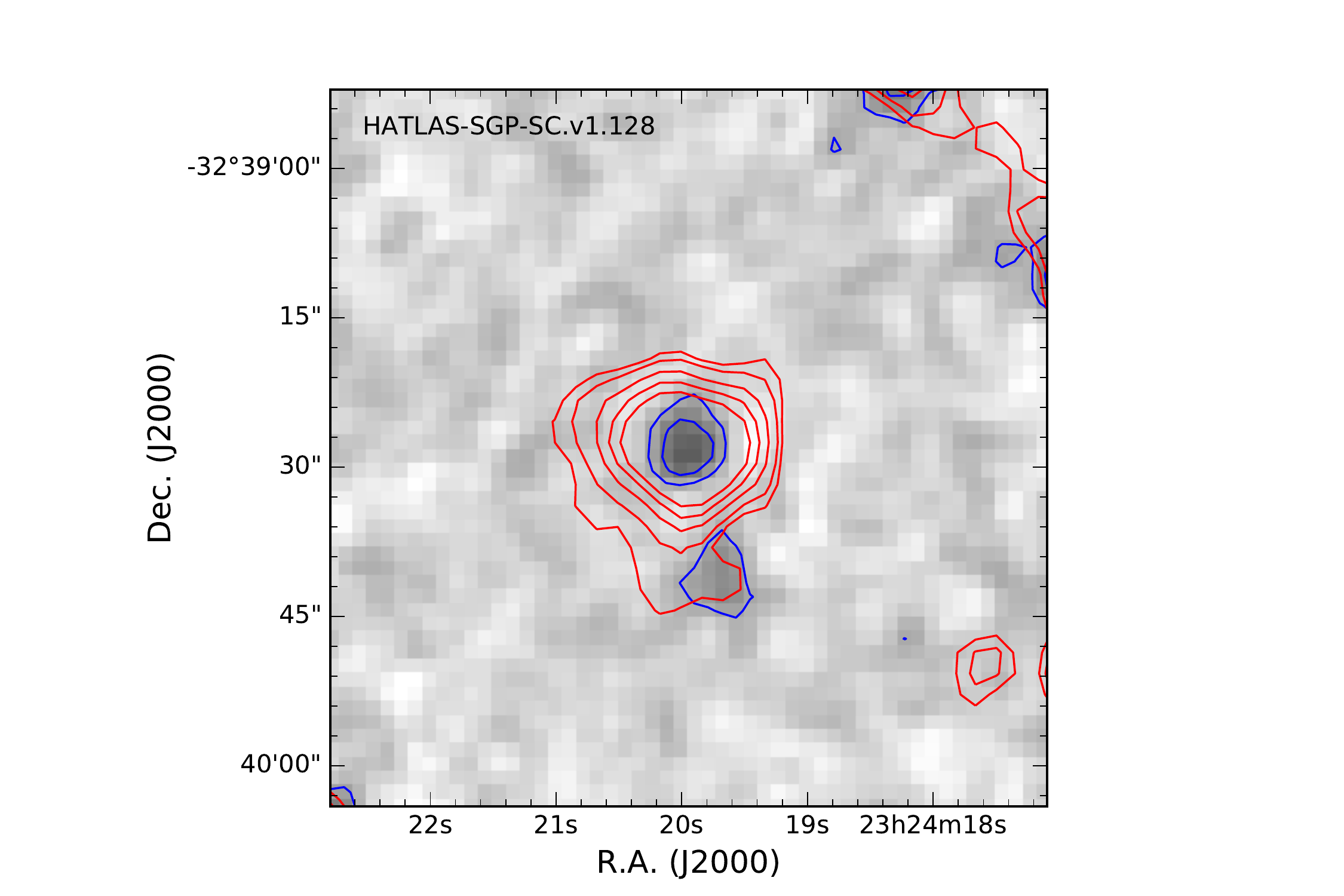}
    \includegraphics[width=0.33\textwidth,trim=60bp 0bp 140bp 40bp ,clip]{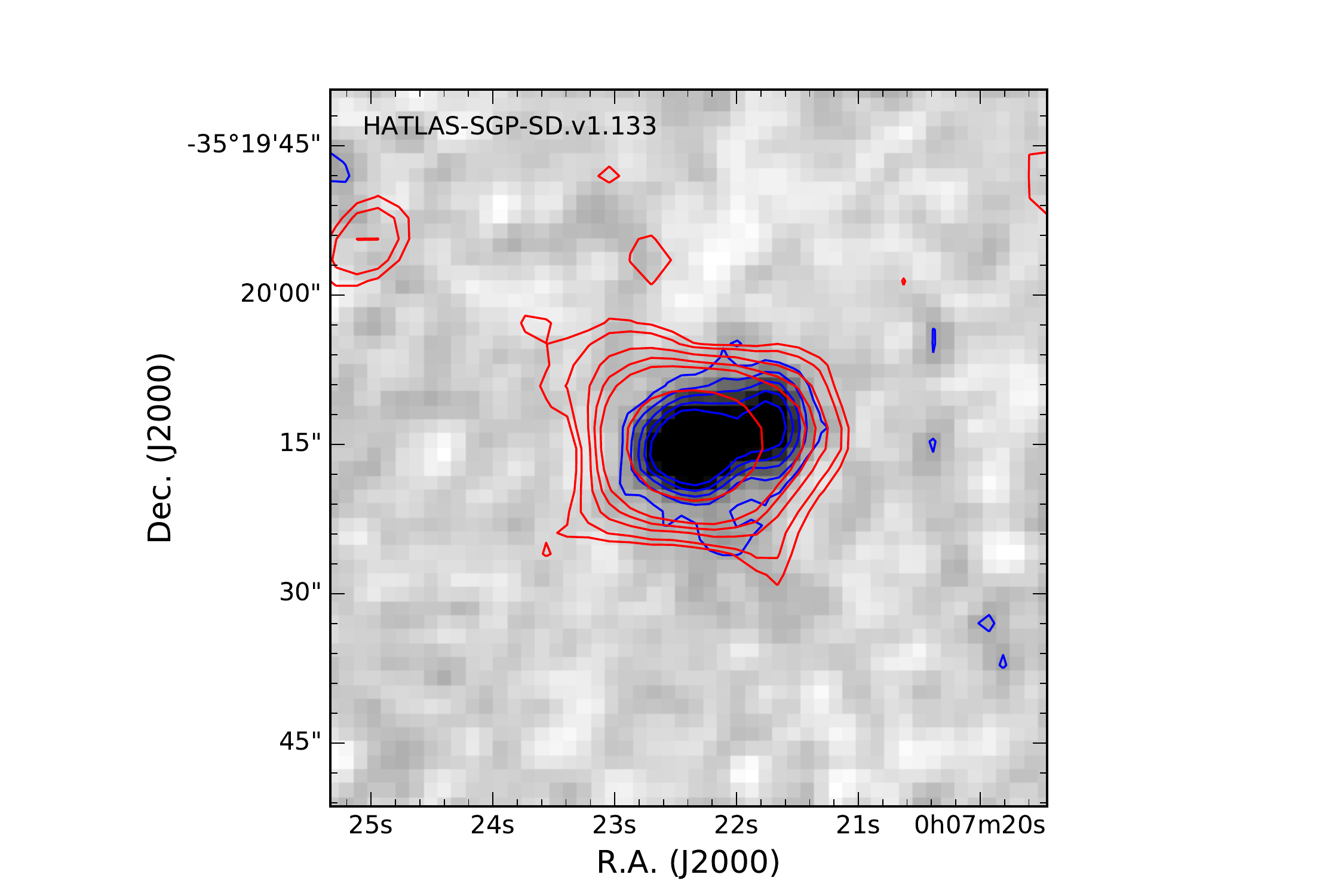}
    \includegraphics[width=0.33\textwidth,trim=60bp 0bp 140bp 40bp ,clip]{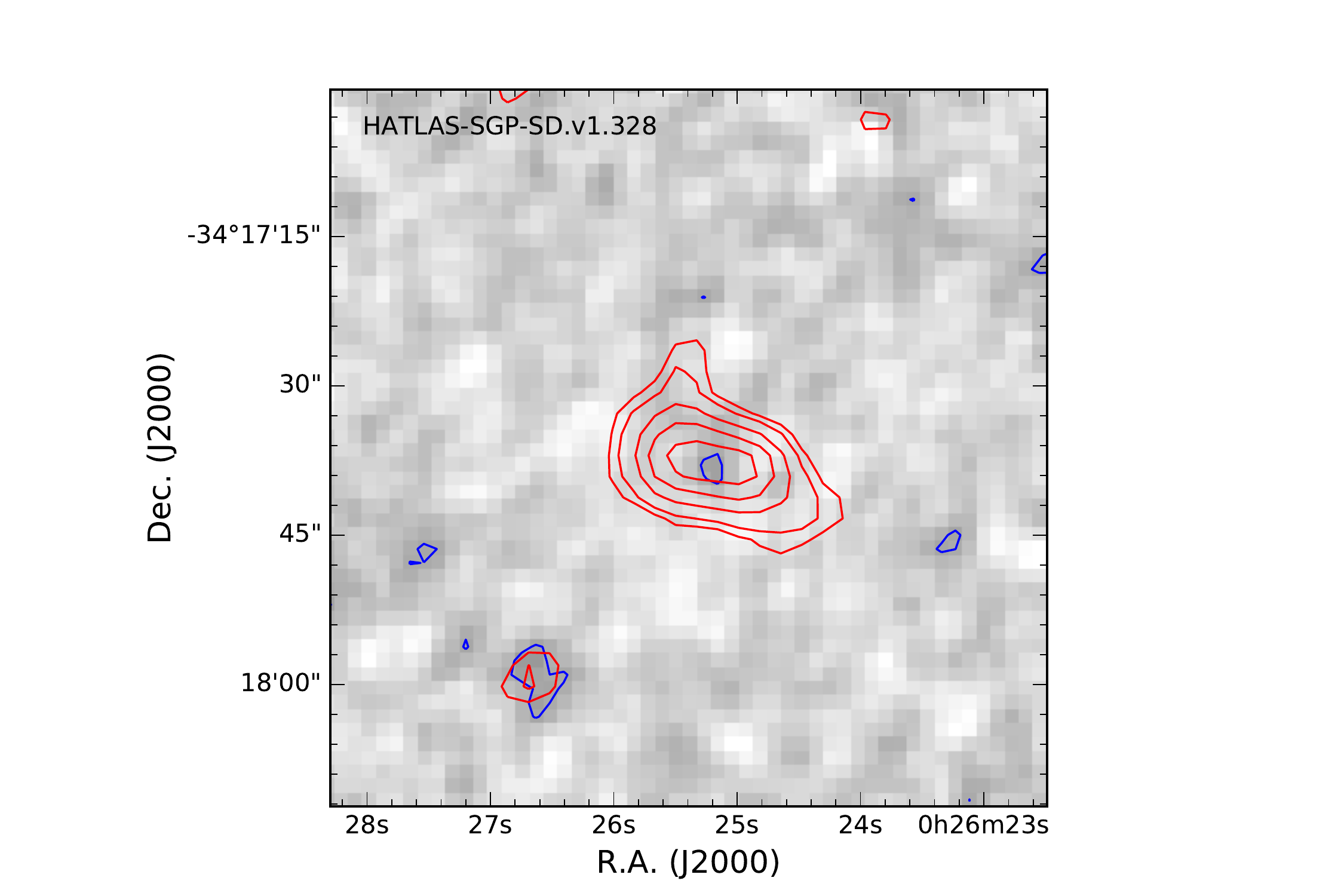}
    \includegraphics[width=0.33\textwidth,trim=60bp 0bp 140bp 40bp ,clip]{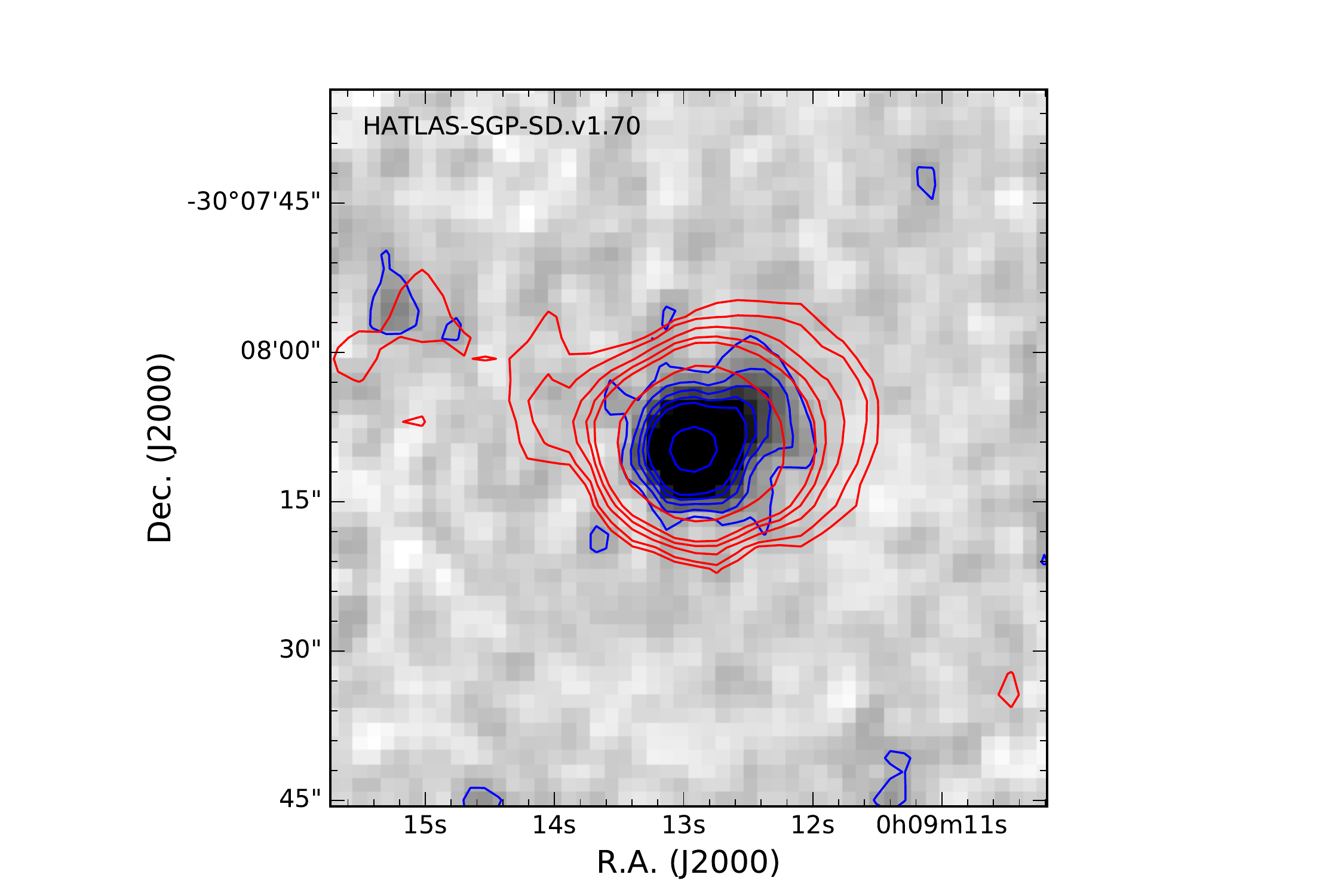}
    \label{fig:all_stamp2}
    \end{figure*}

    \begin{figure*}
    \centering
    \includegraphics[width=0.33\textwidth,trim=60bp 0bp 140bp 40bp ,clip]{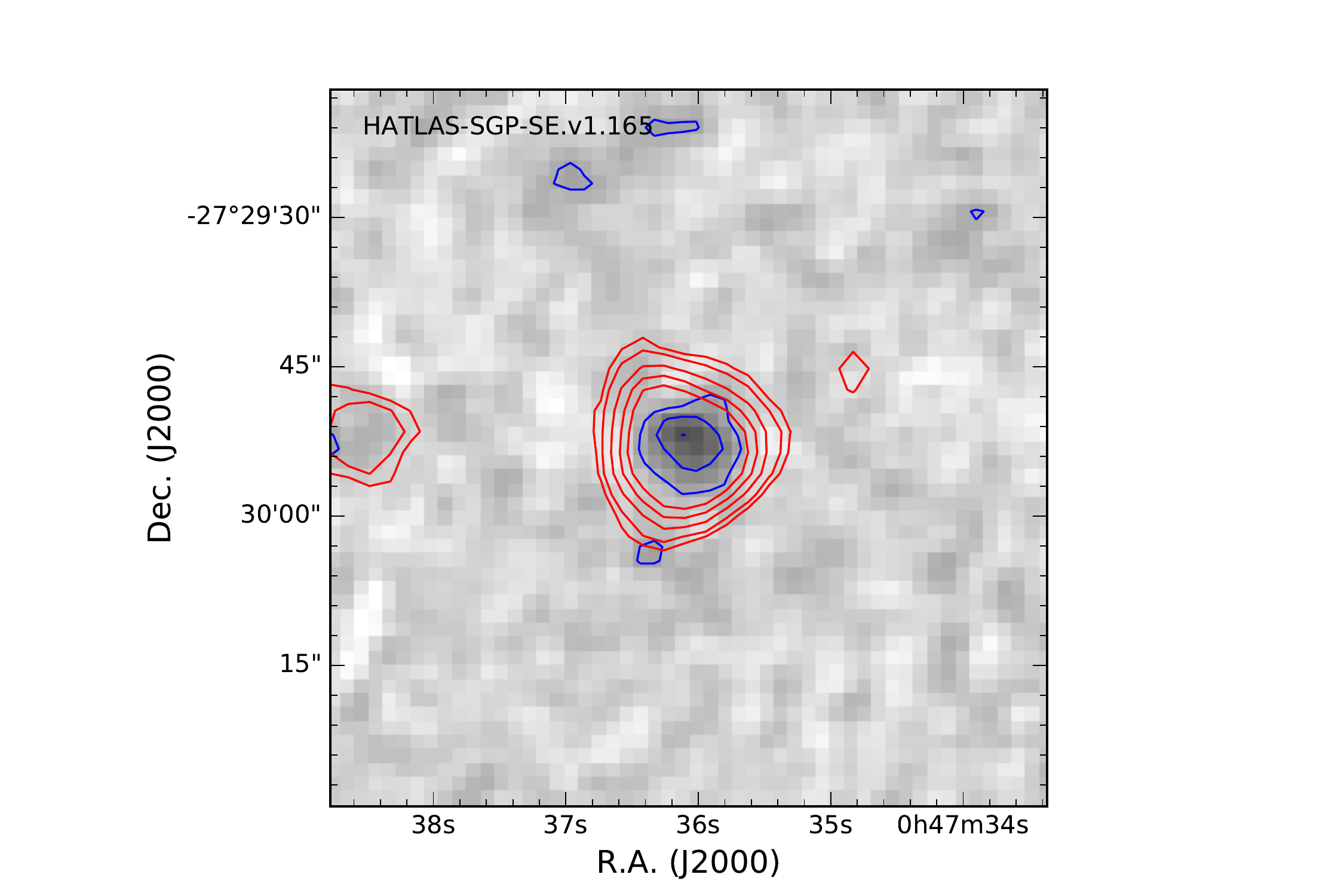}
    \includegraphics[width=0.33\textwidth,trim=60bp 0bp 140bp 40bp ,clip]{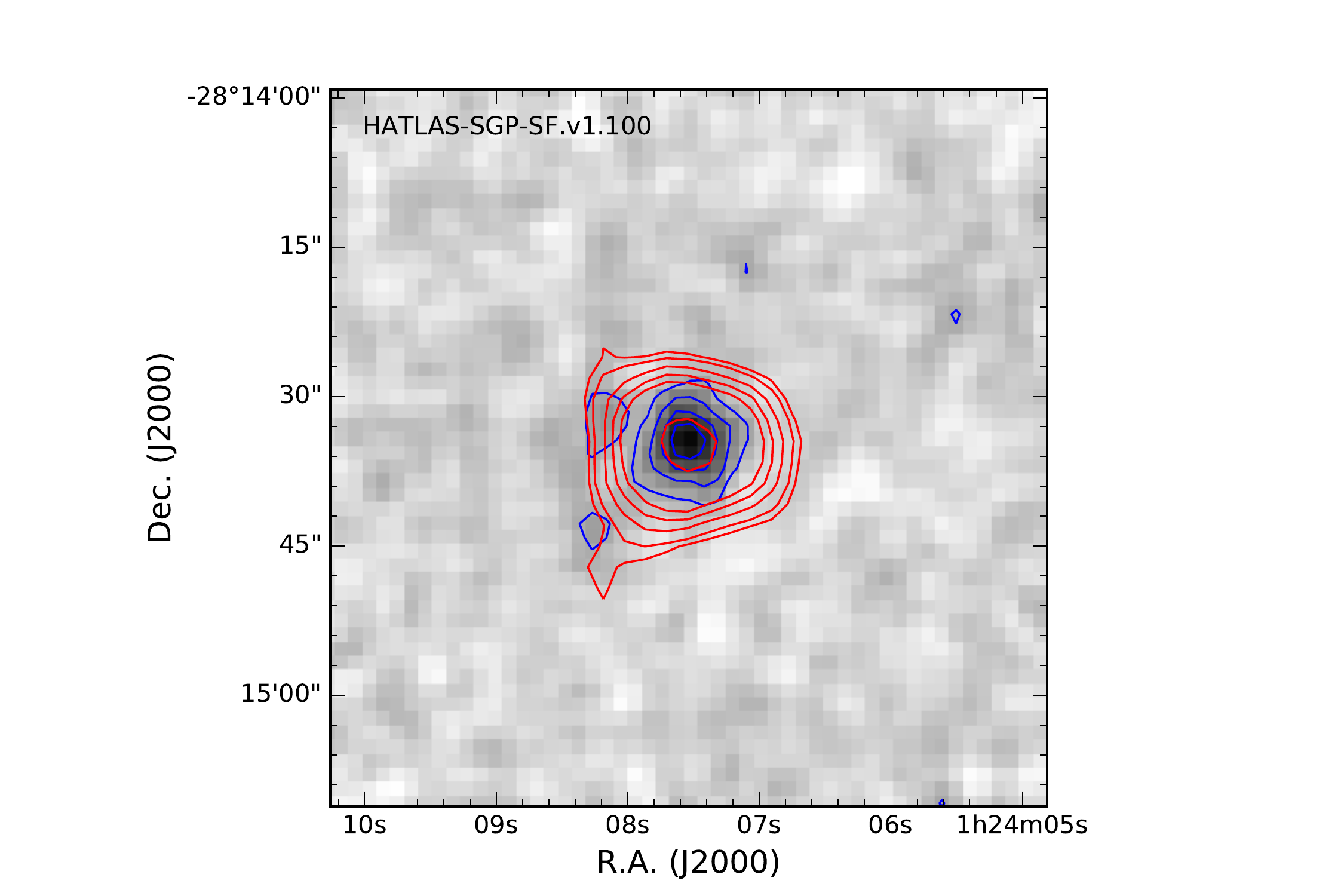}
    \includegraphics[width=0.33\textwidth,trim=60bp 0bp 140bp 40bp ,clip]{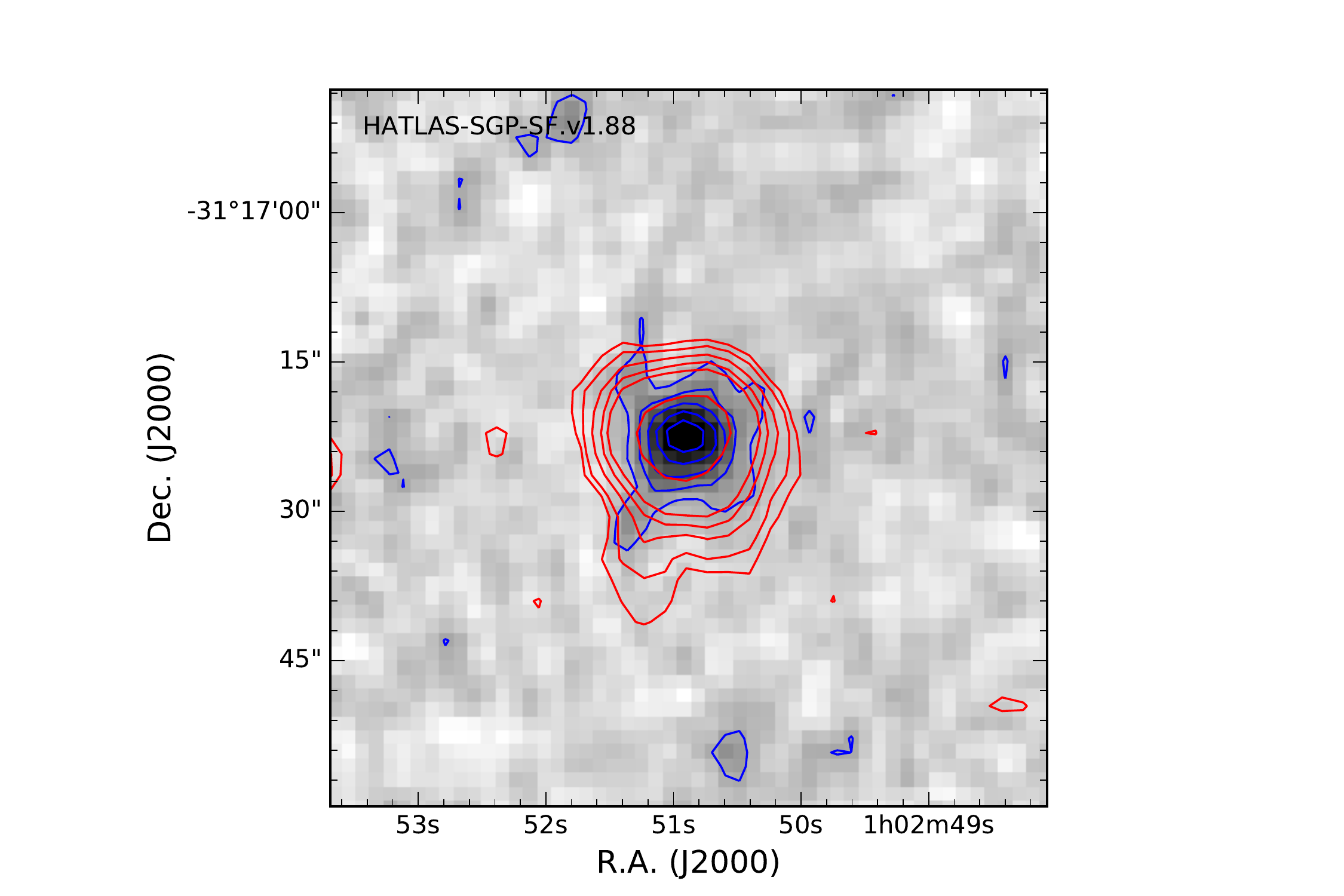}
    \includegraphics[width=0.33\textwidth,trim=60bp 0bp 140bp 40bp ,clip]{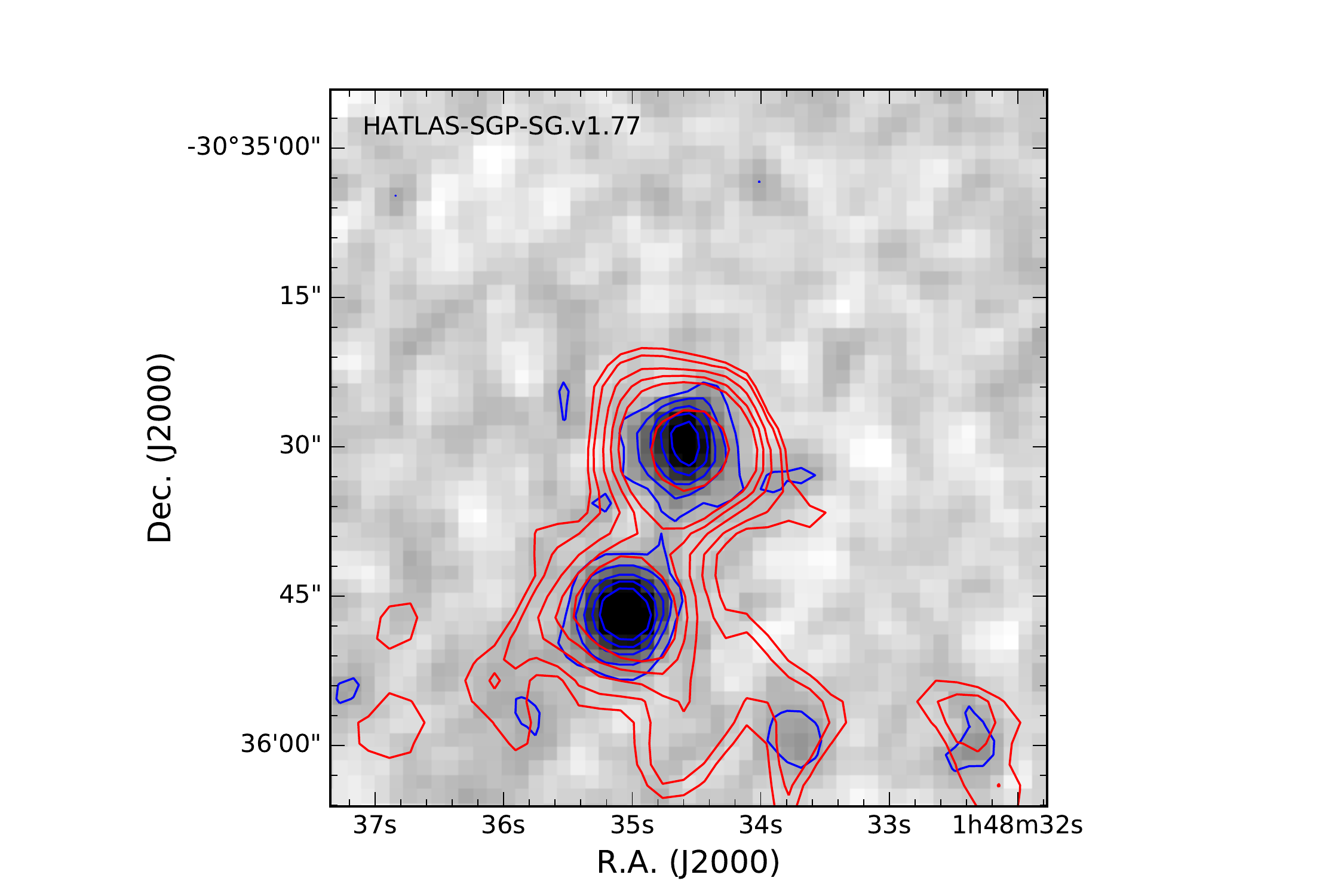}
    \includegraphics[width=0.33\textwidth,trim=60bp 0bp 140bp 40bp ,clip]{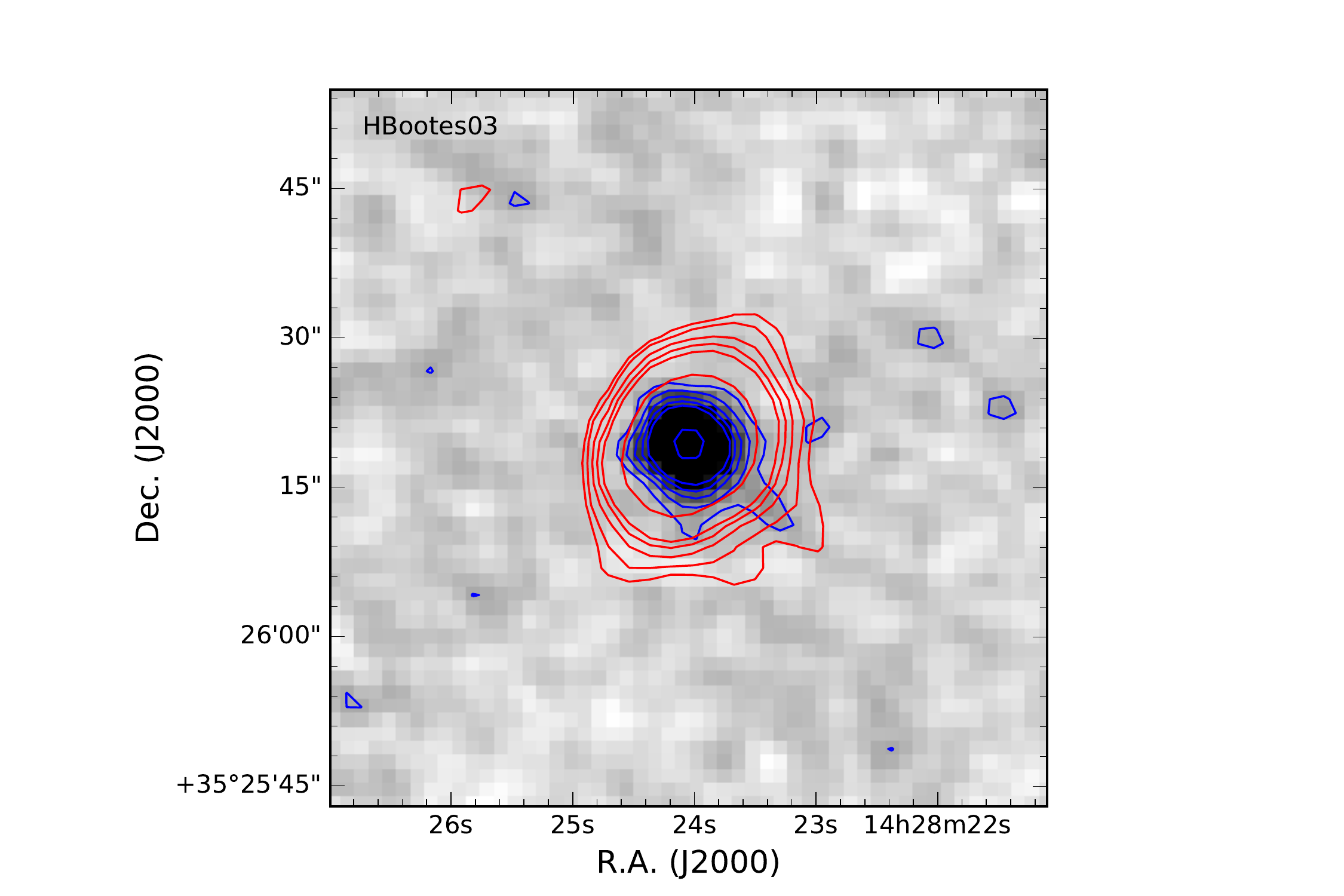}
    \includegraphics[width=0.33\textwidth,trim=60bp 0bp 140bp 40bp ,clip]{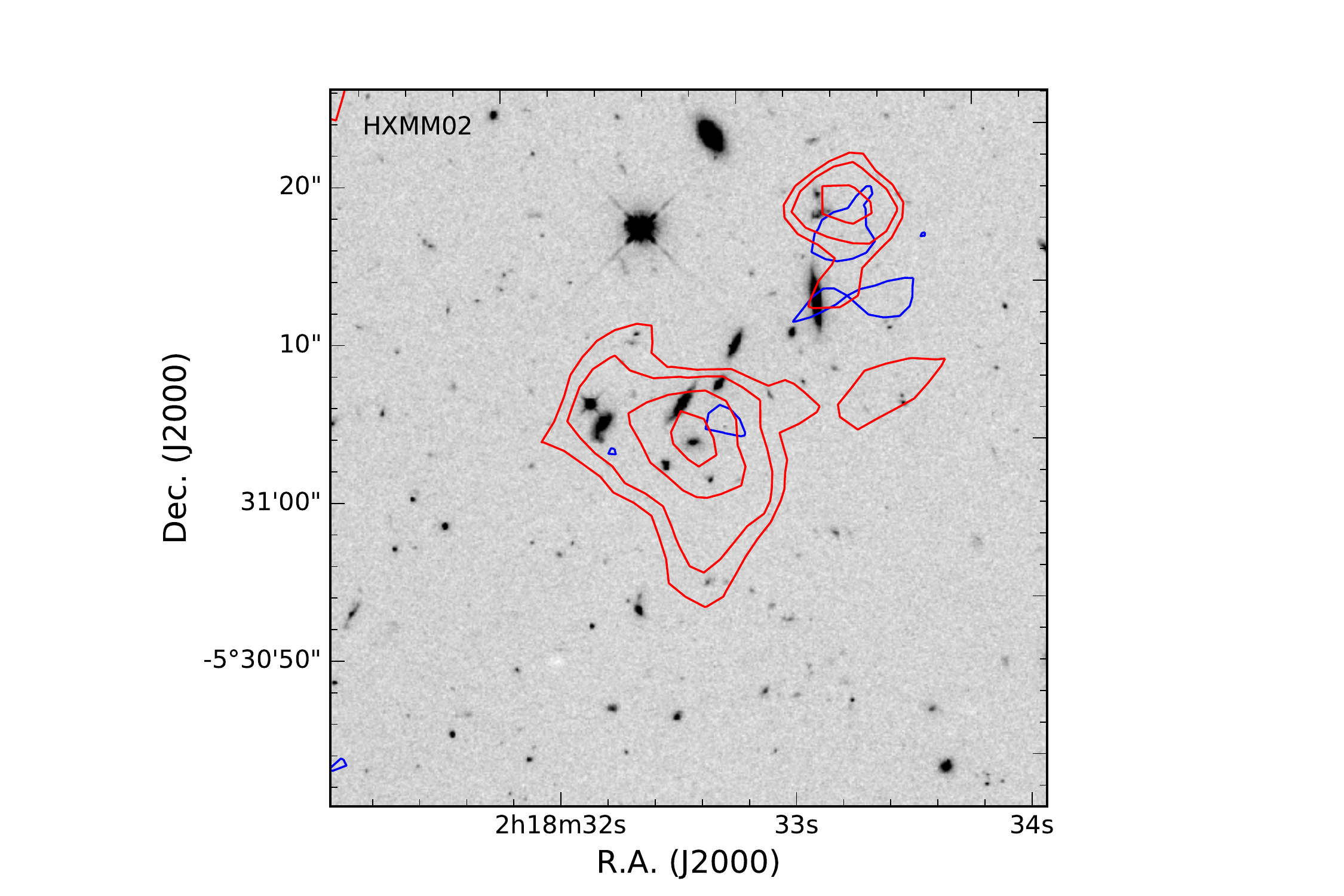}
    \includegraphics[width=0.33\textwidth,trim=60bp 0bp 140bp 40bp ,clip]{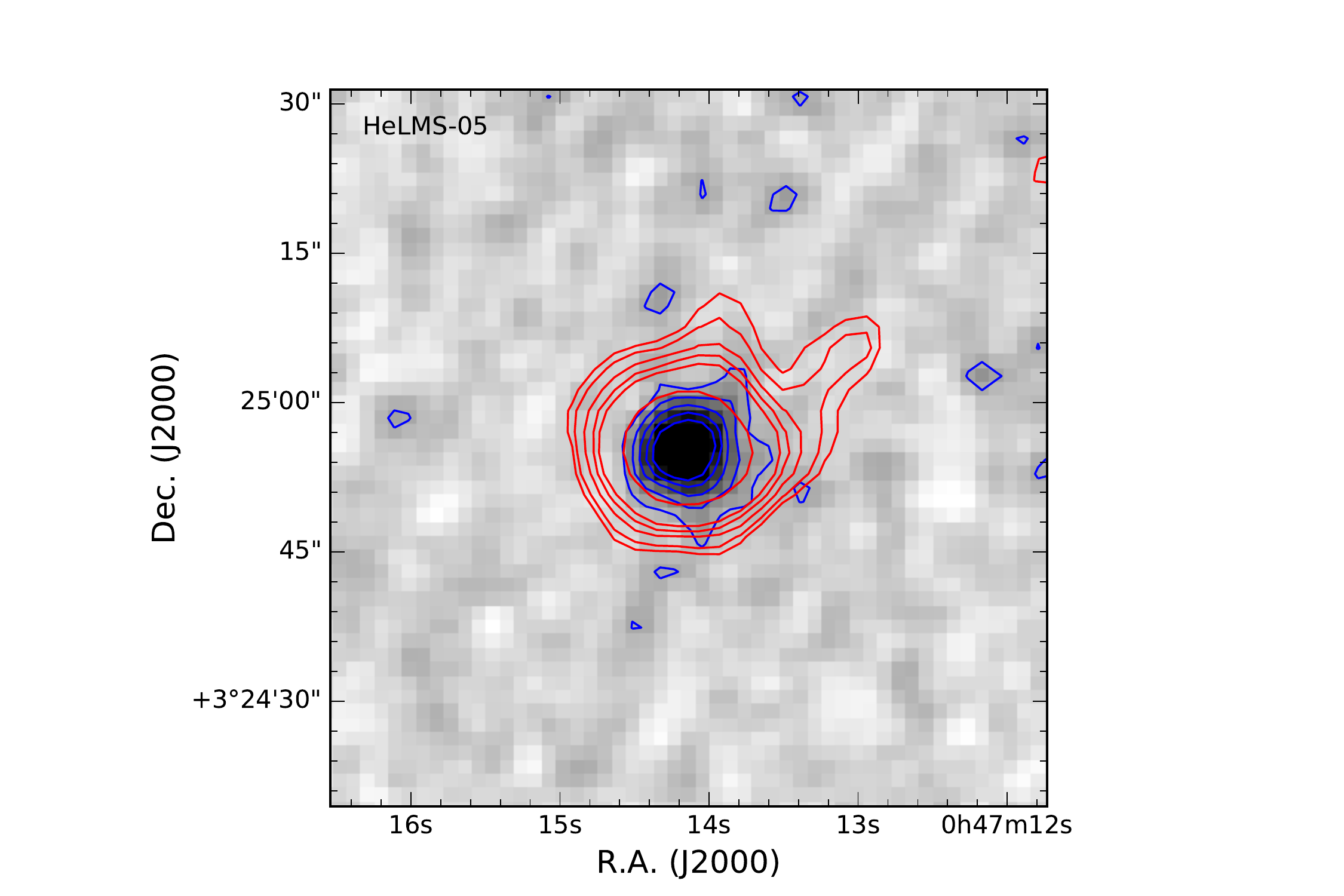}
    \includegraphics[width=0.33\textwidth,trim=60bp 0bp 140bp 40bp ,clip]{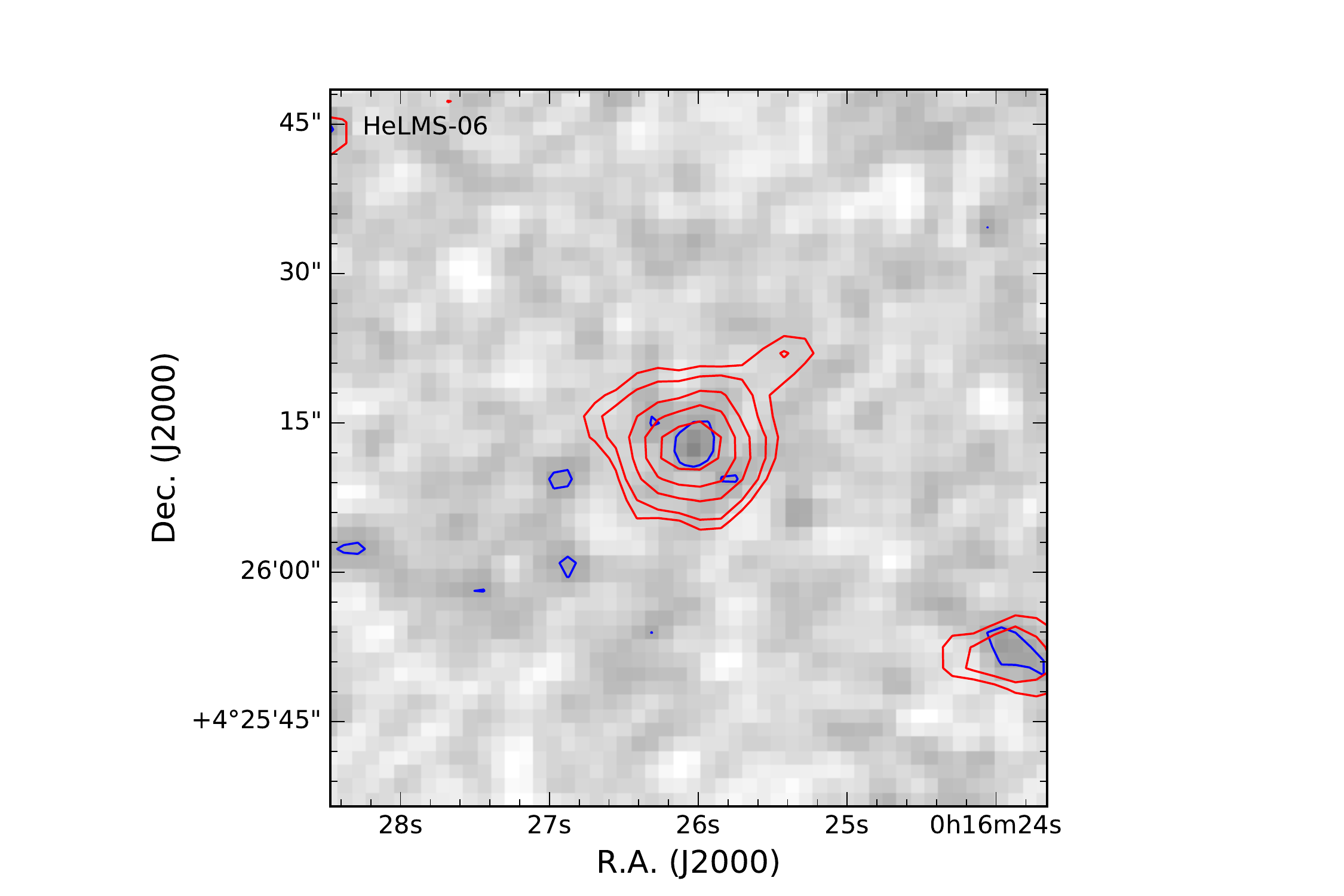}
    \includegraphics[width=0.33\textwidth,trim=60bp 0bp 140bp 40bp ,clip]{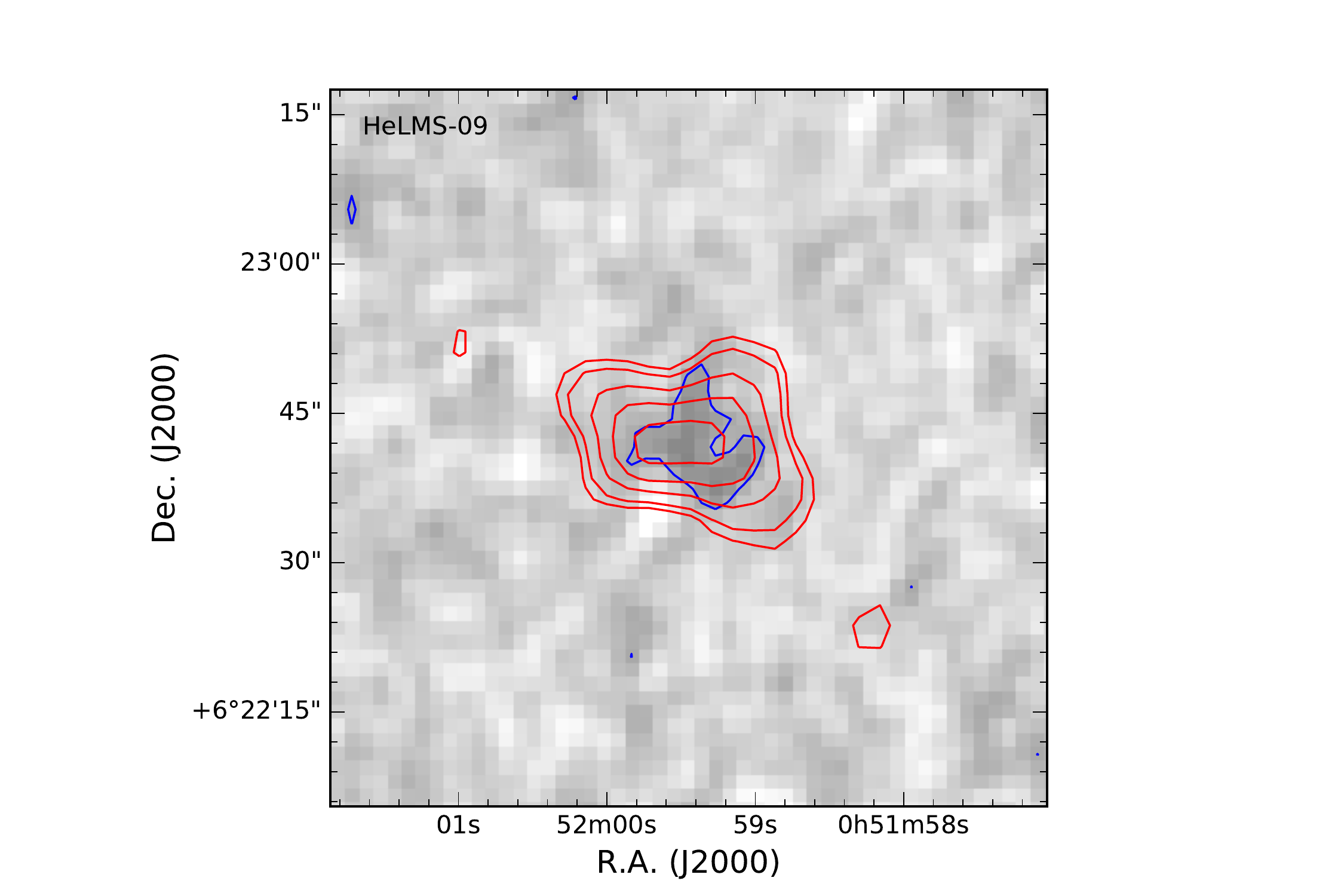}
    \includegraphics[width=0.33\textwidth,trim=60bp 0bp 140bp 40bp ,clip]{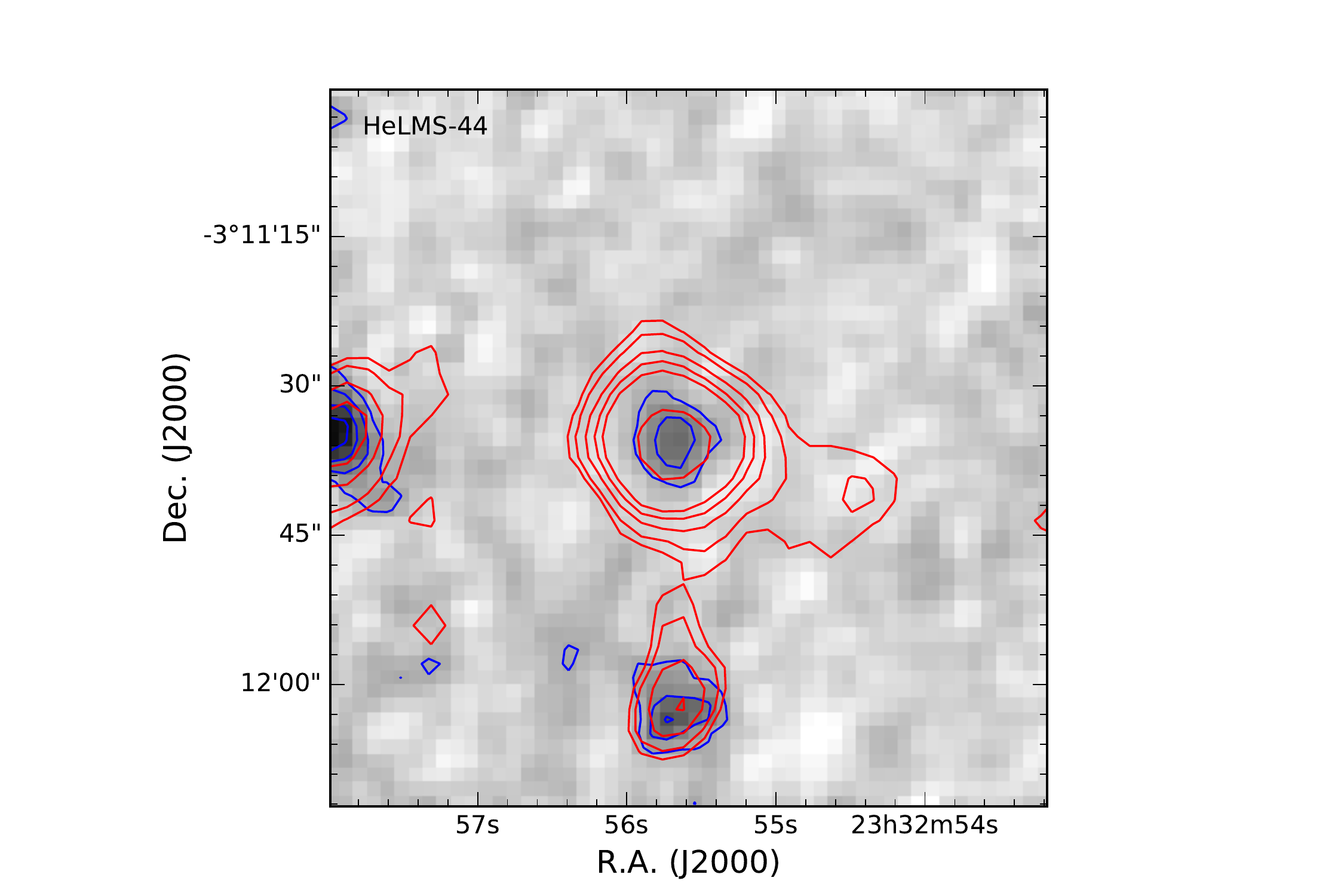}
    \includegraphics[width=0.33\textwidth,trim=60bp 0bp 140bp 40bp ,clip]{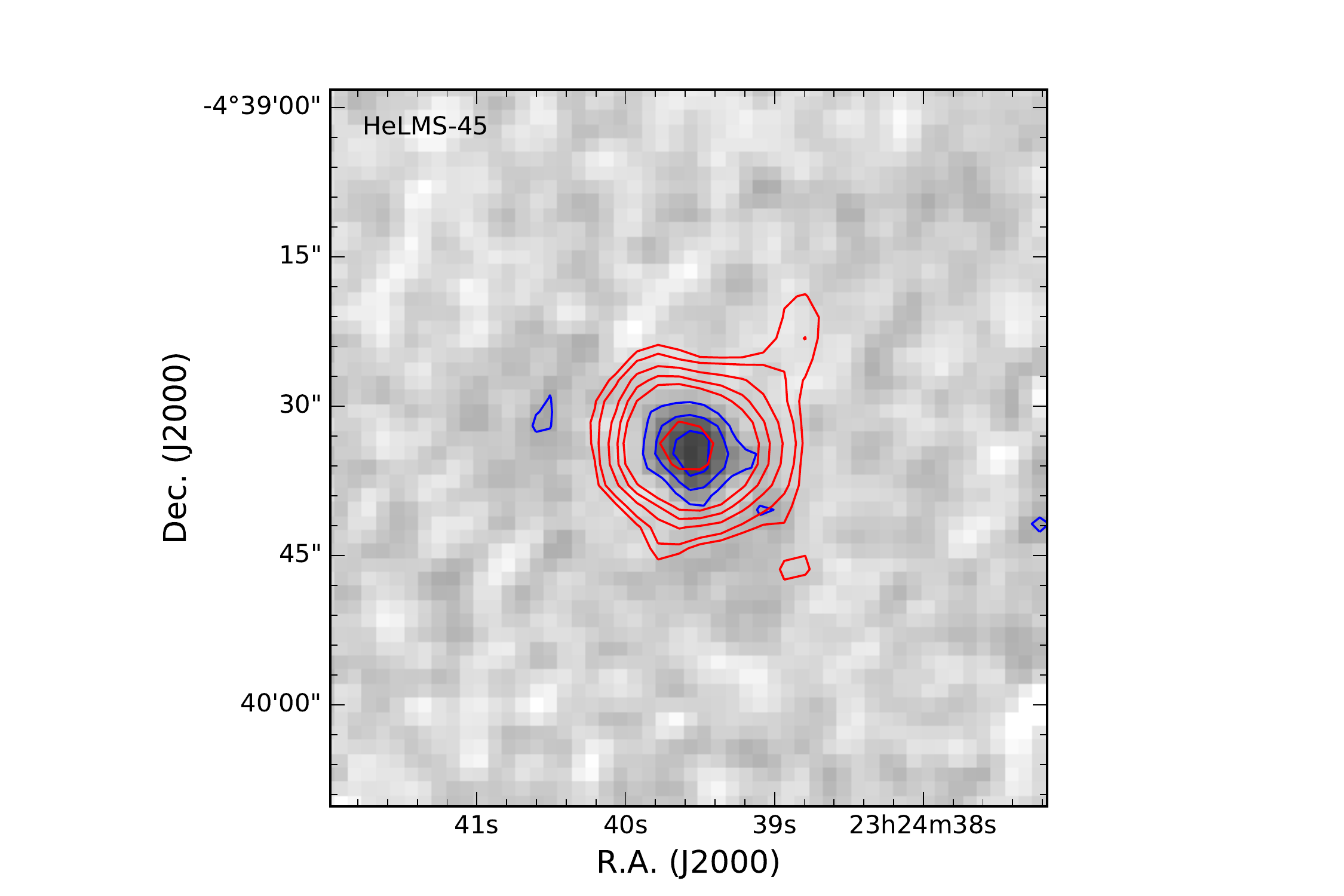}
    \includegraphics[width=0.33\textwidth,trim=60bp 0bp 140bp 40bp ,clip]{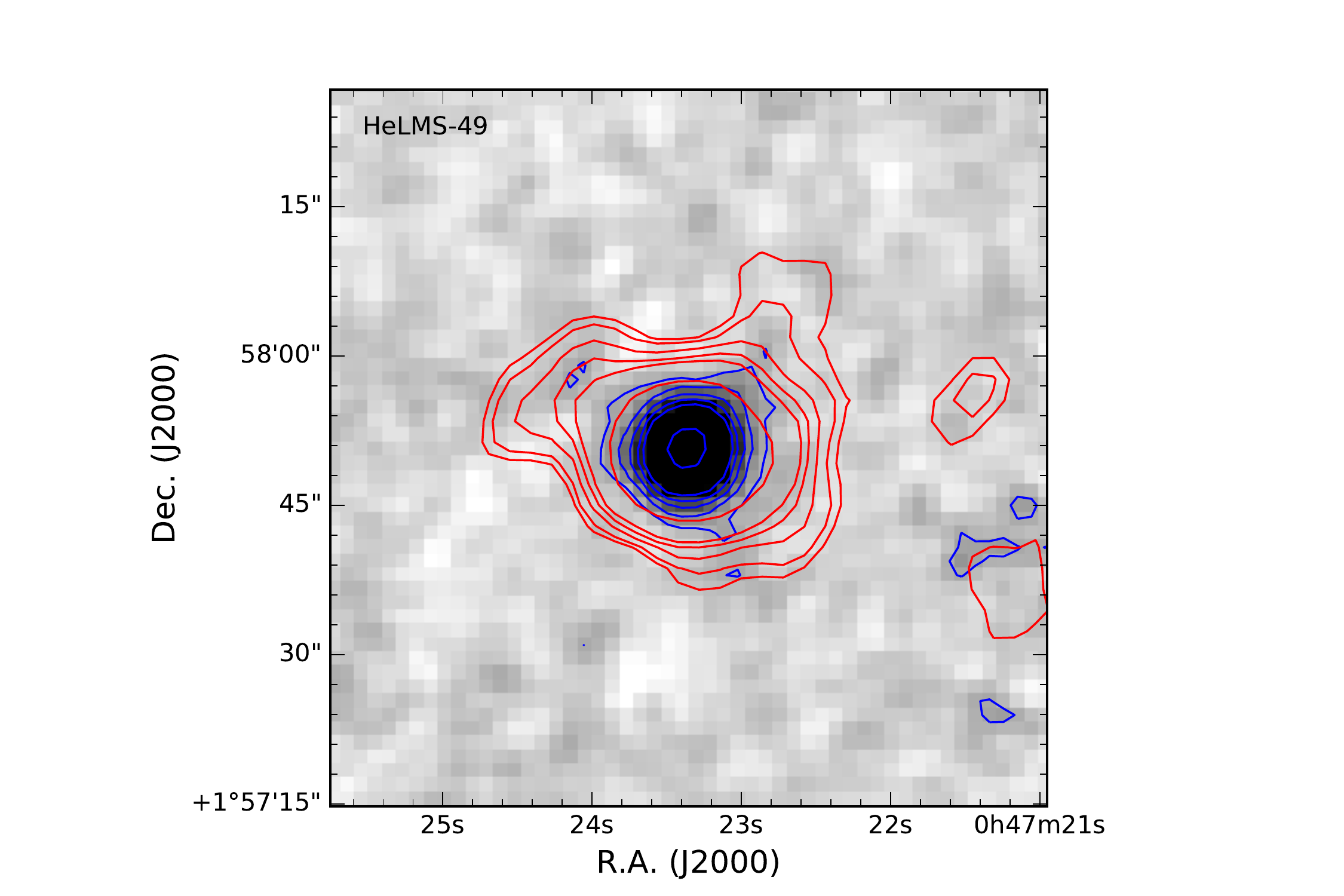}
    \label{fig:all_stamp3}
    \end{figure*}

    \begin{figure*}
    \centering

    \includegraphics[width=0.33\textwidth,trim=60bp 0bp 140bp 40bp ,clip]{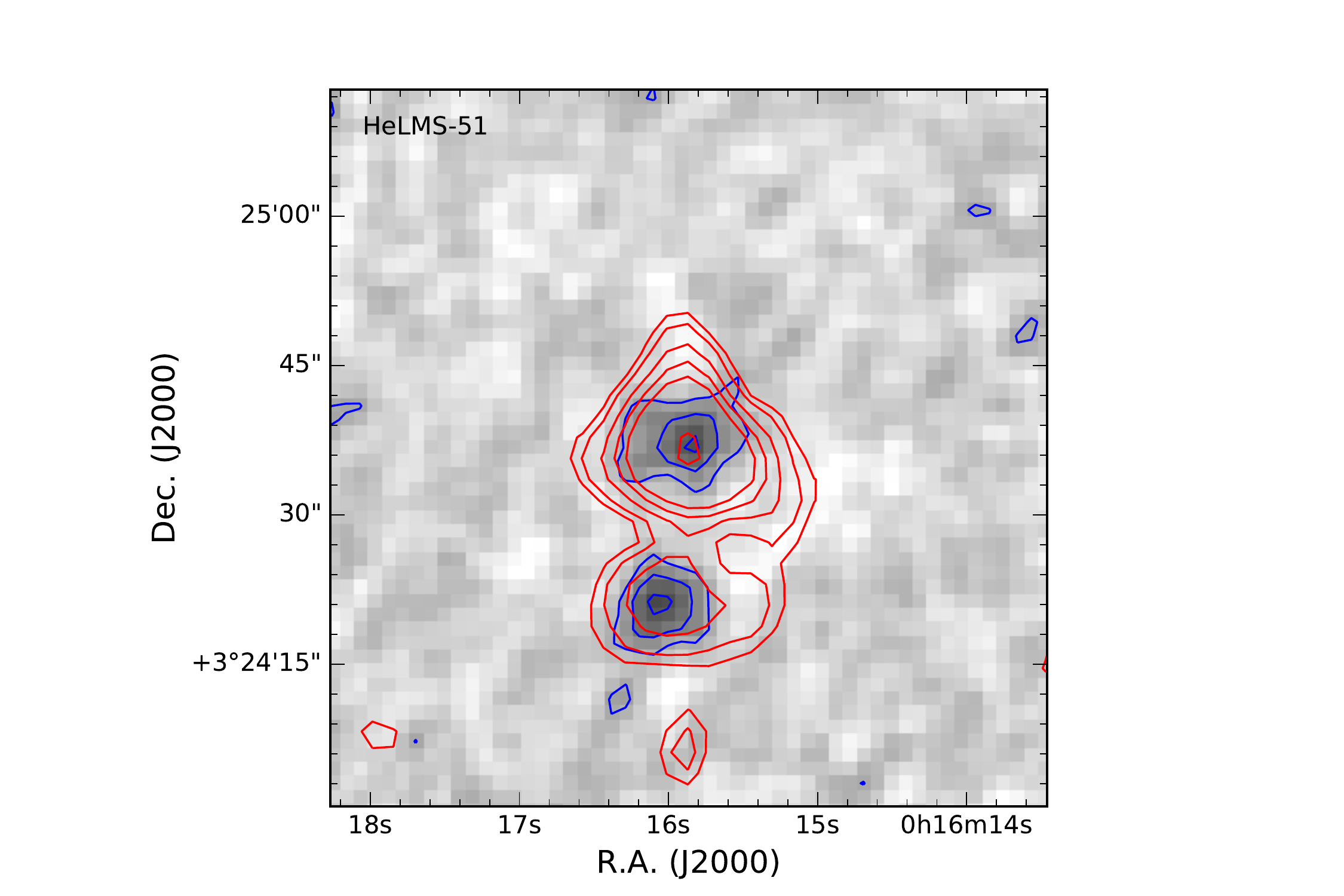}
    \includegraphics[width=0.33\textwidth,trim=60bp 0bp 140bp 40bp ,clip]{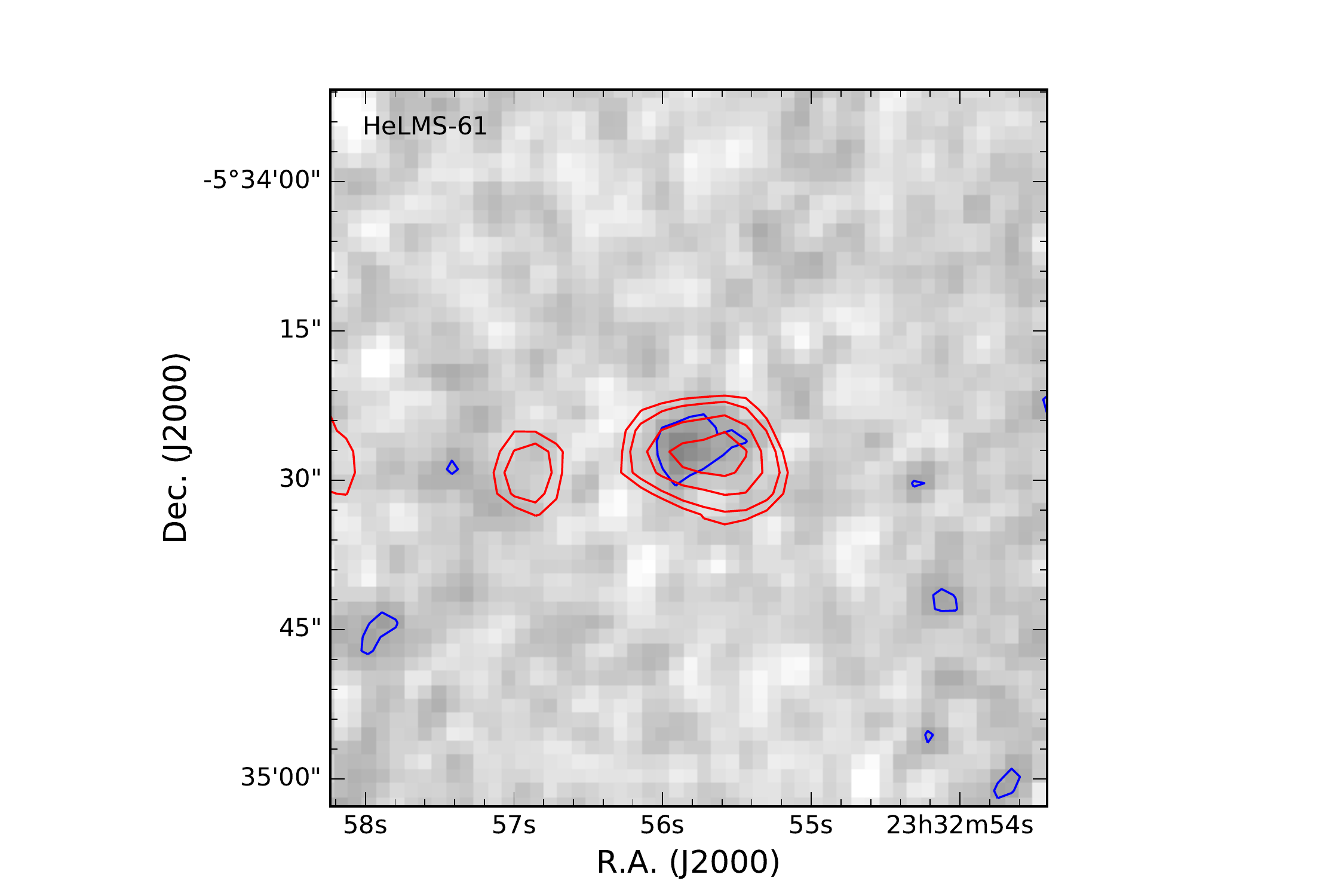}
    \includegraphics[width=0.33\textwidth,trim=60bp 0bp 140bp 40bp ,clip]{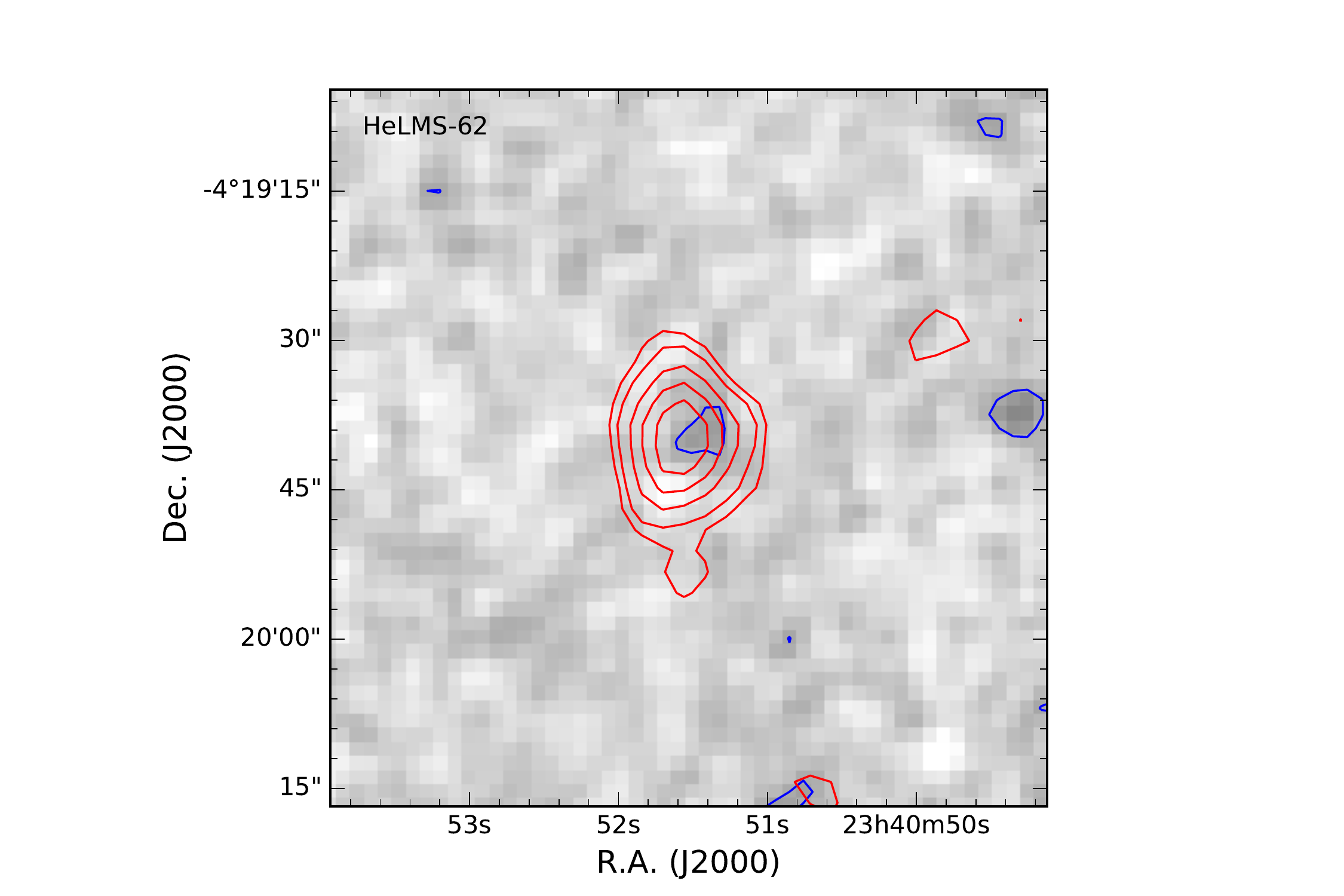}
    \includegraphics[width=0.33\textwidth,trim=60bp 0bp 140bp 40bp ,clip]{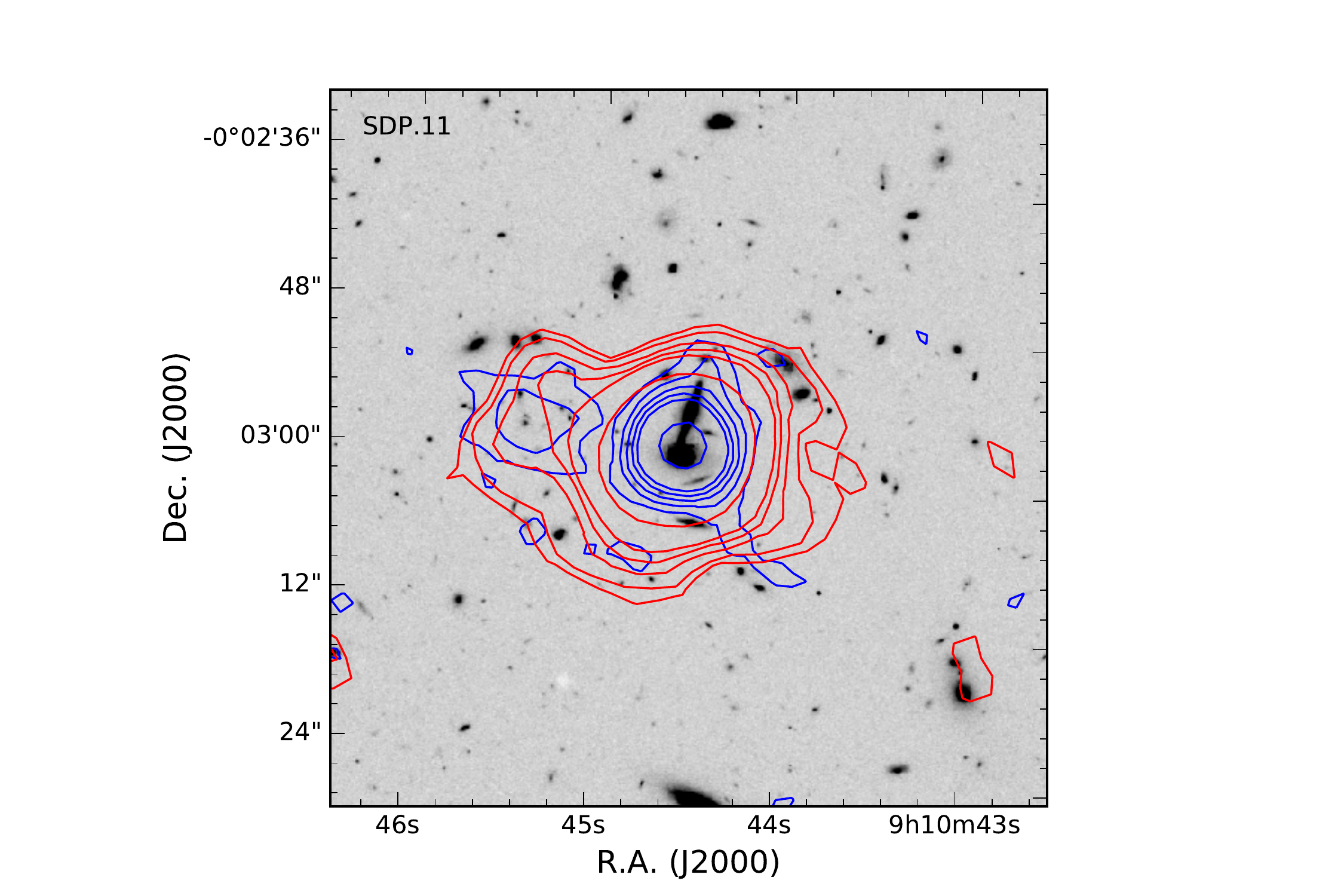}
    \includegraphics[width=0.33\textwidth,trim=60bp 0bp 140bp 40bp ,clip]{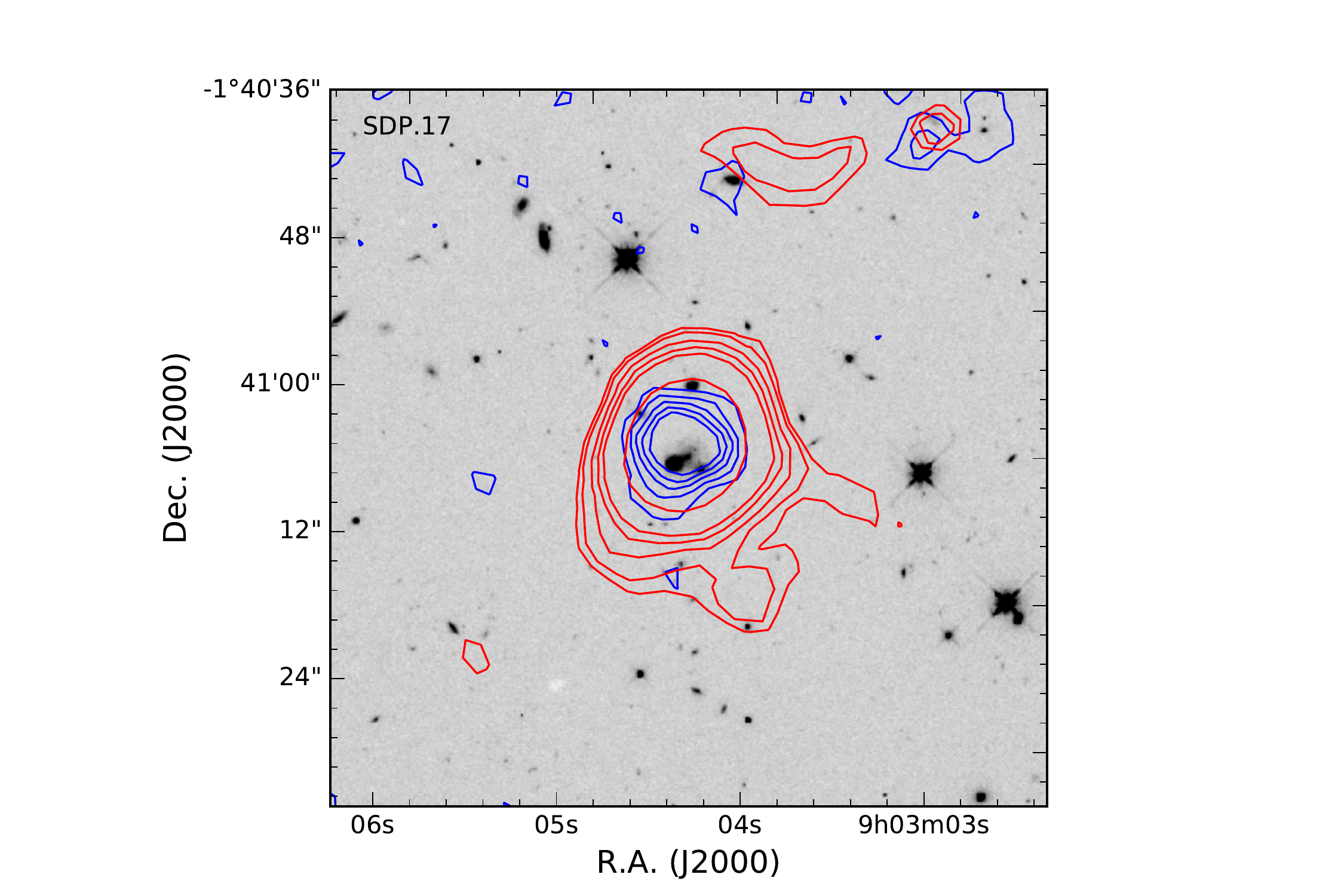}
    \includegraphics[width=0.33\textwidth,trim=60bp 0bp 140bp 40bp ,clip]{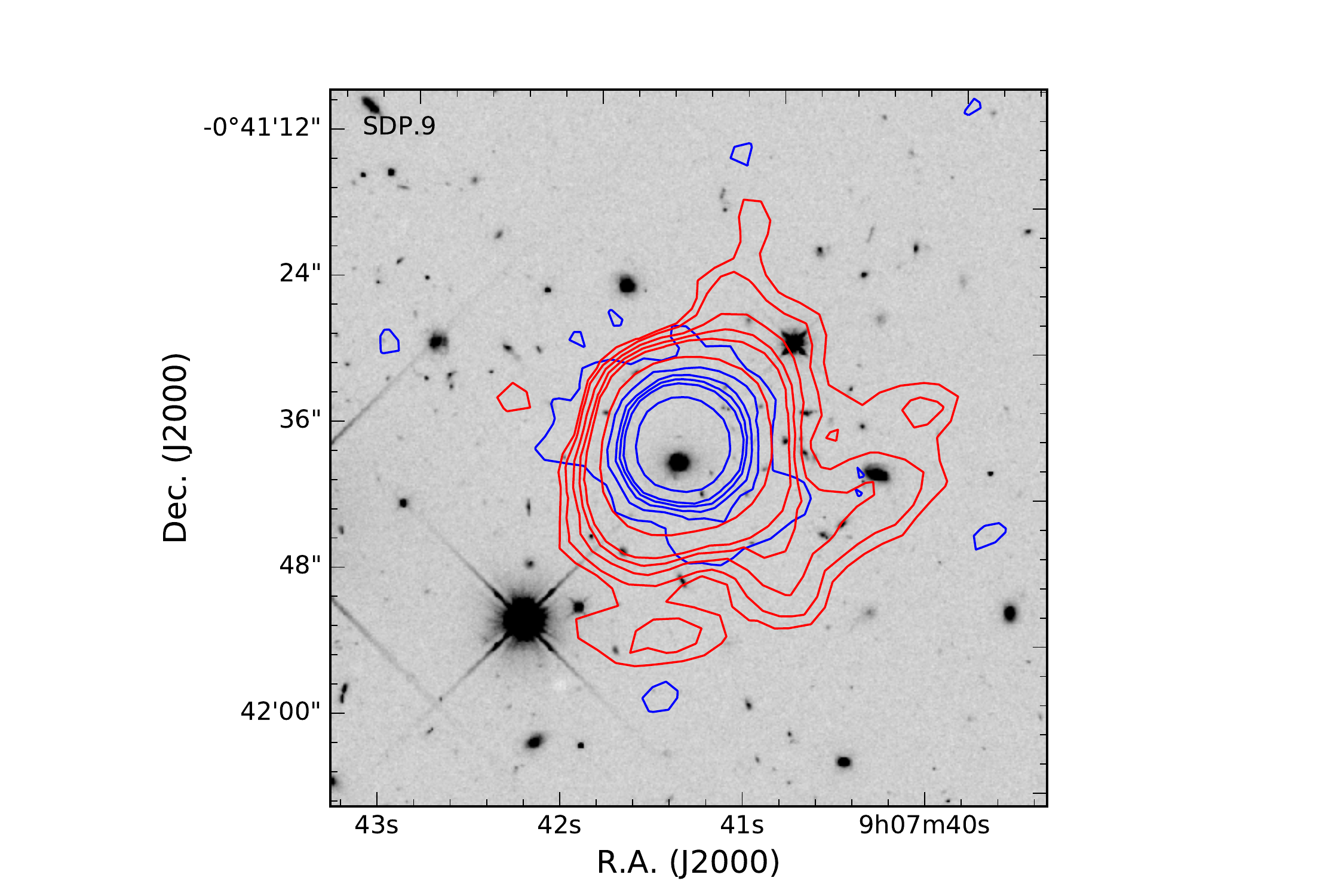}
    \includegraphics[width=0.33\textwidth,trim=60bp 0bp 140bp 40bp ,clip]{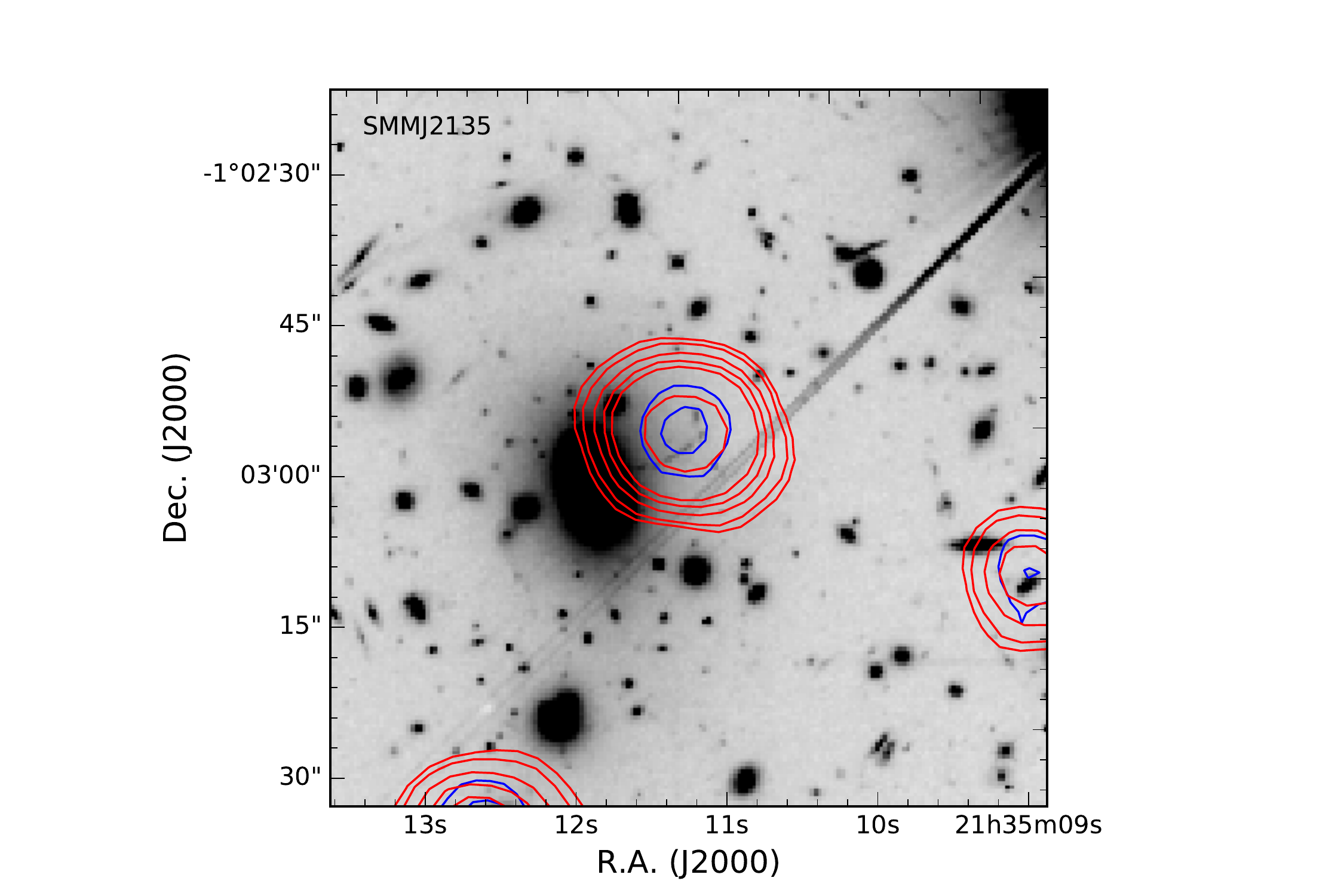}
    \includegraphics[width=0.33\textwidth,trim=60bp 0bp 140bp 40bp ,clip]{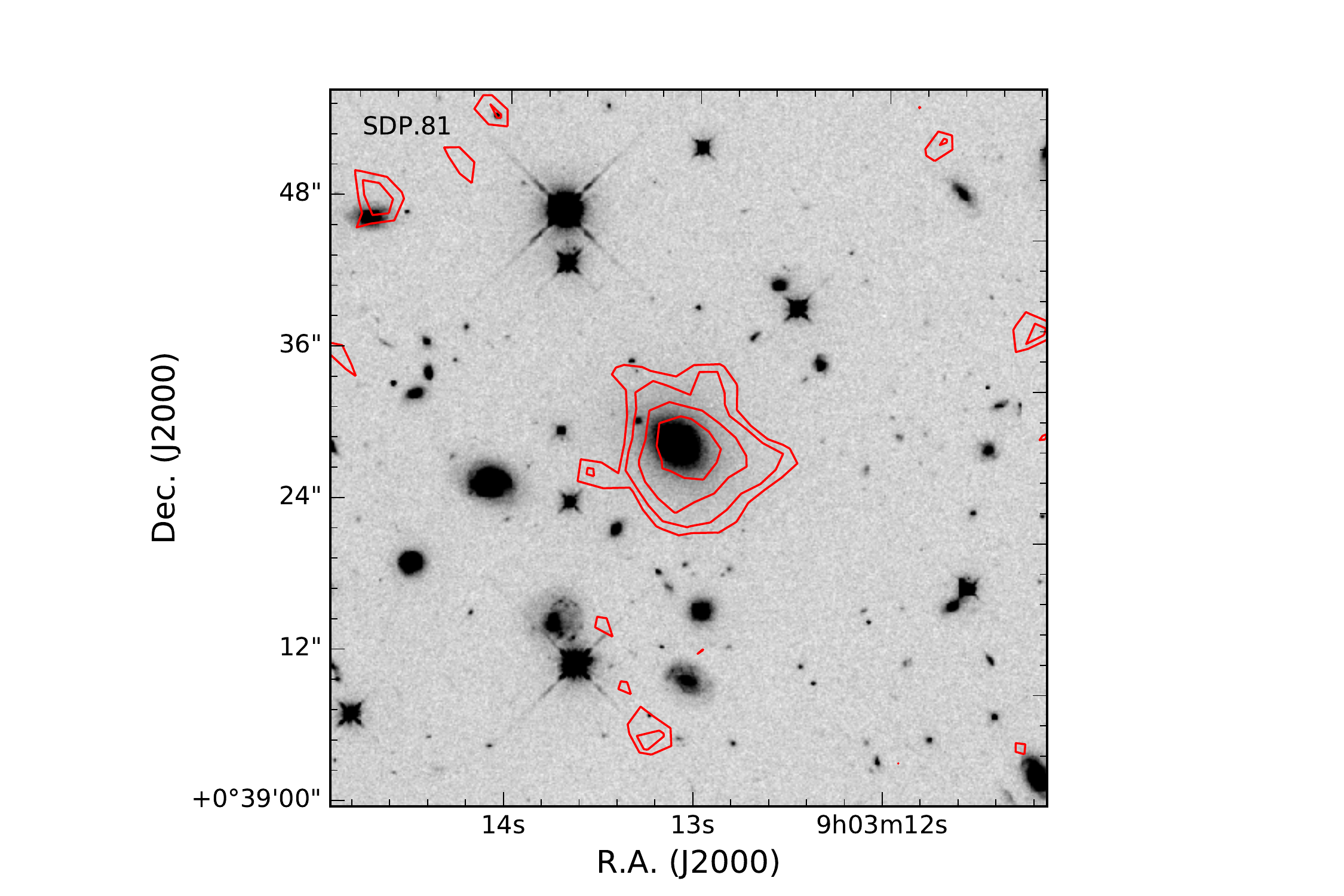}
    \includegraphics[width=0.33\textwidth,trim=60bp 0bp 140bp 40bp ,clip]{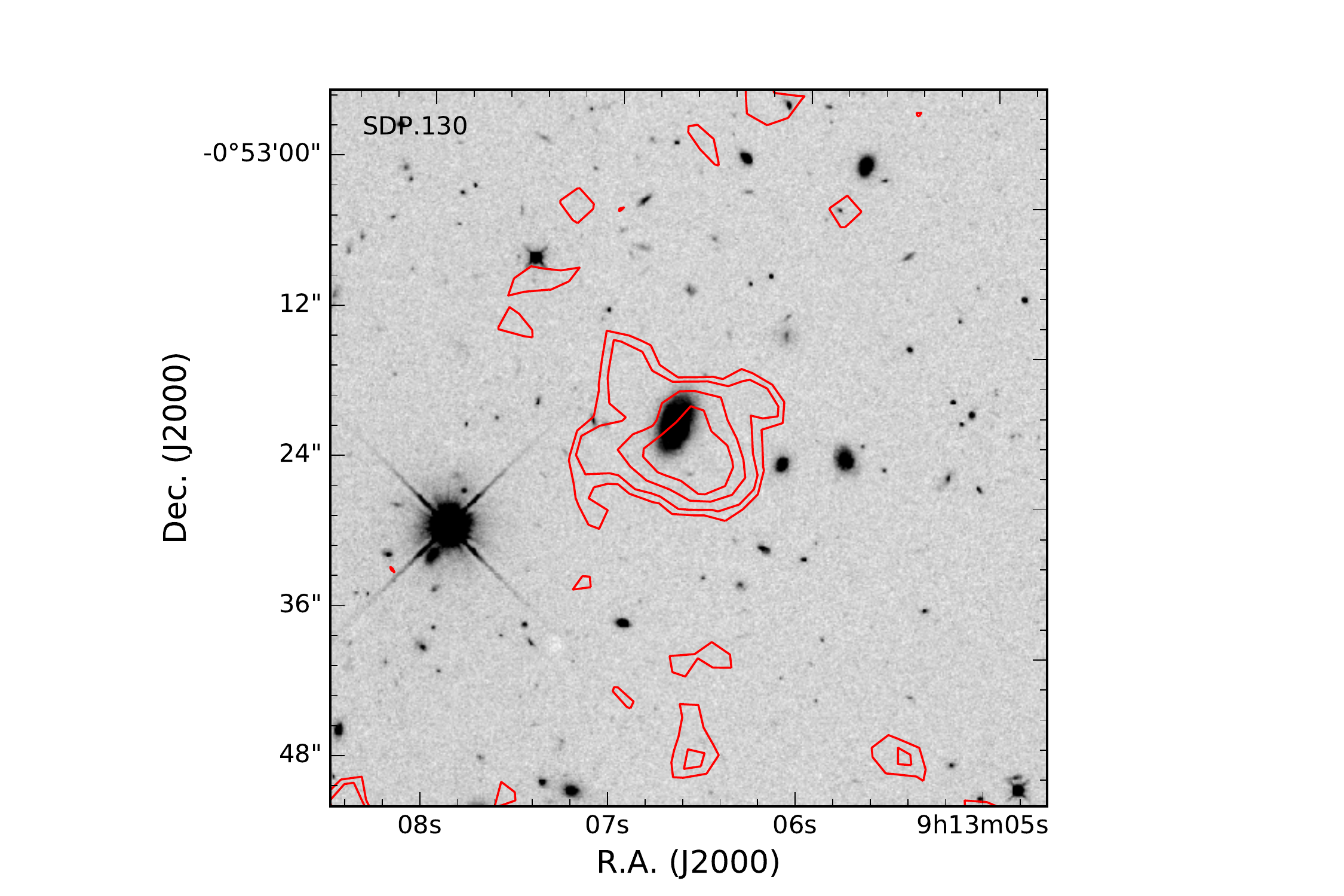}
    \label{fig:all_stamp4}
    \end{figure*}


\section{Baseline subtraction and signal recovery} \label{app:baseline_subtraction}
In this section we compare different background subtraction methods in the data
reduction of \herschel{} SPIRE/FTS data. To remove the overall baseline
profile, and especially the low-frequency ripple features produced by the
imperfect calibration, we perform a convolved baseline fitting.  To do so we
first mask the frequency ranges of a few strong lines (\CII, , \OIII\ 88-\mum,
OH), each with a 2 GHz width, to avoid contaminations from the lines.  Then we
convolve the masked spectra with a Gaussian profile with 15 GHz (FWHM), which
ensures a local baseline feature but without fitting the relatively narrower
line features. We then subtract the baseline profiles from the original spectra
to get the final output spectra for each target.

\begin{figure*}
\centering
\includegraphics[width=0.48\textwidth ]{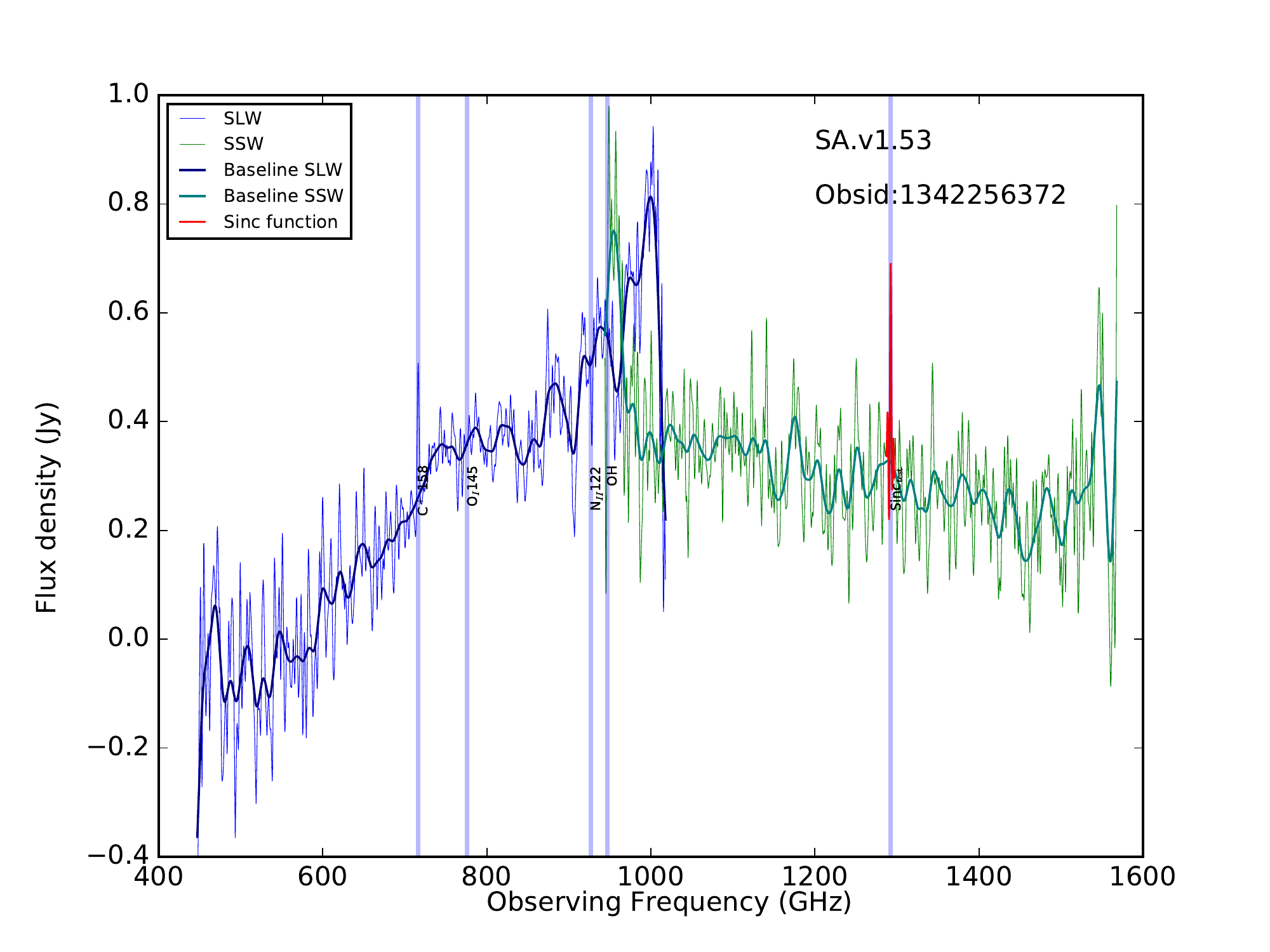}
\includegraphics[width=0.48\textwidth ]{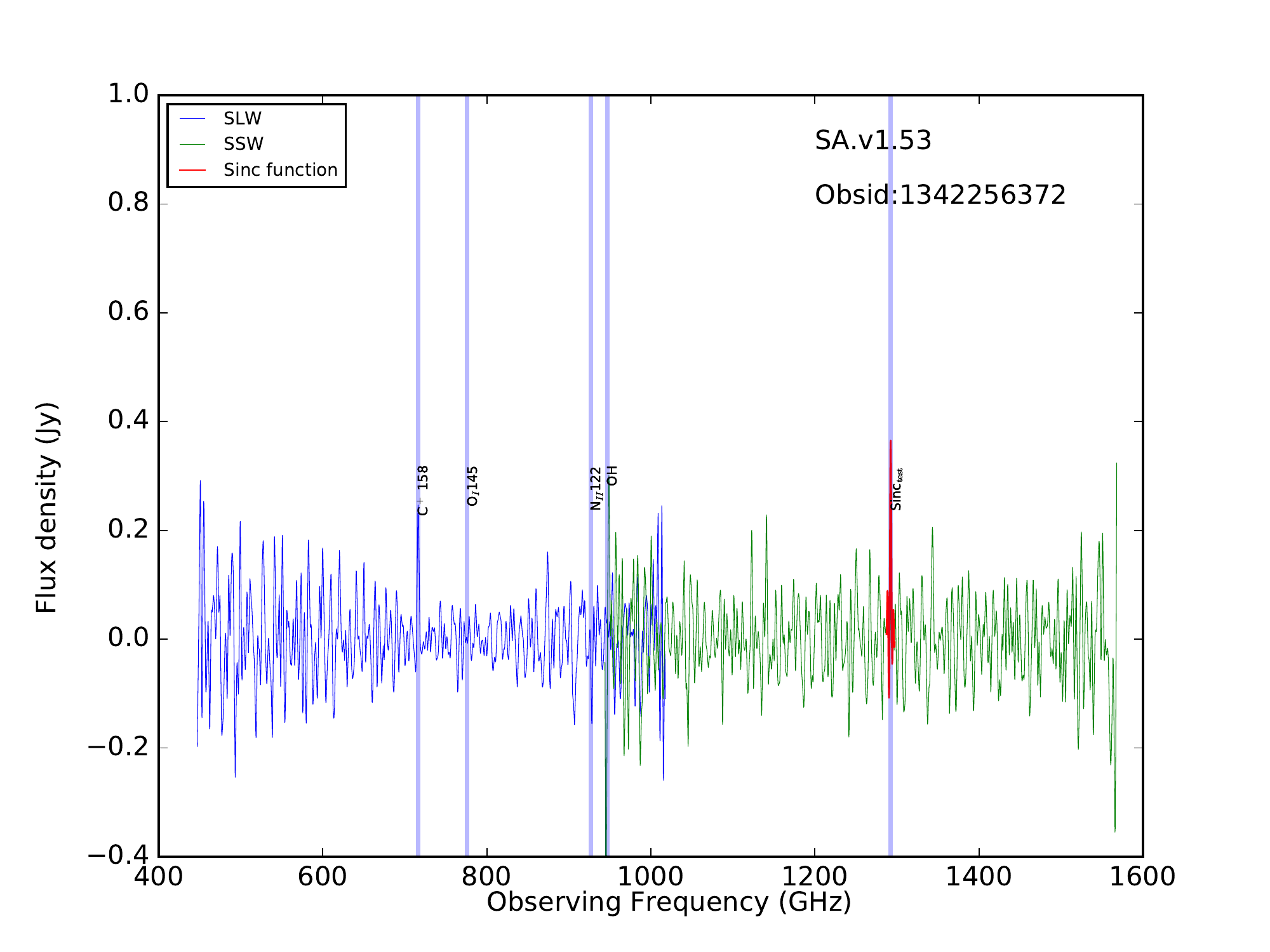}
\caption{An example of our baseline subtraction method. 
{\it Left}: The observed \herschel{} FTS spectra of SA.v1.53, after standard
calibrations. The thin blue line shows the spectrum obtained in the SLW-03
receiver, and the thin green line shows the spectrum obtained in the SSW-04
receiver. The thick lines show the baselines fitted using a 15 GHz Gaussian
width, after masking the frequency ranges of potential line features. The blue
shadowed regions show the masked frequency ranges potentially dominated by line
emission or absorption features. Red line shows an artificial Sinc function
response, which has a peak flux density of 0.3 Jy, a FWHM width of 1.2 GHz, and
is further convolved with a Gaussian profile with a 500 \kms\ FWHM. 
{\it Right}: The FTS spectra after baseline subtraction. The line features (the
\CII\ line and the testing Sinc function) are still robust.}
  \label{fig:sav153}
\end{figure*}

Figure \ref{fig:sav153} presents an example of the baseline subtraction method.
The thin lines show the pipeline produced spectra in the two central pixels of
the \herschel{ } FTS, SLW03 and SSW04. Thick lines show the baselines produced
by convolving the masked spectra (the line free part) and Gaussian profile with
15 GHz in FWMH. The blue shadowed regions show the frequency ranges of the
masked line features, which may potentially bias the baseline fitting. The red
line shows an artificial Sinc function response with a peak of 0.3 Jy and a
FWHM width of 1.2 GHz convolved with a 500 \kms Gaussian FWHM.

\begin{figure*}
\centering
\includegraphics[scale=0.5 ]{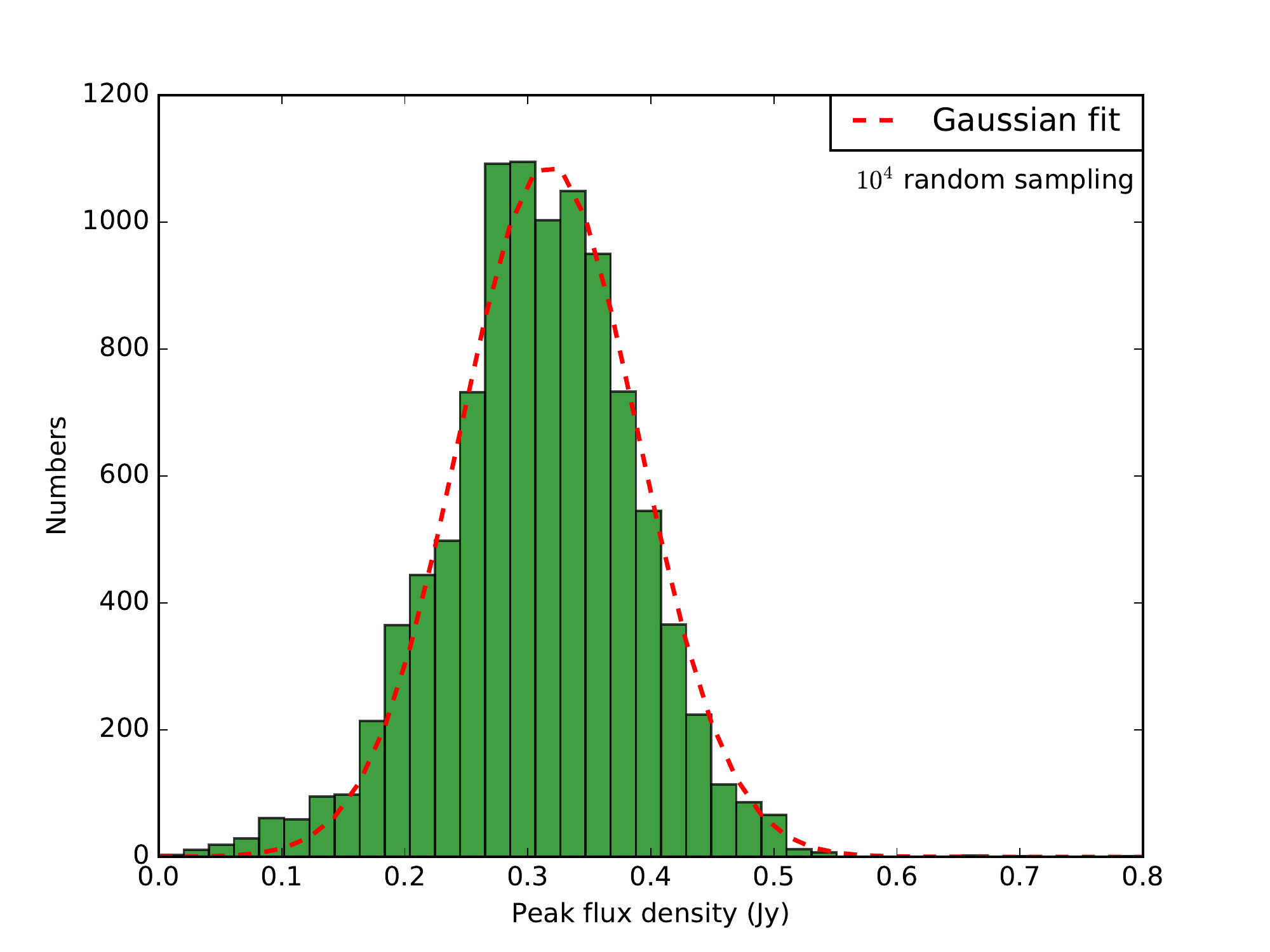}
\caption{
Distribution of the measured line fluxes, for a 10,000 times randomly sampled
signal of 0.3 Jy Sinc function after baseline subtraction.  }
  \label{fig:hist_Gaussian_fit}
\end{figure*}

To fully verify the robustness of this method, we make a sanity test by
randomly insetting artificial spectral signals to the raw data, subtract the
baseline, and test if we can recover the line flux.  We perform the test by
randomly insert artificial signals in the spectra. We find that, if the signal
is higher than 0.5 Jy (peak) and if the line width is less than 2000 \kms, the
line flux can be recovered on very high confidential with an average error
level of less than 20\% of the real fluxes. Figure \ref{fig:hist_Gaussian_fit}
presents a histogram of measured fluxes for 10000 times random test. The
distribution shows a very well centralised flux corresponding to the inserted
artificial noise, with relatively small span.  This indicates that the baseline
subtraction method does not alter the general flux measurement, and the ripple
baseline feature likely not influence much on the high frequency (in Fourier
domain) scales.

We fit a Gaussian to the histogram distribution, and find the $\sigma$ to be
$\sim$0.05 Jy. However we also find that the distribution does not follow a
Gaussian profile very well, especially for both the high and the low ends. We
also examine how much does the recovered flux distribution deviate from normal
distribution, and find that both K-S test and P-test gives $\sim 10^{-8}$,
which mean that the histogram is likely not following a normal distribution.
This is likely because the noise is not uniform across the whole spectral
range, that may bias the noise distribution. Another reason could be that the
noise between channels is correlated, and is not a white noise. It seems that
the noise of the spectra is spiky that enlarges the histogram distribution at
both ends with some extreme values.  Overall, the test indicates a robust flux
measurement can be recovered after the baseline subtraction.

\section{SPIRE FTS spectra of all targets} \label{app:All_spectra}

\begin{figure*}
\centering
\caption{Spectra obtained using \herschel{} SPIRE/FTS, after subtracting the baselines}
\begin{subfigure}{\textwidth}
\caption{  }
\centering
\includegraphics[]{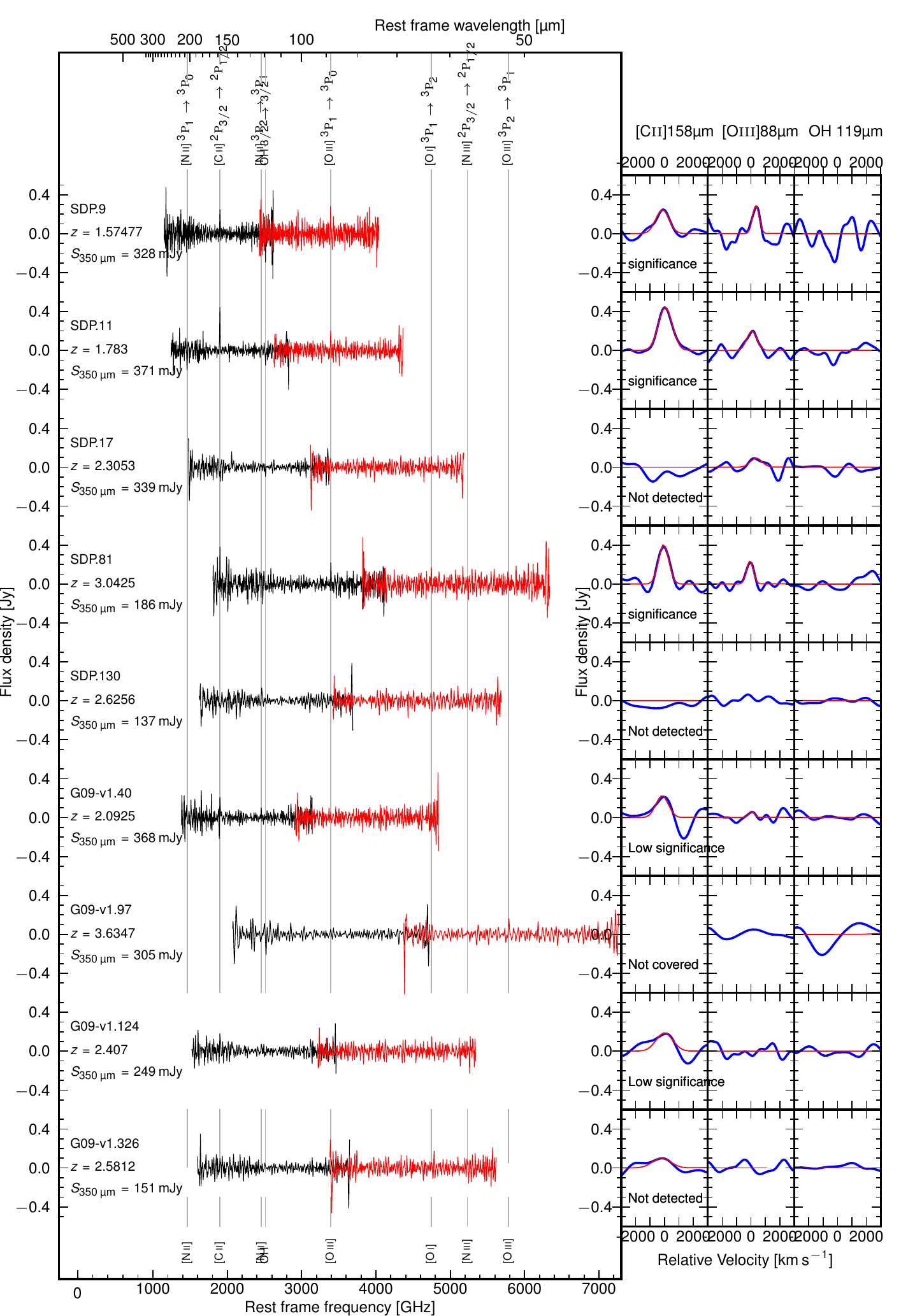}
\label{fig:all_spectra}
\end{subfigure}
\end{figure*}

\newpage

\begin{figure*}
\centering
\ContinuedFloat 
\begin{subfigure}{\textwidth}
\caption{  }
\centering

\includegraphics[]{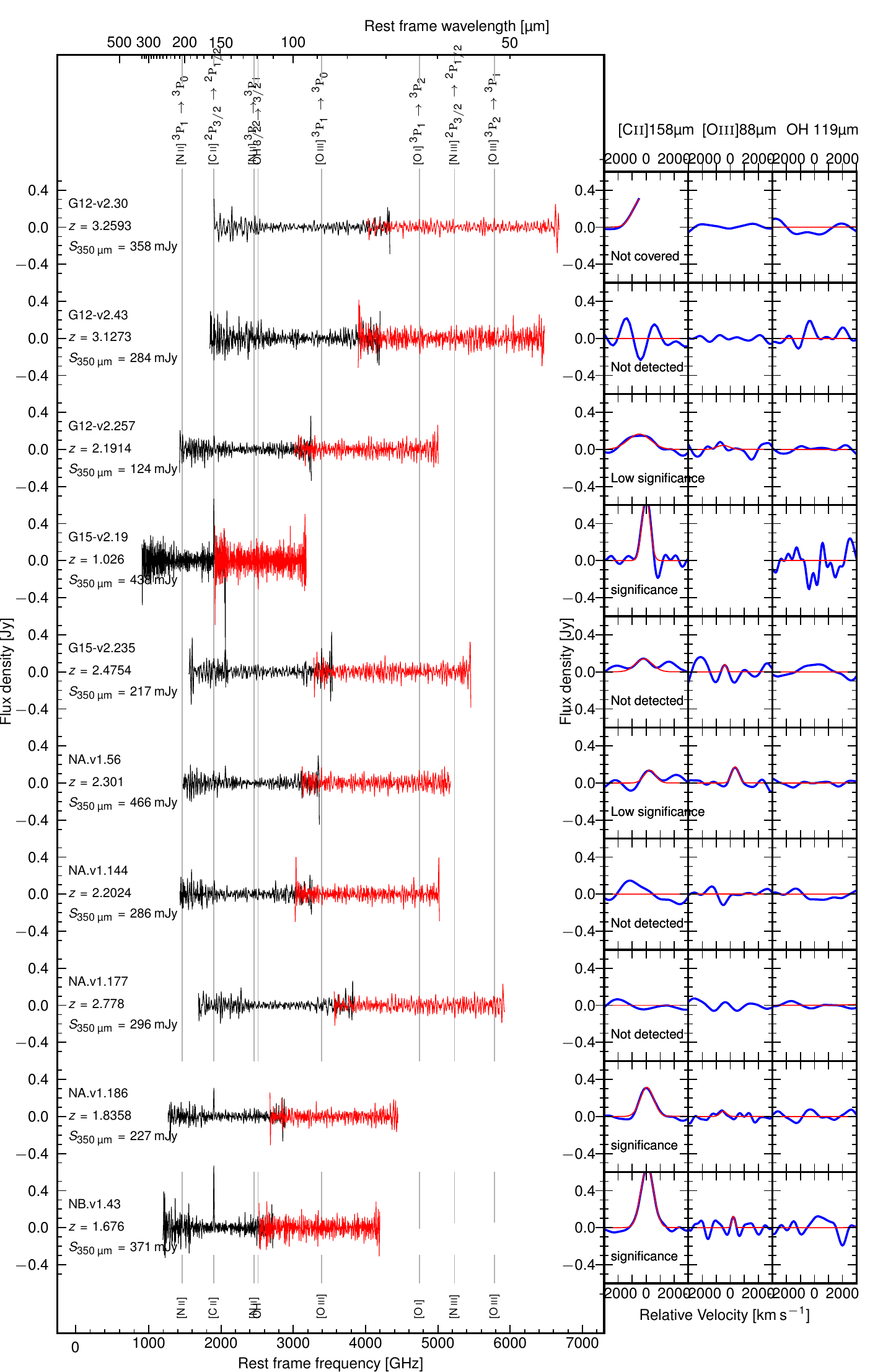}
\label{fig:all_spectra}
\end{subfigure}
\end{figure*}

\begin{figure*}
\centering
\ContinuedFloat 
\begin{subfigure}{\textwidth}
\caption{  }
\centering

\includegraphics[]{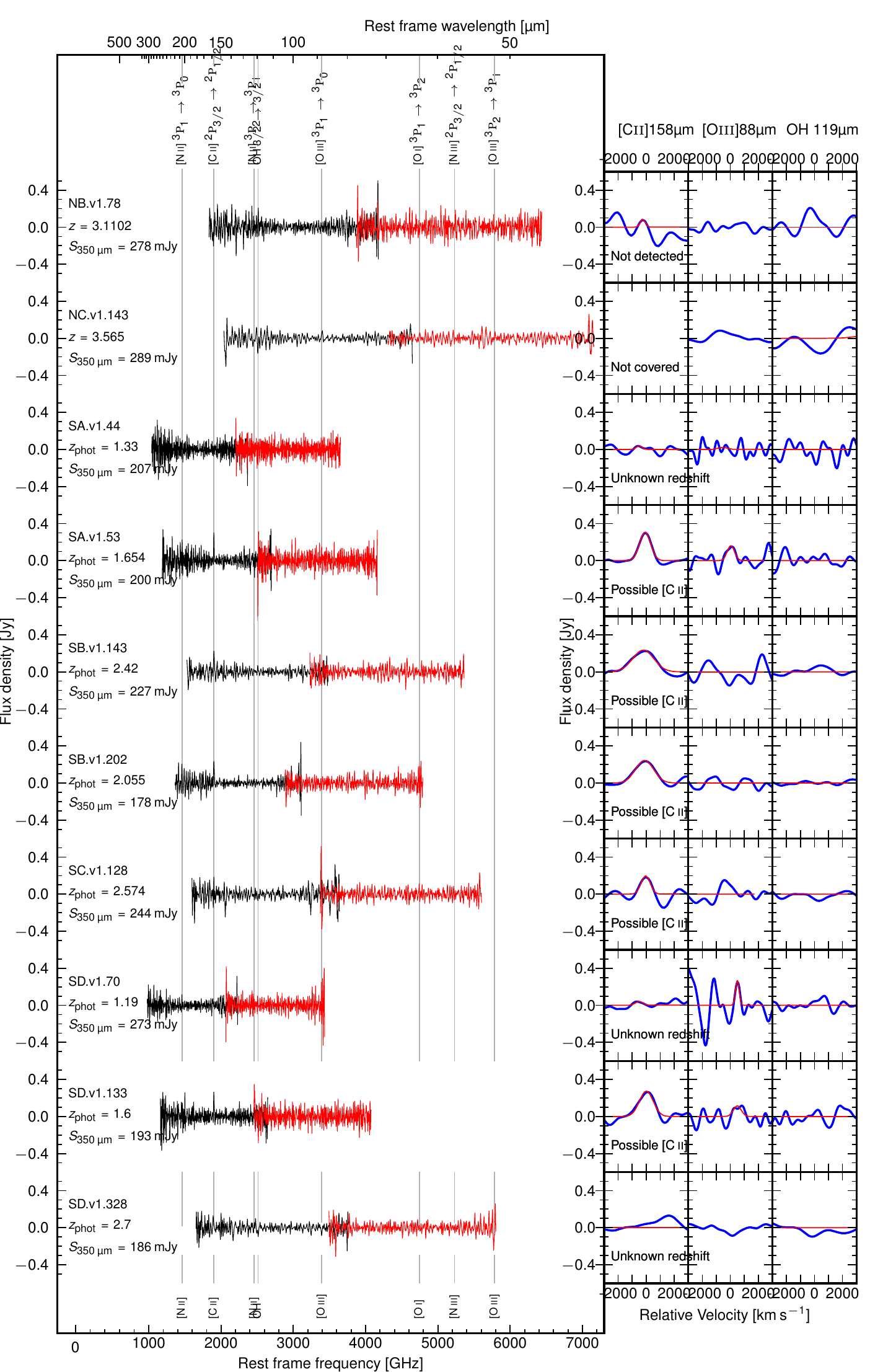}
\label{fig:all_spectra}
\end{subfigure}
\end{figure*}

\begin{figure*}
\centering
\ContinuedFloat 
\begin{subfigure}{\textwidth}
\caption{  }
\centering

\includegraphics[]{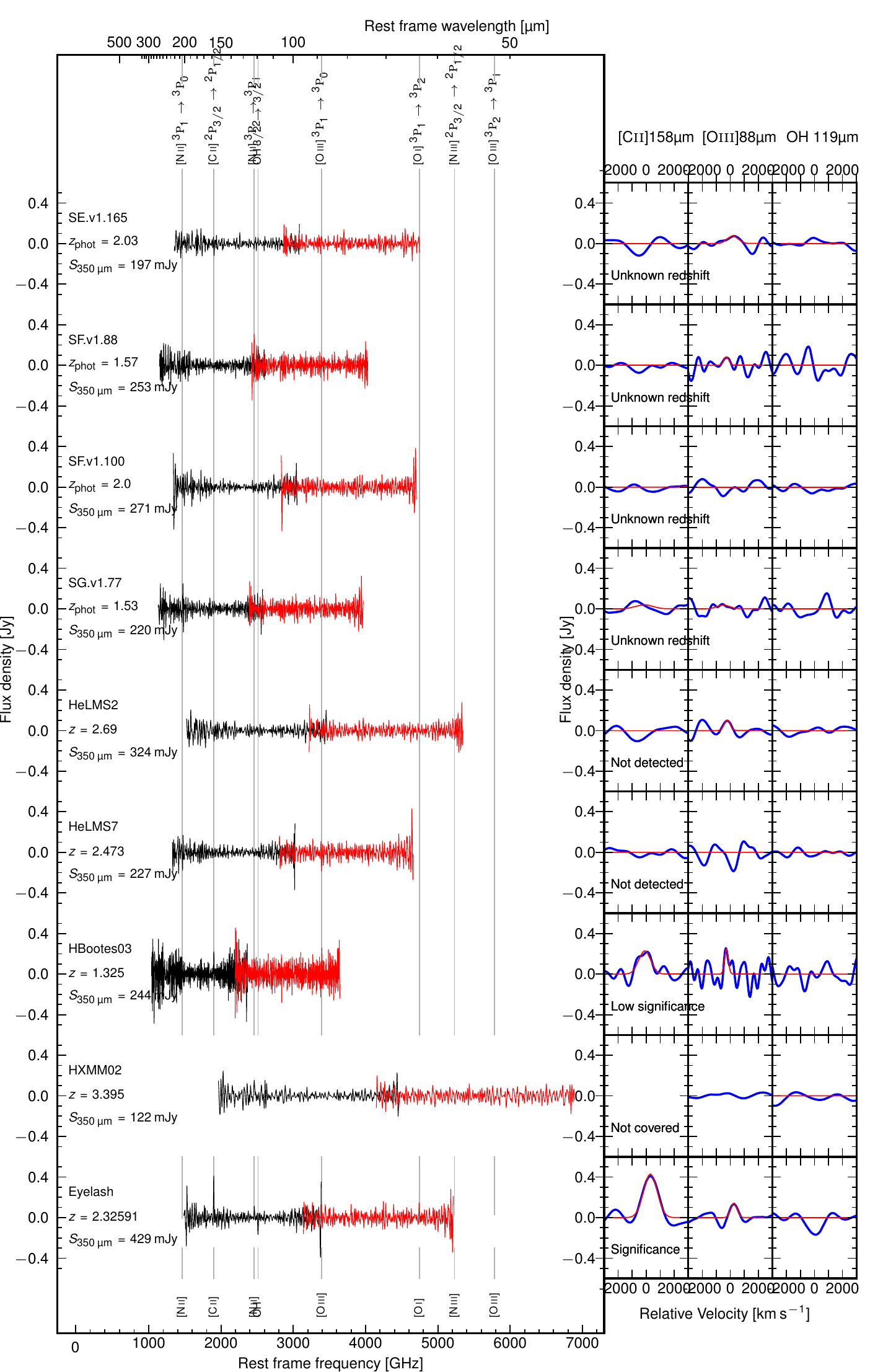}
\label{fig:all_spectra}
\end{subfigure}
\end{figure*}


\section{Comments on individual objects} \label{sec:comments}
SDP.9 (\emph{J090740.0-004200}): This is the lowest redshift and brightest of
the original five strongly lensed DSFGs found in \textit{H}-ATLAS
\citep{Negrello2010}.  NIR imaging suggests a nearly-complete Einstein ring
\citep{Negrello2014}, however the dust emission is dominated by the SE region
\citep{Bussmann2013}.  Despite calculating almost identical dust emission
luminosities, the SFR we derive using our chosen calibration: that presented in
\citet{2011ApJ...737...67M} and recommended by the review of
\citet{Kennicutt2012}; is a factor of 3 higher than value reported by
\citet{Negrello2014}.  We find a relatively low
\lum{\cii{}\,158,\um{}}-to-\lum{FIR} ratio of \(2.5 \pm 0.9 \times 10^{-4}\).

SDP.11 (\emph{J091043.0-000322}): NIR imaging shows this object as a full yet
elongated Einstein ring. 
 Dust emission does not exhibit the same brightness distribution, with
the northern and southern regions dominant.  A \(7.5 \pm 2.3 \times \lineflux\)
\oi\,63-\um{} line has been observed with PACS (Verma et al. in prep.).  We
find the \cii\,158-\um{} line to have a flux \(9.7 \pm 0.9 \times \lineflux\).
50\,\% higher than a recent measurement from ZEUS-2 on APEX
\citep{Ferkinhoff2014}.

  SDP.17 (\emph{J090302.9-014127}):
  With a complex NIR light profile, this system is difficult to
  separate into clear foreground and background components.
  The lensed image is not resolved into multiple components by the SMA
  in even the most extended configuration, suggesting a relatively low
  amplification factor and hence high intrinsic luminosity.
  Despite this, the large emission area suggested by
  \citet{Bussmann2013} would indicate a comparatively modest
  star formation rate surface density, making our \cii\,158-\um{} non-detection
  interesting should that parameter be a key driver of the
  line-to-FIR continuum deficit.

  SDP.81 (\emph{J090311.6+003906}):
  This galaxy is lensed into two arcs, exquisitely revealed by recent
  long-baseline ALMA imaging \citep{2015ApJ...808L...4A}.
  These data have been used to produce a detailed reconstruction of the
  source-plane emission \citep{Dye2015}, which exhibits
  200\,pc-scale clumps.
  We use the amplification derived for the dust emission from this modelling
  effort: \(15.9 \pm 0.7\): which is higher than an estimate from lower
  resolution SMA data
  \citep[\(11.1 \pm 1.1\);][]{Bussmann2013}, and also higher than
  the amplification estimated for the optical emission:
  \(10.2 \pm 0.5\); which appears not to be co-spatial with the rest-frame
  FIR emission \citep{Dye2015}, as is common for starbursts
  (e.g. the Antennae Galaxies).
  An estimate of the source-plane emission radius from those data:
  \(0.5 \pm 0.2\,\text{kpc}\); is also used in our SED fitting.
  The SPIRE FTS spectrum is presented in \citet{Valtchanov2011}
  (programme \gtonefull), but for consistency we reprocess the data here.
  Due to the different method of background subtraction
  (off-axis vs. dark sky) we find a substantially lower \cii\,158-\um{} flux.
  However, as the off-axis method produced reduced noise over the dataset as
  a whole, and is used in the reduction of SPIRE spectra of other
  similarly faint targets
  \citep[e.g.][]{Magdis2014, Gullberg2015}, we use our
  measurement throughout this work.
  Despite the lower line flux, this object still appears to possess the
  highest \(\mu \lum{\cii\,158\,\um}\) and \(\lum{\cii\,158\,\um}/\lum{FIR}\)
  values of our sample.

  SDP.130 (\emph{J091305.0-005343}):
  With a modest \(\sim 2 \times\) amplification this galaxy has significantly
  lower flux densities than the target threshold, and
  \citet{Valtchanov2011} found no lines in the FTS spectrum.
  Using newer calibration data, we find possible \hii{} region emission,
  however the non-detection of \cii\,158-\um{} indicates these are likely
  noise.

  G09-v1.40 (\emph{J085358.9+015537}):
  This galaxy is visible in \(K_\text{s}\) band imaging as an almost-complete
  Einstein ring, consequently experiencing a high magnification factor of
  \(\sim 15\).
  The source-plane object may be highly elongated
  \citep{Calanog2014}.
  We detect \cii\,158\,\um{}.

  G09-v1.97 (\emph{J083051.0+013224}):
  This source is lensed by two foreground galaxies at different redshift.
  While it possesses one of the largest \(\mu \lum{IR}\) values of the
  sample, the high redshift means \cii\,158-\um{} is not covered by the
  FTS.

  G09-v1.124 (\emph{J084933.4+021443}):
  Uniquely within our sample this object is known to be essentially unlensed,
  its broad CO lines suggesting a very high intrinsic luminosity,
  confirmed by a range of follow-up observations which showed a system of
  four separate DSFGs, two of which are individually over the
  HyLIRG limit \citep{Ivison2013}.
  These two brightest are detected as separate point sources in our PACS
  imaging, but together only provide a low significance \cii-158\,\um{}
  feature.
  Recent NIR spectroscopy has revealed a broad-line AGN in this
  system (Oteo et al. in prep.), which will be responsible for some fraction
  of its FIR flux.

  G09-v1.326 (\emph{J091840.8+023047}):
  Available SMA imaging is consistent with a point source and no counterpart is
  found in \textit{HST} or Keck imaging.  Should this be an unlensed DSFG it
  would possess a SFR of \(\sim 5000\,\msun\,\text{yr}^{-1}\), higher than any
  component of the known HyLIRG G09-v1.124, so we instead assume the
  amplification estimate of \(5 \pm 1\) from \citet{Harris2012} derived from
  its CO luminosity and FWHM.  No FIR cooling lines are observed in the FTS
  spectrum, however the flux densities lie below the target threshold.

  G12-v2.30 (\emph{J114638.0-001132}):
  A detailed study of this galaxy is presented in \citet{Fu2012}, indicating a
  complex system in which the existing stellar population is quadruply-lensed
  by a net factor of over \(2 \times\) that of the offset starburst, which
  provides some of the highest 500 and 880-\um{} flux densities in our sample.
  This redshift represents the limit of the FTS coverage of the
  \cii\,158-\um{} line, which therefore lies in the noisiest region of the
  spectrum and is not detected.

  G12-v2.43 (\emph{J113526.3-014606}):
  Multiple lensed images are not resolved in the 880-\um{} SMA data, however a
  faint source visible in \textit{HST} F110W and Keck/NIRC2 \(K_\text{s}\)
  imaging is likely a foreground lensing galaxy.  With an apparent \(\mu \lir >
  10^{14.1}\,\lsun\), among the highest of our sources, we assume the large
  amplification estimate of \citet{Harris2012}: \(17 \pm 11\).  As with
  G12-v2.30, the \cii\,158-\um{} line lies within the high-noise end region of
  the spectrum, and no FIR lines are observed.


  G12-v2.257 (\emph{J115820.1-013753}):
  Two nearby spiral galaxies visible at 100-\um{} will be unresolved within the
  SPIRE beam but likely contribute very little to the measured flux densities
  at the longer wavelengths.  No lensing model is available, though this object
  has one of the lowest 350-\um{} flux densities within our sample indicating a
  low \(\mu \lir\) and likely \(\mu \lesssim 5\).  We however take the
  published amplification estimate of \(13 \pm 7\) from \citet{Harris2012}.  As
  with the other low flux density sources, we do not find any emission line
  features.

  G15-v2.19 (\emph{J142935.3-002836}):
  This object is the brightest (and lowest redshift) within our sample.  A
  detailed study is presented in \citet{Messias2014} indicating a gas-rich
  (25\% of baryonic mass) merger of two source-plane components lensed into an
  almost complete Einstein ring.  The redshift is too low for the spectrum to
  cover \oi\,63-\um{} and \oiii\,88-\um, but a strong \cii\,158-\um{}
  feature is found.  Additionally, we detect 119-\um{} \(\text{OH} \;
  ^2\Pi_{3/2} \; \J{\frac{3}{2}}{\frac{5}{2}}\) absorption.

  With no strong evidence of an AGN \citep{Messias2014} we,
  as with a similar detection within SMM\,J2135 \citep{George2014},
  attribute this to a SNe-driven outflow of molecular gas.

  G15-v2.235 (\emph{J141351.9-000026}):
  This galaxy is weakly lensed by a cluster with only a single elongated image
  visible, suggesting a high intrinsic luminosity.  We find \cii\,158-\um{}
  and the second-highest \lum{\cii\,158-\um}/\lum{FIR} ratio of the sample.

  NA.v1.56 (\emph{J134429.4+303036}):
  This source has the highest 350 and 500\,\um{} flux densities in our sample,
  visible as two well-defined arcs in SMA imaging.  We detect \cii\,158-\um{}
  emission.

  NA.v1.144 (\emph{J133649.9+291801}):
  Available optical--NIR imaging is insufficiently deep to detect a lensing
  galaxy, however the 880-\um{} morphology indicates that this object is
  likely lensed with a small Einstein radius.  A second tentative source at \(z
  = 2.3078\) was suggested by a blind CO line search \citep{Harris2012}.  We do
  not detect any spectral lines.

  NA.v1.177 (\emph{J132859.3+292317}):
  This source is resolved into a single elongated image in SMA data, and likely
  also responsible for a faint arc in Keck/NIRC2 \(K_\text{s}\) imaging.  No
  lens models has been produced so far, and we do not detect any lines.

  NA.v1.186 (\emph{J132504.4+311537}):
  A feature suggestive of \cii\,158-\um{} at \(z = 1.836\) found in our FTS
  spectrum was followed up with CARMA, extending the 3\,mm spectrum below the
  nominal 85\,GHz tuning range to detect CO \J{2}{1}.  A redshift of 1.8358
  was confirmed, negating the previous estimate of \(z = 2.635\).  With no
  published deep NIR or interferometric FIR--mm imaging, a lens model has not
  been produced.

  NB.v1.43 (\emph{J132427.0+284452}):
  NIR images of this galaxy display an \(\approx 10''\) long arc with little
  curvature: much more spatially extended than is visible in submillimetre data
  which arises from a small area towards the centre.  Our FTS spectrum along
  with its detection of strong \cii\,158-\um{} emission was presented in
  \citet{George2013}, however since publication of those data, a lensing model
  has been published.  Lensing bodies are likely either two galaxies or a
  nearby cluster \citep[see][for a discussion]{Bussmann2013}, with a
  (relatively uncertain) lensing amplification factor of \(2.8 \pm 0.4\)
  derived for the dust emission.  Calculations in this work utilise that value.

  NB.v1.78 (\emph{J133008.4+245900}):
  With \(> 200\,\mJy\) flux densities in all SPIRE bands, and at 160-\um
  despite a redshift above 3, this galaxy has the highest \(\mu \lum{IR}\) of
  the sample and a warm dust temperature caused by its small radius and hence
  high SFR surface density.  Coupled with the position of the \cii\,158-\um{}
  line within the spectrum, our non-detection of that transition is therefore
  perhaps unsurprising.

  NC.v1.143 (\emph{J125632.7+233625}):
  The lens has a redshift low enough for it to be visible in SDSS imaging and
  similarly to NB.v1.78, the background galaxy has a high \(\mu \lum{IR}\) and
  an even higher dust temperature and SFR surface density, however the high
  redshift of this source means that \cii\,158\,\um{} is not covered by the FTS
  spectrum.

  \textit{H}-ATLAS SGP sources:
  Without definite redshifts, it is difficult to glean much from these sources.
  Several have low significance SPIRE spectral features potentially
  corresponding to \cii\,158\,\um{}, with the corresponding redshift listed in
  Table~\ref{tab:flux_densities}.

  HeLMS sources: 
  Only two of these galaxies were observed with the FTS, none of them exhibits 
  \cii\,158\,\um{} emission. Further follow-up of these objects needs to be done,
  rendering any conclusive determinations of their properties
  or lensing amplification difficult.

  HBo\"otes03 (\emph{J142824.0+352619}):
  This is a well-studied high-redshift DSFG
  (also known as MIPS J142824.0+352619) discovered before the wide area
  \herschel{} and SPT surveys
  \citep{2006ApJ...636..134B, 2006MNRAS.371..465S}.
  \herschel{} PACS detections of \oiii\,52\,\um{}
  (\(3.7 \pm 0.8 \times \lineflux\))
  and \oi\,63\,\um{} (\(7.8 \pm 1.9 \times \lineflux\)) are presented in
  \citet{2010A&A...518L..36S}, and a \(19.8 \pm 3.0 \times \lineflux\)
  \cii\,158\,\um{} detection with CSO/ZEUS in \citet{HD2010}.
  This latter value is particularly significant, being a factor of 5 higher
  than our measurement (fit shown in Fig.~\ref{fig:sed_fits}).
  We are unsure as to the reason for the magnitude of this discrepancy.

  HXMM02 (\emph{J021830.5\(-\)053124}):
  This object (also known as Orochi or SXDF1100.001) was included due to its
  previous in-depth study \citep{2011MNRAS.415.3081I, 2012PASJ...64L...2I}.
  The small lens Einstein radius and amplification however give it the
  lowest 250 and 350\,\um{} flux densities in our sample, which when
  coupled with its high redshift
  (such that the FTS spectrum does not cover \cii\,158\,\um{}),
  mean no FIR lines are found.
  The dust mass temperature index of 7.2, used to fit all other galaxies is
  insufficient to account for the high PACS flux densities, with the
  best-fitting value being 5.7.
  This suggested additional mass of hot dust, coupled with the low
  FIRRC value (assuming a synchrotron spectral index of -0.75)
  indicates the presence of an AGN.

  SMM\,J2135 (\emph{J213511.6\(-\)010252}):
  The Cosmic Eyelash is one of the most well-studied high-redshift DSFGs, its
  \(37.5 \pm 4.5\) amplification factor enabling the spatial and spectral
  resolution of 3--4 massive star-forming clumps \citep[e.g.][]{Swinbank2010,
  Swinbank2011, Ivison2010,Danielson2013}.
  After a tentative detection of OH\,119-\um{} absorption in the \otone{}
  observation of this source, we included 5 repeat observations of that object
  in \ottwo.  A detailed examination of that data is presented in
  \citet{George2014}, however for consistency we reprocess that data again
  here.  \cii\,158-\um{} appears at a higher velocity in the original spectrum
  than the repeats, however while individual spectra may exhibit slight
  noise-related velocity differences, the fit to the combined spectrum of all
  SMM\,J2135 observations should be more reliable.

\end{appendix}


\end{document}